\documentclass[12pt]{elsart}
\usepackage{feynmp,epsfig,graphics}

\def\tr{{\rm tr} \,}
\def\omx#1{\chi_{#1}}

\def\pslash{\FMslash p}
\def\barpslash{\FMslash{\bar{p}}}
\def\lslash{\FMslash l}
\def\m0{m^{\!\!\!\!^o}}
\def\qslash{\FMslash q}
\def\barqslash{\FMslash{\bar{q}}}

\def\simslash{\FMslash {\sim}}

\def\wslash{\FMslash w}
\def\barwslash{\FMslash{\widetilde{w}}}
\def\kslash{\FMslash k}
\def\barkslash{\FMslash{\bar{k}}}

\def\partialslash{\FMslash \partial}

\begin{document}
GSI-Preprint-2001-12 and ECT*-Preprint-2001-10 
\begin{frontmatter}
\title{Relativistic chiral SU(3) symmetry, large $N_c$ sum rules \\ and \\ meson-baryon scattering}
\author{M.F.M. Lutz$^a$ and E.E. Kolomeitsev$^{a,b}$}
\address{$^a$ Gesellschaft f\"ur Schwerionenforschung (GSI),\\
Planck Str. 1, D-64291 Darmstadt, Germany}
\address{$^b$ ECT$^*$, Villa Tambosi, I-38050 \,Villazzano  (Trento) \\
and INFN, G.C.\ Trento, Italy}
\begin{abstract}
The relativistic chiral $SU(3)$ Lagrangian is used to describe kaon-nucleon
scattering imposing constraints from the pion-nucleon sector and the axial-vector coupling 
constants of the baryon octet states. We solve the covariant coupled-channel Bethe-Salpeter 
equation with the interaction kernel truncated at chiral order $Q^3$ where we include only 
those terms which are leading in the large $N_c$ limit of QCD. The baryon decuplet states 
are an important explicit ingredient in our scheme, because together with the baryon octet 
states they form the large $N_c$ baryon ground states of QCD. Part of our technical 
developments is a minimal chiral subtraction scheme within dimensional regularization, 
which leads to a manifest realization of the covariant chiral counting rules. 
All SU(3) symmetry-breaking effects are well controlled by the combined chiral and large $N_c$ 
expansion, but still found to play a crucial role in understanding the empirical data. 
We achieve an excellent description of the data set typically up to laboratory momenta of 
$p_{\rm lab} \simeq $ 500 MeV.
\end{abstract}
\end{frontmatter}

\section{Introduction}

The meson-baryon scattering processes are an important test for effective field theories 
which aim at reproducing QCD at small energies, where the effective degrees of freedom are 
hadrons rather than quarks and gluons. In this work we focus on the strangeness sector, because there 
the acceptable effective field theories are much less developed and also the empirical data set still 
leaves much room for different theoretical interpretations. In the near future the 
new DA$\Phi$NE facility at Frascati could deliver new data on kaon-nucleon scattering \cite{DAPHNE} and therewith 
help to establish a more profound understanding of the role played by the $SU(3)$ flavor symmetry in hadron 
interactions. At present the low-energy elastic $K^+$-proton scattering data set leads to a rather well 
established $K^+p$ scattering length with $a_{K^+p}\simeq 0.28-0.34$ fm \cite{Dover}. Uncertainties exist, 
however, in the $K^+$-neutron channel where the elastic cross section is extracted 
from the scattering data on the $K^+d\to K^+ p\,n $ and $K^+d \to  K^0 p \,p$ reactions. Since data are 
available only for $p_{\rm lab} > 350$ MeV, the model dependence of the deuteron wave function, the 
final state interactions and the necessary extrapolations down to threshold lead to conflicting values 
for the $K^+$-neutron scattering length \cite{Dumbrajs}. A recent analysis \cite{Barnes} favors 
a repulsive and small value $a_{K^+n}\simeq 0.1$ fm. Since low-energy polarization 
data are not available for $K^+$-nucleon scattering, the separate strength of the various 
p-wave channels can only be inferred by theory at present. This leads to large uncertainties 
in the p-wave scattering volumes \cite{Dumbrajs}. 

The $K^-$-proton scattering length was only recently determined convincingly by a 
kaonic-hydrogen atom measurement \cite{Iwasaki}. In contrast the $K^-$-neutron scattering
length remains model dependent \cite{A.D.Martin,martsakit}. This reflects the fact 
that at low energies there are no $K^-$ deuteron scattering data available 
except for some $K^-d$ branching ratios \cite{Veirs} commonly not included in theoretical 
models of kaon-nucleon scattering. The rather complex multi-channel dynamics of the 
strangeness minus one channel is constrained by not too good quality low-energy elastic and 
inelastic $K^-p$ scattering data \cite{Landoldt} but rather precise $K^-p$
threshold branching ratios \cite{branch-rat}. Therefore the isospin one scattering amplitude 
is constrained only indirectly for instance by the $\Lambda \pi^0$ production data 
\cite{mast-pio}. That leaves much room for different theoretical extrapolations 
\cite{A.D.Martin,martsakit,kim,sakit,gopal,oades,Juelich:2,dalitz,keil}. 
As a consequence the subthreshold $\bar K N$ scattering amplitudes, which determine 
the $\bar K$-spectral function in nuclear matter to leading order in the density expansion, 
are poorly controlled. In the region of the $\Lambda(1405)$ resonance the isospin zero 
amplitudes of different analyses may differ by a factor of two \cite{Kaiser,Ramos}. 

Therefore it is desirable to make use of the chiral symmetry constraints of QCD. First intriguing 
works in this direction can be found in \cite{Kaiser,Ramos,Hirschegg,Oller-Meissner}. The reliability 
of the extrapolated subthreshold scattering amplitudes can be substantially 
improved by including s- {\it and} p-waves in the analysis of the
empirical cross sections, because  the available
data, in particular in the strangeness minus one channel, are much more precise for $p_{\rm lab}>200$ MeV than 
for $p_{\rm lab}<200$ MeV, where one expects s-wave dominance.

In this work we use the relativistic chiral $SU(3)$ Lagrangian including
an explicit baryon decuplet resonance field with $J^P\!= \!\frac{3}{2}^+$. 
The baryon decuplet field is an important ingredient, because it is part of the baryon 
ground state multiplet which arises in the large $N_c$ limit of QCD \cite{Hooft,Witten}. 
We also consider the effects of a phenomenological baryon nonet d-wave resonance field with
$J^P\!=\!\frac{3}{2}^-$, because some of the p-wave strengths in the $\bar K N$ system can be 
extracted from the available data set reliably only via their interference effects with 
the d-wave resonance $\Lambda (1520)$. 
As to our knowledge this is the first application of the chiral $SU(3)$ Lagrangian density to the 
kaon-nucleon and antikaon-nucleon systems including systematically constraints from the pion-nucleon sector. 
We propose a convenient minimal chiral subtraction scheme for relativistic Feynman diagrams 
which complies manifestly with the standard chiral counting rule \cite{Becher,nn-lutz,Gegelia}.
Furthermore it is argued that the relatively large kaon mass necessarily leads to non-perturbative 
phenomena in the kaon-nucleon channels in contrast to the pion-nucleon system where standard chiral 
perturbation theory ($\chi$PT) can be applied successfully \cite{Gasser,Bernard,Meissner}. In the strangeness 
sectors a partial resummation scheme is 
required \cite{Siegel,Kaiser,Ramos}. We solve the Bethe-Salpeter equation for the scattering amplitude
with the interaction kernel truncated at chiral order $Q^3$ where we include only 
those terms which are leading in the large $N_c$ limit of 
QCD \cite{Hooft,Witten,DJM,Carone,Luty}. The s-, p- and 
d-wave contributions with $J=\frac{1}{2},\frac{3}{2}$ in the scattering amplitude 
are considered. As a novel technical ingredient we construct a covariant projector formalism. 
It is supplemented by a subtraction scheme rather than a cutoff scheme as employed previously 
in \cite{Kaiser,Ramos}. The renormalization scheme is an essential input of our chiral $SU(3)$ dynamics, 
because it leads to consistency with chiral counting rules and an approximate crossing symmetry of the 
subthreshold kaon-nucleon scattering amplitudes. Our scheme avoids, in particular, breaking the $SU(3)$ symmetry by 
channel-dependent cutoff parameters as suggested in \cite{Kaiser} and also a sensitivity 
of the $\Lambda(1405)$ resonance structure to the cutoff parameter implicit in \cite{Ramos}.

We successfully adjust the set of parameters to describe the existing low-energy cross section data on 
kaon-nucleon and antikaon-nucleon scattering including angular distributions to good accuracy. At the same 
time we achieve a satisfactory description of the low-energy s- and p-wave pion-nucleon phase shifts as well as the 
empirical axial-vector coupling constants of the baryon octet states.  We make detailed predictions for the 
poorly known meson-baryon coupling constants of the baryon octet and decuplet states. Furthermore the 
many $SU(3)$ reaction amplitudes and cross sections like $\pi \Lambda \to \pi \Lambda, \pi \Sigma , \bar K \,N $,  
relevant for transport model simulations of heavy-ion reactions, will be presented. As a result of our analysis, 
particularly important for any microscopic description of antikaon propagation in dense nuclear matter, we 
predict sizable contributions from p-waves in the subthreshold $\bar K$-nucleon forward scattering amplitude.

In section 2 we construct the parts of the relativistic chiral Lagrangian relevant for this work. 
All interaction terms are analyzed systematically in the $1/N_c$ expansion of QCD. 
In section 3 we develop the formalism required for the proper treatment of the Bethe-Salpeter 
equation including details on the renormalization scheme. In section 4 we derive  
the coupled channel effective interaction kernel in accordance with the scheme presented in section 3. 
The reader interested primarily in the numerical results can go directly to section 5, which can be read 
rather independently.

\section{Relativistic chiral $SU(3)$ interaction terms in large $N_c$ QCD}

In this section we present all chiral interaction terms to be used in the following 
sections to describe the low-energy meson-baryon scattering data set. 
Particular emphasis is put on the constraints implied by the large $N_c$ 
analysis of QCD. The reader should not be discouraged by the many fundamental parameters 
introduced in this section. The empirical data set includes many hundreds of 
data points and will be reproduced rather accurately. Our scheme makes many predictions 
for poorly known or not known observable quantities like for example the p-wave scattering volumes 
of the kaon-nucleon scattering processes or $SU(3)$ reactions like 
$\pi \Lambda \to \pi \Sigma $. In a more conventional 
meson-exchange approach, which lacks a systematic approximation scheme, many parameters 
are implicit in the so-called form factors. In a certain sense the parameters used in the 
form factors reflect the more systematically constructed and controlled quasi-local 
counter terms of the chiral Lagrangian.

We recall the interaction terms of the relativistic chiral
$SU(3)$ Lagrangian density relevant for the meson-baryon scattering process. 
For details on the systematic construction principle, see for example \cite{Krause}. The basic 
building blocks of the chiral Lagrangian are
\begin{eqnarray}
&& U_\mu = \frac{1}{2}\,e^{-i\,\frac{\Phi}{2\,f}} \left( 
\partial_\mu \,e^{i\,\frac{\Phi}{f}}
+ i\,\Big[A_\mu , e^{i\,\frac{\Phi}{f}} \Big]_+
\right) e^{-i\,\frac{\Phi}{2\,f}}  \;,\qquad \!\!
B \;, \qquad \! \!\Delta_\mu \,, \qquad \! \! B^{*}_\mu \;,
\label{def-fields}
\end{eqnarray}
where we include the pseudo-scalar meson octet field 
$\Phi(J^P\!\!=\!0^-)$, the baryon octet field $B(J^P\!\!=\!{\textstyle{1\over2}}^+)$, the baryon 
decuplet field $\Delta_\mu(J^P\!\!=\!{\textstyle{3\over2}}^+)$ and the baryon nonet resonance 
field $B^*_\mu(J^P\!\!=\!{\textstyle{3\over2}}^-)$ (see \cite{Tripp:1,Tripp:2,Plane}). In (\ref{def-fields}) 
we introduce an external axial-vector source function $A_\mu $ which is required for the systematic 
evaluation of matrix elements of the axial-vector current. A corresponding term 
for the vector current is not shown in (\ref{def-fields}) because it will 
not be needed in this work. The axial-vector source function 
$A^\mu =\sum A_a^\mu\,\lambda^{(a)} $, 
the meson octet field $\Phi=\sum \Phi_a\,\lambda^{(a)}$ and the baryon octet fields 
$B= \sum B_a\,\lambda^{(a)}/\sqrt{2}$, 
$B_\mu^*= B_{\mu,0}^*/\sqrt{3}+\sum B_{\mu,a}^*\,\lambda^{(a)}/\sqrt{2}$ are decomposed using the
Gell-Mann matrices $\lambda_a$ normalized by $\tr \lambda_a \,\lambda_b =
2\,\delta_{ab}$. The baryon decuplet field $\Delta^{abc} $ is completely symmetric 
and related to the physical states by
\begin{eqnarray}
\begin{array}{llll}
\Delta^{111} = \Delta^{++}\,, & \Delta^{113} =\Sigma^{+}/\sqrt{3}\,, &
\Delta^{133}=\Xi^0/\sqrt{3}\,,  &\Delta^{333}= \Omega^-\,, \\
\Delta^{112} =\Delta^{+}/\sqrt{3}\,, & \Delta^{123} =\Sigma^{0}/\sqrt{6}\,, &
\Delta^{233}=\Xi^-/\sqrt{3}\,, & \\
\Delta^{122} =\Delta^{0}/\sqrt{3}\,, & \Delta^{223} =\Sigma^{-}/\sqrt{3}\,, &
& \\
\Delta^{222} =\Delta^{-}\,. & & &
\end{array}
\label{dec-field}
\end{eqnarray}
The parameter $f $ in (\ref{def-fields}) is determined by the weak decay 
widths of the charged pions and kaons properly corrected for chiral SU(3) effects. Taking 
the average of the empirical decay parameters $f_\pi = 92.42 \pm 0.33 $ 
MeV  and $f_K \simeq 113.0 \pm 1.3$ MeV \cite{fpi:exp} one obtains the naive estimate 
$f \simeq  104$ MeV. This value is still within reach of the more detailed analysis  
\cite{GL85} which lead to $f_\pi/f = 1.07 \pm 0.12$. As emphasized in \cite{MO01}, 
the precise value of $f$ is subject to large uncertainties. 

Explicit chiral symmetry-breaking effects are included in terms 
of scalar and pseudo-scalar source fields $\chi_\pm $ proportional to the quark-mass 
matrix of QCD
\begin{eqnarray}
\chi_\pm = \frac{1}{2} \left( 
e^{+i\,\frac{\Phi}{2\,f}} \,\chi_0 \,e^{+i\,\frac{\Phi}{2\,f}}
\pm e^{-i\,\frac{\Phi}{2\,f}} \,\chi_0 \,e^{-i\,\frac{\Phi}{2\,f}} 
\right) \,,
\label{def-chi}
\end{eqnarray} 
where $\chi_0 \sim {\rm diag} (m_u,m_d,m_s)$.
All fields in (\ref{def-fields}) and (\ref{def-chi}) have identical properties under 
chiral $SU(3)$ transformations. The chiral Lagrangian consists of all possible interaction 
terms, formed with the fields $U_\mu, B, \Delta_\mu, B^*_\mu$ and $\chi_\pm$ and their 
respective covariant derivatives. Derivatives of the fields must be included in compliance 
with the chiral $SU(3)$ symmetry. This leads to the notion of a covariant derivative 
${\mathcal D}_\mu$ which is identical for all fields in (\ref{def-fields}) and (\ref{def-chi}). 
For example, it acts on the baryon octet field as
\begin{eqnarray}
\Big[{\mathcal D}_\mu , B\Big]_- &=& \partial_\mu \,B + 
\frac{1}{2}\,\Big[ e^{-i\,\frac{\Phi}{2\,f}} \left( 
\partial_\mu \,e^{+i\,\frac{\Phi}{2\,f}}\right)
+e^{+i\,\frac{\Phi}{2\,f}} \left( 
\partial_\mu \,e^{-i\,\frac{\Phi}{2\,f}}\right), B\Big]_- 
\nonumber\\
&+& \frac{i}{2}\,\Big[ e^{-i\,\frac{\Phi}{2\,f}} \,
A_\mu \,e^{+i\,\frac{\Phi}{2\,f}}- e^{+i\,\frac{\Phi}{2\,f}} \, A_\mu 
\,e^{-i\,\frac{\Phi}{2\,f}}, B\Big]_-  \,.
\label{}
\end{eqnarray} 

The chiral Lagrangian is a powerful tool once it is combined with appropriate
power counting rules leading to a systematic approximation strategy. 
One aims at describing hadronic interactions at low-energy by constructing an expansion 
in small momenta and the small pseudo-scalar meson masses. The infinite set 
of Feynman diagrams are sorted according to their chiral powers. The minimal chiral 
power $Q^{\nu }$ of a given relativistic Feynman diagram,  
\begin{eqnarray}
\nu = 2-{\textstyle {1\over2}}\, E_B + 2\, L
+\sum_i V_i \left( d_i +{\textstyle {1\over2}}\, n_i-2 \right) \;,
\label{q-rule}
\end{eqnarray}
is given in terms of the number of loops, $L$, the number, $V_i$, of vertices of type $i$ 
with $d_i$ 'small' derivatives and $n_i$ baryon fields involved, and  
the number of external baryon lines $E_B$ \cite{Weinberg}. Here one calls a derivative small 
if it acts on the pseudo-scalar meson field or if it probes the virtuality of a baryon field. 
Explicit chiral symmetry-breaking effects are perturbative and included in the counting scheme 
with $\chi_0 \sim Q^2$. 
For a discussion of the relativistic chiral Lagrangian and its required systematic regrouping 
of interaction terms we refer to \cite{nn-lutz}. We will encounter explicit 
examples of this regrouping later. The relativistic chiral Lagrangian requires 
a non-standard renormalization scheme. The $MS$ or $\overline{MS}$ minimal subtraction schemes 
of dimensional regularization do not comply with the chiral counting rule \cite{Gasser}. 
However, an appropriately modified subtraction scheme for relativistic Feynman diagrams leads 
to manifest chiral counting rules \cite{Becher,nn-lutz,Gegelia}. 
Alternatively one may work with the chiral 
Lagrangian in its heavy-fermion representation \cite{J&M} where an appropriate frame-dependent 
redefinition of the baryon fields leads to a more immediate  manifestation of the chiral 
power counting rule (\ref{q-rule}). We will return to this issue in section 3.1 where we 
propose a simple modification of the $\overline {MS}$-scheme which leads to 
consistency with (\ref{q-rule}). Further subtleties of the chiral power counting rule (\ref{q-rule}) 
caused by the inclusion of an explicit baryon resonance field $B^*_\mu$ are addressed in section 4.1 
when discussing the u-channel resonance exchange contributions.

In the $\pi N$ sector, the $SU(2)$ chiral Lagrangian was successfully 
applied \cite{Gasser,Bernard} demonstrating good convergence properties of the perturbative
chiral expansion. In the $SU(3)$ sector, the situation is more involved due in part
to the relatively large kaon mass $m_K \simeq m_N/2$. The perturbative
evaluation of the chiral Lagrangian cannot be justified and one must change
the expansion strategy. Rather than expanding directly the scattering amplitude one may 
expand the interaction kernel according to chiral power counting rules \cite{Weinberg,LePage}. 
The scattering amplitude then follows from the solution of a scattering equation like the 
Lipmann-Schwinger or the Bethe-Salpeter equation. This is analogous to the treatment of the 
$e^+\,e^-$ bound-state problem of QED where a perturbative evaluation of the interaction kernel
can be justified. The rational behind this change of scheme lies in the observation that 
reducible diagrams are typically enhanced close to their unitarity threshold.
The enhancement factor $(2\pi)^n$, measured relative to a reducible diagram with the 
same number of independent loop integrations, is given by 
the number, $n$, of reducible meson-baryon pairs in the diagram, i.e. the number of unitary 
iterations implicit in the diagram. In the $\pi N$ sector this enhancement factor does not 
prohibit a perturbative treatment, because the typical expansion parameter  
$m^2_\pi/(8 \pi \,f^2) \sim 0.1 $ remains sufficiently small. In the $\bar K N$ sector, 
on the other hand, the factor $(2\pi)^n$ invalidates a perturbative 
treatment, because the typical expansion parameter would be $m^2_K/(8 \pi\,f^2) \sim 1$. 
This is in contrast to irreducible diagrams. They yield the typical expansion parameters
$m_\pi/(4 \pi \,f)$ and $m_K/(4\pi \,f)$ which justifies the perturbative 
evaluation of the scattering kernels. We will return to this issue later and discuss this 
phenomena in terms of the Weinberg-Tomozawa interaction in more detail. 

In the next section we will develop the formalism to construct the 
leading orders interaction kernel from the relativistic chiral Lagrangian and then to
solve the Bethe-Salpeter scattering equation. 
In the remainder of this section, we collect all interaction terms needed for the construction 
of the Bethe-Salpeter interaction kernel. We consider all terms of chiral order 
$Q^2$ but only the subset of chiral $Q^3$-terms which are leading in the large $N_c$ limit. 
Loop corrections to the Bethe-Salpeter kernel are neglected, because they carry minimal chiral 
order $Q^3$ and are $1/N_c$ suppressed. The chiral Lagrangian 
\begin{eqnarray}
{\mathcal L} = \sum_n \,{\mathcal L}^{(n)}+\sum_n\,{\mathcal L}^{(n)}_\chi
\label{}
\end{eqnarray}
can be decomposed into terms of different classes ${\mathcal L}^{(n)}$ and ${\mathcal L}^{(n)}_\chi$. 
With an upper index $n$ in ${\mathcal L}^{(n)}$ we indicate the number of fields 
in the interaction vertex. The lower index $\chi $ signals terms with explicit chiral 
symmetry breaking. We assume charge conjugation symmetry and parity invariance in this 
work. To leading chiral order the following interaction terms are required:
\begin{eqnarray}
{\mathcal L}^{(2)} &=&
\tr \bar B \left(i\,\partialslash-\m0_{[8]}\right) \, B
+\frac{1}{4}\,\tr (\partial^\mu \,\Phi )\,
(\partial_\mu \,\Phi )
\nonumber\\
&+&\tr \bar \Delta_\mu \cdot \Big(
\left( i\,\partialslash -\m0_{[10]} \right)g^{\mu \nu }
-i\,\left( \gamma^\mu \partial^\nu + \gamma^\nu \partial^\mu\right)
+i\,\gamma^\mu\,\partialslash\,\gamma^\nu
+\,\m0_{[10]} \,\gamma^\mu\,\gamma^\nu
\Big) \, \Delta_\nu
\nonumber\\
&+&\tr \bar B^*_\mu \cdot \Big(
\left( i\,\partialslash -\m0_{[9]} \right)g^{\mu \nu }
-i\,\left( \gamma^\mu \partial^\nu + \gamma^\nu \partial^\mu\right)
+i\,\gamma^\mu\,\partialslash\,\gamma^\nu
+\,\m0_{[9]} \,\gamma^\mu\,\gamma^\nu
\Big) \, B^*_\nu
\nonumber\\
{\mathcal L}^{(3)} &=&\frac{F_{[8]}}{2\,f} \,\tr  \bar B
\,\gamma_5\,\gamma^\mu \,\Big[\left(\partial_\mu\,\Phi\right),B\Big]_-
+\frac{D_{[8]}}{2\,f} \,\tr  \bar B
\,\gamma_5\,\gamma^\mu \,\Big[\left(\partial_\mu\,\Phi\right),B\Big]_+
\nonumber\\
&-&\frac{C_{[10]}}{2\,f}\,
\tr \left\{
\Big( \bar \Delta_\mu \cdot
(\partial_\nu \,\Phi ) \Big)
\Big( g^{\mu \nu}-\half\,Z_{[10]}\, \gamma^\mu\,\gamma^\nu \Big) \,
B +\mathrm{h.c.}
\right\}
\nonumber\\
&+&\frac{D_{[9]}}{2\,f}\,
\tr \left\{  \bar B^*_\mu \cdot \Big[
(\partial_\nu \,\Phi ) \,
\Big( g^{\mu \nu}-\half\,Z_{[9]}\, \gamma^\mu\,\gamma^\nu \Big)\,\gamma_5 ,
B \Big]_++\mathrm{h.c.} \right\}
\nonumber\\
&+&\frac{F_{[9]}}{2\,f}\,
\tr \left\{
 \bar B^*_\mu \cdot \Big[
(\partial_\nu \,\Phi ) \,
\Big( g^{\mu \nu}-\half\,Z_{[9]}\, \gamma^\mu\,\gamma^\nu \Big)\,\gamma_5 ,
B \Big]_-+\mathrm{h.c.}
\right\} 
\nonumber\\
&+&\frac{C_{[9]}}{8\,f}\,
\tr  \left\{ \bar B^*_\mu \,\tr   \Big[
(\partial_\nu \,\Phi ) \,
\Big( g^{\mu \nu}-\half\,Z_{[9]}\, \gamma^\mu\,\gamma^\nu \Big)\,\gamma_5 ,
B \Big]_++\mathrm{h.c.}
\right\} \;,
\nonumber\\
{\mathcal L}^{(4)}&=& \frac{i}{8\,f^2}\,\tr\bar B\,\gamma^\mu \Big[\Big[ \Phi ,
(\partial_\mu \,\Phi) \Big]_-,B \Big]_- \,,
\label{lag-Q}
\end{eqnarray}
where we use the notations $[A,B]_\pm = A\,B\pm B\,A$ for
$SU(3)$ matrices $A$ and  $B$. Note that the complete chiral interaction terms which 
lead to the terms in (\ref{lag-Q}) are easily recovered by replacing 
$i\,\partial_\mu \,\Phi /f \to U_\mu $. A derivative acting on a baryon field in (\ref{lag-Q}) 
must be understood as the covariant derivative 
$\partial_\mu \,B \to [{\mathcal D}_\mu ,B ]_- $  and 
$\partial_\mu \,\Delta_\nu \to [{\mathcal D}_\mu ,\Delta_\nu ]_- $ .

The $SU(3)$ meson and baryon fields are 
written in terms of their isospin symmetric components
\begin{eqnarray}
\Phi &=& \tau \cdot  \pi
+ \alpha^\dagger \!\cdot \! K +  K^\dagger \cdot \alpha  +
\eta \,\lambda_8 \;,
\nonumber\\
\sqrt{2}\,B &=&  \alpha^\dagger \!\cdot \! N(939)+\lambda_8 \,\Lambda(1115)+ \tau \cdot \Sigma(1195) 
 +\Xi^t(1315)\,i\,\sigma_2 \!\cdot \!\alpha   \, ,
\nonumber\\ \nonumber\\
\alpha^\dagger &=&
 {\textstyle{1\over\sqrt{2}}}\left( \lambda_4+i\,\lambda_5 ,
\lambda_6+i\,\lambda_7 \right)
\;,\;\;\;\tau = (\lambda_1,\lambda_2,\lambda_3)\;,
\label{field-decomp}
\end{eqnarray}
with the isospin doublet fields $K =(K^+,K^0)^t $, $N=(p,n)^t$ and
$\Xi = (\Xi^0,\Xi^-)^t$. The isospin Pauli matrix $\sigma_2$ acts exclusively in the space of
isospin doublet fields $(K,N,\Xi)$ and the matrix valued isospin doublet $\alpha$ 
(see Appendix A). 
For our work we chose the isospin basis, because isospin breaking effects are important only 
in the $\bar K N$ channel. Note that in (\ref{field-decomp})  the numbers in the parentheses indicate 
the approximate mass of the baryon octet fields and $(...)^t$ means matrix transposition. 
Analogously we write the baryon resonance field $B_\mu^*$ as 
\begin{eqnarray}
\sqrt{2}\,B^*_\mu &=&  
\Big( \sqrt{{\textstyle{2\over 3}}}\,\cos \vartheta- \lambda_8\,\sin \vartheta\Big) \,\Lambda_\mu(1520)
+\alpha^\dagger \!\cdot\!  N_\mu(1520)
\nonumber\\
&+& \Big( \sqrt{{\textstyle{2\over 3}}}\,\sin \vartheta+ \lambda_8\,\cos \vartheta\Big) \,\Lambda_\mu(1690)
+\tau \!\cdot \!\Sigma_\mu(1670)  +\Xi_\mu^t(1820)\,i\,\sigma_2 \!\cdot\! \alpha \, ,
\label{def-b-stern}
\end{eqnarray}
where we allow for singlet-octet mixing by means of the mixing angle $\vartheta $ (see \cite{Plane}).
The parameters $\m0_{[8]}$, $\m0_{[9]}$ and $\m0_{[10]}$ in (\ref{lag-Q}) denote the baryon masses 
in the chiral $SU(3)$ limit. Furthermore the products of an anti-decuplet field $\bar
\Delta$ with a decuplet field $\Delta$ and an octet field $\Phi$ transform as  $SU(3)$ octets
\begin{eqnarray}
\Big(\bar \Delta \cdot \Delta \Big)^a_b &=&
\bar \Delta_{bcd}\,\Delta^{acd}\,, \qquad
\Big( \bar \Delta \cdot \Phi \Big)^a_b
=\epsilon^{kla}\,\bar \Delta_{knb}\,\Phi_l^n \,,
\nonumber\\
\Big( \Phi \cdot \Delta \Big)^a_b
&=&\epsilon_{klb}\,\Phi^l_n\,\Delta^{kna} \; ,
\label{dec-prod}
\end{eqnarray}
where $\epsilon_{abc}$ is the completely anti-symmetric pseudo-tensor. For the isospin 
decomposition of $\bar \Delta \cdot \Delta$, $\bar \Delta \cdot \Phi$ and 
$\Phi \cdot \Delta$ we refer to Appendix A. 

The parameters  $F_{[8]}\simeq 0.45$ 
and $D_{[8]}\simeq 0.80$ are constrained by the weak decay widths of the baryon octet states 
\cite{Okun} (see also Tab. \ref{weak-decay:tab}) and $C_{[10]}\simeq 1.6$ can be
estimated from the hadronic decay width of the baryon decuplet states. 
The parameter $Z_{[10]}$ in (\ref{lag-Q}) may be best determined in an $SU(3)$ analysis of 
meson-baryon scattering. While in the pion-nucleon sector it can be absorbed into the 
quasi-local 4-point interaction terms to chiral accuracy $Q^2$ \cite{Tang} (see also Appendix H),  
this is no longer possible if the scheme is extended to $SU(3)$. Our detailed analysis reveals that 
the parameter $Z_{[10]}$ is relevant already at order $Q^2$ if a simultaneous chiral analysis of 
the pion-nucleon and kaon-nucleon scattering processes is performed. The resonance parameters may be estimated 
by an update of the analysis \cite{Plane}. That leads to the values $F_{[9]} \simeq 1.8$, $D_{[9]} \simeq 0.84 $ and 
$C_{[9]} \simeq 2.5 $. The singlet-octet mixing angle $\vartheta \simeq $ 28$^\circ$ 
confirms the finding of \cite{Tripp:2} that the $\Lambda(1520)$ resonance is predominantly a flavor singlet 
state. The  value for the background parameter $Z_{[9]}$ of the $J^P\!\!=\!{\textstyle{3\over2}}^-$ resonance 
is expected to be rather model dependent, because it is unclear so far how to incorporate the  
$J^P\!\!=\!{\textstyle{3\over2}}^-$ resonance in a controlled approximation scheme.
As will be explained in detail $Z_{[9]}$ will drop out completely in our scheme (see sections 4.1-4.2).

\subsection{Large $N_c$ counting}

In this section we briefly recall a powerful expansion strategy which follows from QCD if the 
numbers of colors ($N_c$) is considered as a large number. We present a formulation best suited 
for an application in the chiral Lagrangian leading to a significant parameter reduction. 
The large $N_c$ scaling of  a chiral interaction 
term is easily worked out by using the operator analysis proposed in \cite{Dashen}.
Interaction terms involving baryon fields are represented by matrix 
elements of many-body operators in the large $N_c$ ground-state baryon 
multiplet $| {\mathcal B}  \rangle $. A n-body operator is the product of  
n factors formed exclusively in terms of the bilinear quark-field operators 
$J_i, G_i^{(a)}$ and $T^{(a)}$. These operators are characterized fully by their commutation 
relations, 
\begin{eqnarray}
&&[ G^{(a)}_i\,,G^{(b)}_j]_- ={\textstyle{1\over 4}}\,i\,\delta_{ij}\,f^{ab}_{\;\;\;\,c}\,T^{(c)}
+{\textstyle{1\over 2}}\,i\,\epsilon_{ij}^{\;\;\;k}
\left({\textstyle{1\over 3}}\,\delta^{ab}\,J_{k}+d^{ab}_{\;\;\;\,c}\,G^{(c)}_k\right), \;\;
\nonumber\\
&&[ J_i\,,J_j]_- =i\,\epsilon_{ij}^{\;\;\;k}\,J_{k}\, , \quad 
[ T^{(a)}\,,T^{(b)}]_- =i\,f^{ab}_{\;\;\;\,c}\,T^{(c)}\,,\quad 
\nonumber\\
&& [ T^{(a)}\,,G^{(b)}_i]_- =i\,f^{ab}_{\;\;\;c}\,G^{(c)}_i \;,\quad \!
[ J_i\,,G^{(a)}_j]_- =i\,\epsilon_{ij}^{\;\;\;k}\,G^{(a)}_k \;,  \quad \!
 [ J_i\,,T^{(a)}]_- = 0\;. 
\label{comm}
\end{eqnarray}
The algebra (\ref{comm}), which refelcts the so-called contracted spin-flavor symmetry 
of QCD, leads to a transparent derivation of the many sum rules implied by the 
various infinite subclasses of QCD quark-gluon diagrams as collected to a given order in the 
$1/N_c$ expansion. A convenient realization of the algebra (\ref{comm}) is obtained in 
terms of non-relativistic, flavor-triplet and color $N_c$-multiplet field operators 
$q$ and $q^\dagger$
\begin{eqnarray}
&& J_i = q^\dagger \Bigg( \,\frac{\sigma_i^{(q)}}{2} \otimes 1\Bigg) \,q \,, \qquad 
T^{(a)} = q^\dagger \Bigg( 1 \otimes \frac{\lambda^{(a)}}{2} \Bigg) \,q \,, \;\,
\nonumber\\
&& G^{(a)}_i = q^\dagger \Bigg(\, \frac{\sigma_i^{(q)}}{2} \otimes \frac{\lambda^{(a)}}{2} \Bigg) \,q \,.
\label{}
\end{eqnarray}
If the fermionic field operators $q$ and $q^\dagger $ are assigned 
anti-commutation rules, the algebra (\ref{comm}) follows. The Pauli spin matrices 
$\sigma^{(q)}_i$ act on the two-component spinors of the fermion fields $q, q^\dagger $ 
and the Gell-Mann matrices $\lambda_a$ on their flavor components. Here one needs to emphasize 
that the non-relativistic quark-field operators $q$ and $q^\dagger $ should not be identified 
with the quark-field operators of the QCD Lagrangian \cite{DJM,Carone,Luty}. Rather, they 
constitute an effective tool to represent the operator algebra (\ref{comm}) which allows for an 
efficient derivation of the large $N_c$ sum rules of QCD. A systematic truncation scheme 
results in the operator analysis, because a $n$-body operator is assigned the suppression factor 
$N_c^{1-n}$.  The analysis is complicated by the fact that  matrix elements of $ T^{(a)}$ 
and $G_i^{(a)}$ may be of order $N_c$ in the baryon 
ground state $|{\mathcal B} \rangle$. That implies for instance that matrix elements of 
the (2$n$+1)-body operator $(T_a\,T^{(a)})^n\,T^{(c)}$ are not suppressed relative to the matrix 
elements of the one-body operator $T^{(c)}$. The systematic large $N_c$ operator analysis 
relies on the observation that matrix elements of the spin operator $J_i$, on the other hand, 
are always of order $N_c^0$. Then a set of identities shows how to systematically represent the 
infinite set of many-body operators, which one may write down to a given order in the $1/N_c$ 
expansion, in terms of a finite number of operators. This leads to 
a convenient scheme with only a finite number of operators to a given order \cite{Dashen}.
We recall typical examples of the required operator identities
\begin{eqnarray}
&&[T_a,\, T^{(a)}]_+ -  [J_i,\,J^{(i)}]_+  ={\textstyle {1\over 6}}\,N_c\,(N_c+6)\;,
\quad [ T_{a}\,,G^{(a)}_i]_+ ={\textstyle {2\over 3}}\,(3+N_c)\,J_i \;,
\nonumber\\
&& 27\,[T_a,\, T^{(a)}]_+-12\,[G_i^{(a)},\, G_a^{(i)}]_+ 
= 32\,[J_i,\,J^{(i)}]_+ \;,
\nonumber\\
&& d_{abc}\,[T^{(a)},\,T^{(b)}]_+  -2\,[J_{i},\,G_c^{(i)}]_+
= -{\textstyle {1\over 3}}\,(N_c+3)\,T_c \;,
\nonumber\\
&& d^a_{\;\;bc}\,[G_a^{(i)},\,G_i^{(b)}]_+
+{\textstyle{9\over 4}}\, d_{abc}\,[T^{(a)},\,T^{(b)}]_+ 
= {\textstyle {10\over 3}}\,[J_{i},\,G_c^{(i)}]_+ \;,
\nonumber\\
&& d_{ab}^{\;\;\; c}\,[T^{(a)},\,G^{(b)}_i]_+ =  {\textstyle{1\over 3}}\,[J_{i},\,T^{(c)}]_+ 
-{\textstyle{1\over 3}}\,\epsilon_{ijk}\,f_{ab}^{\;\;\;c}\,[G^{(j)}_a\,,G^{(k)}_b]_+ \;.
\label{operator-ex}
\end{eqnarray}
For instance the first identity in (\ref{operator-ex}) shows how to avoid the  
infinite tower $(T_a\,T^{(a)})^n\,T^{(c)}$ discussed above. Note that the 'parameter'
$N_c$ enters in (\ref{operator-ex}) as a mean to classify the possible realizations of 
the algebra (\ref{comm}). 

As a first and simple example we recall the large $N_c$ structure of the 3-point vertices. 
One readily establishes two operators with parameters $g$ and $h$ to leading order 
in the $1/N_c$ expansion \cite{Dashen}:
\begin{eqnarray}
\langle {\mathcal B}' |\, {\mathcal L}^{(3)}\,| {\mathcal B}  \rangle  =\frac{1}{f}\,
\langle {\mathcal B}' |\, g\,G_i^{(c)}+h\,J_i\,T^{(c)}| {\mathcal B}  \rangle \,
\tr \,\lambda_c\,\nabla^{(i)}\,\Phi + {\mathcal O}\left( \frac{1}{N_c}\right) \;.
\label{3-point-vertex}
\end{eqnarray}
Further possible terms in (\ref{3-point-vertex}) are either redundant or suppressed 
in the $1/N_c$ expansion. For example, the two-body operator 
$ i\,f_{abc}\,G_i^{(a)} \,T^{(b)}  \sim N_c^0$ is reduced by applying the relation
\begin{eqnarray}
&& i\,f_{ab}^{\;\;\;c}\,\Big[G_i^{(a)} \,,T^{(b)} \Big]_- = i\,f_{ab}^{\;\;\;c}\,i\,f^{ab}_{\;\;\;\,d}\,G_i^{(d)} =
-3\,G_i^{(c)} \;. \nonumber
\label{}
\end{eqnarray}
In order to make use of the large $N_c$ result, it is necessary to evaluate the matrix elements 
in (\ref{3-point-vertex}) at 
$N_c=3$ where one has a $\bf 56$-plet with 
$| {\mathcal B}  \rangle= |B(a) ,\Delta(ijk) \rangle $. Most 
economically this is achieved with the completeness identity 
$1=|B\rangle \langle B|+ |\Delta\rangle \langle \Delta | $ in conjunction with 
\begin{eqnarray}
&&T_c\,| B_a(\chi)\rangle = i\,f_{abc}\,| B^{(b)}(\chi)\rangle \;,\qquad
J^{(i)} \,| B_a(\chi )\rangle 
= \frac{1}{2}\,\sigma^{(i)}_{\chi' \chi}| B_a(\chi')\rangle \;,
\nonumber\\
&& G^{(i)}_c\,| B_a(\chi)\rangle = 
\left(\frac{1}{2}\,d_{abc}
+ \frac{1}{3}\,i\,f_{abc}\right)\,\sigma^{(i)}_{\chi'\chi}\,| B^{(b)}(\chi')\rangle
\nonumber\\
&& \qquad \qquad \quad \;
+ \frac{1}{\sqrt{2}\,2}\,\Big(\epsilon_{l}^{\;jk}\,\lambda^{(c)}_{mj}
\,\lambda_{nk}^{(a)}\Big)\,S^{(i)}_{\chi' \chi}
| \Delta^{\!(lmn)}(\chi') \rangle \;,
\label{matrix-el}
\end{eqnarray}
where $S_i\,S^\dagger_j=\delta_{ij}-\sigma_i\,\sigma_j/3$ and 
$\lambda_a\,\lambda_b = {\textstyle{2 \over 3}}\,\delta_{ab}
+(i\,f_{abc}+d_{abc})\,\lambda^{(c)}$. In (\ref{matrix-el}) the baryon octet states 
$| B_b(\chi)\rangle $ are labelled according to their $SU(3)$ octet index $a=1,...,8$ with
the two spin states represented by $\chi=1,2$. Similarly the decuplet states 
$| \Delta_{lmn}(\chi') \rangle$ are listed with $l,m,n=1,2,3$ as defined in (\ref{dec-field}).
Note that the expressions (\ref{matrix-el}) may be 
verified using the quark-model wave functions for the baryon octet and decuplet states. The result (\ref{matrix-el}) 
is however much more general than 
the quark-model, because it follows from the structure of the ground-state baryons in the large 
$N_c$ limit of QCD only. Matching the interaction vertices of the relativistic chiral Lagrangian 
onto the static matrix elements arising in the large $N_c$ operator analysis requires a  
non-relativistic reduction. It is standard to decompose the 
4-component Dirac fields $B$ and $\Delta_\mu $ into baryon octet and 
decuplet spinor fields $B(\chi)$ and $\Delta (\chi)$:
\begin{eqnarray}
\Big(B, \Delta_\mu \Big) \to  \left( 
\begin{array}{c}
 \left(\frac{1}{2}+\frac{1}{2}\,\sqrt{1+\frac{\nabla^2}{M^2}} 
\right)^{\frac{1}{2}} \Big(B(\chi ),S_\mu\,\Delta (\chi )\Big) \\
\frac{(\sigma \cdot \nabla )}{\sqrt{2}\,M} 
\left(1+\sqrt{1+\frac{\nabla^2}{M^2} }\,\right)^{-\frac{1}{2}} 
\Big(B(\chi ),S_\mu\,\Delta (\chi )\Big)
\end{array}
\right)  \,,
\label{}
\end{eqnarray}
where $M$ denotes the baryon octet and decuplet mass in the large $N_c$ limit. To  
leading order one finds $S_\mu =(0, S_i) $ with the transition matrices $S_i$ introduced in 
(\ref{matrix-el}). It is then straightforward to expand in powers of $\nabla/M$ and 
achieve the desired matching. This leads for example to the identification $D_{[8]}=g $, 
$F_{[8]}= 2\,g/3+h$ and $C_{[10]}=2\,g$. The empirical values of $F_{[8]},D_{[8]}$ and $C_{[10]}$ are 
quite consistent with those large $N_c$ sum rules \cite{Jenkins}. 
Note that operators at subleading order in 
(\ref{3-point-vertex}) then parameterize the deviation from $C_{[10]}\simeq 2 \,D_{[8]}$. 

\subsection{Quasi-local interaction terms}
 
We turn to the two-body interaction terms at chiral order $Q^2$. 
From phase space consideration it is evident that to this order
there are only terms which contribute to the meson-baryon s-wave scattering lengths, the 
s-wave effective range parameters and the p-wave scattering volumes. Higher partial 
waves are not affected to this order. The various contributions are regrouped according to their 
scalar, vector or tensor nature as
\begin{eqnarray}
{\mathcal L}^{(4)}_2= {\mathcal L}^{(S)}+{\mathcal L}^{(V)}+{\mathcal L}^{(T)}
\,,
\label{l42}
\end{eqnarray}
where the lower index k in ${\mathcal L}^{(n)}_k$ denotes the minimal chiral order 
of the interaction vertex. In the relativistic framework one observes mixing of the 
partial waves in the sense that for instance ${\mathcal L}^{(S)}, {\mathcal L}^{(V)}$ 
contribute to the s-wave channels and ${\mathcal L}^{(S)}, {\mathcal L}^{(T)}$ to the 
p-wave channels. We write
\begin{eqnarray}
{\mathcal L}^{(S)}&=&\frac{g^{(S)}_0}{8\!\,f^2}\,\tr\bar B\,B
\,\tr (\partial_\mu\Phi) \, (\partial^\mu\Phi)
+\frac{g^{(S)}_1}{8\!\,f^2}\,\tr \bar B \,(\partial_\mu\Phi)
\,\tr(\partial^\mu\Phi) \, B
\nonumber\\
&+&\frac{g^{(S)}_F}{16\!\,f^2} \,\tr\bar B \Big[
\Big[(\partial_\mu\Phi),(\partial^\mu\Phi)
\Big]_+ ,B\Big]_-
\!+\frac{g^{(S)}_D}{16\!\,f^2}\,\tr \bar B \Big[
\Big[(\partial_\mu\Phi),(\partial^\mu\Phi)
\Big]_+, B \Big]_+ \;,
\nonumber\\
{\mathcal L}^{(V)}&=&
\frac{g^{(V)}_0}{16\!\, f^2}\,
\Big(\tr \bar B \,i\,\gamma^\mu\,( \partial^\nu B) \,
\tr(\partial_\nu\Phi) \, ( \partial_\mu\Phi)
+\mathrm{h.c.}\Big)
\nonumber\\
&+&\frac{g^{(V)}_1}{32\!\,f^2}\,
\tr \bar B \,i\,\gamma^\mu\,\Big( ( \partial_\mu\Phi)
\,\tr(\partial_\nu\Phi) \, ( \partial^\nu B)
+ ( \partial_\nu\Phi) \,
\tr ( \partial_\mu \Phi) \, ( \partial^\nu B)
+\mathrm{h.c.}\Big)
\nonumber\\
&+&\frac{g_F^{(V)}}{32\!\,f^2}\,\Big(
\tr   \bar B \,i\,\gamma^\mu\,\Big[
\Big[(\partial_\mu \Phi) , (\partial_\nu\Phi)\Big]_+,
( \partial^\nu B) \Big]_-
+\mathrm{h.c.} \Big)
\nonumber\\
&+& \frac{g^{(V)}_D}{32\!\,f^2}\,\Big(\tr
\bar B \,i\,\gamma^\mu\,\Big[
\Big[( \partial_\mu\Phi) , (\partial_\nu\Phi)\Big]_+,
( \partial^\nu B) \Big]_+
+\mathrm{h.c.} \Big)\,,
\nonumber\\
{\mathcal L}^{(T)}&=&
\frac{g^{(T)}_1}{8\!\,f^2}\,\tr\bar B \,( \partial_\mu\Phi)
\,i\,\sigma^{\mu \nu}\,\tr( \partial_\nu \Phi) \, B
\nonumber\\
&+& \frac{g^{(T)}_D}{16\!\,f^2}\,
\tr \bar B \,i\,\sigma^{\mu \nu}\,\Big[
\Big[(\partial_\mu \Phi) ,( \partial_\nu \Phi)  \Big]_-, B \Big]_+
\nonumber\\
&+&\frac{g^{(T)}_F}{16\!\,f^2}\,
\tr \bar B \,i\,\sigma^{\mu \nu}\,\Big[
\Big[( \partial_\mu \Phi) ,( \partial_\nu\Phi) \Big]_- ,B\Big]_-
\,.
\label{two-body}
\end{eqnarray}
It is clear that if the heavy-baryon expansion is applied to (\ref{two-body})
the quasi-local 4-point interactions can be mapped onto corresponding terms
of  the heavy-baryon formalism presented for example in \cite{CH-Lee}. Inherent in 
the relativistic scheme is the presence of redundant interaction terms which requires
that a systematic regrouping of the interaction terms is performed. This will be 
discussed below in more detail when introducing the quasi-local counter terms at 
chiral order $Q^3$.

We apply the large $N_c$ counting rules in order to estimate the 
relative importance of the  quasi-local $Q^2$-terms in (\ref{two-body}). 
Terms which involve a single-flavor trace are enhanced as compared to the 
double-flavor trace terms. This is because a flavor trace in an interaction term 
is necessarily accompanied by a corresponding color trace if visualized in terms of quark 
and gluon lines. A color trace signals a quark loop and therefore provides the announced   
$1/N_c$ suppression factor \cite{Hooft,Witten}. The counting rules are nevertheless 
subtle, because a certain combination of double trace expressions can be rewritten in 
terms of a single-flavor trace term \cite{Fearing}
\begin{eqnarray}
&&\tr \left( \bar B \, B \right)
\,\tr \Big(\Phi \, \Phi  \Big)
+2\,\tr \left( \bar B \, \Phi \right)
\,\tr \Big(\Phi \, B  \Big)
\nonumber\\
=&&\tr \Big[\bar B, \Phi \Big]_-\,\Big[B, \Phi
\Big]_- 
+\frac{3}{2}\,\tr \bar B \Big[
\Big[\Phi , \Phi  \Big]_+, B \Big]_+ \;.
\label{trace-id}
\end{eqnarray}
Thus one expects for example that both parameters $g_{0}^{(S)}$ and $g_{1}^{(S)}$ may be large 
separately but the combination $2\,g_0^{(S)}-g_1^{(S)}$ should be small. A more detailed 
operator analysis  leads to
\begin{eqnarray}
&&\langle {\mathcal B}' | {\mathcal L}^{(4)}_2 | {\mathcal B} \rangle  = \frac{1}{16\,f^2}\,
\langle {\mathcal B}' | \,O_{ab}(g_1,g_2)\, | {\mathcal B} \rangle
\,\tr [(\partial_\mu \Phi),\lambda^{(a)}]_-\,[(\partial^\mu \Phi),\lambda^{(b)}]_-
\nonumber\\
&&  \qquad \qquad +\frac{1}{16\,f^2}\,
\langle {\mathcal B}' | \,O_{ab}(g_3,g_4)\, | {\mathcal B} \rangle
\,\tr [(\partial_0 \Phi),\lambda^{(a)}]_-\,[(\partial_0 \Phi),\lambda^{(b)}]_-
\nonumber\\
&& \qquad \qquad  + \frac{1}{16\,f^2}\,
\langle {\mathcal B}' | \,O^{(ij)}_{ab}(g_5,g_6)\, | {\mathcal B} \rangle
\,\tr [(\nabla_i \Phi),\lambda^{(a)}]_-\,[(\nabla_j \Phi),\lambda^{(b)}]_- \;,
\nonumber\\ \nonumber\\
&& O_{ab}(g,h) = g\,d_{abc}\,T^{(c)} + h\,[T_a,T_b]_+ 
+{\mathcal O}\left(\frac{1}{N_c} \right) \;,
\nonumber\\
&& O_{ab}^{(ij)}(g,h) = i\,\epsilon^{ijk }\,i\,f_{abc}\left(
g\,G_k^{(c)} + h\,J_{k}\,T^{(c)} \right)
+{\mathcal O}\left(\frac{1}{N_c} \right) \;.
\label{Q^2-large-Nc}
\end{eqnarray}
We checked that other forms for the coupling of the operators $O_{ab}$ to the 
meson fields do not lead to new structures. 
It is straight forward to match the coupling constants $g_{1,..,6}$ onto those of
(\ref{two-body}). Identifying the leading terms in the non-relativistic 
expansion, we obtain: 
\begin{eqnarray}
&& g_0^{(S)}= \frac{1}{2}\,g_1^{(S)} = \frac{2}{3}\,g_D^{(S)}
= -2\,g_2\,, \qquad  \; g_F^{(S)}= -3\,g_1 \,,
\nonumber\\
&& g_0^{(V)}= \frac{1}{2}\,g_1^{(V)}=\frac{2}{3}\,g_D^{(V)}
=-2\,\frac{g_4}{M}\,, \qquad  
g_F^{(V)}= -3\,\frac{g_3}{M} \,, 
\nonumber\\
&& g_1^{(T)}= 0 \,, \qquad  
g_F^{(T)}= -g_5-\frac{3}{2}\,g_6 \,,  \qquad g_D^{(T)}= -\frac{3}{2}\,g_5 \;,
\label{Q^2-large-Nc-result}
\end{eqnarray}
where $M$ is the large $N_c$ value of the baryon octet mass. 
We conclude that to chiral order $Q^2$ there are only six leading large $N_c$ 
coupling constants.

We turn to the quasi-local counter terms to chiral order $Q^3$. It is instructive to  
discuss first a set of redundant interaction terms:
\begin{eqnarray}
{\mathcal L}^{(R)}&=&\frac{h^{(1)}_0}{8 f^2}\,
\tr(\partial^\mu\bar B)\,
(\partial_\nu B)\,
\tr( \partial_\mu \Phi) \, ( \partial^\nu\Phi)
\nonumber\\
&+&\frac{h^{(1)}_1}{16 f^2}\,
\tr(\partial^\mu\bar B)\,  ( \partial_\nu\Phi)\,
\tr(\partial_\mu\Phi) \, (\partial^\nu B)
\nonumber\\
&+&\frac{h^{(1)}_1}{16 f^2}\,
\tr(\partial^\mu\bar B)\,( \partial_\mu \,\Phi)\,
\tr(\partial^\nu\Phi) \, (\partial_\nu B) \,,
\nonumber\\
&+&\frac{h^{(1)}_F}{16 f^2}\,
\tr (\partial^\mu\bar B)\, \Big[
\Big[( \partial_\mu\Phi) , (\partial^\nu \Phi) \Big]_+ ,
(\partial_\nu B) \Big]_-
\nonumber\\
&+& \frac{h^{(1)}_D}{16 f^2}\,
\tr (\partial^\mu\bar B)\,  \Big[
\Big[( \partial_\mu\Phi) , (\partial^\nu \Phi)\Big]_+,
(\partial_\nu B )\Big]_+ \,.
\label{redundant}
\end{eqnarray}
Performing the non-relativistic expansion of (\ref{redundant}) one finds that the 
leading moment is of chiral order $Q^2$. Formally the terms in (\ref{redundant}) are 
transformed into terms of subleading order $Q^3$ by subtracting  ${\mathcal L}^{(V)}$ of
(\ref{two-body}) with $g^{(V)} =  \m0_{[8]} \,h^{(1)}$. Bearing this in mind the  
terms (\ref{redundant}) define particular interaction vertices of chiral order $Q^3$.
Note that in analogy to (\ref{Q^2-large-Nc}) and (\ref{Q^2-large-Nc-result})
we expect the coupling constants 
$h_F^{(1)}$ and $h_D^{(1)}$ with $h_1^{(1)}=2\,h_0^{(1)}= 4\,h_D^{(1)}/3$ 
to be leading in the large $N_c$ limit. 
A complete collection of counter terms of chiral order $Q^3$ is presented in 
Appendix B. Including the four terms in (\ref{redundant}) we find ten independent 
interaction terms which all  contribute exclusively to the 
s- and p-wave channels. Here we present the two additional terms with
$h^{(2)}_{F}$ and $h^{(3)}_{F}$ which are leading in the large $N_c$ expansion:
\begin{eqnarray}
{\mathcal L}^{(4)}_3 &=& {\mathcal L}^{(R)}- {\mathcal L}^{(V)} [g^{(V)}\! = \m0_{[8]}\,h^{(1)}]
\nonumber\\
&+& \frac{h^{(2)}_{F}}{32\!\,f^2}\,
\tr \bar B \,i\,\gamma^\mu\,\Big[ \big[ ( \partial_\mu\Phi)
, (\partial_\nu \Phi)\big]_- , ( \partial^\nu B) \Big]_- +{\rm h.c}
\nonumber\\
&+&\frac{h^{(3)}_{F}}{16\!\,f^2}\,
\tr \bar B \,i\,\gamma^\alpha\,\Big[ \big[ ( \partial_\alpha \,\partial_\mu\Phi)
,( \partial^\mu \Phi) \big]_- ,  B \Big]_- \;.
\label{local-q-3}
\end{eqnarray}
The interaction vertices in (\ref{local-q-3}) can be mapped onto corresponding 
static matrix elements of the large $N_c$ operator analysis: 
\begin{eqnarray}
&&\langle {\mathcal B}' | {\mathcal L}^{(4)}_3 | {\mathcal B} \rangle  = \frac{h_2}{16\,f^2}\,
\langle {\mathcal B}' | \,i\,f_{abc}\,T^{(c)}\, | {\mathcal B} \rangle
\,\partial_\mu \,\Big(
\tr [(\partial_0 \Phi),\lambda^{(a)}]_-\,[(\partial^\mu \Phi),\lambda^{(b)}]_- \Big)
\nonumber\\
&& \qquad \qquad \;\; +\frac{h_3}{16\,f^2}\,
\langle {\mathcal B}' | \,i\,f_{abc}\,T^{(c)}\, | {\mathcal B} \rangle
\, \tr [(\partial_0 \,\partial_\mu \Phi),\lambda^{(a)}]_-\,[(\partial^\mu \Phi),\lambda^{(b)}]_- \,,
\label{}
\end{eqnarray}
where $h^{(2)}_{F} \sim h_2$ and $h_{F}^{(3)} \sim h_3$. We summarize our 
result for the quasi-local chiral interaction vertices of order $Q^3$: at leading order 
the  $1/N_c$ expansion leads to four relevant parameters
only. Also one should stress that the $SU(3)$ structure of the 
$Q^3$ terms as they contribute to the s- and p-wave channels differ from 
the $SU(3)$ structure of the $Q^2$ terms. For instance the $g^{(S)}$
coupling constants contribute to the p-wave channels with four independent 
$SU(3)$ tensors. In contrast, at order $Q^3$ the parameters $h^{(2)}_{F}$ and 
$h^{(3)}_{F}$, which are in fact the only parameters contribution to the 
p-wave channels to this order, contribute with a different and independent 
$SU(3)$ tensor. This is to be compared with the static $SU(3)$ prediction 
that leads to six independent tensors:
\begin{equation}
8\otimes 8= 1 
\oplus 8_S\oplus 8_A \oplus 10\oplus \overline{10}\oplus 27 \;.
\label{}
\end{equation}
Part of the predictive power of the chiral Lagrangian results, because 
chiral $SU(3)$ symmetry selects certain subsets of all $SU(3)$ symmetric 
tensors at a given chiral order. 

\subsection{Explicit chiral symmetry breaking}

There remain the interaction terms proportional to $\chi_\pm $ 
which break the chiral $SU(3)$ symmetry explicitly. We collect here all relevant terms of 
chiral order $Q^2$ \cite{Gasser,Kaiser} and $Q^3$ \cite{Mueller}. It is convenient to visualize the symmetry-breaking 
fields $\chi_\pm$ of (\ref{def-chi}) in their expanded forms:
\begin{eqnarray}
\! \!\! \chi_+ = \chi_0 -\frac{1}{8\,f^2}
\Big[ \Phi, \Big[ \Phi ,\chi_0 \Big]_+\Big]_+ \!+{\mathcal O} \left(\Phi^4 \right) \,,
\;\,
\chi_- = \frac{i}{2\,f}\, \Big[ \Phi ,\chi_0 \Big]_+ 
\!+{\mathcal O} \left(\Phi^3 \right) \,.
\label{chi-exp}
\end{eqnarray}
We begin with the 2-point interaction vertices which all follow exclusively from chiral 
interaction terms linear in $\chi_+$. They read
\begin{eqnarray}
{\mathcal L }_{\chi}^{(2)}&=& -\frac{1}{4}\,\tr \Phi\,\Big[ \chi_0, \Phi \Big]_+
+2\,\tr \bar B \left( b_D\,\Big[ \chi_0 , B \Big]_+ +b_F\,\Big[ \chi_0 , B \Big]_-
+b_0 \, B \,\tr \chi_0 \right)
\nonumber\\
&+&2\,d_D\,\tr \Big(\bar \Delta_\mu\cdot \Delta^\mu  \Big)  \,\chi_0
+2\,d_0 \,\tr \left(\bar \Delta_\mu\cdot  \Delta^\mu\right) \,\tr \chi_0 
\nonumber\\
&+&\tr \bar B \left(i\,\partialslash-\m0_{[8]}\right) \, 
\Big( \zeta_0\,B \, \tr  \chi_0
+  \zeta_D\,[B, \chi_0 ]_+ + \zeta_F\, [B, \chi_0 ]_- \Big) \;, 
\nonumber\\ \nonumber\\
\chi_0 &=&\frac{1}{3} \left( m_\pi^2+2\,m_K^2 \right)\,1
+\frac{2}{\sqrt{3}}\,\left(m_\pi^2-m_K^2\right) \lambda_8 \;,
\label{chi-sb}
\end{eqnarray}
where we normalized $\chi_0$ to give the pseudo-scalar mesons their isospin averaged 
masses. The first term in (\ref{chi-sb}) leads to the finite masses of the pseudo-scalar 
mesons. Note that to chiral order $Q^2$ one has $m_\eta^2 = 4\,(m_K^2-m_\pi^2)/3$.
The parameters $b_D$, $b_F$, and $d_D$ are determined to leading order by the
baryon octet and decuplet mass splitting
\begin{eqnarray} 
&&m_{[8]}^{(\Sigma )}-m_{[8]}^{(\Lambda )}=
{\textstyle{16\over 3}}\,b_D\,(m_K^2-m_\pi^2)\,, \quad
m_{[8]}^{(\Xi )}-m_{[10]}^{(N )} =-8\,b_F\,(m_K^2-m_\pi^2)\,,
\nonumber \\
&&m_{[10]}^{(\Sigma )}-m_{[10]}^{(\Delta )}=m_{[8]}^{(\Xi )}-m_{[10]}^{(\Sigma )}
=m_{[10]}^{(\Omega )}-m_{[10]}^{(\Xi )}
=-\textstyle{4\over 3}\, d_D\, (m_K^2-m_\pi^2)\,.
\label{mass-splitting}
\end{eqnarray}
The empirical baryon masses lead to the estimates $b_D \simeq 0.06$~GeV$^{-1}$, 
$b_F \simeq -0.21$~GeV$^{-1}$, and $d_D\simeq -0.49$~GeV$^{-1}$. 
For completeness we recall the leading large $N_c$ operators for  
the baryon mass splitting (see e.g. \cite{Jenkins}):
\begin{eqnarray}
&&\langle {\mathcal B}' |{\mathcal L }_{\chi}^{(2)} | {\mathcal B} \rangle 
=\langle {\mathcal B}' |\,
b_1\,T^{(8)} + b_2\,[J^{(i)}, G^{(8)}_i]_+ \,| {\mathcal B} \rangle
+{\mathcal O}\left(\frac{1}{N_c^{2}}\right)\,,
\nonumber\\
&&b_D = -\frac{\sqrt{3}}{16}\,\frac{3\,b_2}{m_K^2-m_\pi^2} \,, \qquad 
b_F = -\frac{\sqrt{3}}{16}\,\frac{2\,b_1+b_2}{m_K^2-m_\pi^2}\,, 
\nonumber\\
&& d_D = -\frac{3\,\sqrt{3}}{8}\,\frac{b_1+2\,b_2}{m_K^2-m_\pi^2} \,,
\label{}
\end{eqnarray}
where we matched the symmetry-breaking parts with $\lambda_8$. 
One observes that the empirical values for $b_D+b_F$ and $d_D$ are remarkably consistent with 
the large $N_c$ sum rule $b_D+b_F\simeq {\textstyle{1\over 3}}\,d_D$. The parameters $b_0$ 
and $d_0$ are more difficult to access. They determine the deviation of the octet and 
decuplet baryon masses from their chiral $SU(3)$ limit values $\m0_{[8]}$ and $\m0_{[10]}$:
\begin{eqnarray}
&& m_{[8]}^{(N)} = \m0_{[8]}-2\,m_\pi^2\,(b_0+2\,b_F)-4\,m_K^2\,(b_0+b_D-b_F) \;,
\nonumber\\
&& m_{[10]}^{(\Delta )}  = \m0_{[10]}-2\,m_\pi^2 (d_0 +d_D)-4\,m_K^2 \,d_0 \;,
\label{piN-sig-term}
\end{eqnarray}
where terms of chiral order $Q^3$ are neglected. The size of the parameter $b_0$ is 
commonly encoded into the pion-nucleon sigma term 
\begin{equation}
\sigma_{\pi N}= -2\,m_\pi^2\,(b_D+b_F+2\,b_0) +{\mathcal O}\left(Q^3\right)\,.
\label{spin:naive}
\end{equation}
Note that the former standard value $\sigma_{\pi N}=(45\pm 8)$~MeV of 
\cite{piN-sigterm} is currently under debate \cite{pin-news}.

The parameters $\zeta_0,\zeta_D$ and $\zeta_F$ are required to cancel a divergent term in the 
baryon wave-function renormalization as it follows from the one loop self-energy correction or 
equivalently the unitarization of the s-channel baryon exchange term. It will be demonstrated 
explicitly that within our approximation they will not have any observable effect. 
They lead to a renormalization of the three-point vertices only, which can be accounted for by 
a redefinition of the parameters in (\ref{chi-sb-3}). Thus one may simply drop these 
interaction terms. 

The predictive power of the chiral Lagrangian lies in part in the strong correlation of 
vertices of different degrees as implied by the non-linear fields $U_\mu $ and $\chi_\pm $.
A powerful example is given by the two-point vertices introduced in (\ref{chi-sb}). Since 
they result from chiral interaction terms linear in the $\chi_+$-field (see (\ref{chi-exp})),  
they induce particular meson-octet baryon-octet interaction vertices:
\begin{eqnarray}
{\mathcal L }_{\chi}^{(4)}&=& 
\frac{i}{16\,f^2}\,\tr\bar B\,\gamma^\mu \Big[\Big[ \Phi ,
(\partial_\mu \,\Phi) \Big]_-,\zeta_0\,B \, \tr  \chi_0
+  \zeta_D\,[B, \chi_0 ]_+ + \zeta_F\, [B, \chi_0 ]_-  \Big]_- 
\nonumber\\
&+&\frac{i}{16\,f^2}\,\tr \Big[\zeta_0\,\bar B \, \tr  \chi_0
+  \zeta_D\,[\bar B, \chi_0 ]_+ + \zeta_F\, [\bar B, \chi_0 ]_- ,
\Big[ \Phi ,(\partial_\mu \,\Phi) \Big]_- \Big]_- \,\gamma^\mu\,B
\nonumber\\
&-&\frac{1}{4\,f^2}\,\tr \bar B \left(b_D\,
\Big[ \Big[ \Phi, \Big[ \Phi ,\chi_0 \Big]_+\Big]_+ , B \Big]_+ 
+b_F\,\Big[ \Big[ \Phi, \Big[ \Phi ,\chi_0 \Big]_+\Big]_+ , B \Big]_-
\right)  
\nonumber\\
&-&\frac{b_0}{4\,f^2}\,\tr\bar{B}\, B \,\tr \Big[ \Phi, \Big[ \Phi ,\chi_0 \Big]_+\Big]_+ 
\;.
\label{chi-sb-4}
\end{eqnarray}
To chiral order $Q^3$ there are no further four-point interaction terms with explicit 
chiral symmetry breaking.

We turn to the three-point vertices with explicit chiral symmetry breaking. Here 
the chiral Lagrangian permits two types of interaction terms written as  
${\mathcal L }_{\chi }^{(3)}={\mathcal L }_{\chi,\, +}^{(3)}+{\mathcal L }_{\chi, -}^{(3)}$. 
In ${\mathcal L }_{\chi, +}^{(3)}$ we collect 16 axial-vector terms, which 
result form chiral interaction terms linear in the $\chi_+$ field (see (\ref{chi-exp})),
with a priori unknown coupling constants $F_{0,..,9}$ and $C_{0,...,5}$, 
\begin{eqnarray}
{\mathcal L }_{\chi, +}^{(3)}&=&
\frac{1}{4\,f} \,\tr  \bar B
\,\gamma_5\,\gamma^\mu \,\Big( 
\Big[\left(\partial_\mu\,\Phi\right),F^{}_0\,\Big[\chi_0, B \Big]_+
+F^{}_1\,\Big[\chi_0, B \Big]_-\Big]_+  \Big)+{\rm h.c.}
\nonumber\\
&+&\frac{1}{4\,f} \,\tr  \bar B
\,\gamma_5\,\gamma^\mu \,\Big( 
\Big[\left(\partial_\mu\,\Phi\right),F^{}_2\,\Big[\chi_0, B \Big]_+
+F^{}_3\,\Big[\chi_0, B \Big]_-\Big]_-  \Big)+{\rm h.c.}
\nonumber\\
&+&\frac{1}{2\,f} \,\tr  \bar B
\,\gamma_5\,\gamma^\mu \,\Big( F_4\,\Big[
\Big[\chi_0, \left(\partial_\mu\,\Phi\right)\Big]_+ , B \Big]_+
+F^{}_5\,\Big[\Big[\chi_0, \left(\partial_\mu\,\Phi\right)\Big]_+ , B \Big]_-  \Big)
\nonumber\\
&+&\frac{1}{4\,f} \,\tr  \bar B
\,\gamma_5\,\gamma^\mu \,\Big( 
F^{}_{6}\,B \,\tr \Big( \chi_0 \left(\partial_\mu\,\Phi\right)  \Big)
+F^{}_{7} \left(\partial_\mu\,\Phi\right)
\tr \Big(\chi_0 \, B  \Big) \Big) +{\rm h.c.}
\nonumber\\
&+& \frac{1}{2\,f} \,\tr  \bar B
\,\gamma_5\,\gamma^\mu \,\Big( 
F_8\,\Big[\left(\partial_\mu\,\Phi\right),B\Big]_+ \,\tr \chi_0 
+F_9 \,\Big[\left(\partial_\mu\,\Phi\right),B\Big]_-\,\tr \chi_0 \Big)
\nonumber\\
&-& \frac{1}{2\,f}\,
\tr \left\{C_0 \,\Big( \bar \Delta_\mu \cdot 
\Big[ \chi_0 ,(\partial_\mu \,\Phi )\Big]_+ 
+C_1 \,\Big( \bar \Delta_\mu \cdot 
\Big[ \chi_0 ,(\partial_\mu \,\Phi )\Big]_- \Big)\,B +\rm{h.c.}\right\}
\nonumber\\
&-&\frac{1}{2\,f}\,
\tr \left\{
\Big( \bar \Delta_\mu \cdot (\partial_\mu \,\Phi )  \Big)\,
\Big( C_2\,\Big[\chi_0,B\Big]_++C_3\,\Big[\chi_0,B\Big]_- +\rm{h.c.}\right\}
\nonumber\\
&-&\frac{C_4}{2\,f}\,
\tr \left\{ \Big( \bar \Delta_\mu \cdot \chi_0
\Big)\,\big[(\partial_\mu \,\Phi ),B \big]_- +\rm{h.c.}\right\}
\nonumber\\
&-&\frac{C_5}{2\,f}\,
\tr \left\{ \Big( \bar \Delta_\mu \cdot
(\partial_\mu \,\Phi )\Big)\,
B +\mathrm{h.c.}\right\} \tr \, \chi_0  \;.
\label{chi-sb-3}
\end{eqnarray}
Similarly in ${\mathcal L }_{\chi, -}^{(3)}$ we collect the remaining terms 
which result from chiral interaction terms linear in $\chi_-$. There are three 
pseudo-scalar interaction terms with $\bar F_{4,5,6}$ and four additional terms
parameterized by $\delta F_{4,5,6}$ and $\delta C_0$
\begin{eqnarray}
{\mathcal L }_{\chi, -}^{(3)}&=& \frac{1}{2\,f} \,\tr  \bar B
 \Big[
2\,i\,\gamma_5\,\m0_{[8]}\,
\bar F_4\,\Big[\chi_0,\Phi\Big]_+ 
+\gamma_5\,\gamma^\mu \, \delta F_4\,
\Big[\chi_0, \left(\partial_\mu\,\Phi\right)\Big]_+ 
, B \Big]_+
\nonumber\\
&+&\frac{1}{2\,f} \,\tr  \bar B
 \Big[
2\,i\,\gamma_5\,\m0_{[8]}\,
\bar F_5\,\Big[\chi_0,\Phi\Big]_+ 
+\gamma_5\,\gamma^\mu \, \delta F_5\,
\Big[\chi_0, \left(\partial_\mu\,\Phi\right)\Big]_+
, B \Big]_-
\nonumber\\
&+&\frac{1}{2\,f} \,\tr  \bar B\,
 \Big(
2\,i\,\gamma_5\,\m0_{[8]}\,
\bar F_6\,\tr \big(\chi_0\,\Phi\big) 
+\gamma_5\,\gamma^\mu \, \delta F_6\,\tr 
\big( \chi_0\, \left(\partial_\mu \,\Phi\right)\big) \Big)\, B  
\nonumber\\
&-& \frac{\delta C_0}{2\,f}\,
\tr  \,\Big( \bar \Delta_\mu \cdot 
\Big[ \chi_0 ,(\partial_\mu \,\Phi )\Big]_+ \,B +\rm{h.c.} \Big) \,.
\label{chi-sb-3:p}
\end{eqnarray}
We point out that, while the parameters $F_i$ and $C_i$ contribute to matrix elements of 
the $SU(3)$ axial-vector current $A^\mu_a$, none of the terms of (\ref{chi-sb-3:p}) 
contribute. This follows once the external axial-vector current is 
restored. In (\ref{chi-sb-3}) this is achieved with the replacement 
$\partial^\mu\,\Phi_a \to \partial^\mu\,\Phi_a +2\,f\, A^\mu_a$ (see (\ref{def-fields})). 
Though it is obvious that the pseudo-scalar terms in (\ref{chi-sb-3:p}) proportional to 
$\bar F_i$ do not contribute to the axial-vector current, it is less immediate that the terms 
proportional to $\delta F_i$ and $\delta C_i$ also do not contribute. Moreover, the latter 
terms appear redundant, because terms with identical structure at the 3-point level 
are already listed in (\ref{chi-sb-3:p}). Here one needs to realize that the terms proportional 
to $\delta F_i$ and $\delta C_i$ result from chiral interaction terms linear 
in $[{\mathcal D}_\mu ,\chi_-]_- $ while the terms proportional to $F_i, C_i$  
result from chiral interaction terms linear in $U_\mu $. 

The pseudo-scalar parameters $\bar F_i$ and also $\delta F_{i}$ lead to a 
tree-level Goldberger-Treiman discrepancy. For instance, we have
\begin{eqnarray}
f \,g_{\pi N N} -g_A \,m_N = 2\,m_N\,m_\pi^2 \,
(\bar F_4+\delta F_4+\bar F_5+\delta F_5 ) + {\mathcal O} \left( Q^3\right)  \,,
\label{GTD}
\end{eqnarray}
where we introduced the pion-nucleon coupling constant $g_{\pi N N}$ and the axial-vector 
coupling constant of the nucleon $g_A$. The corresponding generalized Goldberger-Treiman 
discrepancies for the remaining axial-vector coupling constants of the baryon octet states 
follow easily from the replacement rule $F_i \to F_i+\bar F_i+\delta F_i $ for $i=4,5,6$ 
(see also \cite{Goity:Lewis}). We 
emphasize that (\ref{GTD}) must not be confronted directly with the Goldberger-Treiman 
discrepancy as discussed in \cite{GTD,GTD:pi,Goity:Lewis}, because it necessarily involves 
the $SU(3)$ parameter $f $ rather than $f_\pi \simeq 92$ MeV or $f_K \simeq 113$ MeV.

The effect of the axial-vector interaction terms in (\ref{chi-sb-3}) is twofold. First, they 
lead to renormalized values of the $F_{[8]}, D_{[8]}$ and $C_{[10]}$ parameters in (\ref{lag-Q}). 
Secondly they induce interesting $SU(3)$ symmetry-breaking effects which are proportional 
to $(m_K^2-m_\pi^2)\,\lambda_8$. Note that the renormalization of the $F_{[8]},D_{[8]}$ and $C_{[10]}$ parameters 
requires care, because it is necessary to discriminate between  the 
renormalization of the axial-vector current and the one of the meson-baryon coupling 
constants. We introduce $F_R, D_R$ and $C_R$  as they enter 
matrix elements of the axial-vector current :
\begin{eqnarray}
&& F_R =F_{[8]}+ (m_\pi^2+2\,m_K^2)\left(F_9
+{\textstyle{2\over 3}}\,\big( F_2+F_5\big) \right) \;, \qquad 
\nonumber\\
&& D_R= D_{[8]}+ (m_\pi^2+2\,m_K^2)
\left(F_8 +{\textstyle{2\over 3}}\,\big( F_0+F_4\big) \right)\;,
\nonumber\\
&& C_R=C_{[10]} +(m_\pi^2+2\,m_K^2)\,\left(C_5 +{\textstyle{2\over 3}}\,\big( C_0+C_2\big)\right)\,,
\label{ren-FDC}
\end{eqnarray}
and  the renormalized parameters $F_{A,P}, D_{A,P}$ and 
$ C_A$ as they are relevant for the meson-baryon 3-point vertices:
\begin{eqnarray}
&& F_A= F_R + {\textstyle{2\over 3}}\,(m_\pi^2 +2\,m_K^2)\,\delta F_5\;, \quad 
D_A= D_R + {\textstyle{2\over 3}}\,(m_\pi^2 +2\,m_K^2)\,\delta F_4\;, \quad 
\nonumber\\
&& F_P= {\textstyle{2\over 3}}\,(m_\pi^2 +2\,m_K^2)\,\bar F_5\;, \quad 
D_P= {\textstyle{2\over 3}}\,(m_\pi^2 +2\,m_K^2)\,\bar F_4\;, \quad
\nonumber\\
&& C_A= C_R + {\textstyle{2\over 3}}\,(m_\pi^2 +2\,m_K^2)\,\delta C_0\;.
\label{ren-FDC-had}
\end{eqnarray}
The index $A$ or $P$ indicates whether the meson couples to the baryon via an axial-vector 
vertex (A) or a pseudo-scalar vertex (P). 
It is clear that the effects of (\ref{ren-FDC}) and (\ref{ren-FDC-had}) break the chiral 
symmetry but do not break the $SU(3)$ symmetry. In this work we will use the renormalized
parameters $F_R, D_R$ and $C_R$. One can always choose the parameters $F_8,F_9$ and $C_5$ as to obtain
$F_R=F_{[8]}$, $D_R=D_{[8]}$ and $C_R=C_{[10]}$. In order to distinguish the renormalized values 
from their  bare values one needs to determine the parameters $F_8, F_9$ and $C_5$  by 
investigating higher-point Green functions. This is beyond the scope of this work. 

The number of parameters inducing $SU(3)$ symmetry-breaking effects can be reduced 
significantly by the large $N_c$ analysis. We recall the five leading operators presented 
in \cite{Dashen}
\begin{eqnarray}
&&\langle {\mathcal B}' | \,{\mathcal L}_\chi^{(3)}\,| {\mathcal B} \rangle 
=\frac{1}{f}\,\langle {\mathcal B}'|  \,O_i^{(a)}(\tilde c) |{\mathcal B}\rangle \,
\tr\,\lambda_a\,\nabla^{(i)}\, \Phi  
+ 4\,\langle {\mathcal B}'|  \,O_i^{(a)}(c) |{\mathcal B}\rangle \,
A^{(i)}_a \;,
\nonumber\\ \nonumber\\
&& O_i^{(a)}(c) = c_1\,\big( d^{8a}_{\;\;\;b}\,G^{(b)}_{i}  
+{\textstyle {1\over 3}}\,\delta^{a8} \,J_{i}\big)
+c_2\,\big( d^{8a}_{\;\;\;b}\,J_{i}\,T^{(b)}+\delta^{a8} \,J_{i}\big)
\nonumber\\
&& \qquad \qquad \quad +\;c_3\,[G^{(a)}_i, T^{(8)}]_+ +c_4 \, [T^{(a)},G^{(8)}_{i}]_+
 +c_5 \, [J^2,[T^{(8)},G^{(a)}_i]_-]_- \;.
\label{ansatz-3}
\end{eqnarray}
It is important to observe that the parameters $c_i$ and $\tilde c_i$ are a priori independent. 
They are  correlated by the chiral $SU(3)$ symmetry only. With (\ref{matrix-el}) it is 
straightforward to map the 
interaction terms (\ref{ansatz-3}), which all break the $SU(3)$ symmetry linearly, 
onto the chiral vertices of (\ref{chi-sb-3}) and (\ref{chi-sb-3:p}). This procedure 
relates the parameters $c_i $ and $\tilde c_i$. One finds that the matching is possible 
for all operators leaving only ten independent parameters $c_{i}$, $\tilde c_{1,2}$,  
$\bar c_{1,2}$ and $a$, rather than the twenty-three $F_i$, $C_i$ and $\delta F_i, \delta C_i$ and 
$\bar F_i$ parameters in (\ref{chi-sb-3},\ref{chi-sb-3:p}). We derive 
$c_i=\tilde c_i$ for $i=3,4,5$ and
\begin{eqnarray}
&& \!\!\!\!F_1 = -\frac{\sqrt{3}}{2}\,\frac{c_3}{m_K^2-m_\pi^2}\;,\quad 
F_2 = -\frac{\sqrt{3}}{2}\,\frac{c_4}{m_K^2-m_\pi^2} \,,\quad 
F_3 =-\frac{1}{\sqrt{3}}\,\frac{c_3+c_4}{m_K^2-m_\pi^2} \;,
\nonumber\\
&& \!\!\!\!F_4 = -\frac{\sqrt{3}}{4}\,\frac{c_1}{m_K^2-m_\pi^2} \;,\quad  
F_5 = -\frac{\sqrt{3}}{4}\,\frac{{\textstyle{2\over 3}}\,c_1+c_2}{m_K^2-m_\pi^2} \;,\quad  
F_6=-\frac{\sqrt{3}}{2}\,\frac{{\textstyle{2\over 3}}\,c_1+c_2}{m_K^2-m_\pi^2} \;,
\nonumber\\
&& \!\!\!\! C_0 = -\frac{\sqrt{3}}{2}\,\frac{c_1}{m_K^2-m_\pi^2} \,, \quad 
C_1 = -\frac{\sqrt{3}}{2}\,\frac{c_3-c_4+3\,c_5}{m_K^2-m_\pi^2} \;, 
\nonumber\\
&& \!\!\!\! C_3 =-\sqrt{3}\,\frac{c_3}{m_K^2-m_\pi^2}\,, \quad 
C_4 =-\sqrt{3}\,\frac{c_4}{m_K^2-m_\pi^2}\,, \quad 
 F_{0,7}=C_{2,5}=0 \,,
\label{large-Nc-FDC}
\end{eqnarray}
and 
\begin{eqnarray}
&& \bar F_4= -\frac{\sqrt{3}}{4}\,\frac{\bar c_1}{m_K^2-m_\pi^2} \;,\quad  
\delta F_4 = -\frac{\sqrt{3}}{4}\,\frac{\delta c_1}{m_K^2-m_\pi^2} \;, 
\nonumber\\
&& \bar F_5 = -\frac{\sqrt{3}}{4}\,\frac{{\textstyle{2\over 3}}\,\bar c_1+\bar c_2+a}{m_K^2-m_\pi^2} 
\,, \quad \delta F_5 = -\frac{\sqrt{3}}{4}\,\frac{{\textstyle{2\over 3}}\,\delta c_1+\delta c_2-a}{m_K^2-m_\pi^2} 
\,, 
\nonumber\\
&& \bar F_6 =-\frac{\sqrt{3}}{2}\,\frac{{\textstyle{2\over 3}}\,\bar c_1+\bar c_2+a}{m_K^2-m_\pi^2}  \,,
\quad \delta F_6 =-\frac{\sqrt{3}}{2}\,\frac{{\textstyle{2\over 3}}\,\delta c_1+\delta c_2-a}{m_K^2-m_\pi^2}   \,,
\nonumber\\
&& \delta C_0=-\frac{\sqrt{3}}{2}\,\frac{ \tilde c_1- c_1}{m_K^2-m_\pi^2} \;, \qquad 
\delta c_i = \tilde c_i -c_i -\bar c_i \;.
\label{large-Nc-FDC:delta}
\end{eqnarray}
In (\ref{large-Nc-FDC:delta})  the pseudo-scalar parameters 
$\bar F_{4,5,6}$ are expressed in terms of the more convenient dimension less 
parameters $\bar c_{1,2}$ and $a$. Here we insist that an expansion analogous to 
(\ref{ansatz-3}) holds also for the pseudo-scalar vertices in (\ref{chi-sb-3:p}).
\tabcolsep=1.4mm
\renewcommand{\arraystretch}{1.5}
\begin{table}[t]\begin{center}
\begin{tabular}{|c|c||c|c|c|c|c|c|c|}
\hline 
& $g_A $ (Exp.) & $ F_R$ & $ D_R$ & $c_1$ & $c_2 $ & $c_3$ & $c_4 $  & $c_5 $\\
\hline \hline
$ n \to p\,e^-\,\bar \nu_e $ & $1.267 \pm 0.004$& 1 & 1 & $\frac{5}{3\,\sqrt{3}}$ &  $\frac{1}{\sqrt{3}}$& $\frac{5}{\sqrt{3}}$&$ \frac{1}{\sqrt{3}}$ & $\;\;0\;\;$\\  \hline

$ \Sigma^- \to \Lambda \,e^-\,\bar \nu_e  $ & $0.601 \pm 0.015$ & 0 & $\sqrt{\frac{2}{3}}$ & $\frac{\sqrt{2}}{3}$ &  0& 0& 0& 0\\  \hline

$ \Lambda \to p \,e^-\,\bar \nu_e $ & $-0.889 \pm 0.015$ & $-\sqrt{\frac{3}{2}}$ & $-\frac{1}{\sqrt{6}}$ & $\frac{1}{2\,\sqrt{2}}$ & $\frac{1}{2\,\sqrt{2}}$ & $-\frac{3}{2\,\sqrt{2}}$& $\frac{1}{2\,\sqrt{2}}$& 0\\  \hline

$ \Sigma^- \to n \,e^-\,\bar \nu_e $ & $0.342 \pm 0.015$ & $-1$ & 1 & $-\frac{1}{6\,\sqrt{3}}$ & $\frac{1}{2\,\sqrt{3}}$  & $\frac{1}{2\,\sqrt{3}}$& $-\frac{\sqrt{3}}{2}$& 0\\  \hline

$ \Xi^- \to \Lambda \,e^-\,\bar \nu_e $ & $0.306 \pm 0.061 $ & $\sqrt{\frac{3}{2}}$ & $-\frac{1}{\sqrt{6}}$  & $-\frac{1}{6\,\sqrt{2}}$ & $-\frac{1}{2\,\sqrt{2}}$ & $-\frac{1}{2\,\sqrt{2}}$& $-\frac{5}{2\,\sqrt{2}}$& 0\\  \hline

$ \Xi^- \to \Sigma^0 \,e^-\,\bar \nu_e $ & $0.929 \pm 0.112 $ & $\frac{1}{\sqrt{2}}$ & $\frac{1}{\sqrt{2}}$ & $-\frac{5}{6\,\sqrt{6}}$ & $-\frac{1}{2\,\sqrt{6}}$ & $-\frac{5}{2\,\sqrt{6}}$& $-\frac{1}{2\,\sqrt{6}}$& 0\\  \hline


\end{tabular}
\end{center}
\caption{Axial-vector coupling constants for the weak decay processes of the baryon octet states. The empirical values 
for $g_A$ are taken from \cite{Dai}. Here we do not consider  $SU(3)$ symmetry-breaking effects of the  
vector current. The last seven columns give the coefficients of the axial-vector 
coupling constants $g_A$ decomposed into the $F_R$, $D_R$ and $c_i$ parameters.}
\label{weak-decay:tab}
\end{table}

The analysis of \cite{Dai}, which considers constraints from the weak decay 
processes of the baryon octet states and the strong decay widths of the decuplet states, obtains 
$c_2 \simeq -0.15$ and $c_3\simeq 0.09$ but finds  values of $c_1$ and $c_4$ which are compatible 
with zero\footnote{The parameters of \cite{Dai} are obtained with $c_i \to -\sqrt{3}\,c_i/2$ and 
$a= D_R+(c_1+3\,c_3)/\sqrt{3}$ and $b=F_R-2\,a/3+(2\,c_1/3+c_2+2\,c_3+c_4)/\sqrt{3}$.
Note that the analysis of \cite{Dai} does not determine the parameter $c_5$.}. In Tab. (\ref{weak-decay:tab}) we 
reproduce the axial-vector coupling constants as given in \cite{Dai} relevant for the various 
baryon octet weak-decay processes. Besides $C_A+2\,\tilde c_1/\sqrt{3}$, the empirical strong-decay widths 
of the decuplet states constrain the parameters $c_{3}$ and $c_4$ only, as is 
evident from the expressions for the decuplet widths
\begin{eqnarray}
&& \Gamma_{[10]}^{(\Delta )} = \frac{m_N+E_N}{2\,\pi \,f^2}\,\frac{p_{\pi N}^3}{12\,m_{[10]}^{(\Delta )}}
\,\Big( C_A +{\textstyle{2\over \sqrt{3}}}\,(\tilde c_1+3\,c_3)\Big)^2 \,,
\nonumber\\
&& \Gamma_{[10]}^{(\Sigma )} = \frac{m_\Lambda+E_\Lambda}{2\,\pi \,f^2}\,
\frac{p_{\pi \Lambda}^3}{24\,m_{[10]}^{(\Sigma )}}
\,\Big( C_A +{\textstyle{2\over \sqrt{3}}}\,\tilde c_1 \Big)^2 
\nonumber\\
&& \qquad +\frac{m_\Sigma+E_\Sigma}{3\,\pi \,f^2}\,
\frac{p_{\pi \Sigma}^3}{24\,m_{[10]}^{(\Sigma )}}
\,\Big( C_A +{\textstyle{2\over \sqrt{3}}}\,(\tilde c_1+6\,c_4)\Big)^2 \,,
\nonumber\\
&& \Gamma_{[10]}^{(\Xi )} = \frac{m_\Xi+E_\Xi}{2\,\pi \,f^2}\,\frac{p_{\pi \Xi}^3}{24\,m_{[10]}^{(\Xi )}}
\,\Big( C_A +{\textstyle{2\over \sqrt{3}}}\,(\tilde c_1-3\,c_3+3\,c_4)\Big)^2 \,,
\label{decuplet-decay}
\end{eqnarray}
where for example $m_\Delta =\sqrt{m_N^2+p_{\pi N}^2}+\sqrt{m_\pi^2+p_{\pi N}^2}$ and 
$E_N= \sqrt{m_N^2+p_{\pi N}^2}$. For instance, the values 
$ C_A+2\,\tilde c_1/\sqrt{3} \simeq 1.7$, 
$c_3 \simeq 0.09$ and $c_4 \simeq 0.0$ together with 
$f\simeq f_\pi \simeq 93$ MeV lead to isospin averaged partial 
decay widths of the decuplet states which are compatible with the present day empirical estimates. It is 
clear that the six data points for the baryon octet decays can be reproduced by a suitable adjustment of the 
six parameters $F_R$, $D_R$ and $c_{1,2,3,4}$. The non-trivial issue is to what extent the 
explicit $SU(3)$ symmetry-breaking pattern in the axial-vector coupling constants is consistent 
with the symmetry-breaking pattern in the meson-baryon coupling constants. Here a possible 
strong energy dependence of the decuplet self-energies 
may invalidate the use of the simple expressions (\ref{decuplet-decay}). A more 
direct comparison with the meson-baryon scattering data may be required. 

Finally we wish to mention an implicit assumption one relies on if Tab. \ref{weak-decay:tab}
is applied. In a strict chiral expansion the $Q^2$ effects included in that table   
are incomplete, because there are various one-loop diagrams which are not considered but 
carry chiral order $Q^2$ also. However, in a combined chiral and $1/N_c$ expansion it is 
natural to neglect such loop effects, because they are suppressed by $1/N_c$. 
This is immediate with the large $N_c$ scaling rules $m_\pi \sim N_c^0$ and 
$f \sim \sqrt{N_c}$ \cite{Hooft,Witten} together with the fact that the one-loop 
effects are proportional to $m_{K,\pi}^2/(4\pi\,f^2) $ \cite{Bijnens:ga}. On the other hand, 
it is evident from (\ref{3-point-vertex}) and (\ref{ansatz-3}) that the 
$SU(3)$ symmetry-breaking contributions are not necessarily suppressed by $1/N_c$. 
Our approach differs from previous calculations 
\cite{Bijnens:ga,Jenkins:ga1,Jenkins:ga2,Luty-2} where emphasis was put on the one-loop 
corrections of the axial-vector current rather than the quasi-local counter terms which were 
not considered. It is clear that part of the one-loop effects, in particular their  
renormalization scale dependence, can be absorbed into the counter terms 
considered in this work.

We will return to the large $N_c$ counting issue when 
discussing the approximate scattering kernel and also when presenting 
our final set of parameters, obtained from a fit to the data set.

\section{Relativistic meson--baryon scattering}

In this section we prepare the ground for our relativistic
coupled-channel effective field theory of meson--baryon scattering. We
first develop the formalism for the case of elastic $\pi N$ scattering for
simplicity. The next section is devoted to the inclusion of inelastic
channels which leads to the coupled-channel approach. Our approach is based on an 
'old' idea present in the literature for many decades. One aims at reducing the
complexity of the relativistic Bethe-Salpeter equation by a suitable reduction scheme 
constrained to preserve the relativistic unitarity cuts \cite{Sugar,Gross,Tjon}. A famous 
example is the Blankenbecler-Sugar scheme \cite{Sugar} which reduces the Bethe-Salpeter 
equation to a 3-dimensional integral equation. For a beautiful variant developed for 
pion-nucleon scattering see the work by Gross and Surya \cite{Gross}. In our work we 
derive a scheme which is more suitable for the relativistic chiral Lagrangian. We do not 
attempt to establish a numerical solution of the four dimensional Bethe-Salpeter equation based 
on phenomenological form factors and interaction kernels \cite{Afnan}. 
The merit of the chiral Lagrangian is that a major part of the complexity is already eliminated by 
having reduced non-local interactions to 'quasi' local interactions involving 
a finite number of derivatives only. Our scheme is therefore constructed to be 
particularly transparent and efficient for the typical case of 'quasi' local 
interaction terms. In the course of developing our approach we suggest a modified subtraction 
scheme within dimensional regularization, which complies manifestly with the chiral counting rule (\ref{q-rule}). 

Consider the on-shell pion-nucleon scattering amplitude
\begin{eqnarray}
\langle \pi^{i}(\bar q)\,N(\bar p)|\,T\,| \pi^{j}(q)\,N(p) \rangle
=(2\pi)^4\,\delta^4(0)\,
\bar u(\bar p)\,
T^{ij}_{\pi N \rightarrow \pi N}(\bar q,\bar p ; q,p)\,u(p) \,,
\label{on-shell-scattering}
\end{eqnarray}
where  $\delta^4(0)$ guarantees energy-momentum conservation and $u(p)$ is the
nucleon isospin-doublet spinor. In quantum field theory the scattering amplitude
$T_{\pi N \rightarrow \pi N}$ follows as the solution of the Bethe-Salpeter matrix
equation
\begin{eqnarray}
T(\bar k ,k ;w ) &=& K(\bar k ,k ;w )
+\int \frac{d^4l}{(2\pi)^4}\,K(\bar k , l;w )\, G(l;w)\,T(l,k;w )\;,
\nonumber\\
G(l;w)&=&-i\,S_N({\textstyle
{1\over 2}}\,w+l)\,D_\pi({\textstyle {1\over 2}}\,w-l)\,,
\label{BS-eq}
\end{eqnarray}
in terms of the Bethe-Salpeter kernel $K(\bar k,k;w )$ to be specified later, 
the nucleon propagator $S_N(p)=1/(\pslash-m_N+i\,\epsilon)$ and the
pion propagator $D_\pi(q)=1/(q^2-m_\pi^2+i\,\epsilon)$. Self energy corrections in the 
propagators are suppressed and therefore not considered in this work. 
We introduced convenient kinematics:
\begin{eqnarray}
w = p+q = \bar p+\bar q\,,
\quad k= \half\,(p-q)\,,\quad
\bar k =\half\,(\bar p-\bar q)\,,
\label{def-moment}
\end{eqnarray}
where $q,\,p,\, \bar q, \,\bar p$ are the initial and final pion and nucleon 4-momenta.
The Bethe-Salpeter equation (\ref{BS-eq}) implements Lorentz invariance 
and unitarity for the two-body scattering process. It involves the off-shell
continuation of the  on-shell scattering amplitude introduced in
(\ref{on-shell-scattering}). We recall that only the on-shell limit
with $ \bar p^2, p^2\rightarrow m_N^2 $ and $\bar
q^2,q^2\rightarrow m_\pi^2 $ carries direct physical information.
In quantum field theory the off-shell form of the
scattering amplitude reflects the particular choice of the pion and
nucleon interpolating fields chosen in the Lagrangian density and
therefore can be altered basically at will by a redefinition
of the fields~\cite{off-shell,Fearing}.

It is convenient to decompose the interaction kernel and the resulting
scattering amplitude in isospin invariant components
\begin{eqnarray}
K_{ij}(\bar k ,k ;w ) =\sum_I\,K_I(\bar k ,k ;w )\,P^{(I)}_{ij}
\,,\,\,
T_{ij}(\bar k ,k ;w ) &=& \sum_I\,T_I(\bar k ,k ;w )\,P^{(I)}_{ij} \,,
\label{isospin-decom}
\end{eqnarray}
with the isospin projection matrices $P^{(I)}_{ij}$.
For pion-nucleon scattering one has:
\begin{eqnarray}
P^{\left(\frac{1}{2}\right)}_{ij} &=& \frac{1}{3}\,\sigma_i\,\sigma_j \;,\;\;\;
P^{\left(\frac{3}{2}\right)}_{ij} = \delta_{ij}-\frac{1}{3}\,\sigma_i\,\sigma_j
\;,\;\;\;
\sum_k\,P^{\left(I\right)}_{ik}\,P^{\left(I'\right)}_{kj}=
\delta_{I\,I'}\,P^{\left(I\right)}_{ij}\;.
\label{iso:proj}
\end{eqnarray}
The Ansatz (\ref{isospin-decom}) decouples the Bethe-Salpeter equation 
into the two isospin channels $I=1/2$ and $I=3/2$.

\subsection{On-shell reduction of the Bethe-Salpeter equation}

For our application it is useful to exploit the ambiguity in the
off-shell structure
and choose a particularly convenient representation. We decompose
the interaction kernel into an 'on-shell irreducible' part $\bar K $
and 'on-shell reducible' terms $K_L$ and $K_R$ which vanish
if evaluated with on-shell kinematics either in the incoming or outgoing channel respectively 
\begin{eqnarray}
&&K=\bar K+K_L+K_R+K_{LR} \;,
\nonumber\\
&&\bar u_N(\bar p)\,K_L \Big|_{\mathrm{on-shell}} = 0 = K_R
\,u_N(p)\Big|_{\mathrm{on-shell}} \;.
\label{k-decomp}
\end{eqnarray}
The term $K_{LR}$ disappears if evaluated with either incoming or outgoing
on-shell kinematics. Note that the notion of an on-shell irreducible kernel $\bar K$ 
is not unique per se and needs further specifications. The precise definition  of our 
particular choice of on-shell irreducibility will be provided when constructing our 
relativistic partial-wave projectors.
In this subsection we study the generic consequences of decomposing the interaction 
kernel according to (\ref{k-decomp}). With this decomposition of the interaction kernel the
scattering amplitude can be written as follows
\begin{eqnarray}
T &=& \bar T 
-\Big(K_L+K_{LR}\Big)\!\cdot\! \Big( 1-G\!\cdot\! K\Big)^{-1}\!\cdot
\!G\! \cdot\!
\Big(K_R+K_{LR}\Big)-K_{LR}
\nonumber\\
&+&\Big(K_L+K_{LR}\Big)\! \cdot\! \Big( 1-G\!\cdot\! K\Big)^{-1}
+\Big( 1-K\!\cdot\! G\Big)^{-1}\!\cdot\!\Big(K_R+K_{LR} \Big)\,,
\nonumber\\
\bar T &=& \Big(1-V\!\cdot\! G \Big)^{-1} \!\cdot\! V \;,
\label{t-eff}
\end{eqnarray}
where we use operator notation with, e.g., $T=K+ K\!\cdot \!G\!\cdot \!T $ representing
the Bethe-Salpeter equation (\ref{BS-eq}). The effective interaction
$V$ in (\ref{t-eff}) is given by
\begin{eqnarray}
V&=&\Big( \bar K+K_R\!\cdot\! G\!\cdot\! X \Big)\!\cdot\!
\Big(1-G\!\cdot\! K_L-G\!\cdot\! K_{LR}\!\cdot\! G\!\cdot\! X  \Big)^{-1} \;,
\nonumber\\
X &=& \Big( 1-(K_R+K_{LR})\!\cdot\! G\Big)^{-1}\!
\cdot\! \Big( \bar K+K_L\Big) 
\label{v-eff}
\end{eqnarray}
without any approximations. We point out that the interaction kernels $V$ and $K$ are 
equivalent on-shell by construction. This follows from (\ref{t-eff}) and (\ref{k-decomp}), which
predict the equivalence of $T$ and $\bar T$ for on-shell kinematics 
\begin{eqnarray}
\bar{u}_N(\bar p)\, T \,u_N(p)\Big|_{\mathrm{on-shell}}
 \equiv \bar{u}_N(\bar
 p)\,\bar T \,u_N(p)\Big|_{\mathrm{on-shell}} .
\nonumber
\end{eqnarray}
As an explicit simple example for the application of the formalism
(\ref{t-eff},\ref{v-eff}) we  consider the s-channel nucleon pole
diagram as a particular contribution to the interaction
kernel $K(\bar k,k;w)$ in (\ref{BS-eq}). In the isospin $1/2$ channel
its contribution  evaluated with the pseudo-vector pion-nucleon vertex reads
\begin{eqnarray}
K(\bar k, k;w ) &=& -\frac{3\,g_A^2}{4\,f^2}\,
\gamma_5\,\left( {\textstyle{1\over 2}}\,\wslash-\barkslash  \right)\,
\frac{1}{\wslash-m_N-\Delta m_N(w)}\,\gamma_5\,
\left( {\textstyle{1\over 2}}\,\wslash- \kslash \right)\;,
\label{nucleon-pole}
\end{eqnarray}
where we included a wave-function and mass renormalization  
$\Delta m_N(w)$ for later convenience\footnote{The corresponding counter terms 
in the chiral Lagrangian (\ref{chi-sb}) are linear combinations of $b_0,b_D$ and $b_F$
and $\zeta_0$, $\zeta_D$ and $\zeta_F$.}. We construct the various components of the kernel 
according to (\ref{k-decomp})
\begin{eqnarray}
\bar K&=&-\frac{3\,g_A^2}{4\,f^2}\,
\frac{\big( \wslash-m_N \big)^2}{\wslash+\bar m_N}\,,\quad
K_{LR}=\frac{3\,g_A^2}{4\,f^2}\,
\Big( {\barpslash}-m_N \Big)\,
\frac{1}{\wslash+\bar m_N}\,\Big( \pslash- m_N\Big)\;,
\nonumber\\
K_L&=&\frac{3\,g_A^2}{4\,f^2}\,
\Big(\barpslash-m_N \Big)\,
\frac{\wslash- m_N}{\wslash+\bar m_N}\,,\quad
K_R=\frac{3\,g_A^2}{4\,f^2}\,
\frac{\wslash-m_N}{\wslash+\bar m_N}\,\Big( \pslash- m_N\Big)
\,,
\label{nucleon-pole-off}
\end{eqnarray}
where $\bar m_N = m_N+\gamma_5\,\Delta m_N(w)\,\gamma_5$. 
The solution of the Bethe-Salpeter equation is derived in two steps. First
solve for the auxiliary object $X$ in (\ref{v-eff})
\begin{eqnarray}
X &=& \frac{3\,g_A^2}{4\,f^2}\,
\Big( \barpslash-\wslash\Big)\,
\frac{1}{\wslash+\bar m_N}\,\Big( \wslash- m_N\Big)
\nonumber\\
&+&\frac{3\,g_A^2}{4\,f^2}\,
\Big( \barpslash+\wslash-2\,m_N\Big)\,
\frac{3\,g_A^2\,I_\pi^{(l)}}
{4\,f^2\,(\wslash+\bar m_N)+3\,g_A^2\,I_\pi}\,
\frac{ \wslash- m_N}{\wslash+\bar m_N} \;,
\label{x-example}
\end{eqnarray}
where one encounters the pionic tadpole integrals:
\begin{eqnarray}
I_\pi &=&i\,
\int  \frac{d^d\,l}{(2\,\pi)^d }
\frac{\mu^{4-d}}{l^2-m_\pi^2+i\,\epsilon }
\,,\quad
I^{(l)}_\pi =i\,
\int  \frac{d^d\,l}{(2\,\pi)^d }
\frac{\mu^{4-d}\,\lslash }{l^2-m_\pi^2+i\,\epsilon }\,,
\label{def-tadpole}
\end{eqnarray}
properly regularized for space-time dimension $d$ in terms of the renormalization scale $\mu $.
Since $I_\pi^{(l)}=0$ and $K_R\cdot  G\cdot
\!(\bar K+K_L)\equiv 0$ for our example the expression (\ref{x-example}) reduces to 
$X=\bar K+K_L $. The effective potential $V(w)$
and the on-shell equivalent scattering amplitude $\bar T $ follow
\begin{eqnarray}
V(w)&=& -\frac{3\,g_A^2}{4\,f^2}\,\frac{ (\wslash- m_N)^2}{\wslash+\bar m_N}
\left(1+
\frac{3\,g_A^2\,I_\pi}{4\,f^2}\,\frac{\wslash- m_N}{\wslash+\bar m_N}
\right)^{-1} \;,
\nonumber\\
\bar T(w) &=& \frac{1}{1-V(w)\,J_{\pi N}(w)}\,V(w) \;.
\label{pin-example}
\end{eqnarray}
The divergent loop function $J_{\pi N}$ in (\ref{pin-example}) defined via 
$\bar{K}\!\cdot\!G\!\cdot\!\bar{K}= \bar K\, J_{\pi N}\,\bar K$ may be decomposed
into scalar master-loop functions $I_{\pi N}(\sqrt{s}\,)$ and $I_{N},I_\pi$ with
\begin{eqnarray}
J_{\pi N}(w)&=& \left(m_N +
\frac{w^2+m_N^2-m_\pi^2}{2\,w^2}\,
\wslash\right)\,I_{\pi N}(\sqrt{s}\,)
+\frac{I_N-I_{\pi}}{2\,w^2}\,\wslash \;,
\nonumber\\
I_{\pi N}(\sqrt{s}\,)&=&-i\,\int
\frac{d^dl}{(2\pi)^d}\,\frac{\mu^{4-d}}{l^2-m_\pi^2}\,\frac{1}{(w-l)^2-m_N^2}
\;,
\nonumber\\
I_N &=&i\,
\int  \frac{d^d\,l}{(2\,\pi)^d }
\frac{\mu^{4-d}}{l^2-m_N^2+i\,\epsilon }
\label{jpin-def}
\end{eqnarray}
where $s=w^2$. The result (\ref{pin-example}) gives an explicit example of the powerful
formula (\ref{t-eff}). The Bethe-Salpeter equation may be solved in two steps. Once the 
effective potential $V$ is evaluated the scattering amplitude $\bar T$ is given in 
terms of the loop function $J_{\pi N}$ which is independent on the form of the interaction.  
In section 3.3 we will generalize the result (\ref{pin-example}) by constructing a complete  
set of covariant projectors which will define our notion of on-shell irreducibility 
explicitly. Before discussing the result (\ref{pin-example}) in more detail we wish to 
consider the regularization and renormalization scheme required for the relativistic loop 
functions in (\ref{jpin-def}). 

\subsection{Renormalization program}

An important requisite of the chiral Lagrangian is a consistent regularization and renormalization 
scheme for its loop diagrams. The regularization scheme should respect all symmetries built 
into the theory but should also comply with the power counting rule (\ref{q-rule}). 
The standard $MS$ or $\overline{MS}$ subtraction scheme of dimensional regularization appears inconvenient, because  
it contradicts standard chiral power counting rules if applied to relativistic Feynman
diagrams \cite{gss,kambor}. We will suggest a modified subtraction scheme for relativistic diagrams, 
properly regularized in space-time dimensions $d$, which complies with the chiral counting rule 
(\ref{q-rule}) manifestly.

We begin with a discussion of the regularization scheme for the one-loop functions $I_\pi$, $I_N$ and 
$I_{\pi N}(\sqrt{s}\,)$ introduced in (\ref{jpin-def}). One encounters some freedom in  
regularizing and renormalizing those master-loop functions, which are typical 
representatives for all one-loop diagrams. We first recall their well-known properties
at $d=4$. The loop function $I_{\pi N}(\sqrt{s}\,)$ is made finite by one 
subtraction, for example at $\sqrt{s}=0$,
\begin{eqnarray}
\!\!\!\!I_{\pi N}(\sqrt{s}\,)&=&\frac{1}{16\,\pi^2}
\left( \frac{p_{\pi N}}{\sqrt{s}}\,
\left( \ln \left(1-\frac{s-2\,p_{\pi N}\,\sqrt{s}}{m_\pi^2+m_N^2} \right)
-\ln \left(1-\frac{s+2\,p_{\pi N}\sqrt{s}}{m_\pi^2+m_N^2} \right)\right)
\right.
\nonumber\\
&+&\left.
\left(\frac{1}{2}\,\frac{m_\pi^2+m_N^2}{m_\pi^2-m_N^2}
-\frac{m_\pi^2-m_N^2}{2\,s}
\right)
\,\ln \left( \frac{m_\pi^2}{m_N^2}\right) +1 \right)+I_{\pi N}(0)\;,
\nonumber\\
p_{\pi N}^2 &=&
\frac{s}{4}-\frac{m_N^2+m_\pi^2}{2}+\frac{(m_N^2-m_\pi^2)^2}{4\,s}  \;.
\label{ipin-analytic}
\end{eqnarray}
One finds $I_{\pi N}(m_N)-I_{\pi N}(0)= (4\pi)^{-2}+{\mathcal O}(m_\pi/m_N)\sim Q^0$  
in conflict with the expected minimal chiral power $Q$. 
On the other hand the leading chiral power of the subtracted loop function  
complies with the prediction of the standard chiral power counting rule (\ref{q-rule}) with 
$I_{\pi N}(\sqrt{s}\,)-I_{\pi N}(\mu_S ) \sim Q$ rather than the anomalous 
power $Q^0$ provided that $\mu_S \sim m_N$ holds. The anomalous contribution 
is eaten up by the subtraction constant $I_{\pi N}(\mu_S)$. This can be seen by expanding 
the loop function
\begin{eqnarray}
I_{\pi N}(\sqrt{s}\,)&=&i\,\frac{\sqrt{\phi_{\pi N}}}{8\,\pi\,m_N}
+\frac{\sqrt{\phi_{\pi N} }}{16\,\pi^2\,m_N}\,
\ln \left( \frac{\sqrt{s}-m_N+\sqrt{\phi_{\pi N} }}
{\sqrt{s}-m_N-\sqrt{\phi_{\pi N} }} \right)
\nonumber\\
&+&\frac{\sqrt{m^2_\pi-\mu_N^2}}{8\,\pi\,m_N}
-\frac{\sqrt{\mu_N^2-m_\pi^2}}{16\,\pi^2\,m_N}\,
\ln \left( \frac{\mu_N+\sqrt{\mu_N^2-m_\pi^2}}
{\mu_N-\sqrt{\mu_N^2-m_\pi^2}} \right)
\nonumber\\
&-&\frac{\sqrt{s}-\mu_S}{16\,\pi^2\,m_N}\,\ln \left( \frac{m_\pi^2}{m_N^2}\right)
+{\mathcal O}\left( \frac{(\mu_S-m_N)^2}{m_N^2},Q^2 \right)+I_{\pi N}(\mu_S)\;,
\label{ipin-expand}
\end{eqnarray}
in powers of $\sqrt{s}-m_N\sim Q$ and $\mu_N=\mu_S -m_N $. Here we
introduced the approximate phase-space  factor $\phi_{\pi N}= (\sqrt{s}-m_N)^2-m_\pi^2$.

It should be clear from the simple example of $I_{\pi N}(\sqrt{s}\,)$ in (\ref{ipin-analytic}) 
that a manifest realization of the chiral counting rule (\ref{q-rule}) is closely linked 
to the subtraction scheme implicit in any regularization scheme. A priori it is unclear 
in which way the subtraction constants of various loop functions are related by the 
pertinent symmetries\footnote{This aspect was not addressed satisfactorily 
in \cite{nn-lutz,Gegelia}.}. Dimensional regularization has proven to be an extremely 
powerful tool how to regularize and how to subtract loop functions in accordance with all 
symmetries. 
Therefore we recall the expressions for 
the master-loop function $I_N$, and $I_\pi$ and $I_{\pi N}(m_N)$ as they follow in 
dimensional regularization:
\begin{eqnarray}
&& I_N = m_N^2\,\frac{\Gamma (1-d/2)}{(4\pi)^2} 
\left(\frac{m_N^2}{4\,\pi \,\mu^2} \right)^{(d-4)/2}
\nonumber\\
&& \quad \;\,= \frac{m_N^2}{(4\,\pi)^2}
\left( -\frac{2}{4-d}+\gamma-1 -\ln (4 \pi)+\ln \left( \frac{m_N^2}{\mu^2}\right)
+ {\mathcal O}\left(4-d \right)\right) \;,
\label{n-tadpole}
\end{eqnarray}
where $d$ is the dimension of space-time and $\gamma $ the Euler constant.  The expression
for the pionic tadpole follows by replacing the nucleon mass $m_N$ in (\ref{n-tadpole})
by the pion mass $m_\pi$. The merit of dimensional regularization is that one is free 
to subtract all poles at $d=4$ including any specified finite term without violating any of the 
pertinent symmetries. In the $\overline{MS }$-scheme the pole $1/(4-d)$ is subtracted 
including the finite part $\gamma- \ln (4 \pi)$. That leads to 
\begin{eqnarray}
&&I_{N,\overline{MS}}= \frac{m_N^2}{(4 \pi)^2}
\left( -1+\ln \left(\frac{m_N^2}{\mu^2}\right) \right)\,,\quad  \;
I_{\pi,\overline{MS}} =\frac{m_\pi^2}{(4\pi)^2}
\left(-1+\ln \left(\frac{m_\pi^2}{\mu^2}\right) \right)\,,
\nonumber\\
&& I_{\pi N,\overline{MS}}\,(m_N)= -\frac{1}{(4\pi)^2} \,
\ln \left(\frac{m_N^2}{\mu^2} \right) 
-\frac{m_\pi}{16\,\pi \,m_N}+
{\mathcal O}\left( \frac{m_\pi^2}{m_N^2}\right) \,.
\label{master-dim}
\end{eqnarray}
The result (\ref{master-dim}) confirms the expected chiral power for the pionic 
tadpole $I_\pi \sim Q^2$. However, a striking disagreement with the chiral counting rule
(\ref{q-rule}) is found for the $\overline {MS}$-subtracted loop functions  
$I_{\pi N}\sim Q^0$ and $I_N \sim Q^0$. 
Recall that for the subtracted loop function $I_{\pi N}(\sqrt{s}\,)-I_{\pi N}(m_N)\sim Q $  
the expected minimal chiral power is manifest (see (\ref{ipin-expand})). 
It is instructive to trace the source of the anomalous chiral powers. By means of the 
identities 
\begin{eqnarray}
&& I_{\pi N}(m_N) = \frac{I_\pi-I_N}{m_N^2-m_\pi^2} + I_{\pi N}(m_N)-I_{\pi N}(0)  \;,
\nonumber\\
&&  I_{\pi N}(m_N)-I_{\pi N}(0)= 
\frac{1}{(4\pi)^2}-\frac{m_\pi}{16\,\pi\,m_N}  \left(1-\frac{m_\pi^2}{8\,m_N^2} \right)
\nonumber\\
&& \qquad \qquad \qquad +\frac{1}{(4 \pi)^2}\left( 1
-\frac{3}{2}\,\ln \left( \frac{m_\pi^2}{m_N^2}\right) \right)
\frac{m_\pi^2}{m_N^2}
+{\mathcal O}\left( \frac{m_\pi^4}{m_N^4}, d-4 \right) \,,
\label{ipin-dim}
\end{eqnarray}
it appears that once the subtraction scheme is specified for the tadpole terms 
$I_\pi$ and $I_N$ the required subtractions for the remaining master-loop functions are  
unique. In (\ref{ipin-dim}) we used the algebraic consistency identity\footnote{Note 
that (\ref{algebra-1}) leads to a well-behaved loop function 
$J_{\pi N}(w)$ at $w=0$ (see (\ref{jpin-def})).}
\begin{eqnarray}
I_\pi-I_N= \Big(m_N^2-m_\pi^2\Big)\,I_{\pi N}(0) \,,
\label{algebra-1}
\end{eqnarray}
which holds for any value of the space-time dimension d, and expanded the finite expression
$I_{\pi N}(m_N)-I_{\pi N}(0)$ in powers of $m_\pi/m_N$ at $d=4$. The result (\ref{ipin-dim}) 
seems to show that one either violates the desired chiral power for the 
nucleonic tadpole, $I_N$, or for the loop function $I_{\pi N}(m_N)$. 
One may for example subtract the pole at $d=4$ including the finite constant 
$$
\gamma-1 -\ln (4 \pi)+\ln \left( \frac{m_N^2}{\mu^2}\right) \,.
$$
That leads to a vanishing 
nucleonic tadpole $I_N \to 0$, which would be consistent with the expectation $I_N\sim Q^3$,
but we find $I_{\pi N}(m_N) \to 1/(4\pi)^2 +{\mathcal O}\left(m_\pi \right)$, in disagreement 
with the expectation $I_{\pi N}(m_N)\sim Q$. 
This problem can be solved if one succeeds in defining a subtraction scheme which 
acts differently on $I_{\pi N}(0)$ and $I_{\pi N}(m_N)$. We stress that this is legitimate, 
because $I_{\pi N}(0)$  probes our effective theory outside its applicability domain. 
Mathematically this can be  achieved most economically and 
consistently by subtracting a pole in $I_{\pi N}(\sqrt{s}\,)$ at $d=3$ which arises in 
the limit $m_\pi /m_N \to 0$.  To be explicit we recall the expression for 
$I_{\pi N}(\sqrt{s}\,)$ at arbitrary space-time dimension $d$ (see eg. \cite{Becher}):
\begin{eqnarray}
&& I_{\pi N}(\sqrt{s}\,) = \left(\frac{m_N}{\mu}\right)^{d-4}
\frac{\Gamma(2-d/2)}{(4\pi)^{d/2}}\,\int_0^1 \,d z \; C^{d/2-2} \;,
\nonumber\\
&& \qquad  C= z^2-
\frac{s-m_N^2-m_\pi^2}{m^2_N}\,z \,(1-z)+\frac{m_\pi^2}{m_N^2}\,(1-z)^2-i\,\epsilon  \,.
\label{}
\end{eqnarray}
We observe that at $d=3$ the loop function $I_{\pi N}(\sqrt{s}\,)$ is finite at $\sqrt{s}=0$ 
but infinite at $\sqrt{s}=m_N$ if one applies the limit $m_\pi /m_N \to 0$. One finds
\begin{eqnarray}
&& I_{\pi N}(m_N\,) = \frac{1}{8\,\pi}\,\frac{\mu}{m_N}\,\frac{1}{d-3}
-\frac{1}{16\,\pi}\,\frac{\mu}{m_N}
\left(\ln (4\,\pi)+ \ln \left(\frac{\mu^2}{m_N^2}\right)+\frac{\Gamma'(1/2)}{\sqrt{\pi}}\right)
\nonumber\\
&&\qquad \qquad \,+\,{\mathcal O}\left(  \frac{m_\pi}{m_N},d-3\right) \;.
\label{}
\end{eqnarray}
We are now prepared to introduce a minimal chiral subtraction scheme which may be viewed as 
a simplified variant of the scheme of Becher and Leutwyler in \cite{Becher}. As usual one 
first needs to evaluate the contributions to an observable quantity at arbitrary space-time 
dimension $d$. The result shows poles at $d=4$ and $d=3$ if considered in the non-relativistic 
limit with $m_\pi/m_N \to 0$. Our modified subtraction scheme is  defined by the 
replacement rules:
\begin{eqnarray}
&& \frac{1}{d-3} \to -\frac{m_N}{2\pi\,\mu} \;, \qquad 
\frac{2}{d-4} \to  \gamma-1 -\ln (4 \pi)+\ln \left( \frac{m_N^2}{\mu^2}\right) \,,
\label{def-sub}
\end{eqnarray}
where it is understood that poles at $d=3$ are isolated in the non-relativistic 
limit with $m_\pi/m_N \to 0$. The $d=4$ poles are isolated with the ratio $m_\pi/m_N$ at its 
physical value. The limit $d\to 4$ is applied after the pole terms are replaced according 
to (\ref{def-sub}). We emphasize that there are no infrared singularities in the residuum 
of the $1/(d-3)$-pole terms. In particular we observe that the anomalous subtraction implied in 
(\ref{def-sub}) does not lead to a potentially troublesome pion-mass dependence of the 
counter terms\footnote{In the scheme of 
Becher and Leutwyler  \cite{Becher} a pion-mass dependent subtraction scheme for the scalar 
one-loop functions is suggested. To be specific the master-loop function $I_{\pi N}(\sqrt{s})$ 
is subtracted by a regular polynomial in $s, m_N$ and $m_\pi$ of infinite order. That may 
lead to a pion-mass dependence of the counter terms in the chiral Lagrangian. In our scheme 
we subtract only a constant which agrees with the non-relativistic limit 
of the suggested polynomial of Becher and Leutwyler. Our subtraction constant does not 
depend on the pion mass.}. To make contact with a non-relativistic scheme one needs to 
expand the loop function in powers of $p/m_N $ where $p$ represents any external three  
momentum.

We collect our results for the loop functions $I_N$, $I_\pi$ and $I_{\pi N}(m_N)$
as implied by the subtraction prescription (\ref{def-sub}):
\begin{eqnarray}
&& \bar I_N = 0 \,, \qquad 
\bar I_\pi = \frac{m_\pi^2}{(4\pi)^2}\,\ln \left( \frac{m_\pi^2}{m_N^2}\right)\;,
\nonumber\\
&& \bar I_{\pi N}(m_N) = -\frac{m_\pi}{16\,\pi\,m_N}  \left(1-\frac{m_\pi^2}{8\,m_N^2} \right)
\nonumber\\
&& \qquad \qquad \; +\frac{1}{(4 \pi)^2}\left( 1
-\frac{1}{2}\,\ln \left( \frac{m_\pi^2}{m_N^2}\right) \right)
\frac{m_\pi^2}{m_N^2}
+{\mathcal O}\left( \frac{m_\pi^4}{m_N^4} \right) \,,
\label{result-bar}
\end{eqnarray}
where the '$\bar{\phantom{X}}$' signals a  subtracted loop functions\footnote{
The renormalized loop function $J_{\pi N}(w)$ is no longer well-behaved at $w=0$. This was 
expected and does not cause any harm, because the point $\sqrt{s}=0$ is far outside the 
applicability domain of our effective field theory.}. The result shows 
that now the loop functions behave according to their expected minimal chiral power (\ref{q-rule}).
Note that the one-loop expressions (\ref{result-bar}) do no longer depend on the 
renormalization scale $\mu$ introduced in dimensional regularization\footnote{One 
may make contact with the non-relativistic so-called PDS-scheme of \cite{KSW} by 
slightly modifying the replacement rule for the pole at $d=3$. With
$$\frac{1}{d-3} \to 1-\frac{m_N}{2\pi\,\mu} \;, \qquad 
I_{\pi N}(m_N) \to  \frac{1}{16\,\pi}\,\frac{2\,\mu-m_\pi}{m_N}
+{\mathcal O}\left( \frac{m_\pi^2}{m_N^2}\right)\,,$$
power counting is manifest if one counts $\mu \sim Q$. This is completely 
analogous to the PDS-scheme}. This should not be too surprising, because the renormalization 
prescription (\ref{def-sub}) has a non-trivial effect on the counter terms of the chiral 
Lagrangian.  The prescription (\ref{def-sub}) defines also a unique subtraction 
for higher loop functions in the same way the $\overline {MS}$-scheme does. The 
renormalization scale dependence will be explicit at the two-loop level, reflecting the 
presence of so-called overall divergences.

We checked that all scalar one-loop functions, subtracted according to (\ref{def-sub}), comply 
with their expected minimal chiral power (\ref{q-rule}). Typically, loop functions which are 
finite at $d=4$ are not affected by the subtraction prescription (\ref{def-sub}). Also,  
loop functions involving pion propagators only do not show singularities at $d=3$ and 
therefore can be  related easily to the corresponding loop functions of the 
$\overline{MS}$-scheme. Though it may be tedious to relate the standard 
$\overline{MS}$-scheme to our scheme in the nucleon sector, in particular when multi-loop 
diagrams are considered, we assert that we propose 
a well defined prescription for regularizing all divergent loop integrals. A prescription 
which is far more convenient than the $\overline {MS}$-scheme, because it complies manifestly 
with the chiral counting rule (\ref{q-rule})\footnote{Note that it is unclear how to 
generalize our prescription in the presence of two massive fields with respective masses 
$m_1$ and $m_2$ and 
$m_1 \gg m_2$. For the decuplet baryons one may count $m_{[8]}-m_{[9]} \sim Q $ as suggested 
by large $N_c$ counting arguments and therefore start with a common mass for the baryon 
octet and decuplet states. The presence of the $B_\mu^*$ field in the chiral Lagrangian does  
cause a problem. Since we include the $B_\mu^*$ field only at tree-level in the interaction 
kernel of pion-nucleon and kaon-nucleon scattering, we do not further investigate the possible 
problems in this work. Formally one may avoid such problems all together if one counts 
$m_{[8]}-m_{[9]} \sim Q$ even though this assignment may not be effective.}.

If we were to perform calculations within standard chiral perturbation theory in terms of the 
relativistic chiral Lagrangian we would be all set for any computation. However, as advocated before 
we are heading towards a non-perturbative chiral theory. That requires a more elaborate renormalization 
program, because we wish to  discriminate carefully reducible and irreducible diagrams and sum the reducible 
diagrams to infinite order. The idea is to take over the renormalization program of standard chiral 
perturbatioon theory to the interaction kernel. In order to apply the standard perturbative 
renormalization program for the interaction kernel one has to move all divergent parts lying  
in reducible diagrams into the interaction kernel. That problem is solved in part by constructing 
an on-shell equivalent interaction kernel according to (\ref{k-decomp},\ref{t-eff},\ref{v-eff}). 
It is evident that the 'moving' of divergences needs to be controlled  by an additional renormalization condition.   
Any such condition imposed should be constructed so as to respect  
crossing symmetry approximatively. While standard chiral perturbation theory leads directly to 
cross symmetric amplitudes at least approximatively, it is not automatically so in a resummation scheme.

Before introducing our general scheme we examine the above issues explicitly with the 
example worked out in detail in the previous section. Taking the s-channel nucleon pole term as the 
driving term in the Bethe-Salpeter equation we derived the explicit result (\ref{pin-example}). 
We first discuss its implicit 
nucleon mass renormalization. The result (\ref{pin-example}) shows a pole at the 
physical nucleon mass with $\wslash =-m_N$, only if the mass-counter term 
$\Delta m_N$ in (\ref{nucleon-pole}) is identified as follows
\begin{eqnarray}
\Delta m_N =  \frac{3\,g_A^2}{4\,f^2}\,2\,m_N\,\Big (m_\pi^2\,I_{\pi N}(m_N)-I_N \Big)\,.
\label{mass-ren}
\end{eqnarray}
In the $\overline{MS}$-scheme the divergent parts of $\Delta  m_N$, or more precisely 
the renormalization scale dependent parts, may be absorbed into 
the nucleon bare mass \cite{gss}. In our scheme we renormalize by simply replacing 
$I_{\pi N}\to \bar I_{\pi N} $, $I_\pi \to \bar I_\pi$  and 
$I_N \to \bar I_N= 0$. The result (\ref{mass-ren}) together 
with (\ref{result-bar}) then reproduces the well known result \cite{gss,kambor}
$$
\Delta m_N = -\frac{3\,g_A^2\,m_\pi^3}{32\,\pi\,f^2}+\cdots \;,
$$ 
commonly derived in terms of the one-loop nucleon self 
energy $\Sigma_N(p)$. In order to offer a more direct comparison of (\ref{pin-example}) 
with the  nucleon self-energy $\Sigma_N(p)$ we recall the one-loop result
\begin{eqnarray}
&&\!\!\!\!\!\! \Sigma_N (p) = \frac{3\,g_A^2}{4\,f^2}\, \Bigg( 
 m_N \,\Big( m_\pi^2\,I_{\pi N}(\sqrt{p^2}\,) -I_N\Big)
-\pslash \,\Bigg(\frac{p^2-m_N^2}{2\,p^2} \, I_\pi 
+\frac{p^2+m_N^2}{2\,p^2} \,I_N
\nonumber\\
&& \qquad \qquad \qquad 
+\left( \frac{(p^2-m_N^2)^2}{2\,p^2}- m_\pi^2\,\frac{p^2+m_N^2}{2\,p^2}
\right) I_{\pi N}(\sqrt{p^2}\,)
\Bigg) \Bigg)\;,
\label{n-self}
\end{eqnarray}
in terms of the convenient master-loop function $I_{\pi N}(\sqrt{p^2})$ and the 
tadpole terms $I_N$ and $I_\pi$. We emphasize that the 
expression (\ref{n-self}) is valid for arbitrary space-time dimension $d$. The wave-function 
renormalization, ${\mathcal Z}_N$, of the nucleon can be read off (\ref{n-self})
\begin{eqnarray}
{\mathcal Z}^{-1}_N &=& 1-\frac{\partial \,\Sigma_N }{\partial \,\pslash}\Bigg|_{\pslash \,=m_N} 
\!\!= 1 -\frac{3\,g_A^2}{4\,f^2} 
\left( 2\,m_\pi^2\,m_N \,\frac{\partial \,I_{\pi N}(\sqrt{p^2})}{\partial \,\sqrt{p^2}}
 \Bigg|_{p^2=m_N^2}
\!\!-I_\pi 
\right) 
\nonumber\\
&=& 1- \frac{3\,g_A^2}{4\,f^2}\,\frac{m_\pi^2 }{(4\pi)^2}
\left( -4 -3\,\ln \left(\frac{m_\pi^2}{m_N^2}\right) 
+3\,\pi\,\frac{m_\pi}{m_N}
+{\mathcal O}\left( \frac{m^2_\pi}{m^2_N} \right)
\right) \,,
\label{def-zn}
\end{eqnarray}
where in the last line of (\ref{def-zn}) we applied our minimal chiral 
subtraction scheme (\ref{def-sub}).
The result (\ref{def-zn}) agrees with the expressions obtained previously in 
\cite{gss,kambor,Becher}. Note that in (\ref{def-zn}) we suppressed the contribution of 
the counter terms $\zeta_0, \zeta_D$ and $\zeta_F$.

It is illuminating to discuss the role played by the pionic tadpole 
contribution, $I_\pi$, from $V(w)$ in (\ref{pin-example}) and from $J_{\pi N}(w)$
in (\ref{jpin-def}). In the mass renormalization (\ref{mass-ren}) both tadpole contributions 
cancel identically. 
If one dropped the pionic tadpole term $I_\pi$ in the effective interaction $V(w)$ one would 
find a mass renormalization $\Delta m_N \sim m_N\,I_\pi/f^2 \sim Q^2 $, in conflict with the
expected minimal chiral power $\Delta m_N \sim Q^3$. Since we would like to evaluate 
the effective interaction $V$ in chiral perturbation theory we take this cancellation as 
the motivation to 'move' all tadpole contributions from the loop function $J_{\pi N}(w)$
to the effective interaction kernel $V(w)$. We split the meson-baryon propagator 
$G={\mathcal Z}_N\,(G_R+\Delta G)$ into two terms $G_R$ and $\Delta G$  which leads to a renormalized 
tadpole-free loop function $G_R$. The renormalized effective potential $V_R(w)$ follows 
\begin{eqnarray}
\bar T_R= \frac{{\mathcal Z}_N}{1-V\cdot G}\cdot V=\frac{1}{1-V_R\cdot G_R}\cdot V_R \;,
\quad
V_R = \frac{{\mathcal Z}_N}{1-{\mathcal Z}_N\,V\cdot\Delta G}\cdot V \;,
\label{ren-v}
\end{eqnarray}
where we introduced the renormalized scattering amplitude
$\bar T_R = {\mathcal Z}^{\frac{1}{2}}_N \,\bar T\, {\mathcal Z}^{\frac{1}{2}}_N$ as implied by the LSZ 
reduction scheme. 
Now the cancellation of the pionic tadpole contributions is easily implemented by applying 
the chiral expansion to $V_R(w)$. The renormalized interaction kernel $V_R(w)$ 
is real by construction. 

Our renormalization scheme is still incomplete\footnote{We 
discuss here the most general case not necessarily imposing the minimal chiral subtraction 
scheme (\ref{def-sub}).}. We need 
to specify how to absorb the remaining logarithmic divergence in $I_{\pi N}(\sqrt{s}\,)$.
The strategy is to move all divergences from the unitarity loop function 
$J_{\pi N}(w)$ into the renormalized potential $V_R(w)$ via (\ref{ren-v}). 
For the effective potential one may then apply the standard perturbative renormalization 
program. We impose the renormalization condition that the effective potential
$V_R(w)$ matches the scattering amplitudes at a subthreshold energy $\sqrt{s}=\mu_S$:
\begin{eqnarray}
\bar T_R (\mu_S) = V_R(\mu_S) \;.
\label{ren-cond}
\end{eqnarray}
We argue that the choice for the subtraction point $\mu_S$ is rather well determined
by the crossing symmetry constraint. In fact one is lead almost uniquely 
to the convenient point $\mu_S =m_N$. For the case of pion-nucleon 
scattering one observes that the renormalized effective interaction $V_R$ is real only if 
$m_N-m_\pi<\mu_S < m_N+m_\pi$ holds. The first condition reflects the s-channel 
unitarity cut with  $\Im I_{\pi N}(\mu_S) =0 $ only if $\mu_S < m_N+m_\pi$. The 
second condition signals the u-channel unitarity cut. 
A particular convenient choice for the subtraction point is $\mu_S=m_N$, because it protects 
the nucleon pole term contribution. It leads to
\begin{eqnarray}
\bar T_R (w)&=& V_R(w)+{\mathcal O}\Big(\left( \wslash+m_N\right)^0\Big)
\nonumber\\
&=&-\frac{3\,({\mathcal Z}_N\,g_A)^2}{4\,f^2}\,
\frac{4\,m^2_N }{\wslash+m_N}
+{\mathcal O}\Big(\left( \wslash+m_N\right)^0\Big) \,,
\label{pole-protection}
\end{eqnarray}
if the  renormalized loop function is subtracted in such a way that $I_{\pi N,R}(m_N) =0 $ and 
$I_{\pi N,R}'(m_N)=0$ hold. 

Note that we insist on a minimal subtraction for the 
scalar loop function $I_{\pi N}(\sqrt{s}\,)$ 'inside' the full loop function $J_{\pi N}(w)$. 
According to (\ref{ipin-analytic}) one subtraction suffices to render $I_{\pi N}(\sqrt{s}\,)$ 
finite. A direct subtraction for $J_{\pi N}(w)$ would require an infinite subtraction polynomial
which would not be specified by the simple renormalization condition (\ref{ren-cond}). 
We stress that the double subtraction in $I_{\pi N}(\sqrt{s}\,)$ is necessary in order 
to meet the condition (\ref{ren-cond}). Only with $I_{\pi N,R}'(m_N)=0$
the renormalized effective potential $V_R(w)$ in (\ref{pole-protection}) represents 
the s-channel nucleon pole term in terms of the physical coupling constant ${\mathcal Z}_N\,g_A$ 
properly. For example, it is evident that if $V_R$ is truncated at chiral order $Q^2$  
one finds ${\mathcal Z}_N=1$. The one-loop wave-function renormalization (\ref{def-zn}) is needed 
if one considers the $Q^3$-terms in $V_R$.

We observe that the subtracted loop function $I_{\pi N}(\sqrt{s}\,)-I_{\pi N}(\mu_S)$ 
is in fact independent of the subtraction point to order $Q^2$ if one counted 
$\mu_S-m_N\sim Q^2$. In this case one derives from (\ref{ipin-expand}) the expression
\begin{eqnarray}
I_{\pi N}(\sqrt{s}\,)-I_{\pi N}(\mu_S)
&=&i\,\frac{\sqrt{\phi_{\pi N}}}{8\,\pi\,m_N}
+\frac{m_\pi}{16\,\pi\,m_N}
-\frac{\sqrt{s}-m_N}{16\,\pi^2\,m_N}\,\ln \left( \frac{m_\pi^2}{m_N^2}\right)
\nonumber\\
&+&\frac{\sqrt{\phi_{\pi N}}}{16\,\pi^2\,m_N}\,
\ln \left( \frac{\sqrt{s}-m_N+\sqrt{\phi_{\pi N}}}
{\sqrt{s}-m_N-\sqrt{\phi_{\pi N}}} \right)
+{\mathcal O}\left( Q^2 \right) \;,
\label{ipin-expand:b}
\end{eqnarray}
where $\phi_{\pi N}= (\sqrt{s}-m_N)^2-m_\pi^2$.  The result (\ref{ipin-expand:b})
suggests that one may set up a systematic expansion scheme in powers of 
$\mu_S-m_N \sim Q^2$. The subtraction-scale independence of physical
results then implies that all powers $(\mu_S-m_N)^n$ cancel except for the leading term 
with $n=0$. In this sense one may say that the scattering amplitude is independent of the 
subtraction point $\mu_S$. Note that such a scheme does not necessarily require a  
perturbative expansion of the scattering amplitude. It may be advantageous instead to 
restore that minimal $\mu_S$-dependence in the effective potential $V$, leading to a 
subtraction scale independent scattering amplitude at given order in $(\mu_S-m_N)$. 
Equivalently, it is legitimate to directly insist on the 'physical'
subtraction with $\mu_S=m_N$. There is a further important point to be made: we would reject 
a conceivable scheme in which the inverse effective potential $V^{-1}$ is expanded in 
chiral powers, even though it would obviously facilitate the construction of that minimal 
$\mu_S$-dependence. As examined in detail in \cite{nn-lutz} expanding the inverse effective 
potential requires a careful analysis to determine if the effective potential has a zero within 
its applicability domain. If this is the case one must reorganize the expansion scheme. 
Note that this is particularly cumbersome in a coupled-channel scenario where one must 
ensure that $\det V \neq 0$ holds. Since in the $SU(3)$ limit the Weinberg-Tomozawa interaction 
term, which is the first term in the chiral expansion of $V $, leads 
to $\det V_{WT}=0$ (see (\ref{WT-k})) one should not pursue this path\footnote{There is a 
further strong indication that the expansion of the inverse effective potential indeed 
requires a reorganization. The effective p-wave interaction kernel is troublesome, because 
the nucleon-pole term together with a smooth background term will lead necessarily to a 
non-trivial zero.}. 

We address an important aspect related closely to our renormalization scheme, the
approximate crossing symmetry. At first, one may insist on either a 
strict perturbative scheme or an approach which performs a simultaneous iteration of 
the s- and u-channel in order to meet the crossing symmetry constraint. We point out, 
however, that a simultaneous iteration of the s- and u-channel is not required, if the 
s-channel iterated and u-channel iterated amplitudes match at subthreshold 
energies $\sqrt{s}\simeq \mu_S \simeq m_N $ to high accuracy. This is a sufficient 
condition in the chiral framework, because the overlap of the applicability domains of   
the s- and u-channel iterated amplitudes are restricted to a small matching window 
at subthreshold energies in any case. This will be discussed in more detail in section 4.3. 
Note that crossing symmetry is not necessarily observed in a cutoff regularized approach. 
In our scheme the meson-baryon scattering process is {\it perturbative} below the s and 
u-channel unitarity thresholds by construction and therefore meets the crossing symmetry 
constraints approximatively. {\it Non-perturbative} effects as implied by the unitarization 
are then expected at energies outside the matching window.  

Before turning to the coupled channel formalism we return to the question of
the convergence of standard $\chi$PT. The poor convergence of the standard $\chi$PT scheme in
the $\bar KN$ sector is illustarted by comparing the effect of the iteration
of the Weinberg-Tomozawa term in the $\bar KN$ and $\pi N$ channels.

\subsection{Weinberg-Tomozawa interaction and convergence of $\chi$PT}

We consider the Weinberg-Tomozawa interaction term $K^{}_{WT}(\bar k,k;w)$ as the driving term in the 
Bethe-Salpeter equation. This is an instructive example because it exemplifies the 
non-perturbative nature of the strangeness channels and it also serves as a transparent first 
application of our renormalization scheme. The on-shell equivalent scattering amplitude 
$\bar T_{WT}(w)$ is
\begin{eqnarray}
&& \!\! V_{WT}(w) =\left( 1-\frac{c}{4 f^2}\,I_L\right)^{-1} 
\!\! \frac{c}{4 f^2}  \left(2 \;\wslash -2\!\,M 
-I_{LR}\,\frac{c}{4 f^2} \right)
\left(1-I_R\,\frac{c}{4 f^2} \right)^{-1},
\nonumber\\
&&\!\! \bar T_{WT}(w) = \frac{1}{1-V_{WT}(w)\,J(w)}\,V_{WT}(w) \;,\quad 
K_{WT}(\bar k,k;w) = c^{}_{}\,\frac{\qslash+\barqslash}{4\,f^2}\;,
\label{wt-tadpole}
\end{eqnarray}
where the Bethe-Salpeter scattering equation (\ref{BS-eq}) was solved algebraically
following the construction 
(\ref{k-decomp},\ref{t-eff}). Here we identify $K_R\cdot G \to I_R$, $G\cdot K_L \to I_L$
and $K_R\cdot G\cdot K_L \to I_{LR}$ with the mesonic tadpole loop $I_M$ (see (\ref{def-tadpole})) with 
$I_L=I_R=I_M$ and $I_{LR}=(\wslash-M)\,I_{M}$.  The meson-baryon loop function $J(w)$ is 
defined for the $\pi N$ system in (\ref{jpin-def}). 
The coupling constants $c^{(I)}_{MB\to MB}$ specifies the strength of the 
Weinberg-Tomozawa interaction in a given meson-baryon channel with isospin $I$. 
Note that (\ref{wt-tadpole}) is written in a way such that coupled channel effects 
are easily included by identifying the proper matrix structure of its building blocks. 
A detailed account of these effects will be presented in subsequent sections.  
We emphasize that the mesonic tadpole $I_M$ cannot be absorbed systematically in $f$, 
in particular when the coupled channel generalization of (\ref{wt-tadpole}) with its  
non-diagonal matrix $c$ is considered. At leading chiral order $Q$ the tadpole 
contribution should be dropped in the effective interaction $V$ in any case. 

It is instructive to consider the s-wave $\bar K N $ scattering lengths up 
to second order in the Weinberg-Tomozawa interaction vertex. 
The coefficients $c^{(0)}_{\bar K N\to \bar KN}=3$ and 
$c^{(0)}_{\bar K N\to \bar KN}=3$ and $c^{(1)}_{\bar K N\to \bar KN}=1$ lead to  
\begin{eqnarray}
4\,\pi \left(1+\frac{m_K}{m_N} \right) a_{\bar  K N}^{(I=0)} &=&
\frac{3}{2}\,\frac{m_K}{f^2} \left(
1+\frac{3\,m^2_K}{16\,\pi^2\,f^2}
\left(\pi-\ln \left(\frac{m_K^2}{m_N^2}\right) \right)\right)+\cdots \;,
\nonumber\\
4\,\pi \left(1+\frac{m_K}{m_N} \right) a_{\bar  K N}^{(I=1)} &=&
\frac{1}{2}\,\frac{m_K}{f^2} \left(
1+\frac{m^2_K}{16\,\pi^2\,f^2}
\left(\pi-\ln \left(\frac{m_K^2}{m_N^2}\right) \right)\right)+\cdots \;,
\label{wt-square-KN}
\end{eqnarray}
where we included only the s-channel iteration effects following
from (\ref{wt-tadpole}). The loop function $I_{KN}(\sqrt{s})$ was subtracted at the nucleon mass 
and all tadpole contributions are dropped. Though the expressions (\ref{wt-square-KN}) are 
incomplete in the $\chi $PT framework (terms of chiral order $Q^2$ and $Q^3$ terms are 
neglected) it is highly instructive to investigate the convergence property of a reduced 
chiral Lagrangian with the Weinberg-Tomozawa interaction only. According to 
(\ref{wt-square-KN}) the relevant expansion parameter $m_K^2/(8\pi f^2) \simeq 1$ is about one
in the $\bar KN$ sector. One observes the enhancement factor $2 \pi$ as compared to irreducible
diagrams which would lead to the typical factor $ m^2_K/(4\pi \,f)^2$. 
The perturbative treatment of the Weinberg-Tomozawa
interaction term is therefore unjustified and a change in approximation
scheme is required. In the isospin zero $\bar KN$ system the Weinberg-Tomozawa interaction 
if iterated to all orders in the s-channel (\ref{wt-tadpole}) leads to a pole in the 
scattering amplitude at subthreshold
energies $\sqrt{s}< m_N+m_K$. This pole is a precursor of
the $\Lambda(1405)$ resonance \cite{dalitz-1,Siegel,Kaiser,Ramos,Hirschegg}.

\begin{figure}[t]
\begin{center}
\includegraphics[width=12cm,clip=true]{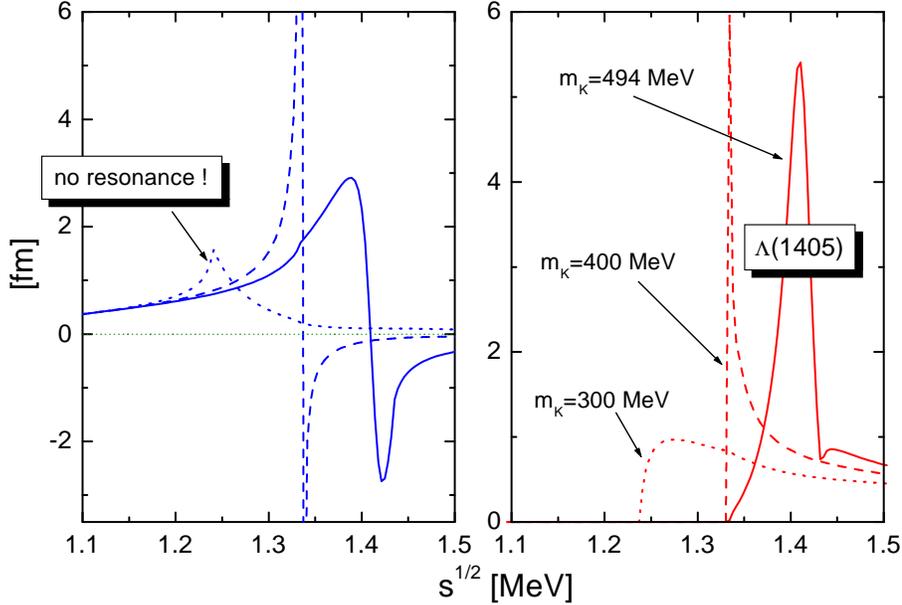}
\end{center}
\caption{Real (l.h.s.) and imaginary (r.h.s.) part of the isospin zero
s-wave $K^-$-nucleon 
scattering amplitude as it follows from the $SU(3)$ Weinberg-Tomozawa
interaction term in 
a coupled channel calculation. We use $f = 93 $ MeV and identify the
subtraction point 
with the $\Lambda(1116)$ mass.}
\label{fig:wt}
\end{figure}

In Fig.~\ref{fig:wt} we anticipate our final result for the leading interaction term of the
chiral $SU(3)$ Lagrangian density suggested by Tomozawa and Weinberg. 
If taken as input for the multi-channel Bethe-Salpeter equation,
properly furnished with a renormalization scheme leading to a subtraction point close to
the baryon octet mass, a rich structure of the scattering amplitude arises. 
Details for the coupled channel generalization of (\ref{wt-tadpole}) are presented in the 
subsequent sections. Fig.~\ref{fig:wt} shows the s-wave solution of the multi-channel Bethe-Salpeter as a 
function of the kaon mass. For physical kaon masses the isospin zero scattering amplitude 
exhibits a resonance structure at energies where one would expect the $\Lambda(1405)$ 
resonance. We point out that the resonance structure disappears as the kaon mass is decreased.
Already at a hypothetical kaon mass of $300$ MeV the $\Lambda(1405)$ resonance is no longer 
formed. Fig.~\ref{fig:wt} demonstrates that the chiral $SU(3)$ Lagrangian is
necessarily non-perturbative in the strangeness sector. This confirms the findings of
\cite{Kaiser,Ramos}. In previous works however the $\Lambda(1405)$ resonance
is the result of a fine tuned cutoff parameter which gives rise to a different kaon mass
dependence of the scattering amplitude \cite{Ramos}. In our scheme the choice of subtraction
point close to the baryon octet mass follows necessarily from the 
compliance of the expansion scheme with approximate crossing symmetry. 
Moreover, the identification of the subtraction point with the $\Lambda$-mass in the 
isospin zero channel protects the hyperon exchange s-channel pole contribution and 
therefore avoids possible pathologies at subthreshold energies.

We turn to the pion-nucleon sector. The chiral $SU(2)$ Lagrangian has been 
successfully applied to pion-nucleon scattering in standard chiral
perturbation theory \cite{Bernard,Meissner,pin-q4}. Here the typical expansion parameter 
$m^2_\pi/(8\,\pi\,f^2) \ll 1$ characterizing the unitarization  is sufficiently 
small and one would expect good convergence properties. The application of the chiral $SU(3)$
Lagrangian to pion-nucleon scattering on the other hand is not completely worked out so far. 
In the SU(3) scheme the $\pi N$ channel couples for example to the $K \Sigma $ channel. Thus 
the slow convergence of the unitarization in the $K \Sigma$ channel suggests 
to expand the interaction kernel rather than the scattering amplitude also in the 
strangeness zero channel. This may improve the convergence properties of the chiral 
expansion and extend its applicability domain to larger energies. Also,  
if the same set of parameters are to be used in the pion-nucleon and kaon-nucleon 
sectors the analogous partial resummation of higher order counter terms included  
by solving the Bethe-Salpeter equation should be applied. 
We illustrate such effects for the case of the Weinberg-Tomozawa interaction. 
With $c^{(\frac{3}{2})}_{\pi N\to \pi N}=-1$ and 
$c^{(\frac{3}{2})}_{\pi N\to K \Sigma}=-1$  the isospin three half s-wave pion-nucleon 
scattering lengths $a_{\pi N}^{(\frac{3}{2})}$ receive the typical correction terms
\begin{eqnarray}
&&4\,\pi \left(1+\frac{m_\pi}{m_N} \right) a_{\pi N}^{(\frac{3}{2})} =
-\frac{m_\pi}{2\,f^2}\, \Bigg(
1-\frac{m^2_\pi}{16\,\pi^2\,f^2}
\left(\pi-\ln \left(\frac{m_\pi^2}{m_N^2}\right) \right) 
\nonumber\\
&&\qquad \qquad - \frac{(m_\pi+m_K)^2}{32\,\pi^2\,f^2} 
\Bigg( -1-\frac{1}{2}\,\ln \left(\frac{m_K^2}{m_\Sigma^2}\right) 
\nonumber\\
&& \qquad \qquad 
+\pi \left( \frac{3\,m_K}{4\,m_N}-\frac{m_\Sigma-m_N}{2\,m_K} \right) 
+{\mathcal O}\left( Q^2\right)\Bigg)\Bigg)
+ {\mathcal O}\left( m_\pi^2\right)\;,
\label{wt-square-piN}
\end{eqnarray}
where we again considered exclusively the unitary correction terms. Note that the 
ratio $(m_\Sigma -m_N)/m_K$  arises in (\ref{wt-square-piN}), because we first expand 
in powers of $m_\pi$ and only then expand further with $m_\Sigma-m_N \sim Q^2$ and 
$m_K^2 \sim Q^2 $. The correction terms in (\ref{wt-square-piN}) induced by the 
kaon-hyperon loop, which is subtracted at the nucleon 
mass, exemplify the fact that the parameter $f$ is renormalized by the strangeness sector 
and therefore must not be identified with the chiral limit value of $f$ as derived for the 
$SU(2)$ chiral Lagrangian. This is evident if one confronts the Weinberg-Tomozawa
theorem of the chiral $SU(2)$ symmetry with (\ref{wt-square-piN}). 
The expression (\ref{wt-square-piN}) demonstrates further
that this renormalization of $f$ appears poorly convergent in the kaon mass. Note in 
particular the anomalously large term $\pi \,m_K/m_N $. Hence it is advantageous to consider 
the partial resummation induced by a unitary coupled channel treatment of pion-nucleon 
scattering.

\subsection{Partial-wave decomposition of the Bethe-Salpeter equation}

The Bethe-Salpeter equation (\ref{BS-eq}) can be solved analytically for
quasi-local interaction terms which typically arise in the chiral Lagrangian.
The scattering equation is decoupled by introducing relativistic projection
operators ${Y}^{(\pm)}_n(\bar q,q;w)$ with good total angular momentum:
\begin{eqnarray}
&&{Y}^{(\pm )}_n(\bar q,q;w)=\frac{1}{2}\,\Bigg(\frac{\wslash}{\sqrt{w^2}}\pm 1\Bigg)\,
\bar Y_{n+1}(\bar q,q;w)
\nonumber\\
&&\;\;\;\;\;\;\;\;\;\;-\frac{1}{2}\,\Bigg(  \barqslash -\frac{w\cdot \bar q}{w^2}\,\wslash \Bigg)
\Bigg(\frac{\wslash}{\sqrt{w^2}} \mp 1\Bigg)\,
\Bigg( \qslash -\frac{w\cdot q}{w^2}\,\wslash \Bigg)
\bar Y_{n}(\bar q,q;w)\;,
\nonumber\\
&&\bar Y_{n}(\bar q,q;w)= \sum_{k=0}^{[(n-1)/2]}\,\frac{(-)^k\,(2\,n-2\,k) !}{2^n\,k !\,(n-k)
!\,(n-2\,k -1) !}\,Y_{\bar q \bar q}^{k}\,Y_{\bar q q}^{n-2\,k-1}\,Y_{q q}^{k}\;,
\nonumber\\
&&Y_{\bar q \bar q}=\frac{(w\cdot \bar q )\,(\bar q\cdot w)}{w^2}
-\bar q \cdot \bar q
\;,\;\;\;
Y_{q  q}=\frac{(w\cdot  q )\,( q\cdot w)}{w^2} -q \cdot  q \;,
\nonumber\\
&&Y_{\bar q q}=\frac{(w\cdot \bar q )\,(q\cdot w)}{w^2} -\bar q \cdot q  \;.
\label{cov-proj}
\end{eqnarray}
For the readers convenience we provide the leading order projectors
${Y}_n^{(\pm)}$ relevant for the $J={\textstyle{1\over 2}}$ and $J={\textstyle{3\over 2}}$ 
channels explicitly:
\begin{eqnarray}
{Y}_0^{(\pm )}(\bar q,q;w) &=& \frac{1}{2}\,\left( \frac{\wslash}{\sqrt{w^2}}\pm 1 \right)\;,
\nonumber\\
{Y}_1^{(\pm)}(\bar q,q;w) &=& \frac{3}{2}\,\left( \frac{\wslash}{\sqrt{w^2}}\pm 1 \right)
\left(\frac{(\bar q \cdot w )\,(w \cdot q)}{w^2} -\big( \bar q\cdot q\big)\right)
\nonumber\\
&-&\frac{1}{2}\,\Bigg(  \barqslash -\frac{w\cdot \bar q}{w^2}\,\wslash \Bigg)
\Bigg(\frac{\wslash}{\sqrt{w^2}}\mp 1\Bigg)\,
\Bigg( \qslash -\frac{w\cdot q}{w^2}\,\wslash \Bigg)\;.
\label{}
\end{eqnarray}

The objects ${Y}^{(\pm)}_n(\bar q,q;w)$ are constructed to have the following
convenient property: Suppose that the interaction kernel $K$ in (\ref{BS-eq}) can be
expressed as linear combinations of the ${Y}^{(\pm)}_n(\bar q,q;w)$ with a set of
coupling functions $V^{(\pm)}(\sqrt{s}\,, n)$, which may depend on the variable $s$,
\begin{eqnarray}
K(\bar k ,k ;w ) &=& \sum_{n=0}^\infty \left(
V^{(+)}(\sqrt{s};n)\,{Y}^{(+)}_n(\bar q,q;w)
+V^{(-)}(\sqrt{s};n)\,{Y}^{(-)}_n(\bar q,q;w) \right) \, ,
\label{k-sum}
\end{eqnarray}
with $ w = p+q$, $ k= (p-q)/2 $ and $\bar k =(\bar p-\bar q)/2 $.
Then in a given isospin channel the unique solution reads
\begin{eqnarray}
&& T(\bar k ,k ;w ) = \sum_{n=0}^\infty \left(
M^{(+)}(\sqrt{s};n )\,{Y}^{(+)}_n(\bar q,q;w)
+ M^{(-)}(\sqrt{s};n)\,{Y}^{(-)}_n(\bar q,q;w)\right) \;,
\nonumber\\
&& M^{(\pm)}(\sqrt{s};n) = \frac{V^{(\pm )}(\sqrt{s};n)}{1-V^{(\pm)}(\sqrt{s};n)
\,J^{(\pm)}_{\pi N }(\sqrt{s};n)} \;,
\label{t-sum}
\end{eqnarray}
with a set of divergent loop functions $J^{(\pm)}_{\pi N }(\sqrt{s};n)$ defined by
\begin{eqnarray}
J^{(\pm)}_{\pi N}(\sqrt{s};n)\,{Y}^{(\pm)}_n(\bar q,q;w)
&&= -i\int \frac{d^4l}{(2\pi)^4}\,{Y}^{(\pm)}_n(\bar q,l;w)\,
S_N(w-l)\,
\nonumber\\
&& \;\;\;\;\;\;\;\;\;\;\;\;\;\;\;\;\;\;\;\;\;
\times D_\pi(l)\,{Y}^{(\pm)}_n(l,q;w)\;.
\label{jpin-n-def}
\end{eqnarray}
We underline that the definition of the loop functions in (\ref{jpin-n-def}) is non 
trivial, because it assumes that $Y_n^{(\pm )}\cdot G \cdot Y_n^{(\pm )} $ is indeed 
proportional to $Y_n^{(\pm)}$. An explicit derivation of this property, which is in fact 
closely linked to our renormalization scheme, is given in Appendix C. We recall 
that the loop functions $J^{(\pm)}_{\pi N}(\sqrt{s};n)$, which are badly divergent, have a 
finite and well-defined imaginary part
\begin{eqnarray}
\Im\,J^{(\pm)}_{\pi N }(\sqrt{s};n) &=& \frac{p^{2\,n+1}_{\pi N}}{8\,\pi\,\sqrt{s}}
\left(  \frac{\sqrt{s}}{2}+ \frac{m_N^2-m_\pi^2}{2\,\sqrt{s}}\pm m_N \right)\;.
\label{}
\end{eqnarray}
We specify how to renormalize the loop functions. In dimensional regularization
the loop functions can be written as linear combinations of scalar one
loop functions
$I_{\pi N}(\sqrt{s}\,)$, $I_\pi$, $I_N $ and $I^{(n)}$, 
\begin{eqnarray}
I^{(n)}=i\,\int \frac{d^4l}{(2\pi)^4}\,\Big(l^2 \Big)^n\;,
\label{tadpole:a}
\end{eqnarray}
According to our renormalization procedure we drop $I^{(n)}$, the tadpole
contributions $I_\pi$, $I_N $ and replace
$I_{\pi N}(s) $ by $I_{\pi N}(\sqrt{s}\,)-I_{\pi N}(\mu_S )$. This leads to
\begin{eqnarray}
J^{(\pm)}_{\pi N }(\sqrt{s}; n) &=&
p_{\pi N}^{2\,n}(\sqrt{s}\,)
\left( \frac{\sqrt{s}}{2}+ \frac{m_N^2-m_\pi^2}{2\,\sqrt{s}}\pm m_N \right)
\Delta I_{\pi N}(\sqrt{s}\,)\;,
\nonumber\\
\Delta I_{\pi N}(\sqrt{s}\,) &=& I_{\pi N}(\sqrt{s}\,)-I_{\pi N}(\mu_S ) \;,
\label{result-loop}
\end{eqnarray}
with the master loop function $I_{\pi N}(\sqrt{s}\,)$ and $p_{\pi N}(\sqrt{s}\,)$ 
given in (\ref{ipin-analytic}). In the center of mass frame $p_{\pi N}$ represents 
the relative momentum. We emphasize 
that the loop functions $J^{(\pm)}_{\pi N }(\sqrt{s}; n)$ are renormalized in accordance
with (\ref{ren-cond}) and (\ref{ren-v}) where $\mu_S=m_N$\footnote{ 
Note that consistency with the renormalization condition (\ref{ren-cond}) requires
a further subtraction in the loop function $J^{(-)}_{\pi N}(\sqrt{s},0)$ 
if the potential $V^{(-)}_{\pi N}(\sqrt{s},0) \sim 1/(s-m_N^2+i\,\epsilon)$ exhibits the 
s-channel nucleon pole (see (\ref{pole-protection})).}. 
This leads to tadpole-free loop functions and also to $M^{(\pm )}(m_N,n)=V^{(\pm )}(m_N,n)$. 
The behavior of the loop functions $J^{(\pm )}_{\pi N}(\sqrt{s},n)$ close to threshold
\begin{eqnarray}
\Im \,J^{(+)}_{\pi N}(\sqrt{s},n) \sim p_{\pi N}^{2\,n+1} \;, \qquad 
\Im \,J^{(-)}_{\pi N}(\sqrt{s},n) \sim p_{\pi N}^{2\,n+3} \,,
\label{}
\end{eqnarray}
already tells the angular momentum, $l$, of a given channel with $l=n$ for the $'+'$ and 
$l=n+1$ for the $'-'$ channel.

The Bethe-Salpeter equation (\ref{t-sum}) decouples into reduced scattering amplitudes
$M^{(\pm)}(\sqrt{s},n)$ with well-defined angular momentum. In order to unambiguously
identify the total angular momentum $J$ we recall the partial-wave decomposition of
the on-shell scattering amplitude $T$ \cite{Landoldt}. The amplitude $T$ is decomposed
into invariant amplitudes $F^{(I)}_\pm(s,t)$ carrying good isospin $I$
\begin{eqnarray}
T &=& \sum_{I}\,\left(
\frac{1}{2}\,\Bigg(\frac{\wslash}{\sqrt{w^2}}+1
\Bigg)\, F^{(I)}_+(s,t)+ \frac{1}{2}\,\Bigg(\frac{\wslash}{\sqrt{w^2}}-1
\Bigg)\,F^{(I)}_-(s,t)\right) P_{I}
\label{}
\end{eqnarray}
where $s= (p+q)^2=w^2$ and $t=(\bar q-q)^2$ and $P_I$ are the isospin projectors introduced in (\ref{iso:proj}).
Note that the choice of invariant amplitudes is not unique. Our choice is particularly
convenient to make contact with the covariant projection operators (\ref{cov-proj}).
For different choices, see \cite{Landoldt}.
The amplitudes $F_{\pm}(s,t)$ are decomposed into partial-wave amplitudes
$f^{(l)}_{J=l\pm \frac{1}{2}}(\sqrt{s}\,)$ \cite{Landoldt}
\begin{eqnarray}
F_+(s,t) &=& \frac{8\,\pi\,
\sqrt{s}}{E+m_N}\,\sum_{n=1}^\infty
\,\Big( f_{J=n+\frac{1}{2}}^{(n-1)}(\sqrt{s}\,)-f_{J=n- \frac{1}{2}}^{(n+1)}(\sqrt{s}\,)\Big)\,P'_n(\cos \theta)
\;,
\nonumber\\
F_-(s,t) &=&\frac{8\,\pi\,
\sqrt{s}}{E-m_N}\,\sum_{n=1}^\infty
\,\Big( f_{J=n-\frac{1}{2}}^{(n)}(\sqrt{s}\,)-f_{J=n+ \frac{1}{2}}^{(n)}(\sqrt{s}\,)\Big)\,P'_n(\cos \theta)\;,
\nonumber\\
P'_n (\cos \theta)&=& \sum_{k=0}^{[(n-1)/2]}\,\frac{(-)^k\,(2\,n-2\,k) !}{2^n\,k !\,(n-k)
!\,(n-2\,k -1) !}\,\Big(\cos  \theta\Big)^{n-2\,k-1} \,,
\label{t-on-decomp}
\end{eqnarray}
where $[n/2]= (n-1)/2$ for $n$ odd and $[n/2]= n/2$ for $n$
even. $P'_n(z)$ is the derivative
of the Legendre polynomials. In the center of mass frame $E$ represents the nucleon
energy and $\theta$ the scattering angle:
\begin{eqnarray}
&&E=\frac{1}{2}\,\sqrt{s}+\frac{m_N^2-m_\pi^2}{2\,\sqrt{s}} \;,\;\;\;
t=(\bar q-q)^2=-2\,p_{\pi N}^2\,\Big(1-\cos \theta \Big)  \;.
\label{}
\end{eqnarray}
The unitarity condition formulated for the partial-wave amplitudes
$f_J^{(l)}$ leads to their representation in terms of the scattering phase 
shifts $\delta^{(l)}_J$
\begin{eqnarray}
&&p_{\pi N}\,f^{(l)}_{J=l\pm \frac{1}{2}}(\sqrt{s}\,) =
\frac{1}{2\,i}\left( e^{2\,i\,\delta^{(l)}_{J=l\pm \frac{1}{2}} (s)}-1 \right)
=\frac{1}{\cot \delta^{(l)}_{J=l\pm \frac{1}{2}} (s)-i}\;.
\label{}
\end{eqnarray}
One can now match the reduced amplitudes 
$M^{(\pm)}_n(s)$ of (\ref{t-sum}) and the partial-wave amplitudes 
$f^{(l)}_{J=l\pm \frac{1}{2}}(s)$
\begin{eqnarray}
f^{(l)}_{J=l\pm\frac{1}{2}}(\sqrt{s}\,) &=& \frac{p^{2\,J-1}_{\pi N}}{8\,\pi\,\sqrt{s}}
\left( \frac{\sqrt{s}}{2}+\frac{m_N^2-m_\pi^2}{2\,\sqrt{s}} \pm m_N\right)
M^{(\pm )}_{}(\sqrt{s},J-{\textstyle{1\over2}})\;.
\label{match}
\end{eqnarray}

It is instructive 
to consider the basic building block $\bar Y_n(\bar q,q;w)$ of the covariant projectors
${Y}_n^{(\pm)}(\bar q,q;w)$
in (\ref{cov-proj}) and observe the formal similarity with $P'_n(\cos
\theta )$
in (\ref{t-on-decomp}). In fact in the center of mass frame with
$w_{cm}=(\sqrt{s},0)$ one
finds $p_{\pi N}^{2\,n-2}\,P'_n(\cos \theta )=Y_n(\bar q,q;w_{cm})$.
This observation leads to a straightforward proof of (\ref{t-sum})
and (\ref{jpin-n-def}). It is sufficient to prove the orthogonality of the projectors
${Y}_n^{(\pm)}(\bar q,q;w)$ in the center of mass
frame, because the projectors are free from kinematical singularities. One readily finds
that the imaginary part of the unitary products
${Y}^{(\pm)}_n\,G\,{Y}^{(\pm)}_m$ vanish
unless both projectors are the same. It follows that the unitary product of
projectors which are expected to be orthogonal can at most be a real polynomial involving
the tadpole functions $I_\pi, I_N$ and $I^{(n)}$. Then our renormalization procedure
as described in section 3.2 leads to (\ref{t-sum}) and (\ref{jpin-n-def}). We emphasize that our argument relies 
crucially on the fact that the projectors ${Y}_n^{(\pm)}(\bar q,q;w)$ are free 
from kinematical singularities in $q$ and $\bar q$. This implies in particular that the object 
$Y_n(\bar q,q;w)$ must not be identified with $p_{\pi N}^{2\,n-2}\,P'_n(\cos \theta )$ as 
one may expect naively. 

We return to the assumption made in (\ref{k-sum}) that the interaction kernel $K$ can be 
decomposed in terms of the projectors $Y_n^{(\pm)}$. Of course this is not possible 
for a general interaction kernel $K$. We point out, however, that the on-shell equivalent 
interaction kernel $V$ in (\ref{t-eff}) can be decomposed into the $Y_n^{(\pm)}$ if 
the on-shell irreducible kernel $\bar K$ and the on-shell reducible kernels 
$K_{L,R}$ and $K_{LR}$ of (\ref{k-decomp}) are identified properly. 
The on-shell irreducible kernel $\bar K$ of (\ref{k-decomp}) is defined by
decomposing the  interaction kernel according to
\begin{eqnarray}
&&\bar K ^{(I)}(\bar q, q;w)=
\sum_{n=0}^\infty \left(\bar K^{(I)}_+(\sqrt{s};n)\,{Y}_n^{(+)}(\bar q, q;w)
+\bar K^{(I)}_-(\sqrt{s};n)\,{Y}_n^{(-)}(\bar q, q;w)\right) \;,
\nonumber\\
&&\bar K^{(I )}_\pm(\sqrt{s};n)= \int_{-1}^1 \frac{dz}{2} \,
\frac{K^{(I)}_\pm(s,t)}{\,p^{2\,n}_{\pi N}}\,P_n(z)
\nonumber\\
&&\;\;\;\;\;\;\;\;\;\;\;\;\;\;\;\;+ \int_{-1}^1 \frac{dz}{2} \,
\left( \frac{1}{2}\,\sqrt{s}+\frac{m_N^2-m_\pi^2}{2\,\sqrt{s}}\mp m_N\right)^2
\frac{K^{(I)}_\mp(s,t)}{p^{2\,n+2}_{\pi N}}\,P_{n+1}(z) \;,
\label{bark-def-pin}
\end{eqnarray}
where $t= -2\,p_{\pi N}^2\,(1-x)$ and $K_\pm^{(I)}(s,t)$ follows from the decomposition of the 
interaction kernel $K$:
\begin{eqnarray}
&&K^{(I)}(\bar q, q; w) \Big|_{\rm{on-shell}}=
\frac{1}{2}\,\Bigg(\frac{\wslash}{\sqrt{w^2}}+1
\Bigg)\, K_+^{(I)}(s,t) + \frac{1}{2}\,\Bigg(\frac{\wslash}{\sqrt{w^2}}-1
\Bigg)\,K_-^{(I)}(s,t) \;.
\label{}
\end{eqnarray}
Then $K-\bar K $ is on-shell
reducible by construction and therefore can be decomposed into $K_L,K_R,K_{LR}$.
Note that it does not yet follow that the induced effective interaction $V$ can 
be decomposed into the $Y_n^{(\pm)}$. This may need an iterative procedure in particular 
when the interaction kernel shows non-local structures induced for example by a $t$-channel
meson-exchange. The starting point of the iteration is given with $ K_0 = K$ and 
$V_0 = \bar K  $ as defined via (\ref{bark-def-pin}). Then $K_{n+1} = V[K_n]$,   
where $V[K_n]$ is defined in (\ref{v-eff}) with respect to 
$\bar K_n$ as given in (\ref{bark-def-pin}). The effective 
interaction $V$ is then identified with 
$V=^{\rm \;\;\;\,lim}_{\;n \to \infty}\,  \bar K_n $. 
In our work we will not encounter this complication, because the effective interaction kernel 
is treated to leading orders of chiral perturbation theory.

\section{$SU(3)$ coupled-channel dynamics}

The Bethe-Salpeter equation (\ref{BS-eq}) is readily generalized for a coupled-channel 
system. The chiral  $SU(3)$ Lagrangian with baryon octet and
pseudo-scalar meson octet couples the $\bar K N$ system  to five inelastic channels
$\pi \Sigma $, $\pi \Lambda $, $\eta \Lambda$, $\eta \Sigma$ and $K \Xi $ and the 
$\pi N$ system to the three channels $K \Sigma$, $\eta N$ and $K\Lambda$. The strangeness
plus one sector with the $K N$ channel is treated separately in the next  
section when discussing constraints from crossing symmetry. For simplicity we assume in the 
following discussion good isospin symmetry. Isospin symmetry 
breaking effects are considered in Appendix D. In order to establish our convention
consider for example the two-body meson-baryon interaction terms in (\ref{two-body}).
They can be rewritten in the following form
\begin{eqnarray}
{\mathcal L}(\bar k ,k ;w)&=&
\sum_{I=0,\frac{1}{2},1,\frac{3}{2}}\,R^{(I)\,\dagger }(\bar q,\bar p)\,\gamma_0
\,K^{(I)}(\bar k ,k ;w )\,R^{(I)}(q,p) \,,\;\;\;\;
\nonumber\\
R^{(0)}&=& \left(
\begin{array}{c}
\textstyle{1\over\sqrt{2}}\,K^\dagger\,N \\
\textstyle{1\over\sqrt{3}}\,\vec{\pi}_c \; \vec{\Sigma} \\
\eta_c \,\Lambda\\
\textstyle{1\over\sqrt{2}}\,K^t\,i\,\sigma_2\,\Xi
\end{array}
\right) \;,\;\;\;
\vec{R}^{(1)}= \left(
\begin{array}{c}
\textstyle{1\over\sqrt{2}}\,K^\dagger\,\vec{\sigma}\,N \\
\textstyle{1\over i\sqrt{2}}\,\vec{\pi}_c \, \times \vec{\Sigma} \\
\vec{\pi}_c \,\Lambda\\
\eta_c \,\vec{\Sigma} \\
\textstyle{1\over\sqrt{2}}\,K^t\,i\,\sigma_2\,\vec{\sigma}\,\Xi
\end{array}
\right) \;,
\nonumber\\
R^{(\frac{1}{2})} &=&
\left(
\begin{array}{c}
\textstyle{1\over\sqrt{3}}\,\pi_{c} \cdot \sigma\,N \\
\textstyle{1\over\sqrt{3}}\,\Sigma \cdot \sigma  \,K \\
\eta_c \,N \\
K\,\Lambda 
\end{array}
\right) \;,\;\;\;\;
R^{(\frac{3}{2})}= \left(
\begin{array}{c}
\pi_c \cdot S\,N \\
\Sigma \cdot S \,K 
\end{array}
\right) \;,
\label{r-def}
\end{eqnarray}
where $R(q,p)$ in (\ref{r-def}) is defined by $R(q,p)= \int d^4x \, d^4y\,e^{-i\,q\,x-i\,p\,y}\,\Phi(x)\,B(y)$, 
and $\Phi(x)$ and $B(y)$ denoting the meson and baryon fields respectively. In (\ref{r-def}) we decomposed 
the pion field $\vec \pi = \vec \pi_c+\vec \pi_c^\dagger $ and the 
eta field $\eta = \eta_c+\eta_c^\dagger$\footnote{
For a neutral scalar field $\phi(x)=\phi_c(x)+\phi^\dagger_c(x)$ with mass $m$ we write
$$\phi_c(0,\vec x)= \int \frac{d^3 k}{(2\pi)^3} \,\frac{e^{i\,\vec k \cdot \vec x}}{2\,\omega_k}\,a(\vec k)
\;, \qquad \phi^\dagger_c(0,\vec x)= \int \frac{d^3 k}{(2\pi)^3} \,
\frac{e^{-i\,\vec k \cdot \vec x}}{2\,\omega_k}\,a^\dagger(\vec k)\,,$$
where $\omega_k=(m^2+\vec k^2\,)^{\frac{1}{2}}$ and 
$[a(k),a^\dagger(k')]_-= (2\pi)^3\,2\,\omega_k\,\delta^3(k-k')$.  
In (\ref{r-def}) we suppress terms which do not contribute to the two-body scattering
process at tree-level. For example terms like $\bar N\,\eta_c \,N\,\eta_c$ 
or $\bar N \,\eta_c^\dagger\,N\,\eta_c^{\dagger }$ are dropped.}. Also we apply the isospin decomposition 
of (\ref{field-decomp}). The isospin 
$1/2$ to $3/2$ transition matrices 
$S_i$ in (\ref{r-def}) are normalized by $S^\dagger_i\,S_j=\delta_{ij}-\sigma_i\,\sigma_j/3$. 
The Lagrangian density ${\mathcal L}(x)$ in coordinate space
is related to its momentum space representation through 
\begin{eqnarray}
\int d^4x\,{\mathcal L}(x) &=& \int \frac{d^4k}{(2\,\pi)^4}\,
\frac{d^4\bar k}{(2\,\pi)^4}\,\frac{d^4w}{(2\,\pi)^4}\,
{\mathcal L}(\bar k ,k ;w)\;.
\label{lag-momentum}
\end{eqnarray}
The merit of the notation (\ref{r-def}) is threefold. Firstly, the phase convention for the
isospin states is specified. Secondly, it defines the convention for the interaction kernel 
$K$ in the Bethe-Salpeter equation. Last it provides also a convenient scheme to  read 
off the isospin decomposition for the interaction kernel $K$ directly 
from the interaction Lagrangian (see Appendix A). The coupled channel Bethe-Salpeter matrix equation reads
\begin{eqnarray}
T^{(I)}_{ab}(\bar k ,k ;w ) &=& K^{(I)}_{ab}(\bar k ,k ;w )
+\sum_{c,d}\int\!\! \frac{d^4l}{(2\pi)^4}\,K^{(I)}_{ac}(\bar k , l;w )\,
G^{(I)}_{cd}(l;w)\,T^{(I)}_{db}(l,k;w )\;,
\nonumber\\
G^{(I)}_{cd}(l;w)&=&-i\,D_{\Phi(I,d)}(\half\,w-l)\,S_{B(I,d)}(
\half\,w+l)\,\delta_{cd} \,,
\label{BS-coupled}
\end{eqnarray}
where $D_{\Phi(I,d)}(q)$ and $S_{B(I,d)}(p)$ denote the
meson propagator and baryon propagator respectively for a given channel $d$ with
isospin $I$. The matrix structure of the coupled-channel interaction kernel $K_{ab}(\bar k ,k ;w )$ is defined
via (\ref{r-def}) and
\begin{eqnarray}
&&\Phi(0,a)=(\bar K,\pi,\eta , K)_a\;,\;\;\;\;\;\;\ B(0,a)=(N,\Sigma,\Lambda,\Xi)_a\;,
\nonumber\\
&&\Phi(1,a)=(\bar K,\pi,\pi,\eta , K)_a\;,\;\;\ B(1,a)=(N,\Sigma,\Lambda,\Sigma ,\Xi)_a\;,
\nonumber\\
&&\Phi({\textstyle{1\over 2}},a)=(\pi,K, \eta, K)_a\;,\;\;\;\;\;\;\ 
B({\textstyle{1\over 2}},a)=(N,\Sigma,N, \Lambda )_a\;,
\nonumber\\
&&\Phi({\textstyle{3\over 2}},a)=(\pi,K)_a\;,\;\;\qquad  \quad \;\;
B({\textstyle{3\over 2}},a)=(N,\Sigma )_a\;.
\label{def-channel}
\end{eqnarray}

We proceed and identify the on-shell equivalent coupled channel interaction kernel $V$ 
of (\ref{v-eff}). At leading chiral orders it is legitimate to identify $V^{(I)}_{ab}$ with
$\bar K^{(I)}_{ab}$ of (\ref{k-decomp}), because the loop corrections in (\ref{v-eff}) are 
of minimal chiral power $Q^3$ (see (\ref{q-rule})). To chiral order $Q^3$ the interaction 
kernel receives additional terms from one loop diagrams involving the on-shell reducible 
interaction kernels $K_{L,R}$ as well as from irreducible one-loop diagrams. In the notation 
of (\ref{k-decomp}) one finds
\begin{eqnarray}
V = \bar K + K_R \cdot G \cdot \bar K +\bar K\,\cdot G \cdot K_L +
K_R \cdot G \cdot K_L +{\mathcal O}\left(Q^4 \right) \;. 
\label{v-idef}
\end{eqnarray}
Typically the $Q^3$ terms induced by $K_{L,R}$ in (\ref{v-idef}) are tadpoles 
(see e.g. (\ref{pin-example},\ref{wt-tadpole})). The only non-trivial contribution arise 
from the on-shell reducible parts of the u-channel baryon octet terms. However, by construction,  
those contributions have the same form as the irreducible $Q^3$ loop-correction terms of 
$\bar K$. In particular they do not show the typical enhancement factor of $2 \pi$ associated 
with the s-channel unitarity cuts. In the large $N_c$ limit all loop correction terms to 
$V$ are necessarily suppressed by $1/N_c$. This follows, because any hadronic loop 
function if visualized in terms of quark-gluon diagrams involves at least one quark-loop, 
which in turn is $1/N_c$ suppressed \cite{Hooft,Witten}. Thus it is legitimate to take  
$V = \bar K$ in this work.

We do include those correction terms of suppressed order in the $1/N_c$ 
expansion which are implied by the physical baryon octet and decuplet exchange contributions.
Individually the baryon exchange diagrams are of forbidden order $N_c$. 
Only the complete large $N_c$ baryon ground state multiplet with 
$J=({\textstyle {1 \over 2}}, ...,{\textstyle {N_c\over 2}} )$ leads to an exact cancellation 
and a scattering amplitude of order $N_c^0$ \cite{Manohar}. However this cancellation persists 
only in the limit of degenerate baryon octet $\m0_{[8]}$ and decuplet mass $\m0_{[10]}$. 
With $m_\pi < \m0_{[10]}-\m0_{[8]}$ the cancellation is incomplete and thus leads to an 
enhanced sensitivity of the scattering amplitude to the physical baryon-exchange contributions.
Therefore one should sum the $1/N_c$ suppressed contributions of the form 
$(\m0_{[10]}-\m0_{[8]})^n/m_\pi^n \sim N_c^{-n}$. We take this into account by evaluating the 
baryon-exchange contributions to subleading chiral orders but avoid the expansion in either of  
$(\m0_{[10]}-\m0_{[8]})/m_\pi $ or $m_\pi/(\m0_{[10]}-\m0_{[8]})$. We also include the $SU(3)$ 
symmetry-breaking counter terms of the 3-point meson-baryon vertices and the quasi-local 
two-body counter terms of chiral order $Q^3$ which are leading in the $1/N_c$ expansion. 
Note that the quasi-local two-body counter terms of large chiral order 
are not necessarily suppressed by $1/N_c$ relatively to the terms of 
low chiral order. This is plausible, because for example a t-channel 
vector-meson exchange, which has a definite large $N_c$ scaling behavior, leads to 
contributions in all partial waves. Thus quasi-local counter terms with different partial-wave 
characteristics may have identical large $N_c$ scaling behavior even though they 
carry different chiral powers. Finally we argue that it is justified to perform the 
partial $1/N_c$ resummation of all 
reducible diagrams implied by solving the Bethe-Salpeter equation (\ref{t-eff}). In section 3 
we observed that reducible diagrams are typically enhanced close to their unitarity threshold.
The typical enhancement factor of $2 \pi$ per unitarity cut, measured relatively to irreducible 
diagrams (see (\ref{wt-square-KN})), is larger than the number of colors $N_c=3$ of our world.

By analogy with (\ref{t-sum}) the coupled-channel scattering amplitudes 
$T^{(I)}_{ab} $ are decomposed into their on-shell equivalent 
partial-wave amplitudes $M^{(I,\pm)}_{ab} $
\begin{eqnarray}
\bar T^{(I)}_{ab}(\bar k,k; w)&=& \sum_{n=0}^\infty\,M^{(I,+)}_{ab}(\sqrt{s};n)\,{Y}_n^{(+)}(\bar q, q;w)
\nonumber\\
&+&\sum_{n=0}^\infty\,M^{(I,-)}_{ab}(\sqrt{s};n)\,{Y}_n^{(-)}(\bar q, q;w) \;,
\label{tttt}
\end{eqnarray}
where $k=\! {\textstyle{1\over 2}}\,w -q$ and 
$\bar k=\! {\textstyle{1\over 2}}\,w -\bar q$ and $s= w_\mu\, w^\mu$. The covariant 
projectors ${Y}_n^{(\pm )}(\bar q, q;w)$ were defined in (\ref{cov-proj}). 
Expressions for the differential cross sections given in 
terms of the partial-wave amplitudes $M_{ab}^{(\pm)}$ can be found in Appendix F.
The form of the scattering amplitude (\ref{tttt}) follows, because 
the effective interaction kernel $V^{(I)}_{ab}$ of (\ref{v-eff}) is decomposed 
accordingly 
\begin{eqnarray}
V^{(I)}_{ab}(\bar k,k;w)&=&
\sum_{n=0}^\infty\,V^{(I,+)}_{ab}(\sqrt{s};n)\,{Y}_n^{(+)}(\bar q, q;w)
\nonumber\\
&+&\sum_{n=0}^\infty\,V^{(I,-)}_{ab}(\sqrt{s};n)\,{Y}_n^{(-)}(\bar q, q;w) \;.
\label{not-def}
\end{eqnarray}
The coupled-channel Bethe-Salpeter equation (\ref{BS-coupled}) reduces to 
a convenient matrix equation for the effective interaction kernel $V^{(I)}_{ab}$
and the invariant amplitudes $M^{(I,\pm)}_{ab} $
\begin{eqnarray}
M^{(I,\pm)}_{ab}(\sqrt{s};n) &=&  V^{(I,\pm )}_{ab}(\sqrt{s};n) 
\nonumber\\
&+& \sum_{c,d}
V^{(I,\pm)}_{ac}(\sqrt{s};n)\,J^{(I,\pm )}_{cd}(\sqrt{s};n)\,M^{(I,\pm)}_{db}(\sqrt{s};n) \;,
\label{}
\end{eqnarray}
which is readily solved with:
\begin{eqnarray}
M^{(I,\pm)}_{ab}(\sqrt{s};n)&=& \Bigg[\left( 1- V^{(I,\pm )}_{}(\sqrt{s};n)\,J^{(I,\pm )}_{}(\sqrt{s};n)\right)^{-1} 
V^{(I,\pm )}_{}(\sqrt{s};n)\Bigg]_{ab} \,.
\label{}
\end{eqnarray}
It remains to specify the coupled-channel loop matrix function $J^{(I,\pm )}_{ab} \!\! \sim \delta_{a b }$, which 
is diagonal in the coupled-channel space. We write
\begin{eqnarray}
&& J^{(I,\pm )}_{aa}(\sqrt{s};n)=
\left( \frac{\sqrt{s}}{2}+ \frac{m_{B(I,a)}^2
-m_{\Phi(I,a)}^2}{2\,\sqrt{s}}\pm m_{B(I,a)} \right)
\Delta I^{(k)}_{\Phi(I,a)\,B(I,a)}(\sqrt{s}\, )
\nonumber\\
&& \qquad \qquad \qquad \! \times  
\left(\frac{s}{4}-\frac{m_{B(I,a)}^2+m_{\Phi(I,a)}^2}{2}
+\frac{\big(m_{B(I,a)}^2-m_{\Phi(I,a)}^2\big)^2}{4\,s} \right)^n
\;,
\nonumber\\
&& \Delta I^{(k)}_{\Phi(I,a)\,B(I,a)}(\sqrt{s}\,) = I_{\Phi(I,a)\,B(I,a)}
(\sqrt{s}\,)
\nonumber\\
&& \qquad \qquad \qquad \quad -
\sum_{l=0}^k\,\frac{1}{l}\left(\frac{\partial }{\partial \,\sqrt{s}} \right)^l\,
\Bigg|_{\sqrt{s}= \mu_S}\,I_{\Phi(I,a)\,B(I,a)}(\sqrt{s}\,)\;.
\label{result-loop:ab}
\end{eqnarray}
The index $a$ labels a specific channel consisting of a  meson-baryon pair of given 
isospin ($\Phi(I,a)$, $B(I,a)$). In particular $m_{\Phi(I,a)}$ and $m_{B(I,a)}$ denote 
the empirical isospin averaged meson and baryon octet mass respectively.
The scalar master-loop integral, $I_{\Phi(I,a)\,B(I,a)}(\sqrt{s}\,)$,
was introduced explicitly in (\ref{ipin-analytic}) for the pion-nucleon channel. 
Note that we do not use the expanded form of (\ref{ipin-expand}).
The subtraction point $\mu_S^{(I)}$ is identified with $\mu_S^{(0)}\!=\!m_\Lambda$ and 
$\mu_S^{(1)}\!=\!m_\Sigma$ in the isospin zero and isospin one channel respectively so as to 
protect the hyperon s-channel pole structures. Similarly we use 
$\mu_S^{(\frac{1}{2})}=\mu_S^{(\frac{3}{2})}=m_N$ in the pion-nucleon sector. 
We emphasize that in the p-wave loop functions $J^{(-)}(\sqrt{s},0)$ and 
$J^{(+)}(\sqrt{s},1)$ we perform a double subtraction of the internal master-loop 
function with $k=1$ in (\ref{result-loop:ab}) 
whenever a large $N_c$ baryon ground state manifests itself with a 
s-channel pole contribution in the associated partial-wave scattering amplitude.
In all remaining channels we use $k=0$ in (\ref{result-loop:ab}).
This leads to consistency with the renormalization condition
(\ref{ren-cond}), in particular in the large $N_c$ limit with $m_{[8]}=m_{[10]}$.

We proceed by identifying the on-shell irreducible interaction kernel $\bar K^{(I)}_{ab}$
in (\ref{v-idef}).
The result (\ref{bark-def-pin}) is generalized to 
the case of inelastic scattering. This leads to the matrix structure of the interaction 
kernel $K_{ab}$ defined in (\ref{r-def}). In a given partial wave the effective interaction kernel
$\bar K^{(I,\pm )}_{ab}(\sqrt{s};n)$ reads
\begin{eqnarray}
&&\bar K ^{(I)}_{ab}=
\sum_{n=0}^\infty \left(\bar K^{(I,+)}_{ab}(\sqrt{s};n)\,{Y}_n^{(+)}(\bar q, q;w)
+\bar K^{(I,-)}_{ab}(\sqrt{s};n)\,{Y}_n^{(-)}(\bar q, q;w)\right) \;,
\nonumber\\
&&\bar K^{(I,\pm)}_{ab}(\sqrt{s};n)= \int_{-1}^1 \frac{dz}{2} \,
\frac{K^{(I,\pm)}_{ab}(s,t^{(I)}_{ab})}{\big( p^{(I)}_{a}\,
p^{(I)}_{b}\big)^{n}}\,P_n(z)
\nonumber\\
&&\;\;\;\;\;\;\;\;\;\;\;+\int_{-1}^1 \frac{dz}{2} \,
\,\Big(E^{(I)}_a\mp m_{B(I,a)} \Big) \,\Big(E^{(I)}_b \mp m_{B(I,b)} \Big)
\frac{K^{(I,\mp)}_{ab}(s,t^{(I)}_{ab})}{\big( p^{(I)}_{a}\,
p^{(I)}_{b}\big)^{n+1}}\,P_{n+1}(z)\;,
\label{bark-def}
\end{eqnarray}
where $P_n(z)$ are the Legendre polynomials and 
\begin{eqnarray}
&&t^{(I)}_{ab}= m_{\Phi(I,a)}^2+m_{\Phi(I,b)}^2 -2\,\omega^{(I)}_a\,\omega^{(I)}_b
+2\,p^{(I)}_a\,p^{(I)}_b\,z \;,\;\;\;
E^{(I)}_a=\sqrt{s}-\omega^{(I)}_a\;,
\nonumber\\
&&\omega^{(I)}_a=\frac{s+m^2_{\Phi(I,a)}-m^2_{B(I,a)}}{2\,\sqrt{s}}\;,\;\;\;
\Big(p^{(I)}_{a}\Big)^2 =\Big(\omega^{(I)}_a\Big)^2-m_{\Phi(I,a)}^2\;.
\label{}
\end{eqnarray}
The construction of the on-shell irreducible interaction kernel requires the identification of the 
invariant amplitudes $K_{ab}^{(I,\pm )}(s,t) $ in a given channel $ab$:
\begin{eqnarray}
&&K^{(I)}_{ab} \Big|_{\rm{on-sh.}}
= \frac{1}{2}\,\Bigg(\frac{\wslash}{\sqrt{w^2}}+1
\Bigg)\, K_{ab}^{(I,+)}(s,t) + \frac{1}{2}\,\Bigg(\frac{\wslash}{\sqrt{w^2}}-1
\Bigg)\,K_{ab}^{(I,-)}(s,t) \;.
\label{}
\end{eqnarray}
Isospin breaking effects are easily incorporated by constructing super matrices 
$V^{(I' I)}$, $J^{(I'I)}$ and $T^{(I'I)}$ which couple different isospin  states. 
We consider isospin breaking effects induced by isospin off-diagonal loop function 
$J^{(I'I)}$ but impose $V^{(I'I)}\sim \delta_{I'I}$. For technical details we refer 
to Appendix D.

We underline that our approach deviates here from the common chiral expansion scheme as implied by 
the heavy-fermion representation of the chiral Lagrangian. A strict chiral expansion of the 
unitarity loop function $I_{\Phi(I,a)\,B(I,a)}(\sqrt{s}\,)$ does not reproduce the correct 
s-channel unitarity cut. One must perform an infinite summation of interaction terms in 
the heavy-fermion chiral Lagrangian to recover the correct threshold behavior. 
This is achieved more conveniently  by working directly with the manifest relativistic scheme,  
where it is natural to write down the loop functions in terms of the physical masses. 
In this work we express results systematically in terms of physical parameters avoiding the 
use of bare parameters like $\m0_{[8]}$ whenever possible. 

\subsection{Construction of effective interaction kernel}

We collect all terms of the chiral Lagrangian contributing to order
$Q^3$ to the interaction kernel $K^{(I)}_{ab}(\bar k ,k ;w )$ of (\ref{BS-coupled}):
\begin{eqnarray}
K^{(I)}(\bar k,k;w) &=& K^{(I)}_{WT}(\bar k,k;w)+K^{(I)}_{s-[8]}(\bar k,k;w)
+K^{(I)}_{u-[8]}(\bar k,k;w)
\nonumber\\
&+&K^{(I)}_{s-[10]}(\bar k,k;w)
+K^{(I)}_{u-[10]}(\bar k,k;w)+K^{(I)}_{s-[9]}(\bar k,k;w)
\nonumber\\
&+&K^{(I)}_{u-[9]}(\bar k,k;w)
+K^{(I)}_{[8][8]}(\bar k,k;w)+K^{(I)}_{\chi}(\bar k,k;w) \,. 
\label{k-all}
\end{eqnarray}
The various contributions to the interaction kernel will be  written in a form 
which facilitates the derivation of $K^{(I,\pm)}_{ab}(s,t)$ in (\ref{bark-def}). It is then 
straightforward to derive the on-shell irreducible interaction kernel $V$.

We begin with a discussion of the Weinberg-Tomozawa interaction term 
$K^{(I)}_{WT}$ in (\ref{lag-Q}). To chiral order 
$Q^3 $ it is necessary to consider the effects of the baryon wave-function 
renormalization factors ${\mathcal Z}$ and of the further counter terms 
introduced in (\ref{chi-sb-4}). We have
\begin{eqnarray}
\Big[\,K^{(I)}_{W\,T\,}(\bar k,k;w)\Big]_{ab}&=&
\Big[C_{WT}^{(I)}\Big]_{ab}\,
\Big(1+{\textstyle{1\over 2}}\,\Delta \zeta_{B(I,a)}
+{\textstyle{1\over 2}}\,\Delta \zeta_{B(I,b)} \Big)
\,\frac{\barqslash+\qslash}{4\,f^2}\;.
\label{WT-k}
\end{eqnarray}
The dimensionless coefficient matrix
$C_{WT}^{(I)}$ of the Weinberg-Tomozawa interaction is given in Tab. \ref{tabpi-1} for the 
strangeness zero channels and in the Appendix E for the strangeness minus one channels. 
The  constants $\Delta \zeta_{B(c)}$ in (\ref{WT-k}) receive 
contributions from the baryon octet wave-function renormalization factors ${\mathcal Z}$ and 
the parameters $\zeta_0,\zeta_D$ and $\zeta_F$ of (\ref{chi-sb-4}):
\begin{eqnarray}
&&\Delta \zeta = \zeta +{\mathcal Z}-1 +\cdots \,,
\nonumber\\
&&\zeta_N = \big( \zeta_0+2\,\zeta_F  \big)\, m_\pi^2 
+ \big( 2\,\zeta_0 +2\,\zeta_D-2\,\zeta_F\big) \,m_K^2 \;,
\nonumber\\
&& \zeta_\Lambda =\big( \zeta_0-{\textstyle{2\over 3}}\,\zeta_D  \big)\, m_\pi^2 
+ \big( 2\,\zeta_0 +{\textstyle{8\over 3}}\,\zeta_D\big) \,m_K^2 \;,
\nonumber\\
&& \zeta_\Sigma =\big( \zeta_0+2\,\zeta_D  \big)\, m_\pi^2 + 2\,\zeta_0  \,m_K^2 \;,
\nonumber\\
&& \zeta_\Xi = \big( \zeta_0-2\,\zeta_F  \big)\, m_\pi^2 
+ \big( 2\,\zeta_0 +2\,\zeta_D+2\,\zeta_F\big) \,m_K^2 \;.
\label{zeta-par}
\end{eqnarray}
The dots in (\ref{zeta-par}) represent corrections terms of order 
$Q^3$ and further contributions from irreducible one-loop diagrams not considered here.
We point out that the coupling constants $\zeta_0$, $\zeta_D$ and $\zeta_F$, which 
appear to renormalize the strength of the Weinberg-Tomozawa interaction term, also 
contribute to the baryon wave-function factor ${\mathcal Z}$ as is 
evident from (\ref{chi-sb}). In fact, it will be demonstrated that the explicit and 
implicit dependence via the wave-function renormalization factors ${\mathcal Z}$ cancel 
identically at leading order. Generalizing (\ref{def-zn}) we derive the 
wave-function renormalization constants for the $SU(3)$ baryon octet fields
\begin{eqnarray}
&& {\mathcal Z}^{-1}_{B(c)}-1 = \zeta_{B(c)}
-\frac{1}{4\,f^2}\,\sum_{a} \,\xi_{B(I,a)}\,m_{\Phi(I,a)}^2 
\left( G^{(B(c))}_{\Phi (I,a)\, B(I,a)} \right)^2 \,,
\label{zeta-result}\\
&& \xi_{B(I,a)}=  2\,m_{B(c)}\,
\frac{\partial \,I_{\Phi (I,a)\, B(I,a)}(\sqrt{s}\,)}{\partial \sqrt{s}} 
\Bigg|_{\sqrt{s}=m_{B(c)}} -\frac{I_{\Phi (I,a)}}{m_{\Phi(I,a)}^2 } 
+{\mathcal O}\left( Q^4\right)\,,
\nonumber
\end{eqnarray}
where the sum in (\ref{zeta-result}) includes all $SU(3)$ channels as listed in (\ref{r-def}). 
The result (\ref{zeta-result}) is  expressed in terms of the dimension less
coupling constants $G_{\Phi B}^{(B)}$. For completeness we collect here all 
required 3-point coupling coefficients:
\begin{eqnarray}
&&G_{\pi N}^{(N)} = \sqrt{3}\,\Big( F+D \Big) \;, \quad 
G_{K \,\Sigma}^{(N)} = -\sqrt{3}\,\Big( F-D \Big) \;,\quad 
G_{\eta \,N}^{(N)} = {\textstyle{1\over\sqrt{3}}}\,\Big( 3\,F-D \Big)\;,
\nonumber\\
&& G_{K \,\Lambda}^{(N)} = -{\textstyle{1\over\sqrt{3}}}\,\Big( 3\,F+D \Big) \;,\quad 
G_{\bar K N}^{(\Lambda)} = \sqrt{2}\,G_{K \,\Lambda}^{(N)} 
\;, \quad G_{\pi \Sigma}^{(\Lambda)} = 2\,D \;,
\nonumber\\
&& G_{\eta \,\Lambda}^{(\Lambda)} = -{\textstyle{2\over\sqrt{3}}}\,D \;,\quad 
G_{K \Xi}^{(\Lambda)} = -\sqrt{{\textstyle{2\over 3}}}\,\Big( 3\,F-D\Big)  \;,
\quad  G_{\bar K N}^{(\Sigma)} = 
\sqrt{{\textstyle{2\over 3}}}\,G_{K \Sigma}^{(N)}\;, 
\nonumber\\
&& G_{\pi \Sigma}^{(\Sigma)} = -\sqrt{8}\,F \;, \quad 
G_{\pi \Lambda}^{(\Sigma)} = 
{\textstyle{1\over\sqrt{3}}}\,G_{\pi \Sigma}^{(\Lambda)} \;,\quad
G_{\eta \Sigma }^{(\Sigma)} = {\textstyle{2\over\sqrt{3}}}\,D\;, 
\nonumber\\
&&G_{K\Xi}^{(\Sigma)} = \sqrt{2}\,\Big( F+D\Big) \;,\quad 
G_{\bar K \Lambda}^{(\Xi)} = -{\textstyle{1\over\sqrt{2}}}\,
G_{K \Xi}^{(\Lambda)}\;, \quad 
G_{\bar K \Sigma}^{(\Xi)} = -\sqrt{{\textstyle{3\over2}}}\,
G_{K \Xi}^{(\Sigma)}\;,
\nonumber\\
&&G_{\eta \,\Xi}^{(\Xi)} = -{\textstyle{1\over\sqrt{3}}}\,\Big( 3\,F+D \Big)\;,\quad 
G_{\pi \,\Xi}^{(\Xi)} = \sqrt{3}\,\Big( F-D \Big) \;,
\label{G-explicit}
\end{eqnarray}
at leading chiral order $Q^0$. The loop function $I_{\Phi (I,a)\, B(I,a)}(\sqrt{s}\,)$ and 
the mesonic tadpole term $I_{\Phi(I,a)}$ are the obvious generalizations of 
$I_{\pi N}(\sqrt{s}\,)$ and $I_\pi$ introduced in (\ref{ipin-analytic}). 
We emphasize that the contribution $\zeta_{B(c)}$ in (\ref{zeta-result}) is identical
to the corresponding contribution in $ \Delta \zeta_{B(c)}$. 
Therefore, if all one-loop effects are dropped, one finds the result
\begin{eqnarray}
{\mathcal Z}^{-1}=1 +\zeta \;, \qquad 
\Delta \zeta =0 +{\mathcal O} \left(Q^3, \frac{1}{N_c}\,Q^2 \right) \;.
\label{delta-zeta}
\end{eqnarray}
Given our renormalization condition (\ref{ren-cond}) that requires the 
baryon s-channel pole contribution to be represented by the renormalized effective 
potential $V_R$, we conclude that (\ref{delta-zeta}) holds in our approximation. Note 
that the loop correction to ${\mathcal Z}$ as given in (\ref{zeta-result}) must be 
considered as part of the renormalized effective potential $V_R$ and consequently  
should be dropped as they are suppressed by the factor $Q^3/N_c$.  
It is evident that the one-loop contribution to the baryon ${\mathcal Z}$-factor 
is naturally moved into $V_R$ by that double subtraction with $k=1$ in 
(\ref{result-loop:ab}), as explained previously. 

\tabcolsep=0.7mm
\renewcommand{\arraystretch}{1.45}
\begin{table}[t]
\begin{tabular}{|r||c||c|c|c||c|c||c|c|c|c||c|c|c|c|}
 \hline   
& $ C_{WT}^{(\frac{1}{2})} $  &  $ C_{N_{[8]}}^{(\frac{1}{2})}$  & $C_{\Lambda_{[8]}}^{(\frac{1}{2})}$  &  
$C_{\Sigma_{[8]}}^{(\frac{1}{2})}$ 
& $C_{\Delta_{[10]}}^{(\frac{1}{2})}$ & $C_{\Sigma_{[10]} }^{(\frac{1}{2})}$ 
    &$\widetilde{C}_{N_{[8]}}^{(\frac{1}{2})}$ & $\widetilde{C}_{\Lambda_{[8]} }^{(\frac{1}{2})}$ &
    $\widetilde{C}_{\Sigma_{[8]}}^{(\frac{1}{2})}$ & $\widetilde {C}_{\Xi_{[8]} }^{(\frac{1}{2})}$   &
$\widetilde {C}_{\Delta_{[10]} }^{(\frac{1}{2})}$ &
    $\widetilde {C}_{\Sigma_{[10]}}^{(\frac{1}{2})}$& $\widetilde C_{\Xi_{[10]}}^{(\frac{1}{2})}$
\\ \hline \hline 
$11$& 2 & 1& 0  & $0$ & 0 & 0  & -$\frac{1}{3}$ &0  & 0 &0 &$\frac{4}{3}$ &0 & 0\\ \hline 
$12$& $\frac{1}{2}$ & 1& 0  & $0$ & 0 & 0  &0  &$\frac{1}{\sqrt{6}}$  & -1 &0 &0 &-1 & 0\\ \hline 
$13$& 0 & 1 & 0  &  0  & 0 & 0  & 1  &0  & 0 &0 &0 &0 & 0\\ \hline 
$14$& -$\frac{3}{2}$ & 1& 0 & $0$  & 0 & 0  &0  & 0 & $\sqrt{\frac{3}{2}}$ &0 &0&$\sqrt{\frac{3}{2}}$  & 0\\ \hline \hline

$22$& 2 & 1& 0  & $0$  & 0 & 0  &0  &0  & 0 &-$\frac{1}{3}$ &0 &0 & -$\frac{1}{3}$\\ \hline 
$23$& -$\frac{3}{2}$& 1 & 0  & $0$  & 0& 0  &0  &0  & $\sqrt{\frac{3}{2}}$ &0 &0 &$\sqrt{\frac{3}{2}}$ & 0\\ \hline 
$24$& 0 & 1 & 0  & $0$ & 0 & 0  &0  &0  & 0 &-1 &0 & 0 & -1\\ \hline  \hline

$33$& 0 & 1 & 0  & 0  & 0& 0  &1  &0  & 0 &0 &0 &0 & 0\\ \hline 
$34$& -$\frac{3}{2}$ & 1 & 0 & $0$ & 0 & 0  &0  &$\frac{1}{\sqrt{2}}$  & 0 &0 &0 &0 & 0\\ \hline \hline

$44$& 0& 1  & 0  & $0$  & 0 & 0  &0  &0  & 0 &1 &0 &0 & 1\\ \hline \hline

\hline   
& $ C_{WT}^{(\frac{3}{2})} $  &  $ C_{N_{[8]}}^{(\frac{3}{2})}$  & $C_{\Lambda_{[8]}}^{(\frac{3}{2})}$  &  
$C_{\Sigma_{[8]}}^{(\frac{3}{2})}$ 
& $C_{\Delta_{[10]}}^{(\frac{3}{2})}$ & $C_{\Sigma_{[10]} }^{(\frac{3}{2})}$ 
    &$\widetilde{C}_{N_{[8]}}^{(\frac{3}{2})}$ & $\widetilde{C}_{\Lambda_{[8]} }^{(\frac{3}{2})}$ &
    $\widetilde{C}_{\Sigma_{[8]}}^{(\frac{3}{2})}$ & $\widetilde {C}_{\Xi_{[8]} }^{(\frac{3}{2})}$   &
$\widetilde {C}_{\Delta_{[10]} }^{(\frac{3}{2})}$ &
    $\widetilde {C}_{\Sigma_{[10]}}^{(\frac{3}{2})}$& $\widetilde C_{\Xi_{[10]}}^{(\frac{3}{2})}$
\\ \hline \hline 
$11$& -1 & 0& 0  & 0 & 1 & 0  &$\frac{2}{3}$  &0  & 0 &0 &$\frac{1}{3}$ &0 & 0\\ \hline 
$12$& -1 & 0& 0  & 0 & 1 & 0  &0  &$\frac{1}{\sqrt{6}}$  & $\frac{1}{2}$ &0 &0 &$\frac{1}{2}$ & 0\\ \hline 
\hline

$22$& -1 & 0 & 0  & 0 & 1 & 0  &0  & 0  & 0 & $\frac{2}{3}$ &0 &0 & $\frac{2}{3}$\\ \hline 
\end{tabular}
\vspace*{2mm} \caption{Weinberg-Tomozawa interaction strengths and baryon exchange 
coefficients in the strangeness zero channels as defined in (\ref{k-nonlocal}).} 
\label{tabpi-1}
\end{table}

It is instructive to afford a short detour and explore to what extent the parameters 
$\zeta_0, \zeta_D $ and $\zeta_F $ can be dialed to give ${\mathcal Z}=1$. In 
the $SU(3)$ limit with degenerate meson and also degenerate baryon masses 
one finds a degenerate wave-function renormalization factor ${\mathcal Z}$
\begin{eqnarray}
&&{\mathcal Z}^{-1}-1=3\,\zeta_0+2\,\zeta_D 
- \frac{m_\pi^2}{(4\pi\,f)^2}
\left( \frac{5}{3}\,D^2+3\,F^2\right) 
\Bigg( -4 -3\,\ln \left(\frac{m_\pi^2}{m_N^2}\right) 
\nonumber\\
&& \qquad \qquad \qquad \qquad \qquad \qquad \qquad 
+3\,\pi\,\frac{m_\pi}{m_N}
+{\mathcal O}\left( \frac{m^2_\pi}{m^2_N} \right)
\Bigg) \;,
\label{su3-limit:z}
\end{eqnarray}
where we used the minimal chiral subtraction prescription (\ref{def-sub}). 
The result (\ref{su3-limit:z}) agrees identically with the $SU(2)$-result (\ref{def-zn})
if one formally replaces  $g^2_A $ by $4\, F^2+ {\textstyle{20\over 9}}\,D^2 $ 
in (\ref{def-zn}). It is clear from (\ref{su3-limit:z}) that in the $SU(3)$ limit the 
counter term $3\,\zeta_0+2\,\zeta_D$ may be dialed such so as to impose ${\mathcal Z}=1$. This 
is no longer possible once the explicit $SU(3)$ symmetry-breaking effects are included. 
Note however that consistency of the perturbative renormalization procedure suggests that 
for hypothetical $SU(3)$-degenerate $\xi$-factors in (\ref{zeta-result}) it should be 
possible to dial $\zeta_0$, $\zeta_D$ and $\zeta_F$ so as to find ${\mathcal Z}_B =1$ 
for all baryon octet fields. This is expected, because for example in the 
$\overline {MS}$-scheme 
one finds a $SU(3)$-symmetric renormalization scale\footnote{Note that in the  
minimal chiral subtraction scheme the counter terms $\zeta_{0,D,F}$ are already renormalization 
scale independent. The consistency of this procedure follows from the symmetry conserving 
property of dimensional regularization.} dependence of $\xi_{B(c)}$ 
in (\ref{zeta-result}) with $\xi_{B(c)}  \sim  \ln \mu^2$. Indeed in this case the choice 
\begin{eqnarray}
&&\zeta_0 = \frac{\xi}{4\,f^2}\left(\frac{26}{9}\,D^2+2\,F^2\right)\, , \qquad 
\zeta_D = \frac{\xi}{4\,f^2}\left(-D^2+3\,F^2\right)\,, \quad 
\nonumber\\
&& \zeta_F = \frac{\xi}{4\,f^2}\,\frac{10}{3}\,D\,F \,,
\label{}
\end{eqnarray}
would lead to ${\mathcal Z}=1$ for all baryon octet wave functions. We will return to the $SU(3)$
symmetry-breaking effects in the baryon wave-function renormalization factors 
when discussing the meson-baryon 3-point vertices at subleading orders.

We proceed with the s-channel and u-channel exchange diagrams of the baryon octet. They  
contribute as follows
\begin{eqnarray}
\Big[K^{(I)}_{s-[8]}(\bar k,k;w)\Big]_{ab} &=& -\sum_{c\,=\,1}^3\,
\Big(\barqslash -R^{(I,c)}_{L,ab}\Big)\,
\frac{\Big[C^{(I,c)}_{[8]}\Big]_{ab}}
{4\,f^2\,\big(\wslash+m^{(c)}_{[8]}\big)}\,\Big( \qslash -R^{(I,c)}_{R,ab}\Big)\;,
\nonumber\\
\Big[K^{(I)}_{u-[8]}(\bar k,k;w)\Big]_{ab}  &=&
\sum_{c\,=\,1}^4\,
\frac{\Big[\widetilde C^{(I,c)}_{[8]}\Big]_{ab} }{4\,f^2 }\,\Bigg(
\wslash+m^{(c)}_{\,[8]}+\tilde R^{(I,c)}_{L,ab}+ \tilde R^{(I,c)}_{R,ab}
\nonumber\\ 
&-& \Big(\barpslash+m^{(c)}_{\,[8]}+ \tilde R^{(I,c)}_{L,ab} \Big)\, 
\frac{1}{\barwslash+m^{(c)}_{\,[8]} }
\, \Big(\pslash+m^{(c)}_{\,[8]}+ \tilde R^{(I,c)}_{R,ab}\Big)\Bigg)\;,
\label{wave-ren}
\end{eqnarray}
where $\widetilde w_\mu=p_\mu-\bar q_\mu$. 
The index $c$ in (\ref{wave-ren}) labels the baryon octet exchange with 
$c\to (N,\Lambda, \Sigma ,\Xi)$. In particular $m^{(c)}_{[8]}$ denotes   
the physical baryon octet masses and $R^{(I,c)}$ and  $\tilde R^{(I,c)}$ characterize 
the ratios of pseudo-vector to pseudo-scalar terms in the meson-baryon vertices.  
The dimensionless matrices $C^{}_{[8]}$ and $\widetilde C^{}_{[8]}$
give the renormalized strengths of the s-channel and 
u-channel baryon exchanges respectively. They are expressed most conveniently 
in terms of s-channel $C^{}_{B(c)}$ and u-channel $ \tilde C^{}_{B(c)}$ coefficient matrices
\begin{eqnarray}
&& \Big[C^{(I,c)}_{[8]}\Big]_{ab} = \Big[C^{(I)}_{B(c)}\Big]_{ab}\,
\bar A_{\Phi (I,a) B(I,a)}^{(B(c))}\,\bar A_{\Phi (I,b) B(I,b)}^{(B(c))} \;,
\nonumber\\
&& \Big[\widetilde C^{(I,c)}_{[8]}\Big]_{ab}=
\Big[\widetilde C^{(I)}_{B(c)}\Big]_{ab}\,
\bar A_{\Phi (I,b) B(I,a)}^{(B(c))}\,\bar A_{\Phi (I,a) B(I,b)}^{(B(c))}
 \;,
\label{def-not}
\end{eqnarray}
and the renormalized three-point coupling constants  $\bar A$.
According to the LSZ-scheme the bare coupling constants, $A$, are related to the renormalized 
coupling constants, $\bar A$, by means of the  baryon octet ${\mathcal Z}$-factors
\begin{eqnarray}
\bar A_{\Phi (I,a) B(I,b)}^{(B(c))}= {\mathcal Z}^{\frac{1}{2}}_{B(c)}\,A_{\Phi (I,a) B(I,b)}^{(B(c))}
\,{\mathcal Z}_{B(I,b)}^{\frac{1}{2}}\,.
\label{g-ren:z}
\end{eqnarray}
To leading order $Q^0$ the bare coupling constants $A = G(F,D)+{\mathcal O}\left(Q^2 \right)$ are 
specified in (\ref{G-explicit}). The chiral correction terms to the coupling constants $A$ and 
$R $ will be discussed in great detail subsequently.
In our convention the s-channel coefficients $C_{B(c)}$ are either one 
or zero and the u-channel coefficients $\tilde C^{}_{B(c)}$ represent 
appropriate Fierz factors resulting from the interchange of initial and final meson states. 
To illustrate our notation explicitly the coefficients are listed in Tab. \ref{tabpi-1} for the 
strangeness zero channels but relegated to Appendix E for the strangeness minus one channel. 
We stress that the expressions for the s- and u-channel exchange contribution (\ref{wave-ren}) 
depend on the result (\ref{delta-zeta}), because there would otherwise be a factor 
${\mathcal Z}^{-1}/(1+\zeta)$ in front of both contributions. That would necessarily lead to 
an asymmetry in the treatment of s- versus u-channel exchange contribution, because the
s-channel contribution would be further renormalized by the unitarization.
In contrast, we observe in our scheme the proper balance of s- and u-channel exchange 
contributions in the scattering amplitude, as is required for the realization of the 
large $N_c$ cancellation mechanism \cite{Manohar}. 

We turn to the $SU(3)$ symmetry-breaking effects in the meson-baryon coupling constants. 
The 3-point vertices have pseudo-vector and pseudo-scalar components determined by 
$F_i+\delta F_i$ and $\bar F_i$ of (\ref{chi-sb-3},\ref{chi-sb-3:p}) 
respectively. We first collect the symmetry-breaking terms in the axial-vector coupling 
constants $A$ introduced in (\ref{g-ren:z}). They are of chiral order $Q^2$ but lead to 
particular $Q^3$-correction terms in the scattering amplitudes. One finds
\begin{eqnarray}
&& A = G(F_A,D_A) -\frac{2}{\sqrt{3}}\,(m_K^2-m_\pi^2)\,\Delta A \;,\quad 
\tilde F_i = F_i + \delta F_i \;,
\nonumber\\
&&\Delta A_{\pi N}^{(N)} = 
3\,(F_1+F_3)-F_0-F_2+2\,(\tilde F_4+\tilde F_5) \;, 
\nonumber\\
&&\Delta A_{K \Sigma}^{(N)} = 
{\textstyle{3\over 2}}\,(F_1-F_3) +{\textstyle{1\over 2}}\,(F_0-F_2)-\tilde F_4+\tilde F_5 \;, 
\nonumber\\
&&\Delta A_{\eta \,N}^{(N)} = 
-F_1+3\,F_3+{\textstyle{1\over3}}\,F_0-F_2
+{\textstyle{2\over3}}\,\tilde F_4-2\,\tilde F_5 +2\,(\tilde F_6+{\textstyle{4\over3}}\,\tilde F_4) \;, 
\nonumber\\
&&\Delta A_{K \Lambda}^{(N)} = 
-{\textstyle{1\over 2}}\,F_1-{\textstyle{3\over 2}}\,F_3+{\textstyle{1\over 2}}\,F_0+
{\textstyle{3\over2}}\,F_2 +{\textstyle{1\over3}}\,\tilde F_4+\tilde F_5
+(F_7+{\textstyle{4\over 3}}\,F_0)\;, 
\nonumber\\
&& \Delta A_{\bar K N}^{(\Lambda)} = \sqrt{2}\,\Delta A_{K \,\Lambda}^{(N)} \,,\quad 
\nonumber\\
&& \Delta A_{\pi \Sigma}^{(\Lambda)} = 
 {\textstyle{4\over \sqrt{3}}}\,\tilde F_4 +\sqrt{3}\,(F_7+{\textstyle{4\over3}}\,F_0)  \;, 
\nonumber\\
&&\Delta A_{\eta \,\Lambda}^{(\Lambda)} = 
{\textstyle{4\over 3}}\,(F_0 +\tilde F_4)+2\,(\tilde F_6+{\textstyle{4\over3}}\,\tilde F_4)
+2\,(F_7+{\textstyle{4\over3}}\,F_0) \;, 
\nonumber\\
&& \Delta A_{K \Xi}^{(\Lambda)} = 
-{\textstyle{1\over \sqrt{2}}}\,(F_0+F_1)
+{\textstyle{3\over \sqrt{2}}}\,(F_2+F_3) 
-{\textstyle{\sqrt{2}\over 3}}\,(\tilde F_4-3\,\tilde F_5) 
-\sqrt{2}\,(F_7 +{\textstyle{4\over3}}\,F_0) \;, 
\nonumber\\
&& \Delta A_{\bar K N}^{(\Sigma)} = 
\sqrt{{\textstyle{2\over 3}}}\,\Delta A_{K \Sigma}^{(N)} \;, \quad 
\Delta A_{\pi \Sigma}^{(\Sigma)} = -4\,\sqrt{{\textstyle{2\over 3}}}\,(F_2+\tilde F_5) \;, \quad 
\Delta A_{\pi \Lambda}^{(\Sigma)} = 
{\textstyle{1\over\sqrt{3}}}\,\Delta A_{\pi \Sigma}^{(\Lambda)} \;,
\nonumber\\
&& \Delta A_{\eta \Sigma }^{(\Sigma)} =  {\textstyle{4\over 3}}\,(F_0-\tilde F_4)
+2\,(\tilde F_6+{\textstyle{4\over3}}\,\tilde F_4) \;,
\nonumber\\
&& \Delta A_{K \Xi}^{(\Sigma)} =  -\sqrt{{\textstyle{3\over 2}}}\,(F_1+F_3)+
\sqrt{{\textstyle{1\over 6}}}\,(F_0+F_2)
-\sqrt{{\textstyle{2\over 3}}}\,(\tilde F_4+\tilde F_5)\;,
\nonumber\\ 
&& \Delta A_{\bar K \Lambda}^{(\Xi)} = -{\textstyle{1\over\sqrt{2}}}\,
\Delta A_{K \Xi}^{(\Lambda)} \;, \quad 
\Delta A_{\bar K \Sigma}^{(\Xi)} = -\sqrt{{\textstyle{3\over2}}}\,
\Delta A_{K \Xi}^{(\Sigma)} \;,
\nonumber\\
&& \Delta A_{\eta \,\Xi}^{(\Xi)} = 
F_1+3\,F_3+{\textstyle{1\over3}}\,F_0+F_2
+{\textstyle{2\over3}}\,\tilde F_4+2\,\tilde F_5+2\,(\tilde F_6+{\textstyle{4\over3}}\,\tilde F_4) \;,
\nonumber\\
&& \Delta A_{\pi \Xi}^{(\Xi)} = F_0+3\,F_1-F_2-3\,F_3+2\,(\tilde F_5-\tilde F_4) \;,
\label{G-explicit-new}
\end{eqnarray}
where the renormalized coupling constants $D_A$ and $F_A$ 
are given in (\ref{ren-FDC-had}) and $G(F,D)$ in (\ref{G-explicit}).
We emphasize that to order $Q^2$ the effect of the ${\mathcal Z}$-factors in (\ref{g-ren:z}) 
can be accounted for by the following redefinition of the $F_R,D_R$ and 
$F_i$ parameters in (\ref{G-explicit-new})  
\begin{eqnarray}
&& F_R \to F_R +\, \Big( 2\,m_K^2+m_\pi^2\Big)\,
\Big(\zeta_0+{\textstyle{2\over 3}}\,\zeta_D\Big) \,F_R \;, \qquad 
\nonumber\\
&& D_R \to D_R + \Big( 2\,m_K^2+m_\pi^2\Big)\,
\Big( \zeta_0+{\textstyle{2\over 3}}\,\zeta_D\Big) \,D_R \;, 
\nonumber\\
&& F_0 \to F_0 +\zeta_D\,D_R\,, \qquad F_1 \to F_1+\zeta_F \,F_R
\,, \qquad F_2 \to F_2 +\zeta_D \,F_R\,,
\nonumber\\
&& F_3 \to F_3 + \zeta_F \,F_R\,,\qquad \,F_7 \to F_7 -{\textstyle{4\over 3}}\,\zeta_D\,\,D_R\,.
\label{zeta-redundant}
\end{eqnarray}
As a consequence the parameters $\zeta_0$, $\zeta_D$ and $\zeta_F$ are redundant in 
our approximation and can therefore be dropped. It is legitimate to identify the
bare coupling constants $A$ with the renormalized coupling constants $\bar A$. Note 
that the same renormalization holds for matrix elements of the axial-vector current.

We recall the large $N_c$ result (\ref{large-Nc-FDC}). The many parameters $F_i$ and 
$\delta F_i$ in (\ref{G-explicit-new}) are expressed in terms of the seven parameters 
$c_{i}$ and $\delta c_i= \tilde c_i-c_i-\bar c_i$. For the readers convenience we provide 
the axial-vector coupling constants, which are most relevant for the pion-nucleon and 
kaon-nucleon scattering processes, in terms of those large $N_c$ parameters
\begin{eqnarray}
&& A_{\pi N}^{(N)} = \sqrt{3}\,\Big( F_A+D_A \Big) +
{\textstyle{5\over 3}}\, (c_1+\delta c_1) + c_2+\delta c_2 -a+ 5\, c_3 + c_4\;, 
\nonumber\\
&& A_{\bar K N}^{(\Lambda)} = -\sqrt{{\textstyle{2\over3}}}\,\Big( 3\,F_A+D_A \Big)
+{\textstyle{1\over\sqrt{2}}}\,\Big(c_1+\delta c_1+c_2+\delta c_2-a-3\,c_3+c_4 \Big)\,,
\nonumber\\
&& A_{\bar K N}^{(\Sigma)} = -\sqrt{2 }\,\Big( F_A-D_A \Big)
-{\textstyle{1\over\sqrt{6}}}\,\Big({\textstyle{1\over 3}}\,(c_1+\delta c_1)-c_2-\delta c_2+a-c_3+3\,c_4 \Big)\,,
\nonumber\\
&& A_{\pi \Sigma}^{(\Lambda)} = 2\,D_A +{\textstyle{2\over\sqrt{3}}}\,(c_1+\delta c_1)\;,
\label{c1234}
\end{eqnarray}
where
\begin{eqnarray}
F_A &=& F_R 
-\frac{\beta}{\sqrt{3}}\,\Big({\textstyle{2\over 3}}\,\delta c_1 +\delta c_2 -a\Big) \,, 
\qquad \; \;D_A = D_R - \frac{\beta}{\sqrt{3}}\,\delta c_1 \,,
\nonumber\\
\beta &=& \frac{m_K^2+m_\pi^2/2}{m_K^2-m_\pi^2}  \,.
\label{c1234-b}
\end{eqnarray}
The result (\ref{c1234}) shows that the axial-vector coupling 
constants $A_{\pi N}^{(N)}$, $A_{\bar K N}^{(\Lambda)}$, $A_{\bar K N}^{(\Sigma)}$ and 
$A_{\pi \Sigma}^{(\Lambda)}$, can be fine tuned with $c_i $ in (\ref{large-Nc-FDC}) to be 
off their $SU(3)$ limit values. Moreover, the SU(3) symmetric
contribution proportional to $ F_A$ or $ D_A$ deviate from their corresponding 
$F_R$ and $D_R$ values relevant for matrix elements of the axial-vector current, once 
non-zero values for $\delta c_{1,2}$ are established. We recall, however, that 
the values of $F_R, D_R$ and $c_i$ are strongly constrained by the weak decay widths of the 
baryon-octet states (see Tab. \ref{weak-decay:tab}). Thus the SU(3) symmetry-breaking pattern 
in the axial-vector current and the one in the meson-baryon axial-vector coupling constants 
are closely linked.

It is left to specify the pseudo-scalar part, $P$, of the meson-baryon vertices introduced
in (\ref{chi-sb-3:p}). In (\ref{wave-ren}) their effect was encoded 
into the $R$ and $\tilde R $ parameters with 
$R, \tilde R \sim \bar F_{}\sim \bar c_{},a_{}$. Applying the large $N_c$ result 
of (\ref{large-Nc-FDC:delta}) we obtain
\begin{eqnarray}
&& R_{L,ab}^{(I,c)} \,A_{\Phi (I,a)\, B(I,a)}^{(B(c))} = 
\Big(m_{B(I,a)}+m_{B(c)} \Big)\,P_{\Phi (I,a)\, B(I,a)}^{(B(c))} 
\;,
\nonumber\\
&& R_{R,ab}^{(I,c)} \,A_{\Phi (I,b)\, B(I,b)}^{(B(c))} = 
\Big(m_{B(I,b)}+m_{B(c)} \Big)\,P_{\Phi (I,b)\, B(I,b)}^{(B(c))} 
\;,
\nonumber\\
&& \tilde R_{L,ab}^{(I,c)} \,A_{\Phi (I,b)\, B(I,a)}^{(B(c))} = 
\Big(m_{B(I,a)}+m_{B(c)} \Big)\,P_{\Phi (I,b)\, B(I,a)}^{(B(c))} 
\;,
\nonumber\\
&& \tilde R_{R,ab}^{(I,c)} \,A_{\Phi (I,a)\, B(I,b)}^{(B(c))} = 
\Big(m_{B(I,b)}+m_{B(c)} \Big)\,P_{\Phi (I,a)\, B(I,b)}^{(B(c))} 
\;,
\nonumber\\ \nonumber\\
&&P_{\pi N}^{(N)} = -\Big({\textstyle{5\over3}}\, \bar c_1 +\bar c_2+a\Big)\, \Big(\beta -1\Big)\;, \quad 
P_{K \Sigma}^{(N)} = -\Big( {\textstyle{1\over3}}\,\bar c_1 -\bar c_2-a \Big)\, 
\Big( \beta +{\textstyle{1\over2}} \Big) \;, 
\nonumber\\
&&P_{\eta \,N}^{(N)} = {\textstyle{7\over3}}\, \bar c_1 +\bar c_2+a
- \beta \,\Big( {\textstyle{1\over3}}\,\bar c_1 +\bar c_2+a \Big)\;, \qquad 
\nonumber\\
&& P_{K \Lambda}^{(N)} = \Big( \bar c_1 +\bar c_2+a\Big)\,\Big(\beta +{\textstyle{1\over2}}\Big)\;, 
\qquad 
P_{\bar K N}^{(\Lambda)} = \sqrt{2}\,P_{K \,\Lambda}^{(N)} \,,\qquad 
\nonumber\\
&&P_{\pi \Sigma}^{(\Lambda)} = - {\textstyle{2\over \sqrt{3}}}\,\bar c_1\,\Big( \beta -1 \Big)\;, 
\quad P_{\eta \,\Lambda}^{(\Lambda)} = {\textstyle{2\over3}}\,\bar c_1\,\Big( 5+\beta \Big) +2\,\bar c_2+2\,a \;, \qquad 
\nonumber\\
&&  P_{K \Xi}^{(\Lambda)} = {\textstyle{1\over \sqrt{2}}}\,\Big( {\textstyle{1\over3}}\,\bar c_1+\bar c_2+a \Big)\,
\Big(2\, \beta +1 \Big)\;, \quad 
P_{\bar K N}^{(\Sigma)} = 
\sqrt{{\textstyle{2\over 3}}}\,P_{K \Sigma}^{(N)} \;, \quad \!\!
\nonumber\\
&& P_{\pi \Sigma}^{(\Sigma)} = \sqrt{{\textstyle{2\over3}}}\,\Big( {\textstyle{4\over 3}}\,\bar c_1 +2\,\bar c_2+2\,a \Big)\,
 \Big(\beta -1\Big) \;, \quad 
P_{\pi \Lambda}^{(\Sigma)} = 
{\textstyle{1\over\sqrt{3}}}\,P_{\pi \Sigma}^{(\Lambda)} \;,
\nonumber\\
&& P_{\eta \Sigma }^{(\Sigma)} = 2\,\bar c_1 \,\Big(1-{\textstyle{1\over 3}}\,\beta \Big)+2\,\bar c_2+2\,a \;,\qquad
\nonumber\\ 
&&  P_{K \Xi}^{(\Sigma)} = -{\textstyle{1\over \sqrt{6}}}\,\Big( {\textstyle{5\over 3}}\,\bar c_1 +\bar c_2+a\Big) \,
\Big(2\,\beta +1\Big)  \;,\quad 
P_{\bar K \Lambda}^{(\Xi)} = -{\textstyle{1\over\sqrt{2}}}\,
P_{K \Xi}^{(\Lambda)} \;, \quad 
\nonumber\\
&& P_{\bar K \Sigma}^{(\Xi)} = -\sqrt{{\textstyle{3\over2}}}\,
P_{K \Xi}^{(\Sigma)} \;,\quad 
P_{\eta \,\Xi}^{(\Xi)} = {\textstyle{11\over3}}\,\bar c_1 +3\,\bar c_2 + 3\,a
+\beta \,\Big(  \bar c_1 +\bar c_2+a\Big)
\;,\quad \!\! 
\nonumber\\
&&  P_{\pi \Xi}^{(\Xi)} = \Big({\textstyle{1\over3}}\,\bar c_1 -\bar c_2-a\Big)\,\Big( \beta -1 \Big)\;,
\label{P-result}
\end{eqnarray}
where $\beta \simeq 1.12$ was introduced in (\ref{c1234-b}). The pseudo-scalar 
meson-baryon vertices show $SU(3)$ symmetric contribution linear in  
$\beta \,\bar c_i$ and symmetry-breaking terms proportional to $\bar c_i$. 
We emphasize that, even though the physical  meson-baryon coupling constants, $G=A+P$, are the 
sum of their axial-vector and pseudo-scalar components, it is important to carefully 
discriminate both types of vertices, because they give rise to quite different behavior 
for the partial-wave amplitudes off the baryon-octet pole. We expect such effects to be 
particularly important in the strangeness sectors since there (\ref{P-result}) leads to 
$P^{(\Lambda, \Sigma )}_{K N} \sim m_K^2$ to be compared to $P_{\pi N}^{(N)} \sim m_\pi^2$.

We turn to the decuplet exchange terms $K^{(I)}_{s-[10] }$ and $K^{(I)}_{u-[10] }$ 
in (\ref{k-all}). Again we write their respective interaction kernels in a form which facilitates 
the identification of their on-shell irreducible parts
\begin{eqnarray}
K^{(I)}_{s-[10] }(\bar k,k;w) &=&\sum_{c\,=\,1}^2\,
\frac{C^{(I,c)}_{[10]}}{4\,f^2 }\,\Bigg(
\frac{\bar q\cdot q}{\wslash-m_{[10]}^{(c)}}\,
-\frac{(\bar q\cdot w)\,(w\cdot q)}
{(m_{[10]}^{(c)})^2\,\big(\wslash-m_{[10]}^{(c)}\big) }
\nonumber\\
&+&\frac{1}{3}\left(\barqslash +\frac{\bar q \cdot w}{m_{[10]}^{(c)}}\right)
\frac{1}{\wslash+m_{[10]}^{(c)}}
\left(\qslash +\frac{w\cdot q}{m_{[10]}^{(c)}}\right)  -\frac{Z_{[10]}}{3\,m_{[10]}^{(c)}}\,\barqslash \,\qslash
\nonumber\\
&-&\frac{Z_{[10]}^2}{6\,(m_{[10]}^{(c)})^2}\,\barqslash \,\Big(\wslash-2\,m_{[10]}^{(c)}\Big)\,\qslash
+Z_{[10]}\,\barqslash \,\frac{w\cdot q}{3\,(m_{[10]}^{(c)})^2}
\nonumber\\
&+&Z_{[10]} \,\frac{\bar q \cdot w}{3\,(m_{[10]}^{(c)})^2}\,\qslash \Bigg)\;,
\nonumber\\
K^{(I)}_{u-[10] }(\bar k,k;w) &=&\sum_{c\,=\,1}^3\,
\frac{\widetilde C^{(I,c)}_{[10]}}{4\,f^2 }\,\Bigg(
\frac{\bar q\cdot q}{\barwslash-m_{[10]}^{(c)}}
-\frac{(\bar q\cdot \widetilde w)\,(\widetilde w\cdot q)}
{(m_{[10]}^{(c)})^2\,\big(\barwslash-m_{[10]}^{(c)}\big)}
\nonumber\\
&+&\frac{1}{3}\left(\barpslash +m_{[10]}^{(c)}+\frac{q \cdot \widetilde w}{m_{[10]}^{(c)}}\right)
\frac{1}{\barwslash+m_{[10]}^{(c)}}
\left(\pslash +m_{[10]}^{(c)}+\frac{\widetilde w\cdot \bar q}{m_{[10]}^{(c)}}\right)
\nonumber\\
&-&\frac{\widetilde w\cdot (q+\bar q)}{3\,m_{[10]}^{(c)}}
-\frac{1}{3}\,\Big(\wslash+m_{[10]}^{(c)}\Big)
\nonumber\\
&+&\frac{Z_{[10]}}{3\,(m_{[10]}^{(c)})^2}\,\Big(
\barqslash \,\big(\widetilde w\cdot q\big)
+\big(\bar q \cdot \widetilde w\big) \,\qslash
+m_{[10]}^{(c)}\,\big(
\barqslash\,\qslash-2\,(\bar q\cdot q) \big) \Big)
\nonumber\\
&-&\frac{Z_{[10]}^2}{6\,(m_{[10]}^{(c)})^2}\,
\Big( \barpslash \;\barwslash\,\pslash -\widetilde w^2\,\wslash
+2\,m_{[10]}^{(c)}\,\big(\barqslash\,\qslash -2\,q\cdot q\big) \Big)\Bigg) \;.
\label{k-nonlocal}
\end{eqnarray}
The index $c$ in (\ref{k-nonlocal}) labels the decuplet exchange with 
$c\to (\Delta_{[10]} , \Sigma_{[10]}, \Xi_{[10]})$. The dimensionless matrices $C^{(I,c)}_{[10]}$  
and  $\widetilde C^{(I,c)}_{[10]}$ characterize the strength of the s-channel and 
u-channel resonance exchanges respectively. Here we apply the notation introduced in (\ref{def-not}) 
also for the resonance-exchange contributions. In particular the decuplet coefficients 
$C^{(I,c)}_{[10]}$ and $\tilde C^{(I,c)}_{[10]}$ are determined by 
\begin{eqnarray}
&& A = G( C_A) -\frac{2}{\sqrt{3}}\,(m_K^2-m_\pi^2)\,\Delta A \;,\qquad 
\tilde C_0 = C_0 +\delta C_0 \;,
\nonumber\\
&&\Delta A_{\pi N}^{(\Delta_{[10]} )} = \sqrt{2}\,
\big( -{\textstyle{1\over \sqrt{3}}}\,C_2+\sqrt{3}\,C_3+{\textstyle{2\over \sqrt{3}}}\,\tilde C_0 
\big) \;, \quad 
\nonumber\\
&&\Delta A_{K \,\Sigma}^{(\Delta_{[10]} )} = -\sqrt{2}\,
\big({\textstyle{2\over \sqrt{3}}}\,C_2
- {\textstyle{1\over \sqrt{3}}}\,\tilde C_0+\sqrt{3}\,C_1\big) \;,\quad 
\nonumber\\
&& \Delta A_{\bar K N}^{(\Sigma_{[10]})} =\sqrt{{\textstyle{2\over 3}}}\,
\big(-{\textstyle{1\over \sqrt{3}}}\,C_2+\sqrt{3}\,C_3 
- {\textstyle{1\over \sqrt{3}}}\,\tilde C_0-\sqrt{3}\,C_1 \big)-\sqrt{2}\,C_4  \;, \quad  
\nonumber\\
&&\Delta A_{\pi \Sigma}^{(\Sigma_{[10]})} = -\sqrt{{\textstyle{2\over 3}}}\,
\big({\textstyle{2\over \sqrt{3}}}\,C_2
+ {\textstyle{2\over \sqrt{3}}}\,\tilde C_0\big)-2\,\sqrt{2}\,C_4 \;, \qquad
\nonumber\\
&& \Delta A_{\pi \Lambda}^{(\Sigma_{[10]})} = {\textstyle{2\over \sqrt{3}}}\,(C_2-\tilde C_0) \;,  \qquad 
\Delta A_{\eta \Sigma}^{(\Sigma_{[10]})} = {\textstyle{2\over \sqrt{3}}}\,C_2
- {\textstyle{2\over \sqrt{3}}}\,\tilde C_0\;,\quad 
\nonumber\\
&& \Delta A_{K \Xi}^{(\Sigma_{[10]})} = -\sqrt{{\textstyle{2\over 3}}}\,
\big( -{\textstyle{1\over \sqrt{3}}}\,C_2-\sqrt{3}\,C_3 
- {\textstyle{1\over \sqrt{3}}}\,\tilde C_0 +\sqrt{3}\, C_1\big)+\sqrt{2}\,C_4\;, 
\nonumber\\
&&\Delta A_{\bar K \Lambda}^{(\Xi_{[10]})} = 
-{\textstyle{1\over \sqrt{3}}}\, \tilde C_0 -\sqrt{3}\,C_1
-{\textstyle{1\over \sqrt{3}}}\,(3\,C_4+2\,C_2) \;, \quad 
\nonumber\\
&& \Delta A_{\bar K \,\Sigma}^{(\Xi_{[10]})} = {\textstyle{2\over \sqrt{3}}}\,C_2
- {\textstyle{1\over \sqrt{3}}}\,\tilde C_0- \sqrt{3}\, C_1 +\sqrt{3}\,C_4  \;,\quad 
\nonumber\\
&&\Delta A_{\eta \,\Xi}^{(\Xi_{[10]})} = {\textstyle{1\over \sqrt{3}}}\,C_2+\sqrt{3}\,C_3
+ {\textstyle{2\over \sqrt{3}}}\,\tilde C_0 +\sqrt{3}\,C_4 \;,\quad 
\nonumber\\
&&\Delta A_{\pi \,\Xi}^{(\Xi_{[10]})} = {\textstyle{1\over \sqrt{3}}}\,C_2+\sqrt{3}\,C_3
- {\textstyle{2\over \sqrt{3}}}\,\tilde C_0 -\sqrt{3}\,C_4  \;,
\label{}
\end{eqnarray}
where the $SU(3)$ symmetric contributions $G(C)$ are
\begin{eqnarray}
&&G_{\pi N}^{(\Delta_{[10]} )} = \sqrt{2}\,C \;, \quad G_{K \,\Sigma}^{(\Delta_{[10]} )} = -\sqrt{2}\,C \;,
\quad 
G_{\bar K N}^{(\Sigma_{[10]})} =\sqrt{{\textstyle{2\over 3}}}\,C  \;, 
\nonumber\\
&& G_{\pi \Sigma}^{(\Sigma_{[10]})} = -\sqrt{{\textstyle{2\over 3}}}\,C \;,\quad  
G_{\pi \Lambda}^{(\Sigma_{[10]})} = -C \;,  \quad G_{\eta \Sigma}^{(\Sigma_{[10]})} = C\;,
\quad G_{K \Xi}^{(\Sigma_{[10]})} = -\sqrt{{\textstyle{2\over 3}}}\,C\;, 
\nonumber\\
&&G_{\bar K \Lambda}^{(\Xi_{[10]})} = C \;, \quad 
G_{\bar K \,\Sigma}^{(\Xi_{[10]})} = C \;,\quad 
G_{\eta \,\Xi}^{(\Xi_{[10]})} = -C\;,\quad 
G_{\pi \,\Xi}^{(\Xi_{[10]})} = -C \;.
\label{}
\end{eqnarray}
We recall that at leading order in the $1/N_c$ expansion the five symmetry-breaking 
parameters $C_i$ introduced in (\ref{chi-sb-3}) are all given in terms of the 
$c_i$ parameters (\ref{large-Nc-FDC}). The parameter $\delta C_0 \sim \tilde c_1-c_1$ 
and also implicitly $\tilde C_R $ (see (\ref{ren-FDC-had})) probe the $\tilde c_1$ parameter 
introduced in (\ref{ansatz-3}). We emphasize that all $c_i$ parameters but $c_5$ are 
determined to a large extent by the decay widths of the baryon octet states and  
also $\tilde c_1$ is constrained strongly by the $SU(3)$ symmetry-breaking pattern of 
the meson-baryon-octet coupling constants (\ref{ansatz-3}).

The baryon octet resonance contributions $K^{(I)}_{s-[9] }$ and $K^{(I)}_{u-[9] }$
require special attention, because the way how to incorporate systematically these resonances 
in a chiral SU(3) scheme is not clear. We present first the s-channel and u-channel 
contributions as they follow from (\ref{lag-Q}) at tree-level
\begin{eqnarray}
K^{(I)}_{s-[9] }(\bar k,k;w) &=&\sum_{c\,=\,0}^4\,
\frac{C^{(I,c)}_{[9]}}{4\,f^2 }\,\Bigg(
-\frac{\bar q\cdot q}{\wslash+m_{[9]}^{(c)}}\,
+\frac{(\bar q\cdot w)\,(w\cdot q)}
{(m_{[9]}^{(c)})^2\,\big(\wslash+m_{[9]}^{(c)}\big) }
\nonumber\\
&-&\frac{1}{3}\left(\barqslash -\frac{\bar q \cdot w}{m_{[9]}^{(c)}}\right)
\frac{1}{\wslash-m_{[9]}^{(c)}}
\left(\qslash -\frac{w\cdot q}{m_{[10]}^{(c)}}\right)  \Bigg)\;,
\nonumber\\
K^{(I)}_{u-[9] }(\bar k,k;w) &=&\sum_{c\,=\,0}^4\,
\frac{\widetilde C^{(I,c)}_{[9]}}{4\,f^2 }\,\Bigg(
-\frac{\bar q\cdot q}{\barwslash+m_{[9]}^{(c)}}
+\frac{(\bar q\cdot \widetilde w)\,(\widetilde w\cdot q)}
{(m_{[9]}^{(c)})^2\,\big(\barwslash+m_{[9]}^{(c)}\big)}
\nonumber\\
&-&\frac{1}{3}\left(\barpslash -m_{[9]}^{(c)}-\frac{q \cdot \widetilde w}{m_{[9]}^{(c)}}\right)
\frac{1}{\barwslash-m_{[9]}^{(c)}}
\left(\pslash -m_{[9]}^{(c)}-\frac{\widetilde w\cdot \bar q}{m_{[9]}^{(c)}}\right)
\nonumber\\
&-&\frac{\widetilde w\cdot (q+\bar q)}{3\,m_{[9]}^{(c)}}
+\frac{1}{3}\,\Big(\wslash-m_{[9]}^{(c)}\Big)\Bigg) 
\nonumber\\
&-&\frac{Z_{[9]}}{3\,(m_{[9]}^{(c)})^2}\,\Big(
\barqslash \,\big(\widetilde w\cdot q\big)
+\big(\bar q \cdot \widetilde w\big) \,\qslash
-m_{[9]}^{(c)}\,\big(
\barqslash\,\qslash-2\,(\bar q\cdot q) \big) \Big)
\nonumber\\
&+&\frac{(Z_{[9]})^2}{6\,(m_{[9]}^{(c)})^2}\,
\Big( \barpslash \;\barwslash\,\pslash -\widetilde w^2\,\wslash
-2\,m_{[9]}^{(c)}\,\big(\barqslash\,\qslash -2\,q\cdot q\big) \Big)\Bigg)
\label{8stern}
\end{eqnarray}
where the index $c\to (\Lambda(1520),N(1520), \Lambda (1690), \Sigma(1680) ,\Xi(1820))$ extends over
the nonet resonance states.  The coefficient matrices $C^{(I,c)}_{[9]}$ and 
$\widetilde C^{(I,c)}_{[9]}$ are constructed by analogy with those of the octet and decuplet contributions
(see (\ref{G-explicit},\ref{def-not})). In particular one has $C_{B^*(c)}^{(I)}=C_{B(c)}^{(I)}$ and 
$\tilde C_{B^*(c)}^{(I)}=\tilde C_{B(c)}^{(I)}$. The coupling constants 
$A^{(B^*)}_{\Phi B}=G^{(B)}_{\Phi B}(F_{[9]},D_{[9]})$  are given in  
terms of the $F_{[9]}$ and $D_{[9]}$ parameters introduced in (\ref{lag-Q}) for all 
contributions except those for the $\Lambda (1520)$ and $\Lambda (1690)$ resonances 
\begin{eqnarray}
&& G_{\bar K N}^{(\Lambda (1520))} = \sqrt{{\textstyle{2\over 3}}}\,\Big( 3\,F_{[9]}+D_{[9]} \Big)\,\sin \vartheta
+\sqrt{3}\,C_{[9]}\,\cos \vartheta 
 \;,\quad 
\nonumber\\
&& G_{\pi \Sigma}^{(\Lambda(1520))} = -2\,D_{[9]} \,\sin \vartheta 
+{\textstyle{3\over \sqrt{2}}}\,C_{[9]}\,\cos \vartheta\;,\quad 
\nonumber\\
&& G_{\eta \,\Lambda}^{(\Lambda(1520))} = {\textstyle{2\over\sqrt{3}}}\,D_{[9]}\,\sin \vartheta
+\sqrt{{\textstyle{3\over 2}}}\,C_{[9]}\,\cos \vartheta\;,\quad 
\nonumber\\
&& G_{K \Xi}^{(\Lambda(1520))} = \sqrt{{\textstyle{2\over 3}}}\,\Big( 3\,F_{[9]}-D_{[9]}\Big)\,\sin \vartheta  
-\sqrt{3}\,C_{[9]}\,\cos \vartheta\; 
\label{G-9-explicit-1}
\end{eqnarray}
and
\begin{eqnarray}
&& G_{\bar K N}^{(\Lambda (1690))} = -\sqrt{{\textstyle{2\over 3}}}\,\Big( 3\,F_{[9]}+D_{[9]} \Big)\,\cos \vartheta
+\sqrt{3}\,C_{[9]}\,\sin \vartheta 
 \;,\quad 
\nonumber\\
&& G_{\pi \Sigma}^{(\Lambda(1690))} = 2\,D_{[9]} \,\cos \vartheta 
+{\textstyle{3\over \sqrt{2}}}\,C_{[9]}\,\sin \vartheta\;,\quad 
\nonumber\\
&& G_{\eta \,\Lambda}^{(\Lambda(1690))} = -{\textstyle{2\over\sqrt{3}}}\,D_{[9]}\,\cos \vartheta
+\sqrt{{\textstyle{3\over 2}}}\,C_{[9]}\,\sin \vartheta\;,\quad 
\nonumber\\
&& G_{K \Xi}^{(\Lambda(1690))} = -\sqrt{{\textstyle{2\over 3}}}\,\Big( 3\,F_{[9]}-D_{[9]}\Big)\,\cos \vartheta  
-\sqrt{3}\,C_{[9]}\,\sin \vartheta\;. 
\label{G-9-explicit-2}
\end{eqnarray}

The baryon octet field $B^*_\mu $ asks for special considerations in a chiral SU(3) scheme
as there is no straightforward systematic approximation strategy. Given that the baryon octet and 
decuplet states are degenerate in the large $N_c$ limit, it is natural to impose 
$m_{[10]} -m_{[8]} \sim Q$. In contrast with that there is no fundamental reason  
to insist on $m_{N(1520)}-m_N \sim Q$, for example. But, only with $m_{[9]}-m_{[8]} \sim Q $ is it  
feasible to establish 
consistent power counting rules needed for the systematic evaluation of the chiral Lagrangian. 
Note that the presence of the baryon octet resonance states 
in the large $N_c$ limit of QCD is far from abvious. Our opinion differs here from the one expressed 
in \cite{Carone-1,Carone-2} where the d-wave baryon octet resonance states are considered 
as part of an excited large $N_c$ ${\bf 70}$-plet. Recall that reducible diagrams summed by 
the Bethe-Salpeter equation are typically enhanced by a factor of $2 \pi$ relatively to 
irreducible diagrams. We conclude that there are a priori two possibilities: the baryon resonances are a 
consequence of important coupled-channel dynamics or they are already present in the interaction kernel. 
An expansion in $2 \pi/N_c $, in our world with $N_c=3$, does not appear useful. The 
fact that baryon resonances exhibit large hadronic decay widths may be taken as 
an indication that the coupled-channel dynamics is the driving mechanism for the creation of 
baryon resonances. Related arguments have been put forward 
in \cite{Aaron:1,Aaron:2}. Indeed, for instance the $N(1520)$ resonance, was successfully described in 
terms of coupled channel dynamics, including the vector-meson nucleon channels as an important ingredient 
\cite{QM-lutz}, but without assuming a preformed resonance structure in the interaction 
kernel. The successful description of the $\Lambda (1405)$ resonance in our scheme (see Fig. \ref{fig:wt})
supports the above arguments. For a recent discussion of the competing picture in which the $\Lambda (1405)$ 
resonance is considered as a quark-model state we refer to \cite{Kimura}. 

The description of resonances has a subtle consequence for the treatment of the u-channel baryon resonance 
exchange contribution. If a resonance is formed primarily by the 
coupled channel dynamics one should not include an explicit bare u-channel resonance contribution in the 
interaction kernel. 
The then necessarily strong energy dependence of the resonance self-energy would 
invalidate the use of (\ref{8stern}), because for physical energies $\sqrt{s}> m_N+m_\pi$ the 
resonance self-energy is probed far off the resonance pole. Our discussion has non-trivial 
implications for the chiral SU(3) dynamics. Naively one may object that the effect of 
the u-channel baryon resonance exchange contribution in (\ref{8stern}) 
can be absorbed to good accuracy into chiral two-body interaction terms in any case. However, 
while this is true in  a chiral $SU(2)$ scheme, this is no longer possible in a chiral $SU(3)$
approach. This follows because chiral symmetry selects a specific subset of all possible $SU(3)$-symmetric two-body 
interaction terms at given chiral order. In particular one finds that the effect of the $Z_{[9]}$ parameter in (\ref{8stern}) 
can not be absorbed into the chiral two-body parameters $g^{(S)}, g^{(V)}$ or $g^{(T)}$ of (\ref{two-body}). 
For that reason we discriminate two possible scenarios. In scenario I we conjecture that the 
baryon octet resonance states are primarily generated by the coupled channel dynamics of 
the vector-meson baryon-octet channels. Therefore in this scenario the u-channel resonance 
exchange contribution of (\ref{8stern}) is neglected and only its s-channel contribution 
is included as a reminiscence of the neglected vector meson channels. Note that this is analogous to 
the treatment of the $\Lambda (1405)$ resonance, which is generated dynamically in the chiral $SU(3)$ 
scheme (see Fig. \ref{fig:wt}). Here a s-channel pole term is generated by the coupled channel 
dynamics but the associated u-channel term is effectively left out as a much suppressed contribution.
In scenario II we 
explicitly include the s- and the u-channel resonance exchange contributions as given 
in (\ref{8stern}), thereby assuming that the resonance was preformed already in the 
large $N_c$ limit of QCD and only slightly affected by the coupled 
channel dynamics. Our detailed analyses of the data set clearly favor scenario I. 
The inclusion of the u-channel resonance exchange contributions appears to destroy the subtle  balance of chiral s-wave 
range terms and makes it impossible to obtain a reasonable fit to the data set. Thus all results presented in this work 
will be based on scenario I. Note that in that scenario the background parameter $Z_{[9]}$ drops
out completely (see (\ref{v-result-1})).

\tabcolsep=1.1mm
\renewcommand{\arraystretch}{1.3}
\begin{table}[t]
\begin{tabular}{|r||c||c|c||c|c|c||c|c|c|c||c|c|c|}
\multicolumn{14}{c}{}\\ \multicolumn{14}{c}{}\\
 \hline
    & $C_{\pi,0}^{(\frac{1}{2})}$    &  $C_{\pi,D}^{(\frac{1}{2})}$      &  $C_{\pi,F}^{(\frac{1}{2})}$ & $C_{K,0}^{(\frac{1}{2})}$
    &${C}_{K,D}^{(\frac{1}{2})}$ & ${C}_{K,F}^{(\frac{1}{2})}$ &
    ${C}_{0}^{(\frac{1}{2})}$ & ${C}_{1}^{(\frac{1}{2})}$   &${C}_{D}^{(\frac{1}{2})}$ &
    ${C}_{F}^{(\frac{1}{2})}$& $\bar C_{1}^{(\frac{1}{2})}$ & $\bar C_{D}^{(\frac{1}{2})}$& $\bar C_{F}^{(\frac{1}{2})}$\\
\hline \hline 
$11$& -4 & -2 & -2
    & 0 & 0& 0
    & 2 & 0 & 1 & 1
    & 0 & 2  & 2\\ \hline 
$12$& 0 &  $\frac{1}{2}$ & -$\frac{1}{2}$
    & 0 & $\frac{1}{2}$ & -$\frac{1}{2}$
    & 0 & 1 & -$\frac{1}{2}$ & $\frac{1}{2}$
    &-1 & -$\frac{1}{2}$ &  $\frac{1}{2}$  \\\hline 
$13$& 0 & -2& -2
    & 0  & 0  & 0
    & 0 & 0 & 1  & 1
    & 0 & 0 & 0 \\  \hline
$14$& 0 & $\frac{1}{2}$  & $\frac{3}{2}$
    & 0 & $\frac{1}{2}$  & $\frac{3}{2}$
    & 0 & 0  & -$\frac{1}{2}$  & -$\frac{3}{2}$
    & 0 & -$\frac{1}{2}$  & -$\frac{3}{2}$ \\ \hline\hline
$22$& 0  & 0 & 0
    & -4 & -2 & 4
    & 2  & 0 & 1  & -2
    & 0  & -1 & 2 \\ \hline
$23$&  0 & -$\frac{3}{2}$  & $\frac{3}{2}$
    &  0  & $\frac{5}{2}$  & -$\frac{5}{2}$
    &  0 & 0 & -$\frac{1}{2}$  & $\frac{1}{2}$
    &  0 &  $\frac{3}{2}$ & -$\frac{3}{2}$ \\ \hline
$24$&  0 & 0 & 0
    &  0 & -2 & 0
    &  0 & 0 & 1 & 0
    &  0 &-1 & 0 \\ \hline\hline
$33$& $\frac{4}{3}$ & 2 & -$\frac{10}{3}$
    & -$\frac{16}{3}$& -$\frac{16}{3}$ & $\frac{16}{3}$
    & 2 & 0  & $\frac{5}{3}$ & -1
    & 0 & 0  & 0 \\ \hline
$34$&  0 & $\frac{1}{2}$ & $\frac{3}{2}$
    &  0 & -$\frac{5}{6}$ & -$\frac{5}{2}$
    &  0 & 1 & $\frac{1}{6}$ & $\frac{1}{2}$
    & -1 & -$\frac{1}{2}$ & -$\frac{3}{2}$\\ \hline\hline
$44$& 0 & 0 & 0
    &-4 & -$\frac{10}{3}$ & 0
    &2 & 0 & $\frac{5}{3}$ & 0
    & 0 & 1& 0 \\ \hline \hline

& $C_{\pi,0}^{(\frac{3}{2})}$ & $C_{\pi,D}^{(\frac{3}{2})}$ & $C_{\pi,F}^{(\frac{3}{2})}$ & $C_{K,0}^{(\frac{3}{2})}$
& ${C}_{K,D}^{(\frac{3}{2})}$ & ${C}_{K,F}^{(\frac{3}{2})}$ &
    ${C}_{0}^{(\frac{3}{2})}$ & ${C}_{1}^{(\frac{3}{2})}  $ & ${C}_{D}^{(\frac{3}{2})}$ &
    ${C}_{F}^{(\frac{3}{2})}$ & $\bar C_{1}^{(\frac{3}{2})}$ & $\bar C_{D}^{(\frac{3}{2})}$ & $\bar C_{F}^{(\frac{3}{2})}$\\
\hline \hline 
$11$&-4 & -2 & -2  
    & 0& 0 & 0  
    & 2& 0 & 1  & 1  
    & 0 & -1 & -1 \\ \hline 
$12$& 0 & -1 & 1 
    & 0 & -1 & 1 
    & 0 & 1 & 1 & -1
    & -1  & 1 & -1
\\ \hline \hline 
$22$& 0  &  0 & 0
    & -4 & -2 & -2  
    & 2  & 0  & 1  & 1  
    & 0  & -1 & -1
\\ \hline  
\end{tabular}
\vspace*{2mm} \caption{Coefficients of quasi-local interaction terms in the strangeness zero 
channels as defined in (\ref{def-local-1}).} 
\label{tabpi-2}
\end{table}

We turn to the quasi-local two-body interaction terms $K_{[8][8]}^{(I)}$ and 
$K_{\chi}^{(I)}$ in (\ref{k-all}). 
It is convenient to represent the strength 
of an interaction term in a given channel $(I,a)\to (I,b)$ by means of the dimensionless 
coupling coefficients $\big[C^{(I)}_{..}\big]_{ab}$. The $SU(3)$ structure of the interaction terms 
${\mathcal L}^{(S)}$ and ${\mathcal L}^{(V)}$ in (\ref{two-body}) are characterized by the 
coefficients $C^{(I)}_{0}$, $C^{(I)}_{1}$, $C^{(I)}_{D}$, $C^{(I)}_{F}$ and 
${\mathcal L}^{(T)}$  by $\bar C^{(I)}_{1}$, $\bar C^{(I)}_{D}$ and 
$\bar C^{(I)}_{F}$. Similarly the interaction terms (\ref{chi-sb}) which break chiral symmetry
explicitly are written in terms of the coefficients 
$C^{(I)}_{\pi,0}$, $C^{(I)}_{\pi,D}$, $C^{(I)}_{\pi,F}$ and $C^{(I)}_{K,0}$, 
$C^{(I)}_{K,D}$, $C^{(I)}_{K,F}$. We have
\begin{eqnarray}
K^{(I)}_{[8][8]}(\bar k, k; w)&=&  \frac{\bar q \cdot q }{4 f^2} \,
\Big(g^{(S)}_0\,C^{(I)}_0+g^{(S)}_1\,C^{(I)}_1 + g^{(S)}_D\,C^{(I)}_D
+g^{(S)}_F\,C^{(I)}_F\Big)
\nonumber\\
&+&\frac{1}{16 f^2}\,\Big( \qslash\,\big( p+\bar p\big) \cdot \bar q
+\barqslash \,\big(p+ \bar p\big) \cdot  q\Big)\,
\Big( g^{(V)}_0\,C^{(I)}_0+g^{(V)}_1\,C^{(I)}_1 \Big)
\nonumber\\
&+&\frac{1}{16 f^2}\,\Big( \qslash\,\big( p+\bar p\big) \cdot \bar q
+\barqslash \,\big(p+ \bar p\big) \cdot  q\Big)
\,\Big( g^{(V)}_D\,C^{(I)}_D +g^{(V)}_F\,C^{(I)}_F \Big)
\nonumber\\
&+&\frac{i}{4 f^2}\,\Big( \bar q^\mu\,\sigma_{\mu \nu}\,q^\nu \Big)
\,\Big( g^{(T)}_1\,\bar C^{(I)}_{1}+g^{(T)}_{D}\,\bar C^{(I)}_{D} +g^{(T)}_F\,\bar C^{(I)}_{F}\Big)\;,
\nonumber\\
K^{(I)}_{\chi}(\bar k, k; w)&=&
\frac{b_0}{f^2}\,\Big(
m_\pi^2\,C^{(I)}_{\pi,0} +m_K^2\,C^{(I)}_{K,0} \Big)
+\frac{b_D}{f^2}\,\Big(
m_\pi^2\,C^{(I)}_{\pi,D} +m_K^2\,C^{(I)}_{K,D} \Big)
\nonumber\\
&+&\frac{b_F}{f^2}\,\Big(
m_\pi^2\,C^{(I)}_{\pi,F} +m_K^2\,C^{(I)}_{K,F} \Big)\;,
\label{def-local-1}
\end{eqnarray}
where the coefficients $C^{(I)}_{..,ab}$ are listed in Tab. \ref{tabpi-2} for the 
strangeness zero channel and shown in Appendix E for the strangeness 
minus one channel. A complete listing of the $Q^3$ quasi-local two-body interaction terms
can be found in Appendix B.

\subsection{Chiral expansion and covariance}

We proceed and detail the chiral expansion strategy for the interaction kernel 
in which we wish to keep covariance manifestly. Here it is instructive to  rewrite 
first the meson energy $\omega^{(I)}_a$ and the baryon energy $E_a^{(I)}$  in (\ref{bark-def})
\begin{eqnarray}
&&\omega^{(I)}_a(\sqrt{s}) =\sqrt{s}-m^{(I)}_{B(I,a)}
-\frac{\phi^{(I)}_a(\sqrt{s})}{2\,\sqrt{s}}\,,
\;\;\;E_a^{(I)}(\sqrt{s}) = m_{B(I,a)}+\frac{\phi^{(I)}_a(\sqrt{s})}{2\,\sqrt{s}}\;,
\nonumber\\
&&\phi^{(I)}_{a}(\sqrt{s})= \Big(\sqrt{s}-m_{B(I,a)}\Big)^2-m_{\Phi(I,a)}^2 \,,
\label{rel-power}
\end{eqnarray}
in terms of the approximate phase-space  factor $\phi^{(I)}_{a}(\sqrt{s}\,)$.
We assign $\sqrt{s}\sim Q^0$ and $\phi^{(I)}_a/ \sqrt{s} \sim Q^2$. This implies a
unique decomposition of the meson energy $ \omega^{(I)}_a $ in the leading term
$ \sqrt{s}- m_{B(I,a)} \sim Q$ and the subleading term
$-\phi_a^{(I)}/(2 \sqrt{s})\sim Q^2$.
We stress that our assignment leads to $m_{B(I,a)}$ as the leading chiral moment of the
baryon energy $E_a^{(I)}$.
We differ here from the conventional heavy-baryon formalism which assigns
the chiral power $Q$ to the full meson energy $\omega_a^{(I)}$. Consistency with
(\ref{rel-power}), in particular with $E_a^{(I)}=\sqrt{s}- \omega_a^{(I)}$, then results in 
either a leading chiral moment of the baryon energy $E_a^{(I)}$ not equal to the baryon mass 
or an assignment $\sqrt{s}\; \simslash \;Q^0$.

The implications of our relativistic power counting assignment are first exemplified 
for the case of the quasi-local two-body interaction terms.
We derive the effective interaction kernel relevant for the s- and p-wave channels 
\begin{eqnarray}
V^{(I,+)}_{[8][8]}(\sqrt{s};0)&=& \frac{1}{4 f^2}\,
\Big(g^{(S)}_0+\sqrt{s}\,g_0^{(V)}\Big)\,
\Big( \sqrt{s}-M^{(I)}\Big)\,C_0^{(I)}\,\Big( \sqrt{s}-M^{(I)}\Big)
\nonumber\\
&+& \frac{1}{4 f^2}\,\Big(g^{(S)}_1+\sqrt{s}\,g_1^{(V)}\Big)\,
\Big( \sqrt{s}-M^{(I)}\Big)\,C_1^{(I)}\,\Big( \sqrt{s}-M^{(I)}\Big)
\nonumber\\
&+& \frac{1}{4 f^2}\, \Big(g^{(S)}_D+\sqrt{s}\,g_D^{(V)}\Big)\,
\Big( \sqrt{s}-M^{(I)}\Big)\,C_D^{(I)}\,\Big( \sqrt{s}-M^{(I)}\Big)
\nonumber\\
&+& \frac{1}{4 f^2}\,\Big(g^{(S)}_F+\sqrt{s}\,g_F^{(V)}\Big)\,
\Big( \sqrt{s}-M^{(I)}\Big)\,C_F^{(I)}\,\Big( \sqrt{s}-M^{(I)}\Big)
\nonumber\\
&+&{\mathcal O} \left(Q^3\right)\;,
\nonumber\\
V^{(I,-)}_{[8][8]}(\sqrt{s};0)&=&
-\frac{1}{3 f^2} \,M^{(I)}\,
\Big(g^{(S)}_0\,C_0^{(I)}+g^{(S)}_1\,C_1^{(I)}\Big)\, M^{(I)}
\nonumber\\
&-&\frac{1}{3 f^2} \,M^{(I)}\,
\Big(g^{(S)}_D\,C_D^{(I)} +g^{(S)}_F\,C_F^{(I)}\Big)\, M^{(I)}
\nonumber\\
&+& \frac{1}{4 f^2} \left(\sqrt{s}+M^{(I)}\right)
\Big(g^{(T)}_1\,\bar C_{1}^{(I)}+g^{(T)}_D\,\bar C_{D}^{(I)}\Big)
\left(\sqrt{s}+M^{(I)}\right)
\nonumber\\
&+& \frac{1}{4 f^2} \left(\sqrt{s}+M^{(I)}\right)
g^{(T)}_F\,\bar C_{F}^{(I)} \left(\sqrt{s}+M^{(I)}\right)
\nonumber\\
&-&\frac{1}{3 f^2}\,M^{(I)}\,
\Big(g^{(T)}_1\,\bar C_{1}^{(I)}+g^{(T)}_D\,\bar C_{D}^{(I)}+g^{(T)}_F\,\bar C_{F}^{(I)} \Big)\,
M^{(I)}
+ {\mathcal O} \left(Q\right)\;,
\nonumber\\
V^{(I,+)}_{[8][8]}(\sqrt{s};1)&=& -\frac{1}{12 f^2}\,
\Big(g^{(S)}_0\,C_0^{(I)}+g^{(S)}_1\,C_1^{(I)}+g^{(S)}_D\,C_D^{(I)}
+g^{(S)}_F\,C_F^{(I)}\Big)
\nonumber\\
&-&\frac{1}{12 f^2}\,
\Big(g^{(T)}_1\,\bar C_{1}^{(I)}+g^{(T)}_D\,\bar C_{D}^{(I)}+g^{(T)}_T\,\bar C_{F}^{(I)}\Big)
+ {\mathcal O} \left(Q \right)\;.
\nonumber\\
V^{(I,+)}_{\chi}(\sqrt{s},0)&=& \frac{b_0}{f^2}\,\Big( 
m_\pi^2\,C^{(I)}_{\pi,0} +m_K^2\,C^{(I)}_{K ,0} \Big)
+\frac{b_D}{f^2}\,\Big( 
m_\pi^2\,C^{(I)}_{\pi,D} +m_K^2\,C^{(I)}_{K ,D} \Big)
\nonumber\\
&+&\frac{b_F}{f^2}\,\Big( 
m_\pi^2\,C^{(I)}_{\pi,F} +m_K^2\,C^{(I)}_{K ,F}\Big)
+{\mathcal O}\left(Q^3\right)\;,
\label{local-v}
\end{eqnarray}
where we introduced the diagonal baryon mass matrix $M^{(I)}_{ab}=\delta_{ab}\,m_{B(I,a)}$.
Our notation in (\ref{local-v}) implies a matrix multiplication of the mass
matrix $M^{(I)}$ by the coefficient matrices $C^{(I)}$ in the '$ab$' channel-space.
Recall that the d-wave interaction kernel $V^{(I,-)}_{[8][8]}(\sqrt{s},1)$ 
does not receive any contribution from quasi-local counter 
terms to chiral order $Q^3$. Similarly the chiral symmetry-breaking interaction kernel
$V_\chi$ can only contribute to s-wave scattering to this order. 
We point out that the result (\ref{local-v}) is uniquely determined by expanding the meson 
energy according to (\ref{rel-power}). In particular we keep in $V^{(I,+)}_{\Phi B}(\sqrt{s},0)$
the $\sqrt{s}$ factor in front of $g^{(V)}$.
We refrain from any further expansion in $\sqrt{s}-\Lambda $ with some scale
$\Lambda \simeq m_N$. The relativistic chiral Lagrangian supplied with (\ref{rel-power})
leads to well defined kinematical factors included in (\ref{local-v}).
These kinematical factors, which are natural ingredients of the relativistic chiral
Lagrangian, can be generated also in the heavy-baryon formalism by a proper regrouping of
interaction terms. The $Q^3$-terms induced by the interaction kernel (\ref{def-local-1}) are 
shown in Appendix B as part of a complete collection of relevant $Q^3$-terms.

Next we consider the Weinberg-Tomozawa term and the baryon octet and decuplet s-channel 
pole contributions
\begin{eqnarray}
V^{(I,\pm)}_{WT}(\sqrt{s};0)&=&\frac{1}{2\,f^2}\,\left( \sqrt{s}\,C^{(I)}_{WT}
\mp \frac{1}{2}\,\Big[M^{(I)}\,,C^{(I)}_{WT}\Big]_+ \right) \;,
\nonumber\\
V^{(I,\pm)}_{s-[8]}(\sqrt{s};0)
&=& -\sum_{c=1}^3\,\Big(\sqrt{s}\mp M^{(I,c)}_5 \Big)\,
\frac{C^{(I,c)}_{[8]}}{4\,f^2\,\big(\sqrt{s}\pm  m^{(c)}_{[8]}\big) }\,
\Big(\sqrt{s}\mp M^{(I,c)}_5 \Big)\,,
\nonumber\\
V^{(I,+)}_{s-[10]}(\sqrt{s};0)
&=& -\frac{2}{3}\,\sum_{c=1}^2\,\frac{\sqrt{s}+m^{(c)}_{[10]}}{(m^{(c)}_{[10]})^2}\,
\Big(\sqrt{s}- M^{(I)} \Big)\,\frac{C^{(I,c)}_{[10]}}{4\,f^2 }\,\Big(\sqrt{s}- M^{(I)} \Big)
\nonumber\\
&+&\sum_{c=1}^2\,Z_{[10]}\,\frac{2\,\sqrt{s}-m^{(c)}_{[10]}}{3\,(m^{(c)}_{[10]})^2}\,
\Big(\sqrt{s}- M^{(I)} \Big)\,\frac{C^{(I,c)}_{[10]}}{4\,f^2 }\,\Big(\sqrt{s}- M^{(I)} \Big)
\nonumber\\
&+&\sum_{c=1}^2\,Z_{[10]}^2\,\frac{2\,m^{(c)}_{[10]}-\sqrt{s}}{6\,(m^{(c)}_{[10]})^2}\,
\Big(\sqrt{s}- M^{(I)} \Big)\,\frac{C^{(I,c)}_{[10]}}{4\,f^2 }\,\Big(\sqrt{s}- M^{(I)} \Big)
\nonumber\\
&+&{\mathcal O}\left(Q^3 \right)\;,
\nonumber\\
V^{(I,-)}_{s-[10]}(\sqrt{s};0)
&=&\sum_{c=1}^2\, \frac{Z_{[10]}}{3\,m^{(c)}_{[10]}}\,
\Big(\sqrt{s}+ M^{(I)} \Big)\,\frac{C^{(I,c)}_{[10]}}{4\,f^2 }\,\Big(\sqrt{s}+ M^{(I)} \Big)
\nonumber\\
&-&\sum_{c=1}^2\,Z_{[10]}^2\,\frac{\sqrt{s}+2\,m^{(c)}_{[10]}}{6\,(m^{(c)}_{[10]})^2}\,
\Big(\sqrt{s}+ M^{(I)} \Big)\,\frac{C^{(I,c)}_{[10]}}{4\,f^2 }\,\Big(\sqrt{s}+ M^{(I)} \Big)
\nonumber\\
&+&{\mathcal O}\left(Q \right)\;,
\nonumber\\
V^{(I,+)}_{s-[10]}(\sqrt{s};+1)
&=& -\frac{1}{3}\,\sum_{c=1}^2\,
\frac{C^{(I,c)}_{[10]}}{4\,f^2\,(\sqrt{s}-m^{(c)}_{[10]})}
+{\mathcal O}\left(Q \right)\;,
\nonumber\\
V^{(I,-)}_{s-[9]}(\sqrt{s};-1)
&=& -\frac{1}{3}\,\sum_{c=1}^2\,
\frac{C^{(I,c)}_{[9]}}{4\,f^2\,(\sqrt{s}-m^{(c)}_{[9]})}
+{\mathcal O}\left(Q^0 \right)\;
\label{v-result-1}
\end{eqnarray}
where we suppressed terms of order $Q^3$ and introduced:
\begin{eqnarray}
\Big[M^{(I,c)}_5 \Big]_{ab} = \delta_{ab}\, \Big( m_{B(I,a)}+R^{(I,c)}_{L,aa}\Big) \;.
\label{}
\end{eqnarray}
The Weinberg-Tomozawa interaction term
$V_{WT}$ contributes to the s-wave and p-wave interaction kernels with $J=\frac{1}{2}$ 
to chiral order $Q$ and $Q^2$ respectively but not in the $J=\frac{3}{2}$ channels. Similarly
the baryon octet s-channel exchange $V_{s-[8]}$ contributes only in the $J=\frac{1}{2}$ 
channels and the baryon decuplet s-channel exchange $V_{s-[10]}$ to all considered 
channels but the d-wave channel. The $Q^3$-terms not shown in (\ref{v-result-1}) are given in Appendix B.
In (\ref{v-result-1}) we expanded also the baryon-nonet resonance contribution 
applying in particular the questionable formal rule $\sqrt{s}-m_{[9]}\sim Q$. To order $Q^3$ one then finds 
that only the  d-wave interaction kernels are affected. In fact, the vanishing of all 
contributions except the one in the d-wave channel does not depend on the assumption $\sqrt{s}-m_{[9]}\sim Q$. 
It merely reflects the phase space properties of the resonance field. We observe that our strategy preserves 
the correct pole contribution in $V^{(I,-)}_{s-[9]}(\sqrt{s};-1)$ but discards smooth background terms in 
all partial-wave interaction kernels. That is consistent with our discussion of section 4.1 which 
implies that those background terms are not controlled in any case. We emphasize that we keep the physical 
mass matrix $M^{(I)}$ rather than the chiral $SU(3)$ limit value in the interaction kernel. Since the mass 
matrix follows from the on-shell reduction of the interaction kernel it necessarily involves
the physical mass matrix $M^{(I)}$. Similarly we keep $M^{(I)}_5$ unexpanded since only that 
leads to the proper meson-baryon coupling strengths. This procedure is  analogous to keeping 
the physical masses in the unitarity loop functions $J_{MB}(w)$ in (\ref{jpin-n-def}).

We proceed with the baryon octet and baryon decuplet
u-channel contributions. After performing their proper angular average 
as implied by the partial-wave projection in (\ref{bark-def}) their 
contributions to the scattering kernels are written in terms of  matrix valued functions
${ h}^{(I)}_{n \pm}(\sqrt{s},m),{q}^{(I)}_{n \pm }(\sqrt{s},m)$ and 
${ p}^{(I)}_{n \pm }(\sqrt{s},m)$ as
\begin{eqnarray}
\Big[V^{(I,\pm )}_{u-[8]}(\sqrt{s};n)\Big]_{ab} &=& \sum_{c=1}^4\,
\frac{1}{4\,f^2 }\,\Big[\widetilde C^{(I,c)}_{[8]}\Big]_{ab}\,
\Big[{ h}^{(I)}_{n \pm }(\sqrt{s},m^{(c)}_{[8]})\Big]_{ab} \;,
\nonumber\\
\Big[V^{(I,\pm )}_{u-[10]}(\sqrt{s};n)\Big]_{ab} &=&\sum_{c=1}^3\,
\frac{1}{4\,f^2 }\,\Big[\widetilde C^{(I,c)}_{[10]}\Big]_{ab}\,
\Big[{ p}^{(I)}_{n \pm }(\sqrt{s},m_{[10]}^{(c)})\Big]_{ab}\; ,
\nonumber\\
\Big[V^{(I,\pm )}_{u-[9]}(\sqrt{s};n)\Big]_{ab} &=& \sum_{c=1}^4\,
\frac{1}{4\,f^2 }\,\Big[\widetilde C^{(I,c)}_{[9]}\Big]_{ab}\,
\Big[{q}^{(I)}_{n \pm }(\sqrt{s},m^{(c)}_{[9]})\Big]_{ab} \;.
\label{u-result}
\end{eqnarray}
The functions ${h}^{(I)}_{n \pm}(\sqrt{s},m), { q}^{(I)}_{n \pm }(\sqrt{s},m)$ and  
${p}^{(I)}_{n \pm}(\sqrt{s},m)$ can be expanded in  
chiral powers once we assign formal powers to the typical building blocks
\begin{eqnarray}
\mu_{\pm,ab}^{(I)}({s},m)=  m_{B(I,a)}+m_{B(I,b)}-\sqrt{s}\pm m \;.
\label{mupm-scale}
\end{eqnarray}
We count $\mu_- \sim Q$ and $\mu_+ \sim Q^0$ but refrain from any further expansion. 
Then the baryon octet functions ${h}^{(I)}_{n \pm}(\sqrt{s},m)$ to order $Q^3$ read
\begin{eqnarray}
&&\Big[{h}_{0+}^{(I)} (\sqrt{s},m)\Big]_{ab}=\sqrt{s}+m +\tilde R^{(I,c)}_{L,ab}
+\tilde R^{(I,c)}_{R,ab}-
\frac{m+M_{ab}^{(L)}}{\mu_{+,ab}^{(I)}({s},m)}\,
\Bigg(\sqrt{s}+
\nonumber\\
&& \qquad\qquad
+\frac{\phi^{(I)}_{a}({s})\,(\sqrt{s}-m_{B(I,b)})}
{\mu_{+,ab}^{(I)}({s},m)\,\mu_{-,ab}^{(I)}({s},m)}
+\frac{(\sqrt s-m_{B(I,a)})\,\phi^{(I)}_{b}({s})}
{\mu_{+,ab}^{(I)}({s},m)\,\mu_{-,ab}^{(I)}({s},m)}
\nonumber\\
&& \qquad\qquad
+\frac{1}{3}\,\frac{\phi^{(I)}_{a}({s})\,(4\,\sqrt s -\mu_{+,ab}^{(I)}({s},m))\,\phi^{(I)}_{b}({s}) }
{\big( \mu_{+,ab}^{(I)}({s},m)\big)^2\,\big(\mu_{-,ab}^{(I)}({s},m)\big)^2}
\Bigg)\,\frac{m+M_{ab}^{(R)}}{\sqrt{s}}+ {\mathcal O}\left( Q^3\right)\,,
\nonumber\\
&&\Big[{h}_{0-}^{(I)} (\sqrt{s},m)\Big]_{ab}=
\frac{m+M_{ab}^{(L)}}{\mu_{-,ab}^{(I)}({s},m)}\,
\Bigg(2\,\sqrt{s}+\mu_{-,ab}^{(I)}({s},m)
\nonumber\\
&&\qquad\qquad
-\frac{8}{3}\,\frac{m_{B(I,a)}\,m_{B(I,b)}}
{ \mu_{+,ab}^{(I)}({s},m)} \Bigg)\,
\frac{m+M_{ab}^{(R)}}{\mu_{+,ab}^{(I)}({s},m)}
+ {\mathcal O}\left( Q\right)\,,
\nonumber\\
&&\Big[{h}_{1+}^{(I)} (\sqrt{s},m)\Big]_{ab}=-\frac23\,
\frac{(m+M_{ab}^{(L)})\,(m+M_{ab}^{(R)}) }{\mu_{-,ab}^{(I)}({s},m)\,
\big(\mu_{+,ab}^{(I)}({s},m)\big)^2}
+{\mathcal O}\left( Q\right)
\nonumber\\
\nonumber\\
&&\Big[{h}_{1-}^{(I)} (\sqrt{s},m)\Big]_{ab}=
-\frac{32}{15}\,
\frac{(m+M_{ab}^{(L)})\,m_{B(I,a)}\,m_{B(I,b)}\,(m+M_{ab}^{(R)}) }{\big(\mu_{-,ab}^{(I)}({s},m)\big)^2\,
\big(\mu_{+,ab}^{(I)}({s},m)\big)^3}
\nonumber\\
&&\qquad\qquad
+\frac43\,
\frac{(m+M_{ab}^{(L)})\,\sqrt{s}\,(m+M_{ab}^{(R)}) }{\big(\mu_{-,ab}^{(I)}({s},m)\big)^2\,
\big(\mu_{+,ab}^{(I)}({s},m)\big)^2}
+{\mathcal O}\left( Q^{-1}\right)
\label{u-approx-1}\;,
\end{eqnarray}
where we introduced $M_{ab}^{(L)}=m_{B(I,a)}+\tilde R^{(I,c)}_{L,ab}$ and 
$M_{ab}^{(R)}=m_{B(I,b)}+\tilde R^{(I,c)}_{R,ab}$.
The terms of order $Q^3$ can be found in Appendix G where we present also the 
analogous expressions for the decuplet and baryon-octet resonance
exchanges. We emphasize two important points related to the expansion in 
(\ref{u-approx-1}). First, it leads to a separable interaction kernel. Thus the induced 
Bethe-Salpeter equation is solved conveniently by the covariant projector method of 
section 3.4. Secondly, such an expansion is only meaningful in conjunction with the    
renormalization procedure outlined in section 3.2. The expansion leads necessarily to 
further divergencies which require careful attention. 

We close this section with a more detailed discussion of the u-channel exchange. Its 
non-local nature necessarily leads to singularities in the partial-wave scattering 
amplitudes at subthreshold energies. For instance, the expressions (\ref{u-approx-1}) 
as they stand turn useless at energies $\sqrt{s}\simeq m_{B(I,a)}+m_{B(I,b)}-m$ due to
unphysical multiple poles at $\mu_-=0$. One needs to address this problem, because  
the subthreshold amplitudes are an important input for the many-body treatment of 
the nuclear meson dynamics. We stress that a singular behavior in the vicinity of $\mu_-\sim 0$ 
is a rather general property of any u-channel exchange contribution. It is 
not an artifact induced by the chiral expansion. The partial-wave decomposition 
of a u-channel exchange contribution represents the pole term only for sufficiently 
large or small $\sqrt{s}$. To be explicit we consider the u-channel nucleon pole term 
contribution of elastic $\pi N$ scattering
\begin{eqnarray}
\sum_{n=0}^\infty \,\int_{-1}^1\frac{d x}{2}\,
\frac{P_n(x)}{\mu_{\pi N}^{(+)}\,\mu_{\pi N}^{(-)}-2\,\phi_{\pi N}\,x
+{\mathcal O}\left( Q^3\right)}\;,
\label{pin-example:b}
\end{eqnarray}
where $\mu^{(\pm)}_{\pi N}=2\,m_N -\sqrt{s}\pm m_N$.
Upon inspecting the branch points induced by the angular average one concludes that the partial
wave decomposition (\ref{pin-example:b}) is valid only if
$\sqrt{s} >\Lambda_+$ or $\sqrt{s}<\Lambda_-$ with
$\Lambda_\pm=m_N\pm m^2_\pi/m_N+{\mathcal O}(Q^3)$.
For any value in between, $\Lambda_-<\sqrt{s} <\Lambda_+$, the partial-wave decomposition
is not converging. We therefore adopt the following prescription for the
$u$-channel contributions. The unphysical pole terms in
(\ref{u-approx-1}) are systematically replaced by
\begin{eqnarray}
&& m\,\Lambda^{(\pm)}_{ab} (m)= m\,(m_{B(I,a)}+m_{B(I,b)})-m^2 
\nonumber\\
&& \qquad \qquad \pm \left(\Big((m-m_{B(I,b)})^2 - m_{\Phi(I,a)}^2\Big)\,
\Big((m-m_{B(I,a)})^2-m_{\Phi(I,b)}^2\Big)\right)^{1/2} \;,
\nonumber\\
&& \left(\mu^{-1}_{-,ab}(\sqrt{s},m)\right)^n
 \rightarrow
\left(\mu^{-1}_{-,ab}(\Lambda_{ab}^{(-)},m)\right)^n
\,\frac{\sqrt{s}-\Lambda_{ab}^{(+)}}{\Lambda_{ab}^{(-)}-\Lambda_{ab}^{(+)}}
\nonumber\\
&&\qquad \qquad \qquad \quad \;+\,
\left(\mu^{-1}_{-,ab}(\Lambda_{ab}^{(+)},m)\right)^n
\,\frac{\sqrt{s}-\Lambda_{ab}^{(-)}}{\Lambda_{ab}^{(+)}-\Lambda_{ab}^{(-)}} \;,
\label{prescription}
\end{eqnarray}
for  $\Lambda_{ab}^{(-)} < \sqrt{s}< \Lambda_{ab}^{(+)}$  
but kept unchanged for $\sqrt{s}>\Lambda^{(+)}_{ab}$ or $\sqrt{s}<\Lambda^{(-)}_{ab}$ in
a given channel (a,b). The prescription (\ref{prescription}) properly generalizes the 
$SU(2)$ sector result $\Lambda_\pm\simeq m_N\pm m^2_\pi/m_N $ to the $SU(3)$ sector. 
We underline  that (\ref{prescription}) leads to regular expressions for
$h_{\pm n}(\sqrt{s},m)$ but leaves the u-channel pole contributions unchanged above 
threshold. The reader may ask why we impose such a prescription at all. Outside the 
convergence radius of the partial-wave expansion the amplitudes do not make much sense in any 
case. Our prescription is nevertheless required due to a coupled-channel effect. To be 
specific, consider for example elastic kaon-nucleon scattering in the strangeness minus one channel. 
Since there are no u-channel exchange contributions in this channel the physical 
partial-wave amplitudes  do not show any induced singularity structures. For instance in the 
$\pi \Sigma \to \pi \Sigma $ channel, on the other hand, the u-channel hyperon exchange does 
contribute and therefore leads through the coupling of the $\bar K N$ and $\pi \Sigma $ channels 
to a singularity also in the $\bar K N$ amplitude. Such induced singularities are 
an immediate consequence of the approximate treatment of the u-channel exchange contribution
and must be absent in an exact treatment. 
As a measure for the quality of our prescription we propose  
the accuracy to which the resulting subthreshold forward kaon-nucleon scattering 
amplitudes satisfies a dispersion integral. We return to this issue in the result section.

\subsection{$SU(3)$ dynamics in the $S=1$ channels}

We turn to the $K^+$-nucleon scattering process which is related to  
the $K^-$-nucleon scattering process by crossing symmetry. 
The scattering amplitudes $T^{(I)}_{K N \rightarrow K N}$ follow from the 
$K^-$-nucleon scattering amplitudes $T^{(I)}_{\bar K N \rightarrow \bar K N}$ 
via the transformation
\begin{eqnarray}
&&T^{(0)}_{K N \rightarrow K N}(\bar q, q; w)
=-\frac{1}{2}\, T^{(0)}_{\bar K N \rightarrow \bar K N}(-q, -\bar q; w-\bar q-q)
\nonumber\\
&&\quad \quad \quad \quad \quad
+\frac{3}{2}\,T^{(1)}_{\bar K N \rightarrow \bar K N}(-q, -\bar q; w-\bar q-q) \;,
\nonumber\\
&&T^{(1)}_{K N \rightarrow K N}(\bar q, q; w)
=+\frac{1}{2}\, T^{(0)}_{\bar K N \rightarrow \bar K N}(-q, -\bar q; w-\bar q-q)
\nonumber\\
&&\quad \quad \quad \quad \quad
+\frac{1}{2}\,T^{(1)}_{\bar K N \rightarrow \bar K N}(- q, -\bar q; w-\bar q-q) \;.
\label{cross-symmetry}
\end{eqnarray}
It is instructive to work out the implication of crossing symmetry for the kaon-nucleon
scattering amplitudes in more detail. We decompose the on-shell scattering amplitudes 
into their partial-wave amplitudes:
\begin{eqnarray}
&&T^{(I)}_{K N \rightarrow K N}(\bar q, q; w) =
\frac{1}{2}\,\Bigg(\frac{\wslash}{\sqrt{w^2}}+1
\Bigg)\, F^{(I,+)}_{KN}(s,t)+ \frac{1}{2}\,\Bigg(\frac{\wslash}{\sqrt{w^2}}-1
\Bigg)\,F^{(I,-)}_{KN}(s,t)
\nonumber\\
&&\quad \quad \quad = \sum_{n=0}^\infty \,
\Big( M^{(I,+)}_{KN}(\sqrt{s};n)\,{Y}^{(+)}_n(\bar q,q;w)+
 M^{(I,-)}_{KN}(\sqrt{s};n)\,{Y}^{(-)}_n(\bar q,q;w) \Big)\;,
\nonumber\\
&&T^{(I)}_{\bar K N \rightarrow \bar K N}(\bar q, q; w) =
\frac{1}{2}\,\Bigg(\frac{\wslash}{\sqrt{w^2}}+1
\Bigg)\, F^{(I,+)}_{\bar KN}(s,t)+ \frac{1}{2}\,\Bigg(\frac{\wslash}{\sqrt{w^2}}-1
\Bigg)\,F^{(I,-)}_{\bar KN}(s,t)
\nonumber\\
&&\quad \quad \quad = \sum_{n=0}^\infty \,
\Big( M^{(I,+)}_{\bar KN}(\sqrt{s};n)\,{Y}^{(+)}_n(\bar q,q;w)+
 M^{(I,-)}_{\bar KN}(\sqrt{s};n)\,{Y}^{(-)}_n(\bar q,q;w) \Big)
\label{}
\end{eqnarray}
where we suppressed the on-shell nucleon spinors. The crossing identity (\ref{cross-symmetry})
leads to
\begin{eqnarray}
&&F^{(I,\pm )}_{K N}(s,u) = \sum_{n=0}^\infty\,\Bigg(
\pm\,\frac{1}{2}\left(\frac{2\,m_N\mp\sqrt{s}}{\sqrt{u}}+1 \right) \,
\bar Y^{(c)}_{n+1}(\bar q,q; w)
\nonumber\\
&&\quad \quad \quad  \mp
\frac{1}{2}\left( \frac{2\,m_N\mp\sqrt{s}}{\sqrt{u}}-1\right)
\left(\bar E+ m_N \right)^2
\,\bar Y^{(c)}_{n}(\bar q,q; w)
\Bigg)\,\bar M^{(I,+)}_{\bar K N}(\sqrt{u};n)
\nonumber\\
&&\quad \quad \quad \quad \quad +\sum_{n=0}^\infty\,\Bigg(
\pm \,\frac{1}{2}\left(\frac{2\,m_N\mp\sqrt{s}}{\sqrt{u}}-1 \right) \,
\bar Y^{(c)}_{n+1}(\bar q,q; w)
\nonumber\\
&& \quad \quad \quad 
\mp\,\frac{1}{2}\left( \frac{2\,m_N\mp\sqrt{s}}{\sqrt{u}}+1\right)
\left(\bar E-m_N \right)^2
\,\bar Y^{(c)}_{n}(\bar q,q; w)\Bigg)\,\bar M^{(I,-)}_{\bar K N}(\sqrt{u};n) \,,
\nonumber\\
&& \bar E =\frac{1}{2}\,\sqrt{u}+\frac{m_N^2-m_K^2}{2\,\sqrt{u}}\;,\quad
u = 2\,m_N^2+2\,m_K^2-s+2\,p_{KN}^2\,(1-\cos \theta ) \;,
\label{kplus-amplitudes}
\end{eqnarray}
where
\begin{eqnarray}
&& \bar M^{(0,\pm )}_{\bar K N}(\sqrt{u};n) =-\frac{1}{2}\, M^{(0,\pm )}_{\bar K N }(\sqrt{u};n)+
\frac{3}{2}\,M^{(1,\pm )}_{\bar K N }(\sqrt{u};n)
\;,\quad \quad
\nonumber\\
&&\bar M^{(1,\pm )}_{\bar K N}(\sqrt{u};n) =+\frac{1}{2}\, M^{(0,\pm)}_{\bar K N}(\sqrt{u};n)+
\frac{1}{2}\,M^{(1,\pm )}_{\bar K N}(\sqrt{u};n)\;.
\label{kplus-amplitudes:b}
\end{eqnarray}
Note that $ \bar Y^{(c)}_{n}(\bar q,q; w)=\bar Y_{n}(-q,-\bar q; w-q-\bar q)$ was introduced already 
in (\ref{cov-proj}). The partial-wave amplitudes of the $K^+$-nucleon system can now be deduced from 
(\ref{kplus-amplitudes}) using (\ref{bark-def}). 

We proceed with two important remarks. First, if the solution of the coupled-channel 
Bethe-Salpeter equation of the $K^-$-nucleon system of the previous sections is  
used to construct the $K^+$-nucleon scattering amplitudes via (\ref{kplus-amplitudes}) 
one finds real partial-wave amplitudes in conflict with unitarity. And secondly, in any 
case our $K^-$-nucleon scattering amplitudes must not be applied far below the $K^-$-nucleon
threshold. The first point is obvious because in the $K^+$-nucleon channel only two-particle 
irreducible diagrams are summed. Note that the reducible diagrams in the $KN$ sector 
correspond to irreducible contributions in the $\bar K N$ sector and vice versa.  
Since the interaction kernel of the $\bar K N$ sector is evaluated perturbatively it is clear 
that the scattering amplitude does not include that infinite sum of reducible diagrams 
required for unitarity in the crossed channel. 
The second point follows, because the chiral $SU(3)$ Lagrangian 
is an effective field theory where the heavy-meson exchange contributions are integrated out.
It is important to identify the applicability domain correctly. Inspecting the 
singularity structure induced by the light t-channel vector meson exchange contributions one 
observes that they, besides restricting the applicability domain with 
$\sqrt{s} < \sqrt{m_N^2+m_\rho^2/4}+\sqrt{m_K^2+m_\rho^2/4}\simeq 1640$ MeV from above, 
induce cut-structures in between the $K^+$ and $K^-$-nucleon thresholds. 
This will be discussed in more detail below.
Though close to the $K^+$ and $K^-$-nucleon thresholds the tree-level contributions 
of the vector meson exchange are successfully represented by quasi-local interaction vertices, 
basically the Weinberg-Tomozawa interaction term, it is not justified to 
extrapolate a loop evaluated with the effective $K^-$-nucleon vertex down to the $K^+$- 
nucleon threshold. We conclude that one should not identify our $K^-$-nucleon scattering 
amplitudes with the Bethe-Salpeter kernel of the $K^+$-nucleon 
system. This would lead to a manifestly crossing symmetric approach, 
a so-called 'parquet' approximation, if set up in a self consistent manner. 
We reiterate the fatal drawback of a parquet type of approach within our present chiral 
framework: the $K^-$-nucleon amplitudes would be probed far below the $\bar K N$ threshold 
at $\sqrt{s}\simeq m_N-m_K$ outside their validity domain. A clear signal for the 
unreliability of the $K^-$-nucleon scattering amplitudes at $\sqrt{s}\simeq m_N-m_K$ is the 
presence of unphysical pole structures which typically arise at $\sqrt{s}< 700$ MeV. For 
example, the Fig. \ref{fig:wt} of section 4.3  would show unphysical pole structures if 
extended for $\sqrt{s}< 1$ GeV.

\tabcolsep=0.5mm
\renewcommand{\arraystretch}{1.45}
\begin{table}[t]
\begin{tabular}{|r||c||c|c|c||c|c||c|c|c|c||c|c|c|c|}
 \hline   
& $ C_{WT}^{(I)} $  &  $ C_{N_{[8]}}^{(I)}$  & $C_{\Lambda_{[8]}}^{(I)}$  &  
$C_{\Sigma_{[8]}}^{(I)}$ 
& $C_{\Delta_{[10]}}^{(I)}$ & $C_{\Sigma_{[10]} }^{(I)}$ 
    &$\widetilde{C}_{N_{[8]}}^{(I)}$ & $\widetilde{C}_{\Lambda_{[8]} }^{(I)}$ &
    $\widetilde{C}_{\Sigma_{[8]}}^{(I)}$ & $\widetilde {C}_{\Xi_{[8]} }^{(I)}$   &
$\widetilde {C}_{\Delta_{[10]} }^{(I)}$ &
    $\widetilde {C}_{\Sigma_{[10]}}^{(I)}$& $\widetilde C_{\Xi_{[10]}}^{(I)}$
\\ \hline \hline 
$I=0$& 0 & 0 & 0 & 0 & 0 & 0  
     & 0 & -$\frac{1}{2}$ & $\frac{3}{2}$ & 0 & 0 & $\frac{3}{2}$ & 0\\ \hline \hline 

$I=1$& -2 & 0 & 0 & 0 & 0 & 0  
     & 0  & $\frac{1}{2}$ & $\frac{1}{2}$ & 0 & 0 & $\frac{1}{2}$ & 0\\ \hline 
\end{tabular}
\vspace*{2mm} \caption{Weinberg-Tomozawa interaction strengths and baryon exchange coefficients in the 
strangeness plus one channels as defined in (\ref{k-nonlocal}).} 
\label{tabkp-1}
\end{table}

For the above reasons we evaluate the $K^+$-nucleon interaction kernel 
perturbatively from the chiral Lagrangian. Equivalently the interaction kernels,
$V_{K N}$, could be derived from the $K^-$-nucleon interaction kernels 
$V_{\bar K N}$ by applying the crossing identities (\ref{cross-symmetry}). 
The interaction kernels $V_{K N}(\sqrt{s};n)$ in the 
$K^+$-nucleon channel are given by (\ref{local-v},\ref{v-result-1},\ref{u-result}) with the 
required coefficients listed in Tab. \ref{tabkp-1} and Tab. \ref{tabkp-2}. By analogy with  
the treatment of the 
$K^-$-nucleon scattering process we take the $K^+$-nucleon interaction evaluated to 
chiral order $Q^3$ as input for the Bethe-Salpeter equation. 
Again we consider only those $Q^3$-correction terms which are leading in the $1/N_c$
expansion. The partial-wave scattering 
amplitudes $M_{KN}^{(I,\pm )}(\sqrt{s};n)$ then follow
\begin{eqnarray}
M_{KN}^{(I,\pm )}(\sqrt{s};n) &=& \frac{V_{KN}^{(I,\pm )}(\sqrt{s};n)}
{1-V_{KN}^{(I,\pm)}(\sqrt{s};n)\,J^{(\pm)}_{KN}(\sqrt{s}; n)} \;
\label{M:kplus}
\end{eqnarray}
where the loop functions $J^{(\pm)}_{KN}(\sqrt{s};n)$ are given in (\ref{result-loop:ab}) with 
$m_{B(I,a)}=m_N$ and $m_{\Phi(I,a)}=m_K$. The subtraction point $\mu^{(I)}$ is identified 
with the averaged hyperon mass $\mu^{(I)}=(m_\Lambda+m_\Sigma)/2$.

\tabcolsep=1.1mm
\renewcommand{\arraystretch}{1.3}
\begin{table}[t]
\begin{tabular}{|r||c||c|c||c|c|c||c|c|c|c||c|c|c|}
\multicolumn{14}{c}{}\\ \multicolumn{14}{c}{}\\
 \hline
    & $C_{\pi,0}^{(I)}$ &  $C_{\pi,D}^{(I)}$ &  $C_{\pi,F}^{(I)}$ 
    & $C_{K,0}^{(I)}$   & ${C}_{K,D}^{(I)}$  & ${C}_{K,F}^{(I)}$ 
    & ${C}_{0}^{(I)}$   &  ${C}_{1}^{(I)}$  & 
    ${C}_{D}^{(I)}$ & ${C}_{F}^{(I)}$   
    & $\bar C_{1}^{(I)}$ & $\bar C_{D}^{(I)}$ & $\bar C_{F}^{(I)}$\\
\hline \hline 
$I=0$& 0 & 0 & 0
    & -4 & 0 & 4
    & 2 & -1 & 0& -2  
    & 1 & 2& 0   \\ \hline  \hline
$I=1$&0 & 0 & 0  
    & -4 & -4& 0
    & 2  & 1  & 2& 0 
    & -1& 0 &  -2   \\ \hline 
\end{tabular}
\vspace*{2mm} \caption{Coefficients of quasi-local interaction terms in the strangeness plus one
channels as defined in (\ref{def-local-1}).} 
\label{tabkp-2}
\end{table}

We point out that in our scheme we arrive at a crossing symmetric amplitude by matching 
the amplitudes  $M^{(I,\pm)}_{KN}(\sqrt{s},n)$ and $M^{(I,\pm)}_{\bar KN}(\sqrt{s},n)$
at subthreshold energies. The matching interval must be chosen so that both the $K^+$ 
and $K^-$ amplitudes are still within their validity domains. A complication arises due 
to the light vector meson exchange contributions in the t-channel, which we already 
identified above to restrict the validity domain of the present chiral approach. To be explicit 
consider the t-channel $\omega $ exchange. It leads to a branch point at 
$\sqrt{s}= \Lambda_\omega $ with 
$\Lambda_\omega =(m_N^2-m_\omega^2/4)^{1/2}+(m_K^2-m_\omega^2/4)^{1/2} \simeq $ 1138 MeV.
Consequently, one expects the partial-wave $K^+$ and $K^-$-amplitudes to be reliable
for $\sqrt{s} > \Lambda_\omega $ only. That implies, however, that in the crossed channel 
the amplitude is needed for 
$\sqrt{s}>(2\,m_N^2-\Lambda^2_\omega +2\,m_K^2)^{1/2}\simeq $ 978 MeV $< \Lambda_\omega $.
One may naively conclude that crossing symmetry appears outside the scope of our scheme, 
because the matching window is closed.  Note that the minimal critical point 
$\Lambda_{\rm opt}$ needed to open the matching window is
\begin{eqnarray}
\Lambda_{\rm opt.}\simeq \sqrt{m_N^2+m_K^2}\simeq  1061 {\rm MeV}\;. 
\label{opt-match}
\end{eqnarray}
It is determined by the condition $s=u$ at $\cos \theta =1$. We conclude that matching 
the $K^+$ and $K^-$ amplitudes requires that both amplitudes are within their applicability 
domain at $\sqrt{s}= \Lambda_{\rm opt.}$. The point $\sqrt{s}=\Lambda_{\rm opt.}$ is optimal, 
because it identifies the minimal reliability domain required for the matching of the 
subthreshold amplitudes. 
We note that the complication implied by the t-channel vector meson exchanges 
could be circumvented by reconstructing that troublesome branch point explicitly; 
after all it is determined by a tree-level diagram. 
On the other hand, it is clear that one may avoid this complication altogether if one 
considers the forward scattering amplitudes only. For the latter amplitudes the branch cut at 
$\sqrt{s}=\Lambda_\omega $ cancels identically and consequently the forward scattering 
amplitudes should be reliable for energies smaller than $ \Lambda_\omega$ also.  
For that reason, we demonstrate the matching for our forward scattering 
amplitudes only. We reconstruct the approximate forward scattering amplitudes 
$T^{(I)}_{KN}(s)$ in terms the partial-wave amplitudes included in our work
\begin{eqnarray}
T^{(I)}_{KN}(s) &=&
\frac{1}{2\,\sqrt{s}}\left(\frac{s+m_N^2-m_K^2}{2\,m_N}+\sqrt{s} \right)
M^{(I,+)}_{KN}(\sqrt{s};0)
\nonumber\\
&+&\frac{1}{2\,\sqrt{s}}\left(\frac{s+m_N^2-m_K^2}{2\,m_N} -\sqrt{s}\right)
M^{(I,-)}_{KN}(\sqrt{s};0)
\nonumber\\
&+& \frac{1}{\sqrt{s}}\left(\frac{s+m_N^2-m_K^2}{2\,m_N}+\sqrt{s} \right)
\,p^2_{KN}\,M^{(I,+)}_{KN}(\sqrt{s};1) 
\nonumber\\
&+& \frac{1}{\sqrt{s}}\left(\frac{s+m_N^2-m_K^2}{2\,m_N}-\sqrt{s} \right)
\,p^2_{KN}\,M^{(I,-)}_{KN}(\sqrt{s};1) +\cdots \,,
\label{forward-amplitude}
\end{eqnarray}
where $\sqrt{s}=\sqrt{m_N^2+p_{KN}^2}+\sqrt{m_K^2+p_{KN}^2}$. The analogous 
expression holds for $T^{(I)}_{\bar KN}(s)$.  The crossing identities for 
the forward scattering amplitudes 
\begin{eqnarray}
T^{(0)}_{KN}(s) &=& -\frac{1}{2}\,T^{(0)}_{\bar K N} (2\,m_N^2+2\,m_K^2-s)
+\frac{3}{2}\,T^{(1)}_{\bar K N} (2\,m_N^2+2\,m_K^2-s) \;,
\nonumber\\
T^{(1)}_{KN}(s) &=& +\frac{1}{2}\,T^{(0)}_{\bar K N} (2\,m_N^2+2\,m_K^2-s)
+\frac{1}{2}\,T^{(1)}_{\bar K N} (2\,m_N^2+2\,m_K^2-s)\;
\label{forward-crossing}
\end{eqnarray}
are expected to hold approximatively within some matching window centered around
$\sqrt{s}\simeq \Lambda_{\rm opt}$.   
A further complication arises due to the approximate treatment of the u-channel exchange 
contribution in the $K^+N$-channel. Since the optimal matching point 
$\Lambda_{\rm opt.}$ is not too far away from the hyperon poles with  
$\Lambda_{\rm opt.} \sim  m_\Lambda, m_\Sigma$ it is advantageous to investigate 
approximate crossing symmetry for the hyperon-pole term subtracted scattering 
amplitudes. 

We would like to stress that the expected approximate crossing symmetry is 
closely linked to our renormalization condition (\ref{ren-cond}). Since all 
loop functions $J^{(\pm )}(\sqrt{s},n)$ 
vanish close to $\sqrt{s}= (m_\Lambda+m_\Sigma)/2$ by construction, the 
$K^+$ and $K^-$ amplitudes turn perturbative sufficiently close to the optimal matching point 
$\Lambda_{\rm opt.} $. Therefore the approximate crossing symmetry of our scheme follows 
directly from the crossing symmetry of the interaction kernel. In the result section our final 
$K^\pm N$ amplitudes are confronted with the expected approximate crossing symmetry.

Finally we observe that for pion-nucleon scattering the situation is rather different. Given 
the rather small pion mass, the t-channel vector meson exchange contributions do not induce 
singularities in between the $\pi^+$ and $\pi^-$-nucleon thresholds but restrict the 
applicability domain of the chiral Lagrangian to 
$\sqrt{s} < \sqrt{m_N^2+m_\rho^2/4}+\sqrt{m_\pi^2+m_\rho^2/4}\simeq 1420$ MeV.
The approximate crossing symmetry follows directly from the perturbative character of 
the pion-nucleon sector. 

\section{Results}

In this section we present the many results of our detailed chiral 
$SU(3)$ analysis of the low-energy meson-baryon scattering data. We refer to our theory as the 
$\chi $-BS(3) approach for chiral Bethe-Salpeter dynamics of the flavor $SU(3)$ symmetry. Before delving 
into details we briefly summarize the main features and crucial arguments of our approach. We consider the 
number of colors ($N_c$) in QCD as a large parameter relying on a systematic expansion of the interaction kernel in 
powers of $1/N_c$. The coupled-channel Bethe-Salpeter kernel is evaluated in a combined chiral and 
$1/N_c$ expansion including terms of chiral order $Q^3$. We include contributions of s and u-channel baryon 
octet and decuplet states explicitly but only the s-channel contributions of the d-wave $J^P={\textstyle{3\over 2}}^-$ 
baryon nonet resonance states. 
Therewith we consider the s-channel baryon nonet contributions to the interaction kernel
as a reminiscence of further inelastic channels not included in the present scheme like for example the 
$K\,\Delta_\mu $ or $K_\mu \,N$ channel. We expect all baryon resonances, with the important exception of those resonances 
which belong to the large $N_c$ baryon ground states, to be generated by coupled channel dynamics.
Our conjecture is based on the observation that unitary (reducible) loop diagrams are typically enhanced 
by a factor of $2 \pi $ close to threshold relatively to irreducible diagrams. That factor invalidates the 
perturbative evaluation  of the scattering amplitudes and leads necessarily to a non-perturbative scheme with 
reducible diagrams summed to all orders. 

We are painfully aware of the fact that in our present scheme the explicit inclusion of the baryon 
resonance nonet states is somewhat leaving the systematic chiral framework, 
in the absence of a controlled approximation scheme. An explicit baryon resonance contribution is needed because 
there exist no reliable phase shift analyses of the antikaon-nucleon scattering process so far, in particular at 
low energies. Part of
the empirical information on the p-wave amplitudes stems from the interference effects of the p-wave amplitudes with 
the $\Lambda(1520)$ resonance amplitude. The $\Lambda(1520)$ resonance is interpreted as a $SU(3)$ singlet state with 
a small admixture of a  resonance octet state. We stress that the s- and p-wave channels are not affected by the 
baryon nonet states directly, because we discard the nonet u-channel contributions in accordance with our arguments 
concerning resonance generation. As a consequence all s- and p-channels are treated consistently within the chiral 
framework. We do not include further 
resonance fields in the p-wave channels, because those are not required at low energies and also because 
they would destroy the fundamental chiral properties of our scheme. Ultimately we expect those resonances to be 
generated by an extended coupled channel theory also.

The scattering amplitudes for the meson-baryon scattering processes are obtained from the solution of the coupled channel 
Bethe-Salpeter scattering equation. Approximate crossing symmetry of the amplitudes is guaranteed by a 
renormalization program which leads to the matching of subthreshold amplitudes. 
We first present a complete collection of the parameters as they are adjusted 
to the data set. In the subsequent sections we report on details of the fit strategy and confront our results 
with the empirical data. The result section is closed with a detailed analysis of our scattering amplitudes, demonstrating 
their good analyticity properties as well as their compliance with crossing symmetry.

\subsection{Parameters}

The set of parameters is well determined by the available 
scattering data and weak decay widths of the baryon octet states. 
Typically a data point is included in the analysis if $p_{\rm lab} 
< 350$ MeV. There exist low-energy elastic and inelastic $K^-p$ 
cross section data including in part angular distributions and 
polarizations. We include also the low-energy differential $K^+p \to K^+ p$ cross sections 
in our global fit. The empirical constraints set by the $K^+$-deuteron scattering data above 
$p_{\rm lab} > 350$ MeV are considered by requiring a reasonable matching to the single energy $S_{01}$ and $P_{03}$ 
phase shifts of \cite{Hashimoto}. That resolves an ambiguity in the parameter set. 
Finally, in order to avoid the various multi-energy $\pi N$ phase shifts imperspicuous close to threshold, we 
fit to the single-energy phase shifts of \cite{pion-phases}. This will be discussed in more detail below. We 
aim at a uniform quality of the data description. Therefore we form the $\chi^2$ per data point in each sector 
and add those up to the total $\chi^2$ which is minimized by the search algorithms of Minuit \cite{Minuit}. In cases 
where the empirical error bars are much smaller than the accuracy to which we expect the $\chi$-BS(3) to work to the 
given order, we artificially increase those error bars in our global fit. In Tab. \ref{q1param:tab}-\ref{q3param:tab} 
we present the parameter set of our best fit to the data. Note that part of the parameters are predetermined 
to a large extent and therefore fine tuned in a small interval only.

\tabcolsep=1.4mm
\renewcommand{\arraystretch}{1.5}
\begin{table}[h]\begin{center}
\begin{tabular}{|c|c|c|c|c|}
\hline 
$f$ [MeV] & $ C_R$ & $F_R$ & $D_R$ 
\\ 
\hline \hline
90.04  & 1.734 & 0.418 & 0.748  

\\  \hline 

\end{tabular}
\end{center}
\caption{Leading chiral parameters which contribute to meson-baryon 
scattering at order $Q$.}
\label{q1param:tab}
\end{table}

A qualitative understanding of the typical strength in the various channels can 
be obtained already at leading chiral order $Q$. In particular the 
$\Lambda(1405)$ resonance is formed as a result of the coupled-channel 
dynamics defined by the Weinberg-Tomozawa interaction vertices (see Fig. \ref{fig:wt}).  
There are four parameters relevant at that order $f$, $C_R$, $F_R$ and $D_R$.  
Their respective values as given in Tab. \ref{q1param:tab} are the result 
of our global fit to the data set including all parameters of the $\chi$-BS(3) approach. At leading 
chiral order the parameter $f$ determines the weak pion- and kaon-decay processes and at the same 
time the strength of the Weinberg-Tomozawa interaction vertices. To subleading order $Q^2$ the 
Weinberg-Tomozawa terms and the weak-decay constants of the pseudo-scalar meson octet receive independent 
correction terms. The result $f\simeq  90$ MeV is sufficiently close to the 
empirical decay parameters $f_\pi \simeq 92.4$ MeV and 
$f_K \simeq 113.0$ MeV to expect that the $Q^2$ correction terms lead indeed 
to values rather close to the empirical decay constants. Our value for $f$ is consistent 
with the estimate of \cite{GL85} which lead to $f_\pi/f = 1.07 \pm 0.12$.
The baryon octet and decuplet s- and u-channel exchange contributions to the interaction kernels are determined 
by the $F_R, D_R$ and $C_R$ parameters at leading order. Note that $F_R$ and $D_R$ predict 
the baryon octet weak-decay processes and $C_R$ the strong decay widths of the baryon decuplet states 
to this order also.

A quantitative description of the data set requires the inclusion 
of higher order terms. Initially we tried to establish a consistent 
picture of the existing low-energy meson-baryon scattering data 
based on a truncation of the interaction kernels to chiral order 
$Q^2$. This attempt failed due to the insufficient quality of the 
kaon-nucleon scattering data at low energies. In particular some of the 
inelastic $K^-$-proton differential cross sections are strongly influenced by 
the d-wave $\Lambda(1520)$ resonance at energies where the data points 
start to show smaller error bars.  We conclude that, on the one hand, 
one must include an effective baryon-nonet resonance field and, on the other 
hand, perform minimally a chiral $Q^3$ analysis to extend the 
applicability domain to somewhat higher energies. 
Since the effect of the d-wave resonances is only necessary in the strangeness minus one sector, 
they are only considered in that channel. The resonance parameters will be presented 
when discussing the strangeness minus one sector.

\begin{table}[t]\begin{center}
\begin{tabular}{|c|c||c|c||c|c||c|c|}
\hline 
$g_F^{(V)}$[GeV$^{-2}$] & 0.293 & $g_F^{(S)}$[GeV$^{-1}$] & -0.198 & 
$g_F^{(T)}$[GeV$^{-1}$] & 1.106 & $Z_{[10]}$  &  0.719
\\ 
\hline
$g_D^{(V)}$[GeV$^{-2}$] & 1.240 & $g_D^{(S)}$[GeV$^{-1}$] & -0.853 & 
$g_D^{(T)}$[GeV$^{-1}$] & 1.607 & - & -
\\ 
\hline
\end{tabular}
\end{center}
\caption{Chiral $Q^2$-parameters resulting from a fit to low-energy 
meson-baryon scattering data. Further parameters at this order are determined by the 
large $N_c$ sum rules.}
\label{q2param:tab}
\end{table}

At subleading order $Q^2$ the chiral $SU(3)$ 
Lagrangian predicts the relevance of 12 basically unknown parameters,  
$g^{(S)}, g^{(V)}$, $g^{(T)}$ and $Z_{[10]} $, which all need to be adjusted to the 
empirical scattering data. 
It is important to realize that chiral symmetry is largely predictive in the $SU(3)$ sector 
in the  sense that it reduces the number of parameters  beyond 
the static $SU(3)$ symmetry. For example one should compare the six tensors which 
result from decomposing $8\otimes 8= 1 
\oplus 8_S\oplus 8_A \oplus 10\oplus \overline{10}\oplus 27$ into its 
irreducible components with the subset of SU(3) structures selected 
by chiral symmetry in a given partial wave. Thus static $SU(3)$ 
symmetry alone would predict 18 independent terms for the s-wave 
and two p-wave channels rather than the 11 chiral $Q^2$ background 
parameters, $g^{(S)}, g^{(V)}$ and $g^{(T)}$. In our work the number of parameters was
further reduced significantly by insisting on the large $N_c$ sum rules
\begin{eqnarray}
g_1^{(S)}=2\,g_0^{(S)}= 4\,g_D^{(S)}/3 \,, \qquad 
g_1^{(V)}=2\,g_0^{(V)}= 4\,g_D^{(V)}/3 \,, \qquad g_1^{(T)}=0 \,,
\nonumber
\label{}
\end{eqnarray}
for the symmetry conserving quasi-local two body interaction terms (see (\ref{Q^2-large-Nc-result})).
In Tab. \ref{q2param:tab} we collect the values of all free parameters as they result from our 
best global fit. All parameters are found to have natural size. This is an important result, because 
only then is the application of the chiral power counting rule (\ref{q-rule}) justified.
We point out that the large $N_c$ sum rules derived in section 2 implicitly assume that other inelastic channels 
like $K \,\Delta_\mu $ or $K_\mu\,N $ are not too important. The effect of such 
channels can be absorbed to some extent into the quasi-local counter terms, however possibly at the 
prize that their large $N_c$ sum rules are violated. It is therefore a highly non-trivial result 
that we obtain a successful fit imposing (\ref{q2param:tab}). Note that the only previous analysis \cite{Kaiser},
which truncated the interaction kernel to chiral order $Q^2$ but did not include p-waves, found values for the s-wave 
range parameters largely inconsistent with the large $N_c$ sum rules. This may be due in part to the use of channel 
dependent cutoff parameters and the fact that that analysis missed octet and decuplet exchange contributions, 
which are important for the s-wave interaction kernel already to chiral order $Q^2$.

The parameters $b_0, b_D$ and $b_F$ to this order characterize the 
explicit chiral symmetry-breaking effects of QCD via the finite current 
quark masses.  The parameters $b_D$ and $b_F$ are well estimated 
from the baryon octet mass splitting (see (\ref{mass-splitting})) whereas 
$b_0$ must be extracted directly from the meson-baryon scattering data. 
It drives the size of the pion-nucleon sigma term for which conflicting values are 
still being discussed in the literature \cite{pin-news}. Our values 
\begin{eqnarray}
b_0 = -0.346\, {\rm GeV}^{-1} \,, \quad  b_D = 0.061\, {\rm GeV}^{-1} \,, \quad  b_F =-0.195 \,{\rm GeV}^{-1} \,,
\label{b-result}
\end{eqnarray}
are rather close to values expected from the baryon octet mass splitting (\ref{mass-splitting}). The 
pion-nucleon sigma term $\sigma_{\pi N}$ if evaluated at leading chiral order $Q^2$ (see (\ref{spin:naive})) would 
be $\sigma_{\pi N} \simeq 32$ MeV. That value should not be compared directly with $\sigma_{\pi N}$ as 
extracted usually from pion-nucleon scattering data at the Cheng-Dashen point. The required subthreshold
extrapolation involves further poorly convergent expansions \cite{pin-news}. Here we do not attempt to add 
anything new to this ongoing debate. We turn to the analogous symmetry-breaking parameters $d_0$ and $d_D$ 
for the baryon decuplet states. Like for the baryon octet states we use the isospin averaged empirical values 
for the baryon masses in any u-channel exchange contribution. That means we use $m^{(\Delta )}_{[10]} = 1232.0$ 
MeV, $m^{(\Sigma )}_{[10]} = 1384.5$ MeV  and $m^{(\Xi)}_{10} = 1533.5$ MeV in the decuplet exchange expressions.
In the s-channel decuplet expressions we use the slightly different values $m^{(\Delta )}_{10} = 1223.2$ MeV and 
$m^{(\Sigma )}_{10} = 1374.4$ MeV to compensate for a small mass renormalization induced by the unitarization. 
Those values are rather consistent with $d_D \simeq  -0.49$ GeV$^{-1}$ (see (\ref{mass-splitting})). 
Moreover note that all values used are quite compatible with the large $N_c$ sum rule 
$$b_D+b_F = d_D/3 \,.$$ 
The parameter $d_0$ is not determined by our analysis. Its determination required the study of the meson baryon-decuplet
scattering processes.

\tabcolsep=1.1mm
\begin{table}[t]\begin{center}
\begin{tabular}{|c|c||c|c||c|c||c|c||c|c||}
\hline 
$h_F^{(1)}$[GeV$^{-3}$]  & -0.129 & 
$h_D^{(1)}$[GeV$^{-3}$]  & -0.548 &
$h_{F}^{(2)}$[GeV$^{-2}$]  & 0.174 & 
$h_{F}^{(3)}$[GeV$^{-2}$] & -0.221 
\\ 
\hline
\end{tabular}
\end{center}
\caption{Chiral $Q^3$-parameters resulting from a fit to low-energy 
meson-baryon scattering data. Further parameters at this order are determined by the 
large $N_c$ sum rules.}
\label{q3param:tab}
\end{table}

At chiral order $Q^3$ the number of parameters increases 
significantly unless further constraints from QCD are imposed. 
Recall for example that \cite{q3-meissner} presents a large  
collection of already 102 chiral $Q^3$ interaction terms.  A 
systematic expansion of the interaction kernel in powers of $1/N_c$ 
leads to a much reduced parameter set. For example the $1/N_c$ 
expansion leads to only four further parameters $h^{(1)}_{F}$, 
$h^{(1)}_{D}$, $h^{(2)}_{F}$ and $h^{(3)}_{F}$ describing the refined 
symmetry-conserving two-body interaction vertices. This is to be compared with 
the ten parameters established in Appendix B, which were found to be relevant at order 
$Q^3$ if large $N_c$ sum rules are not imposed. In our global fit we insist on the 
large $N_c$ sum rules 
$$ 
h_1^{(1)}=2\,h_0^{(1)}= 4\,h_D^{(1)}/3  \;, \quad h^{(2)}_1 = h^{(2)}_D=0 
\;, \quad h^{(3)}_1 = h^{(3)}_D=0 \,.
$$ 
Note that at order $Q^3$ there are no symmetry-breaking 2-body interaction vertices. To that order the only 
symmetry-breaking effects result from the refined 3-point vertices. Here a particularly rich picture emerges. 
At order $Q^3$ we  established 23 parameters describing symmetry-breaking effects in the 3-point meson-baryon 
vertices.  For instance, to that order the baryon-octet states may couple to the pseudo-scalar mesons
also via pseudo-scalar vertices rather than only via the leading axial-vector vertices. Out of those 
23 parameters 16 contribute at the same time to matrix elements of the axial-vector current. Thus in order 
to control the symmetry breaking effects, it is mandatory to include constraints from the weak decay widths 
of the baryon octet states also. A detailed analysis of the 3-point vertices in the $1/N_c$ expansion of QCD 
reveals that in fact only ten parameters $c_{1,2,3,4,5}$, $\delta c_{1,2}$ and $\bar c_{1,2}$ and $a$, rather 
than the 23 parameters, are needed at leading order in that expansion. Since the leading parameters $F_R, D_R$ 
together with the symmetry-breaking parameters $c_i$ describe at the same time the 
weak decay widths of the baryon octet and decuplet ground states (see Tab. \ref{weak-decay:tab},\ref{weak-decay:tabb}), 
the number of free parameters does not increase significantly at the $Q^3$ level if 
the large $N_c$ limit is applied.

\tabcolsep=1.2mm
\begin{table}[t]\begin{center}
\begin{tabular}{|c|c||c|c||c|c||c|c||c|c||c|c||}
\hline 
$c_1 $ & -0.0707 & 
$c_2 $ & -0.0443 & 
$c_3 $ & 0.0624  & 
$c_4 $ & 0.0119  & 
$c_5 $ & -0.0434 \\
\hline 
$\bar c_1   $ & 0.0754  & 
$\bar  c_2  $ & 0.1533  &  
$\delta c_1 $ & 0.0328  & 
$\delta c_2 $ & -0.0043 &  
$ a $ & -0.2099  \\
\hline
\end{tabular}
\end{center}
\caption{Chiral $Q^3$-parameters, which break the $SU(3)$ symmetry explicitly, resulting from a fit to low-energy 
meson-baryon scattering data. }
\label{q3param:tab}
\end{table}

We conclude that the parameter reduction achieved in this work by insisting on chiral and large $N_c$ 
sum rules is significant. It is instructive to recall that for instance the analysis by Kim \cite{kim}, 
rather close in spirit to modern effective field theories, required already 44 parameters in the strangeness 
minus one sector only. As was pointed out in \cite{Hurtado} that analysis, even though troubled with severe 
shortcomings \cite{piN:Hurtado}, was the only one so far which included s- and p-waves and still reproduced the 
most relevant features of the subthreshold $\bar K N$ amplitudes. We summarize that a combined chiral and large $N_c$ 
analysis leads to a scheme with a reasonably small number of parameters at the $Q^3$ level.

\subsection{Axial-vector coupling constants}

The result of our global fit for the axial-vector coupling constants of the baryon octet 
states are presented in Tab. \ref{weak-decay:tab}. The six data points, which  
strongly constrain the parameters $F_R, D_R$ and $c_{1,2,3,4}$, are well reproduced.
Note that the recent measurement of the decay process 
$\Xi^0 \to \Sigma^+ \,e^-\,\bar \nu_e$ by the KTeV experiment does not provide a 
further stringent constraint so far \cite{KTeV:1,KTeV:2}. The axial-vector coupling constant of that decay 
$g_A(\Xi^0 \to \Sigma^+ \,e^-\,\bar \nu_e) = \sqrt{2}\,g_A (\Xi^- \to \Sigma^0 \,e^-\,\bar \nu_e )$
is related to the decay process $\Xi^- \to \Sigma^0 \,e^-\,\bar \nu_e$ included in Tab. \ref{weak-decay:tabb}
by isospin symmetry. The value given in \cite{KTeV:2} is $g_A(\Xi^0 \to \Sigma^+ \,e^-\,\bar \nu_e) = 1.23 \pm 0.44 $. 
As emphasized in \cite{KTeV:2} it would be important to reduce the uncertainties by more data taking. 
We confirm the result of \cite{Dai} which favors values for $g_A (\Xi^- \to \Sigma^0 \,e^-\,\bar \nu_e )$
and $g_A (\Xi^- \to \Lambda \,e^-\,\bar \nu_e )$ which are somewhat smaller than the central values given 
in Tab. \ref{weak-decay:tab}. This is a non-trivial result because in our approach the parameters $c_{1,2,3,4}$ 
are constrained not only by the weak decay processes of the baryon octet states but also by the meson-baryon 
scattering data. 

\tabcolsep=1.4mm
\renewcommand{\arraystretch}{1.5}
\begin{table}[t]\begin{center}
\begin{tabular}{|c|c||c|c|c|c|c|c|c|}
\hline 
& $g_A $ (Exp.) & $\chi$-BS(3) & SU(3) \\ 
\hline \hline

$ n \to p\,e^-\,\bar \nu_e $ & $1.267 \pm 0.004$& 1.26 & 1.26 
\\  \hline 

$ \Sigma^- \to \Lambda \,e^-\,\bar \nu_e  $ & $0.601 \pm 0.015$ 
& 0.58 & 0.65 
\\  \hline

$ \Lambda \to p \,e^-\,\bar \nu_e $ & $-0.889 \pm 0.015$ & 
-0.92 & -0.90 
\\  \hline

$ \Sigma^- \to n \,e^-\,\bar \nu_e $ & $0.342 \pm 0.015$ & 0.33 & 0.32 
\\  \hline

$ \Xi^- \to \Lambda \,e^-\,\bar \nu_e $ & $0.306 \pm 0.061 $ & 
0.19  & 0.25  
\\  \hline

$ \Xi^- \to \Sigma^0 \,e^-\,\bar \nu_e $ & $0.929 \pm 0.112 $ & 
0.79 & 0.89
\\  \hline


\end{tabular}
\end{center}
\caption{Axial-vector coupling constants for the weak decay processes of the baryon octet 
states. The empirical values 
for $g_A$ are taken from \cite{Dai}. Here we do not consider  
small $SU(3)$ symmetry-breaking effects of the  vector current. The column labelled by SU(3) 
shows the axial-vector coupling constants as they follow from 
$F_R=0.47$ and $D_R=0.79$ and $c_{i}=0$.} 
\label{weak-decay:tabb}
\end{table}

\subsection{Meson-baryon coupling constants}

We turn to the meson-baryon coupling constants. To subleading order the Goldstone bosons couple to the 
baryon octet states via axial-vector but also via suppressed pseudo-scalar vertices (see (\ref{chi-sb-3:p})). 
In Tab. \ref{meson-baryon:tab} we collect the results for all the axial-vector meson-baryon coupling 
constants, $A^{(B)}_{\Phi B}$, and their respective pseudo-scalar parts $P^{(B)}_{\Phi B}$. 
The $SU(3)$ symmetric part of the axial-vector vertices is characterized by parameters $F_A$ and $D_A$
\begin{eqnarray}
F_A &=& F_R -\frac{\beta}{\sqrt{3}}\,\Big({\textstyle{2\over 3}}\,\delta c_1 +\delta c_2 -a\Big) = 
0.270\,,
\quad \!
\nonumber\\
D_A &=& D_R - \frac{\beta}{\sqrt{3}}\,\delta c_1 = 0.726\,,
\label{FDA:def}
\end{eqnarray}
where $\beta \simeq 1.12$. To subleading order the 
parameters $F_A$ and $D_A$ differ from the corresponding parameters $F_R = 0.418$ and 
$D_R =0.748$ relevant for matrix elements of the axial-vector current (see Tab. \ref{weak-decay:tab}) by 
a sizeable amount. The $SU(3)$ symmetry-breaking effects in the axial-vector coupling constants $A$ 
are determined by the parameters $c_i$, which are already tightly constrained by the weak decay 
widths of the baryon octet states, $\delta c_{1,2}$ and $a$. Similarly 
the four parameters $\bar c_{1,2}$ and $a$ characterize the pseudo-scalar meson-baryon 3-point vertices. 
Their symmetric contributions are determined by $F_P$ and $D_P$
\begin{eqnarray}
F_P =  -\frac{\beta}{\sqrt{3}}\,\Big({\textstyle{2\over 3}}\,\bar c_1 +\bar c_2 +a\Big) = 0.004\,,
\quad \!
D_P =  - \frac{\beta}{\sqrt{3}}\,\bar c_1 = -0.049\,.
\label{FDP:def}
\end{eqnarray}
Note that in the on-shell coupling constants $G=A+P$ the parameter 
$a$ drops out. In that sense that parameter should be viewed as representing an effective 
quasi-local 2-body interaction term, which breaks the $SU(3)$ symmetry explicitly.  

\tabcolsep=0.9mm
\renewcommand{\arraystretch}{1.5}
\begin{table}[t]\begin{center}
\begin{tabular}{|c|c|c|c|c|c|c|c|c|c|c|c|c|}
\hline 
& $A^{(N)}_{\pi N} $  & $A^{(N)}_{\eta N} $  & $A^{(\Lambda)}_{\bar 
K N} $  & $A^{(\Lambda)}_{\pi \Sigma } $  & $A^{(\Lambda)}_{\eta 
\Lambda} $ & $A^{(\Lambda)}_{K 
\Xi} $  & $A^{(\Sigma)}_{\bar  K N} $ & $A^{(\Sigma)}_{\pi \Sigma} $  & 
$A^{(\Sigma)}_{\eta \Sigma} $  & $A^{(\Sigma)}_{K \Xi} $  & 
$A^{(\Xi)}_{\pi \Xi} $  & $A^{(\Xi)}_{\eta \Xi} $ 
\\ 
\hline \hline
$\chi$-BS(3) & 2.15  & 0.20  &-1.29   & 1.41   & -0.64  & 0.12   & 0.73  & 
-1.02  & 1.09  & 1.24   & -0.59  & -0.32 
 \\  \hline

SU(3) & 1.73 & 0.05 & -1.25 & 1.45 & -0.84 & -0.07 &0.65  & -0.76 & 0.84 & 1.41 &-0.79 &-0.89 
 \\  \hline \hline

& $P^{(N)}_{\pi N} $  & $P^{(N)}_{\eta N} $  & 
$P^{(\Lambda)}_{\bar K N} $  & $P^{(\Lambda)}_{\pi \Sigma 
} $  & $P^{(\Lambda)}_{\eta 
\Lambda} $ & $P^{(\Lambda)}_{K 
\Xi} $  & $P^{(\Sigma)}_{\bar  K N} $ & $P^{(\Sigma)}_{\pi \Sigma} $  & 
$P^{(\Sigma)}_{\eta \Sigma} $  & $P^{(\Sigma)}_{K \Xi} $  
& $P^{(\Xi)}_{\pi \Xi} $  & $P^{(\Xi)}_{\eta \Xi} $  
\\ 
\hline \hline

$\chi$-BS(3) & -0.01 & 0.16 & 0.04 & -0.01 & 0.20 & -0.07 & -0.11 &-0.00  &-0.02  & -0.09 & 0.01& 0.13
 \\  \hline 
\end{tabular}
\end{center}
\caption{Axial-vector (A) and pseudo-scalar (P) meson-baryon coupling constants for the baryon octet states. 
The row labelled by SU(3) gives results excluding SU(3) symmetry-breaking 
effects with $F_A =0.270$ and $D_A= 0.726$. The total strength of the on-shell meson-baryon vertex 
is determined by $G=A+P$.} 
\label{meson-baryon:tab}
\end{table}

We emphasize that our meson-baryon coupling constants are strongly constrained by the 
axial-vector coupling constants. In particular we reproduce 
a sum rule derived first by Dashen and Weinstein \cite{Dashen:Weinstein}
\begin{eqnarray}
G_{\pi N}^{(N)}-\sqrt{3}\,g_A (n \to p\,e^-\,\bar \nu_e )
&=& -\frac{m_\pi^2}{\sqrt{6}\,m_K^2}\,\Big( G_{\bar K N }^{(\Sigma)}
-g_A(\Sigma^-\!\to n \,e^-\,\bar \nu_e) \Big)
\nonumber\\
&-& \frac{m_\pi^2}{\sqrt{2}\,m_K^2}\,\Big( G_{\bar K N}^{(\Lambda)}
-2\,g_A (\Lambda \to p \,e^-\,\bar \nu_e ) \Big) \;.
\label{DW-sum}
\end{eqnarray}
We observe that, given the expected range of values for $g_{\pi NN}$, $g_{\bar K N \Lambda }$ and 
$g_{\bar K  N \Sigma }$ together with the empirical axial-vector coupling constants, 
the Dashen-Weinstein relation strongly favors a small $f$ parameter value close to $f_\pi$.

\tabcolsep=1.4mm
\renewcommand{\arraystretch}{1.5}
\begin{table}[t]\begin{center}
\begin{tabular}{|c||c|c|c|c|c|c|c|c|}
\hline 
& $\chi $-BS(3) &  \cite{Stoks} & \cite{Juelich:2,Juelich:1,Juelich:3} & \cite{keil}  
& \cite{A.D.Martin} & \cite{Dalitz:g}     \\ 
\hline \hline
$|g_{\pi NN }|$ & 12.9  & 13.0  &  13.5  &  13.1 &  - &  -    
 \\  \hline

$|g_{\bar K N \Lambda}|$ &10.1  & 13.5 & 14.0 &  10.1 & 13.2 & 16.1 
 \\  \hline

$|g_{\pi \Sigma\, \Lambda}|$ & 10.4 & 11.9 & 9.3 & 6.4 & - & -   
 \\  \hline

$|g_{\bar K N \Sigma }|$ & 5.2 & 4.1 & 2.7 & 2.4 & 6.4 & 3.5   
 \\  \hline

$|g_{\pi \Sigma \,\Sigma}|$ & 9.6 & 11.8 & 10.8 & 0.7 & - & -   
 \\  \hline

\end{tabular}
\end{center}
\caption{On-shell meson-baryon coupling constants for the baryon octet states (see (\ref{trans:tab})). We 
give the central values only because reliable error analyses are not available in most cases.} 
\label{meson-baryon:tab:2}
\end{table}

In Tab. \ref{meson-baryon:tab:2}  we confront our results with a representative 
selection of published meson-baryon coupling constants. For the clarity of this comparison 
we recall here the connection with our convention  
\begin{eqnarray}
&& g_{\pi NN} = \frac{m_N}{\sqrt{3}\,f}\,G^{(N)}_{\pi N} \,, \! \quad 
g_{\bar K N \Lambda } = \frac{m_N+m_\Lambda}{\sqrt{8}\,f}\,G^{(\Lambda)}_{\bar KN} \,, \!\quad 
g_{\pi  \Lambda \Sigma }= \frac{m_\Lambda+m_\Sigma }{\sqrt{12}\,f }\,G^{(\Lambda)}_{\pi \Sigma }\, ,
\nonumber\\
&& g_{\bar K N \Sigma }= \frac{m_N+m_\Sigma}{\sqrt{8}\,f}\,G^{(\Sigma)}_{\bar K N} \,,\! \quad 
g_{\pi \Sigma \Sigma} = \frac{m_\Sigma }{\sqrt{2}\,f}\,G^{(\Sigma)}_{\pi \Sigma } \,. 
\label{trans:tab}
\end{eqnarray}
Further values from previous analyses can be found 
in \cite{Dumbrajs}. Note also the interesting recent results within the QCD sum rule approach 
\cite{Kim:Doi:Oka:Lee,Doi:Kim:Oka} and also \cite{Buchmann:Henley}. We do not confront 
our values with those of \cite{Kim:Doi:Oka:Lee,Doi:Kim:Oka} and \cite{Buchmann:Henley}
because the $SU(3)$ symmetry-breaking effects are not yet fully under control in these 
works. The analysis \cite{Stoks} is based on nucleon-nucleon and hyperon-nucleon 
scattering data where $SU(3)$ symmetry breaking effects in the meson-baryon coupling 
constants are parameterized according to the model of \cite{p3-model}. 
The values given in \cite{Juelich:2,Juelich:1,Juelich:3} do not 
allow for $SU(3)$ symmetry-breaking effects and moreover rely on $SU(6)$
quark-model relations. Particularly striking are the extreme $SU(3)$ symmetry-breaking 
effects claimed in \cite{keil}. The parameters result from a K-matrix fit to the 
phase shifts of $K^-N$ scattering as given in \cite{gopal}. We do not confirm these results. 
Note also the recent analysis \cite{Loiseau:Wycech} which deduces the value
$g_{\pi \Lambda \Sigma } = 12.9 \pm 1.2$ from hyperonic atom data, a value somewhat larger than 
our result of $10.4$. For the most recent and accurate pion-nucleon coupling constant 
$g_{\pi NN} = 13.34 \pm 0.09$ we refer to \cite{gpinn:best}. We do not compete 
with the high precision and elaborate analyses of this work.

In Tab. \ref{meson-baryon-decuplet:tab} we collect our results for the meson-baron coupling constants 
of the decuplet states. Again we find only moderate $SU(3)$ symmetry-breaking effects in the coupling 
constants. This is demonstrated by comparing the two rows of Tab. \ref{meson-baryon-decuplet:tab}. 
The SU(3) symmetric part is determined by the parameter $C_A$
\begin{eqnarray}
C_A = C_R -2\,\frac{\beta}{\sqrt{3}}\,\Big(\bar c_1 +\delta c_1 \Big) = 1.593 \,.
\label{}
\end{eqnarray}
We find that the coupling constants given in (\ref{meson-baryon-decuplet:tab}) should not   
be used in the simple expressions (\ref{decuplet-decay}) for the decuplet widths.
For example with $A_{\pi N}^{(\Delta )} \simeq 2.62 $ one would estimate $\Gamma_\Delta \simeq $ 
102 MeV not too close to the empirical value of $\Gamma_\Delta \simeq $ 120 MeV \cite{fpi:exp}. It will 
be demonstrated below, that nevertheless the $P_{33}$ phase shift of the pion-nucleon scattering process, which 
probes the $\Delta $ resonance width, is reproduced accurately. This reflects an important energy dependence in 
the decuplet self energy. 

\tabcolsep=1.2mm
\renewcommand{\arraystretch}{1.5}
\begin{table}[t]\begin{center}
\begin{tabular}{|c|c|c|c|c|c|c|c|c|c|c|c|}
\hline 
& $A^{(\Delta )}_{\,\pi\, N} $  & $A^{(\Delta )}_{\,K\, \Sigma} $  & $A^{(\Sigma )}_{\,\bar 
K \,N} $  & $A^{(\Sigma)}_{\,\pi \,\Sigma } $  & $A^{(\Sigma )}_{\,\pi \,\Lambda} $ 
& $A^{(\Sigma )}_{\,\eta \,\Sigma } $  
& $A^{(\Sigma )}_{\,K \,\Xi} $ & $A^{(\Xi )}_{\,\bar K \,\Lambda} $  & 
$A^{(\Xi )}_{\,\bar K \,\Sigma} $  & $A^{(\Xi )}_{\,\eta \,\Xi} $  & 
$A^{(\Xi )}_{\,\pi \,\Xi} $  
\\ 
\hline \hline
$\chi$-BS(3) & 2.62 & -2.03  & 1.54  & -1.40  & -1.64  &  1.55  &-0.96   & 1.67 & 1.75  &-1.29  &-1.46   
 \\  \hline

SU(3) & 2.25 & -2.25 & 1.30 & -1.30 & -1.59 & 1.59 & -1.30  & 1.59 & 1.59 & -1.59 &-1.59 
 \\  \hline 
\end{tabular}
\end{center}
\caption{Meson-baryon coupling constants for the baryon decuplet states. The row 
labelled by SU(3) gives results obtained  with $C_A =1.593 $ excluding all symmetry-breaking 
effects.} 
\label{meson-baryon-decuplet:tab}
\end{table}

We summarize the main findings of this section. All established parameters
prove the $SU(3)$ flavor symmetry to be an extremely useful and accurate tool.
Explicit symmetry breaking effects are quantitatively important but sufficiently small to 
permit an evaluation within the $\chi$-BS(3) approach. This confirms a beautiful analysis 
by Hamilton and Oades \cite{Hamilton:Oades} who strongly supported the $SU(3)$ flavor symmetry 
by a discrepancy analysis of kaon-nucleon scattering data.

\subsection{Pion-nucleon scattering}

We begin with a detailed account of the strangeness zero sector. For a review of pion-nucleon 
scattering within the conventional meson-exchange picture we refer to \cite{EW}. The various chiral 
approaches will be discussed  more explicitly below. 
Naively one may want to include the pion-nucleon threshold parameters in a global $SU(3)$ fit. In conventional 
chiral perturbation theory the latter are evaluated perturbatively to subleading orders in the chiral 
expansion \cite{Bernard,pin-q4}. The small pion mass justifies the 
perturbative treatment. Explicit expressions for the threshold parameters are given in Appendix H and 
confirm the results \cite{Bernard}. Unfortunately there is no unique set of threshold 
parameters available. This is due to difficulties in extrapolating the empirical data set 
down to threshold, subtle electromagnetic effects and also some inconsistencies in the data set 
itself \cite{GMORW}.
A collection of mutually contradicting threshold parameters is collected in Tab. 8. In order 
to obtain an estimate of systematic errors in the various analyses we confront the 
threshold values with the chiral sum rules:
\begin{eqnarray}
&&4\,\pi \left( 1+\frac{m_\pi}{m_N}\right)
a^{(\pi N)}_{[S_-]}=\frac{m_\pi}{2\,f^2}+{\mathcal O}\left(Q^3 \right)\; ,
\nonumber\\
&&4\,\pi \left( 1+\frac{m_\pi}{m_N}\right)
b^{(\pi N)}_{[S_-]}=\frac{1}{4\,f^2\,m_\pi}
-\frac{2\,g_A^2+1}{4\,f^2\,m_N}
\nonumber\\
&&\quad \quad \quad \quad \quad \quad \quad \quad 
+\frac{C^2}{18\,f^2}\,\frac{m_\pi}{m_N\,(\mu_\Delta+m_\pi)}
+{\mathcal O}\left(Q \right)
\nonumber\\
&&4\,\pi \left( 1+\frac{m_\pi}{m_N}\right)
\Big( a^{(\pi N)}_{[P_{13}]}-a^{(\pi N)}_{[P_{31}]} \Big)
=\frac{1}{4\,f^2\,m_N}+{\mathcal O}\left( Q\right)\;,
\nonumber\\
&& 4\,\pi\,\left( 1+\frac{m_\pi}{m_N}\right)
\,a^{(\pi N)}_{SF}
= -\frac{3\,g_A^2}{2\,f^2\,m_\pi}\,\left(1+\frac{m_\pi}{m_N}\right)
\nonumber\\
&& \quad \quad \quad \quad \quad \quad \quad \quad 
-\frac23\,\frac{C^2}{f^2}\,\frac{m_\pi\,m_N }
{m_\Delta\,(\mu_\Delta^2-m_\pi^2)} +{\mathcal O}\left( Q\right)\;,
\label{sum-rules}
\end{eqnarray}
where $\mu_\Delta = m_\Delta -m_N $. We confirm the result of \cite{Bernard} that 
the spin-flip scattering volume 
$a_{SF}^{(\pi N)}=a^{(\pi N)}_{[P_{11}]}+2\,a^{(\pi N)}_{[P_{31}]}-a^{(\pi N)}_{[P_{13}]}
-2\, a^{(\pi N)}_{[P_{33}]}$ and the combination 
$a^{(\pi N)}_{[P_{13}]}-a^{(\pi N)}_{[P_{31}]}$ in (\ref{sum-rules}) 
are independent of the quasi-local 4-point interaction strengths at leading order. Confronting
the analyses in Tab. \ref{pin:th} with the chiral sum rules quickly reveals that only the EM98 
analysis \cite{EM98} appears consistent with the sum rules within 20 $\%$. The 
analysis \cite{GMORW} and \cite{SP98} badly contradict the chiral sum rules (\ref{sum-rules}), 
valid at leading chiral orders, and therefore would require unnaturally large correction terms, 
possibly discrediting the convergence of the chiral expansion in the pion-nucleon sector. 
The KA86 analysis \cite{Koch86} is consistent with the two p-wave sum rules but appears inconsistent with 
the s-wave range parameter $b_{[S_-]}$. The recent $\pi^-$ hydrogen atom experiment 
\cite{Schroeder} gives rather precise values for the $\pi^-$-proton scattering lengths
\begin{eqnarray}
&& a_{\pi^-p\to \pi^- p} =a_{S_-}^{(\pi N)} +a_{S_+}^{(\pi N)}= ( 0.124\pm 0.001 )\;{\rm fm} \;,\qquad 
\nonumber\\
&& a_{\pi^-p\to \pi^0 n} = -\sqrt{2}\,a_{S_-}^{(\pi N)}=(-0.180\pm 0.008)\; {\rm fm} \;.
\label{pi-atoms}
\end{eqnarray}
These values are in conflict with the s-wave scattering lengths of the EM98 analysis.
For a comprehensive discussion of further constraints from the pion-deuteron scattering 
lengths as derived from recent pionic atom data we refer to \cite{gpinn:best}.
All together the emerging picture is complicated and inconclusive at present. 
Related arguments are presented by Fettes and Mei\ss ner in their work \cite{pin-q4} 
which considers low-energy pion-nucleon phase shifts at chiral order $Q^4$. The resolution of this 
mystery may be found in the most recent work of Fettes and Mei\ss ner \cite{pin-em} where 
they consider the electromagnetic correction terms within the $\chi $PT scheme to order $Q^3$.

\renewcommand{\arraystretch}{1.4}
\tabcolsep=1.3mm
\begin{table}[t]\begin{center}
\begin{tabular}{|c||c|c|c|c|c|}
\hline\hline
 & $\chi $-BS(3) &  KA86\protect\cite{Koch86} &
EM98\protect\cite{EM98} & SP98\protect\cite{SP98} & GMORW\protect\cite{GMORW}\\
\hline\hline
$a_{[S_-]}^{(\pi N)}$~[fm]& 0.124 & 0.130 &  0.109$\pm$ 0.001 & 0.125$\pm$ 0.001 & 0.116$\pm$ 0.004 \\
$a_{[S_+]}^{(\pi N)}$~[fm]& -0.014 &-0.012 & 0.006$\pm$ 0.001  & 0.000$\pm$ 0.001 & 0.005$\pm$ 0.006\\
$b_{[S_-]}^{(\pi N)}$~[$m_\pi^{-3}]$& -0.007 & 0.008&   0.016 & 0.001$\pm$ 0.001 & -0.009$\pm$ 0.012\\
$b_{[S_+]}^{(\pi N)}$~[$m_\pi^{-3}]$& -0.028 &-0.044&  -0.045 & -0.048$\pm$ 0.001 & -0.050$\pm$ 0.016\\
\hline\hline
$a_{[P_{11}]}^{(\pi N)}$~[$m_\pi^{-3}$]& -0.083 & 
-0.078& -0.078$\pm$ 0.003 & -0.073$\pm$ 0.004 & -0.098$\pm$ 0.005 \\
$a_{[P_{31}]}^{(\pi N)}$~[$m_\pi^{-3}$]& -0.045 &
-0.044& -0.043$\pm$ 0.002 & -0.043$\pm$ 0.002 & -0.046$\pm$ 0.004\\
$a_{[P_{13}]}^{(\pi N)}$~[$m_\pi^{-3}$]& -0.038 &
-0.030& -0.033$\pm$ 0.003 & -0.013$\pm$ 0.004 & 0.000$\pm$ 0.004 \\
$a_{[P_{33}]}^{(\pi N)}$~[$m_\pi^{-3}$]& 0.198 &
0.214& 0.214$\pm$ 0.002  & 0.211$\pm$ 0.002 & 0.203 $\pm$ 0.002\\
\hline
\end{tabular}
\end{center}
\caption{Pion-nucleon threshold parameters. The reader not familiar with the common definition of 
the various threshold parameters is referred to Appendix H.}
\label{pin:th}
\end{table}

We turn to another important aspect to be discussed. Even though the EM98 analysis is 
rather consistent with the chiral sum rules (\ref{sum-rules}), does it imply background terms
of natural size? 
This can be addressed by considering a further combination of p-wave scattering volumes 
\begin{eqnarray}
&&4\,\pi \left( 1+\frac{m_\pi}{m_N}\right)\left( a^{(\pi N)}_{[P_{11}]}
-4\,a^{(\pi N)}_{[P_{31}]}\right) = 
\frac{3}{2\,f^2\,m_N} +B+{\mathcal O}\left(Q \right)\;,
\nonumber\\
&&B=-\frac{5}{12 f^2}\,\Big( 2\,\tilde g_0^{(S)}+\tilde g_D^{(S)}+\tilde g_F^{(S)} \Big)
-\frac{1}{3 f^2}\,\Big( \tilde g_D^{(T)}+\tilde g_F^{(T)} \Big) \,,
\label{}
\end{eqnarray}
where we absorbed the Z-dependence into the tilde couplings for simplicity 
(see (\ref{Z-absorb})). The naturalness assumption would lead to  
$B \sim 1/(f^2\,m_\rho)$, a typical size which is compatible with the background term 
$B\simeq 0.92\,m_\pi^{-3}$ of the EM98 solution.

\begin{figure}[t]
\begin{center}
\includegraphics[width=14cm,clip=true]{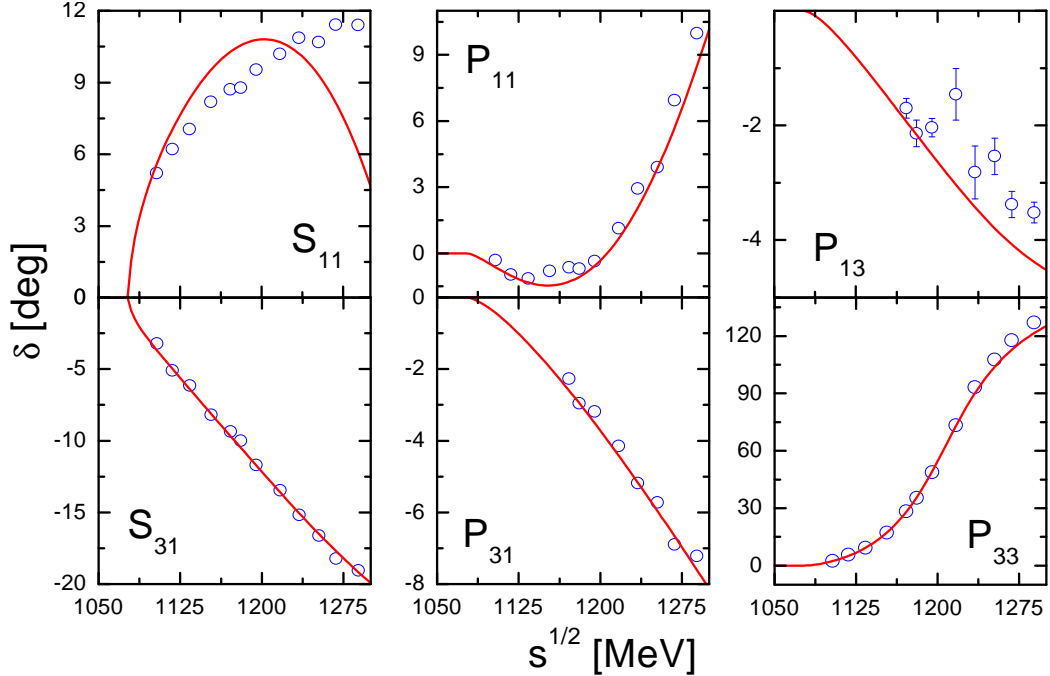}
\end{center}
\caption{S- and p-wave pion-nucleon phase shifts. The single energy phase shifts are taken 
from \cite{pion-phases}.}
\label{fig:pionphases}
\end{figure}

In order to avoid the ambiguities of the threshold parameters we decided to 
include the single energy pion-nucleon phase shifts of \cite{SP98} in our global fit. 
The phase shifts are evaluated in the $\chi$-BS(3) approach including all
channels suggested by the $SU(3)$ flavor symmetry. The single energy phase shifts are fitted 
up to $\sqrt{s} \simeq 1200$ MeV. 
In Fig.~\ref{fig:pionphases} we confront the result of our fit with the empirical phase shifts. 
All s- and p-wave phase shifts are well reproduced up to $\sqrt{s} \simeq 1300$ MeV 
with the exception of the $S_{11}$ phase for which our result agrees with the 
partial-wave analysis less accurately. We emphasize that one 
should not expect quantitative agreement for $\sqrt{s} > m_N+2\,m_\pi \simeq 1215$ MeV where 
the inelastic pion production process, not included in this work, starts. The missing higher 
order range terms in the $S_{11}$ phase are expected to be induced by additional inelastic 
channels or by the nucleon resonances $N(1520)$ and $N(1650)$. 
We confirm the findings of \cite{Kaiser,new-muenchen} that the coupled $SU(3)$ channels, if truncated 
at the Weinberg-Tomozawa level, predict considerable strength in the $S_{11}$ channel around 
$\sqrt{s} \simeq 1500$ MeV where the phase shift shows a resonance-like structure. Note, however 
that it is expected that the nucleon resonances $N(1520)$ and $N(1650)$ couple strongly to each 
other \cite{Sauerman} and therefore one should not expect a quantitative description of the $S_{11}$ 
phase too far away from threshold. Similarly we observe considerable strength in the $P_{11}$ channel 
leading to a resonance-like structure around $\sqrt{s} \simeq 1500 $ MeV. We interpret this phenomenon 
as a precursor effect of the p-wave $N(1440)$ resonance. We stress that our approach differs significantly 
from the recent work \cite{new-muenchen} in which the coupled SU(3) channels are applied to pion induced 
$\eta$ and kaon production which require much larger energies $\sqrt{s} \simeq m_\eta +m_N \simeq$ 1486 MeV 
or $\sqrt{s} \simeq m_K +m_\Sigma \simeq $ 1695 MeV. We believe that such high energies can be 
accessed reliably only by including more inelastic channels. It may be worth mentioning that the 
inclusion of the inelastic channels as required by the $SU(3)$ symmetry leaves the $\pi N$ phase shifts basically 
unchanged for $\sqrt{s}< 1200$ MeV. Our discussion of the pion-nucleon sector is closed by returning to the threshold 
parameters. In Tab. \ref{pin:th} our extracted threshold parameters are presented in the second row. We conclude that 
all threshold parameters are within the range suggested by the various analyses.

\subsection{$K^+$-nucleon scattering}

\begin{figure}[t]
\begin{center}
\includegraphics[width=13cm,clip=true]{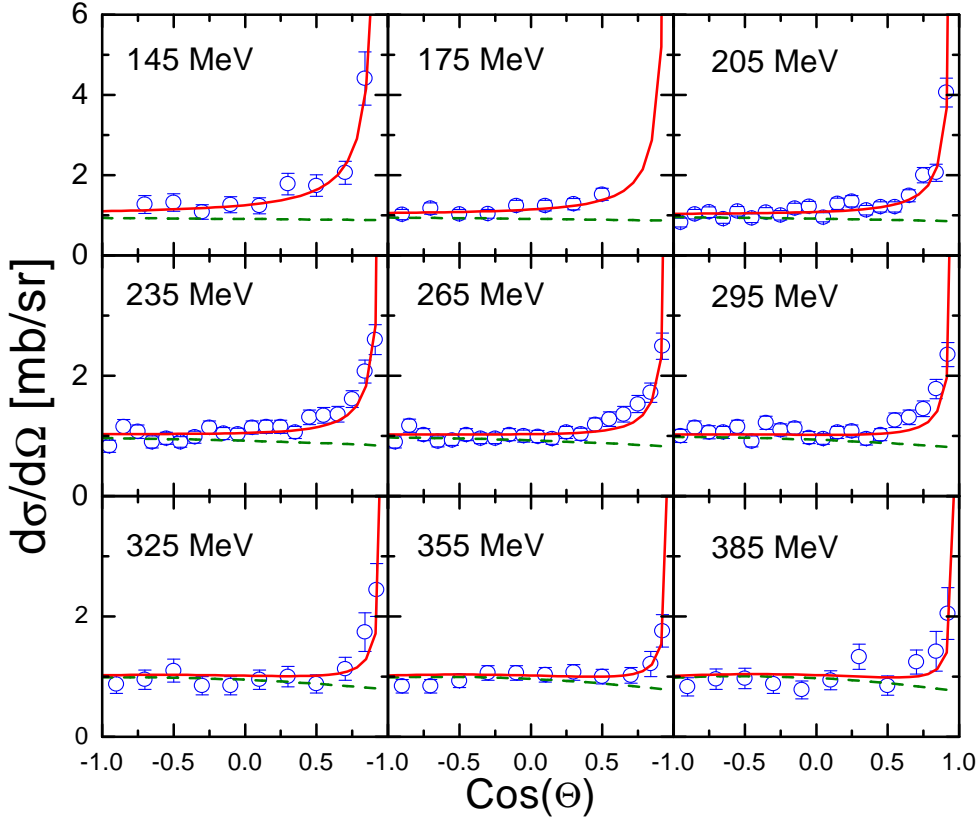}
\end{center}
\caption{Differential cross section of $K^+$-proton scattering. The data points are taken from \cite{kplscat}. The 
solid lines give the result of the $\chi$-BS(3) analysis including Coulomb effects. The 
dashed lines follow for switched off Coulomb interaction.}
\label{fig:Kpdif}
\end{figure}

We turn to the strangeness plus one channel. Since it is impossible to give here a comprehensive 
discussion of the many works dealing with kaon-nucleon scattering we refer to the review 
article by Dover and Walker \cite{Dover} which is still up-to-date in many respects. 
The data situation can be summarized as follows: there exist precise low-energy differential cross 
sections for $K^+ p$ scattering but no scattering data for the $K^+$-deuteron scattering 
process at low energies. Thus all low-energy results in the isospin zero 
channel necessarily follow from model-dependent extrapolations. We include the  
available differential cross section in our global fit. They are nicely reproduced as shown in 
Fig.~\ref{fig:Kpdif}. We include Coulomb interactions which are sizeable in the forward 
direction with $\cos \theta > 0$.

It is instructive to consider the threshold amplitudes in detail. In the $\chi$-BS(3) approach 
the threshold parameters are determined by the threshold values of the effective interaction kernel 
$V^{(\pm)}_{KN}(m_N\!+\!m_K; n)$ 
and the partial-wave loop function $J^{(\pm)}_{KN}(m_N\!+\!m_K; n)$ (see (\ref{result-loop:ab},\ref{M:kplus})). 
Since all loop functions vanish at threshold but the one for the s-wave channel, all p-wave scattering 
volumes remain unchanged by the unitarization and are directly given by the 
threshold values of the appropriate effective interaction kernel $V_{K N}$. The explicit expressions for 
the scattering volumes to leading order can be found in Appendix H. 
In contrast the s-wave scattering lengths are renormalized strongly by the loop function 
$J^{(+)}_{KN}(m_N\!+\!m_K; 0) \neq 0$. 
At leading order the s-wave scattering lengths are
\begin{eqnarray}
4\,\pi\left( 1+\frac{m_K}{m_N}\right) a^{(KN)}_{S_{21}} &=& 
-m_K \left( f^2+\frac{m^2_K}{8\,\pi} 
\left(1-\frac{1}{\pi}\,\ln \frac{m_K^2}{m_N^2} \right)\right)^{-1} \;,
\nonumber\\
4\,\pi\left( 1+\frac{m_K}{m_N}\right) a^{(KN)}_{S_{01}} &=&0 \;,
\label{}
\end{eqnarray}
which lead to $a^{(KN)}_{S_{21}} \simeq -0.22$ fm and $a^{(KN)}_{S01} =0$ fm, 
close to our final values to subleading orders given in Tab. \ref{tab-r-kp}. 
In Tab. \ref{tab-r-kp} we collected  typical results for the p-wave scattering volumes also. 
The large differences in the isospin zero channel reflect the fact that 
this channel is not constrained by scattering data directly \cite{Dover}. 
We find that some of our p-wave scattering volumes, also shown in Tab. \ref{tab-r-kp}, differ 
significantly from the values obtained by previous analyses. Such discrepancies may  be explained 
in part by important cancellation mechanisms among the u-channel baryon octet and decuplet 
contributions (see Appendix H).
An accurate description of the scattering volumes requires a precise input for the meson-baryon 
3-point vertices. Since the $\chi$-BS(3) approach describes the 3-point vertices in accordance 
with all chiral constraints and large $N_c$ sum rules of QCD we believe our values for 
the scattering volumes to be rather reliable.

\tabcolsep=0.5mm
\begin{table}[t]\begin{center}
\begin{tabular}{|c|c|c||c|c|c|c|c|}
\hline 
& $a^{(K N)}_{S_{01}}$ [fm] & $a^{(K N)}_{S_{21}}$ [fm]  & 
$a^{(K N)}_{P_{01}}$ $[m_\pi^{-3}]$ & $a^{(K N)}_{P_{21}}$ $[m_\pi^{-3}]$ &
$a^{(K N)}_{P_{03}}$ $[m_\pi^{-3}]$ & $a^{(K N)}_{P_{23}}$ $[m_\pi^{-3}]$  \\
\hline \hline
$\chi$-BS(3)  & 0.06 & -0.30 & 0.033 & -0.017 & -0.003 & 0.012 \\
\hline
\protect\cite{Hyslop} & 0.0 & -0.33 & 0.028 & -0.056 & -0.046 & 0.025 \\
\hline
\protect\cite{BR:Martin} & -0.04 & -0.32 & 0.030 & -0.011 & -0.007 & 0.007 \\
\hline
\end{tabular}
\end{center}
\caption{$K^+$-nucleon threshold parameters. The values of the $\chi$-BS(3) analysis are given in the first row.
The last two rows recall the threshold parameters as given in \cite{Hyslop} and \cite{BR:Martin}.}
\label{tab-r-kp}
\end{table}

\begin{figure}[t]
\begin{center}
\includegraphics[width=14cm,clip=true]{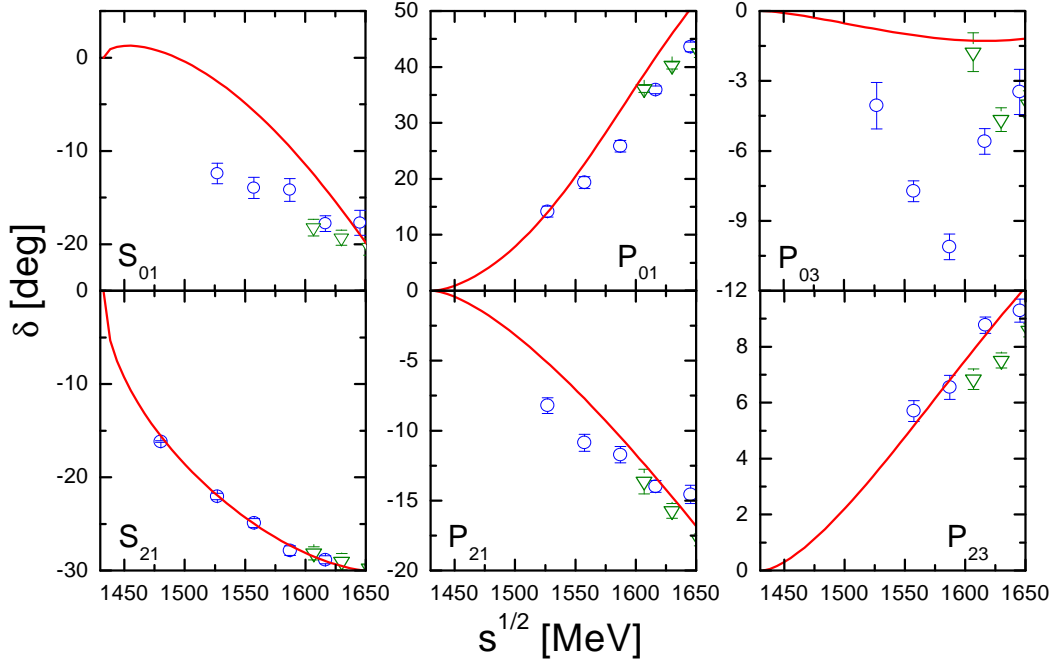}
\end{center}
\caption{S- and p-wave $K^+$-nucleon phase shifts. The solid lines represent the results of the 
$\chi$-BS(3) approach. The open circles are from the Hyslop analysis \cite{Hyslop} and the 
open triangles from the Hashimoto analysis \cite{Hashimoto}.}
\label{fig:K+phases}
\end{figure}

In Fig.~\ref{fig:K+phases} we confront our s- and p-wave $K^+$-nucleon 
phase shifts with the most recent analyses by Hyslop et al. \cite{Hyslop} and Hashimoto \cite{Hashimoto}.
We find that our partial-wave phase shifts are reasonably close to the single energy phase shifts of 
\cite{Hyslop} and \cite{Hashimoto} except the $P_{03}$ phase for which we obtain much smaller strength. 
Note however, that at higher energies we smoothly reach the single energy phase shifts of Hashimoto \cite{Hashimoto}. 
A possible ambiguity in that phase shift is already suggested by the conflicting scattering volumes found in that 
channel by earlier works (see Tab. \ref{tab-r-kp}). The isospin one channel, on the other hand, seems well-established 
even though the data set does not include polarization measurements close to threshold, which are needed 
to unambiguously determine the p-wave scattering volumes.

\subsection{$K^-$-nucleon scattering}

We now turn to our results in the strangeness minus one sector. The antikaon-nucleon scattering 
process shows a large variety of intriguing phenomena. Inelastic channels are already open at threshold 
leading to a rich coupled-channel dynamics. Also the $\bar K N$ state couples to many of the observed 
hyperon resonances for which competing dynamical scenarios are conceivable. 
We fit directly the available data set rather than any partial wave 
analysis. Comparing for instance the energy dependent analyses \cite{gopal} and \cite{garnjost}
one finds large uncertainties in the s- and p-waves in particular at low energies. This reflects on the 
one hand a model dependence of the analysis and on the other hand an insufficient data set. A partial wave 
analysis of elastic and inelastic antikaon-nucleon scattering data without further constraints from theory is 
inconclusive at present \cite{Dover,Gensini}. For a detailed overview of former theoretical 
analyses, we refer to the review article by Dover and Walker \cite{Dover}.

As motivated above we include the d-wave baryon resonance nonet 
field. An update of the analysis \cite{Plane} leads to the 
estimates $F_{[9]} \simeq 1.8$, $D_{[9]} \simeq 0.84$ and $C_{[9]} 
\simeq 2.5$ for the resonance parameters. The singlet-octet mixing angle $\vartheta \simeq 28^\circ $ 
confirms the finding of \cite{Tripp:2}, that the 
$\Lambda(1520)$ resonance is predominantly a flavor singlet state. 
By analogy with the expressions for the decuplet decay widths (\ref{decuplet-decay}) of section 2
we apply the simple expressions
\begin{eqnarray}
&& \Gamma_{N(1520)} = \frac{E_N-m_N}{16\,\pi \,f^2}\,\frac{p_{\pi N}^3}{m_{N(1520)}}
\,\Big( F_{[9]}+ D_{[9]} \Big)^2 
\nonumber\\
&& \qquad \quad \;\,+ \frac{E_N-m_N}{16\,\pi \,f^2}\,\frac{p_{\eta N}^3}{m_{N(1520)}}
\,\Big( F_{[9]}- {\textstyle{1\over \sqrt{3}}}\,D_{[9]} \Big)^2 \,,
\nonumber\\
&& \Gamma_{\Sigma (1680)} = \frac{E_N -m_N}{24\,\pi \,f^2}\,\frac{p_{\bar K N}^3}{m_{\Sigma (1680)}}
\,\Big(F_{[9]}- D_{[9]}  \Big)^2 
+ \frac{E_\Sigma -m_\Sigma}{6\,\pi \,f^2}\,\frac{p_{\pi \Sigma}^3}{m_{\Sigma (1680)}}
\,F_{[9]}^2 
\nonumber\\
&& \qquad \quad \;\, +\frac{E_\Lambda -m_\Lambda}{36\,\pi \,f^2}\,\frac{p_{\pi \Lambda}^3}{m_{\Sigma (1680)}}
\,D_{[9]}^2 \,,
\nonumber\\
&& \Gamma_{\Xi (1820)} = \frac{E_\Lambda-m_\Lambda}{16\,\pi \,f^2}\,
\frac{p_{\bar K \Lambda}^3}{m_{\Xi(1820)}}
\,\Big( F_{[9]}- {\textstyle{1\over \sqrt{3}}}\,D_{[9]} \Big)^2 
\nonumber\\
&& \qquad \quad \;\, +\frac{E_\Sigma-m_\Sigma }{16\,\pi \,f^2}\,
\frac{p_{\bar K \Sigma}^3}{m_{\Xi(1820)}}
\,\Big( F_{[9]}+ D_{[9]} \Big)^2 
\nonumber\\
&& \qquad \quad \;\, +\frac{E_\Xi-m_\Xi }{16\,\pi \,f^2}\,
\frac{p_{\pi \Xi}^3}{m_{\Xi(1820)}}
\,\Big( F_{[9]}- D_{[9]} \Big)^2 \,,
\label{nonet-decay}
\end{eqnarray}
where for example $E_N= \sqrt{m_N^2+p_{\pi N}^2}$ with the relative momentum $p_{\pi N}$ defined in the rest frame 
of the resonance. Our values for $F_{[9]}$ and $D_{[9]} $
describe the decay widths and branching ratios of the $\Sigma (1670)$ and $\Xi (1820)$
reasonably well within their large empirical uncertainties. Note that we put less emphasis on the properties of 
the $N(1520)$ resonance since that resonance is strongly influenced by the $\pi \Delta_\mu $ channel not 
considered here. In our global fit we take $F_{[9]}$ and $D_{[9]}$ fixed as given above but fine-tune the 
mixing angle $\vartheta =27.74^\circ$ and $C_{[9]} =2.509 $. To account for further small inelastic 
three-body channels we assign the 'bare' $\Lambda(1520)$ resonance an energy independent decay width of 
$\Gamma_{\Lambda(1520)}^{(3-{\rm body})}\simeq $ 1.4 MeV. The total cross sections are included in the fit 
for $p_{{\rm lab.}}< $ 500 MeV. For the bare masses of the d-wave resonances we use the values 
$m_{\Lambda (1520)} \simeq 1528.2$ MeV, $m_{\Lambda (1690)}\simeq 1705.3$ MeV and $m_{\Sigma (1680)} \simeq 1690.7 $ MeV.  

\begin{figure}[t]
\begin{center}
\includegraphics[width=13cm,clip=true]{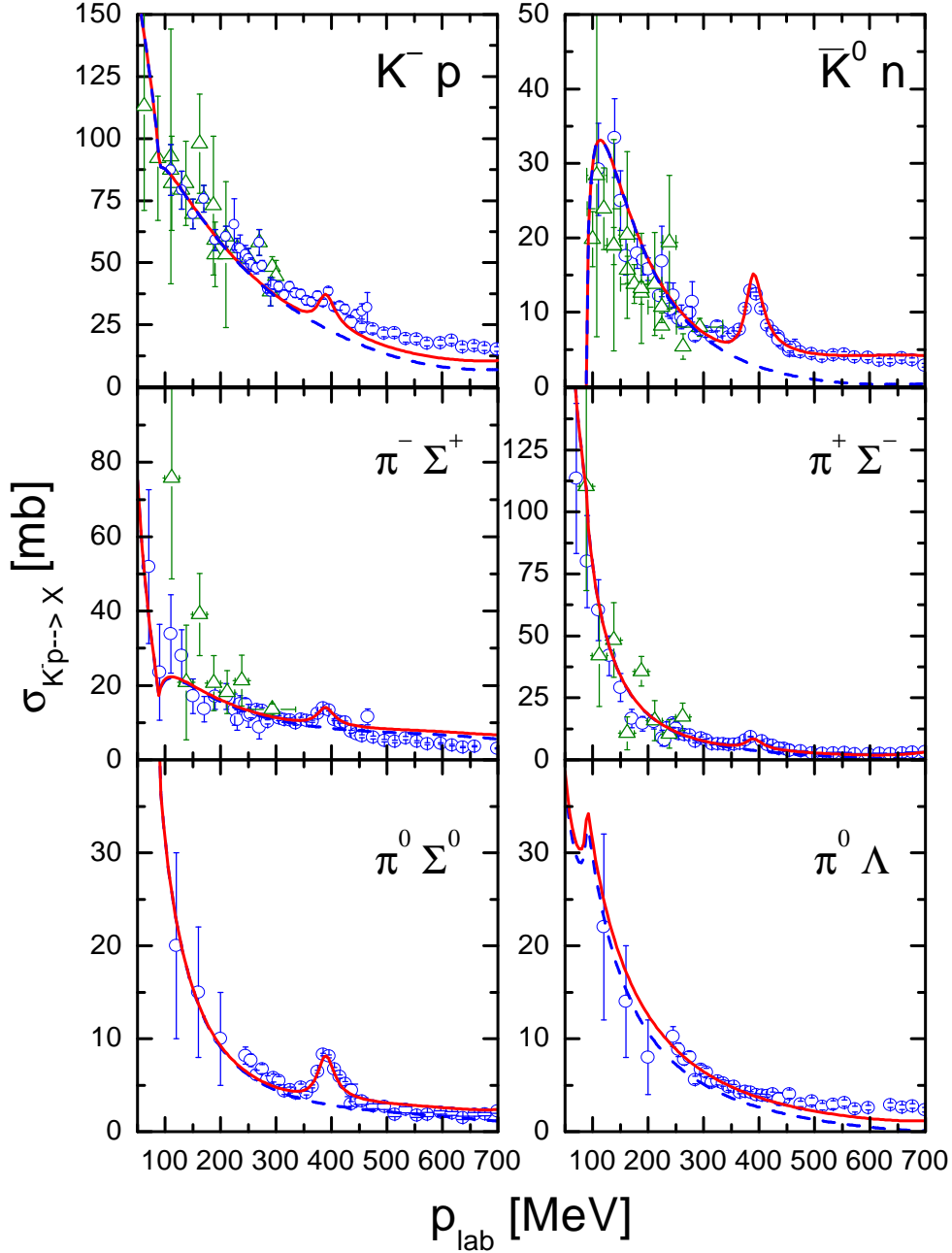}
\end{center}
\caption{$K^-$-proton elastic and inelastic cross sections. The data 
are taken from \cite{mast-pio,sakit,Evans,Ciborowski,mast-ko,bangerter-piS,Armenteros,old-scat}. The solid lines 
show the results of our $\chi$-BS(3) theory including 
all effects of s-, p- and d-waves. The dashed lines represent the s-wave contributions only. We fitted 
the data points given by open circles~\cite{mast-pio,sakit,Evans,Ciborowski,mast-ko,bangerter-piS,Armenteros}. 
Further data points represented by open triangles~\cite{old-scat} were not considered in the global fit.} 
\label{fig:totcross}
\end{figure}

In Fig.~\ref{fig:totcross} we present the result of our fit for the
elastic and inelastic $K^-p$ cross sections. The data set is nicely reproduced including 
the rather precise data points for laboratory momenta 250 MeV$<p_{\rm lab}<$ 500 MeV. 
In Fig.~\ref{fig:totcross} the s-wave contribution to the total cross section is shown with a dashed line. 
Important p-wave contributions are found at low energies only in the $\Lambda \pi^0$ production cross section. 
Note that the $\Lambda \pi^0$ channel carries isospin one and therefore provides valuable
constraints on the poorly known $K^-$-neutron interaction. The deviation of our result 
from the empirical cross sections above $p_{\rm lab} \simeq 500$ MeV in some channels may be due in part 
to the fact that we do not consider the p-wave  $\Lambda(1600)$ and $\Sigma(1660) $ resonances 
quantitatively in this work. As will be demonstrated below when presenting the partial-wave amplitudes, 
there is, however, a strong tendency that those resonances are generated in the $\chi$-BS(3) scheme. We checked that,  
by giving up some of the large $N_c$ sum rules and thereby increasing the number of free parameters we can easily 
obtain a fit with much improved quality beyond $p_{lab} = 500$ MeV. We refrained from presenting those 
results because it is not clear that this procedure leads to the correct partial wave interpretation 
of the total cross sections. Note that the inelastic channel $K^-p \to \Lambda \pi \pi$, not 
included in this work, is no longer negligible at a quantitative level for $p_{\rm lab} > 300$ MeV \cite{watson}. 

\begin{figure}[t]
\begin{center}
\includegraphics[width=10cm,clip=true]{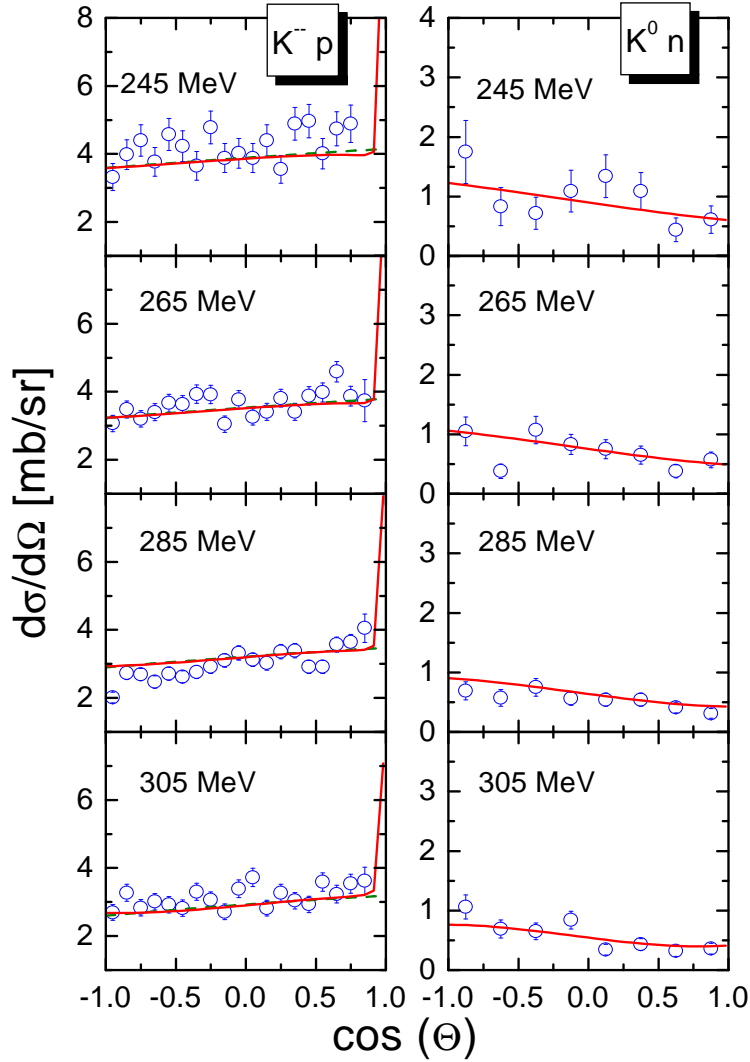}
\end{center}
\caption{$K^-p \to K^-p,\,\bar K^0\,n $ differential cross sections at $p_{\rm lab}=245$ MeV,
$265$ MeV, $285$ MeV and $305$ MeV. The data are taken form \cite{mast-ko}. The solid lines represent 
our $\chi$-BS(3) theory including s-, p- and d-waves as well as Coulomb effects. The dashed lines follow if 
Coulomb interactions are switched off. }
\label{fig:difkm}
\end{figure}

In Fig.~\ref{fig:difkm} we confront our result with a selection of  
differential cross sections from elastic $K^-$-proton scattering \cite{mast-ko}. The angular 
distribution patterns are consistent with weak p-wave interactions.  
The almost linear slope in $\cos \theta $ reflects the interference of
the p-wave contribution with a strong s-wave. Coulomb effects are small in these channels. 
The differential cross sections provide valuable constraints on the p-wave 
interaction strengths.

\begin{figure}[t]
\begin{center}
\includegraphics[width=14.0cm,clip=true]{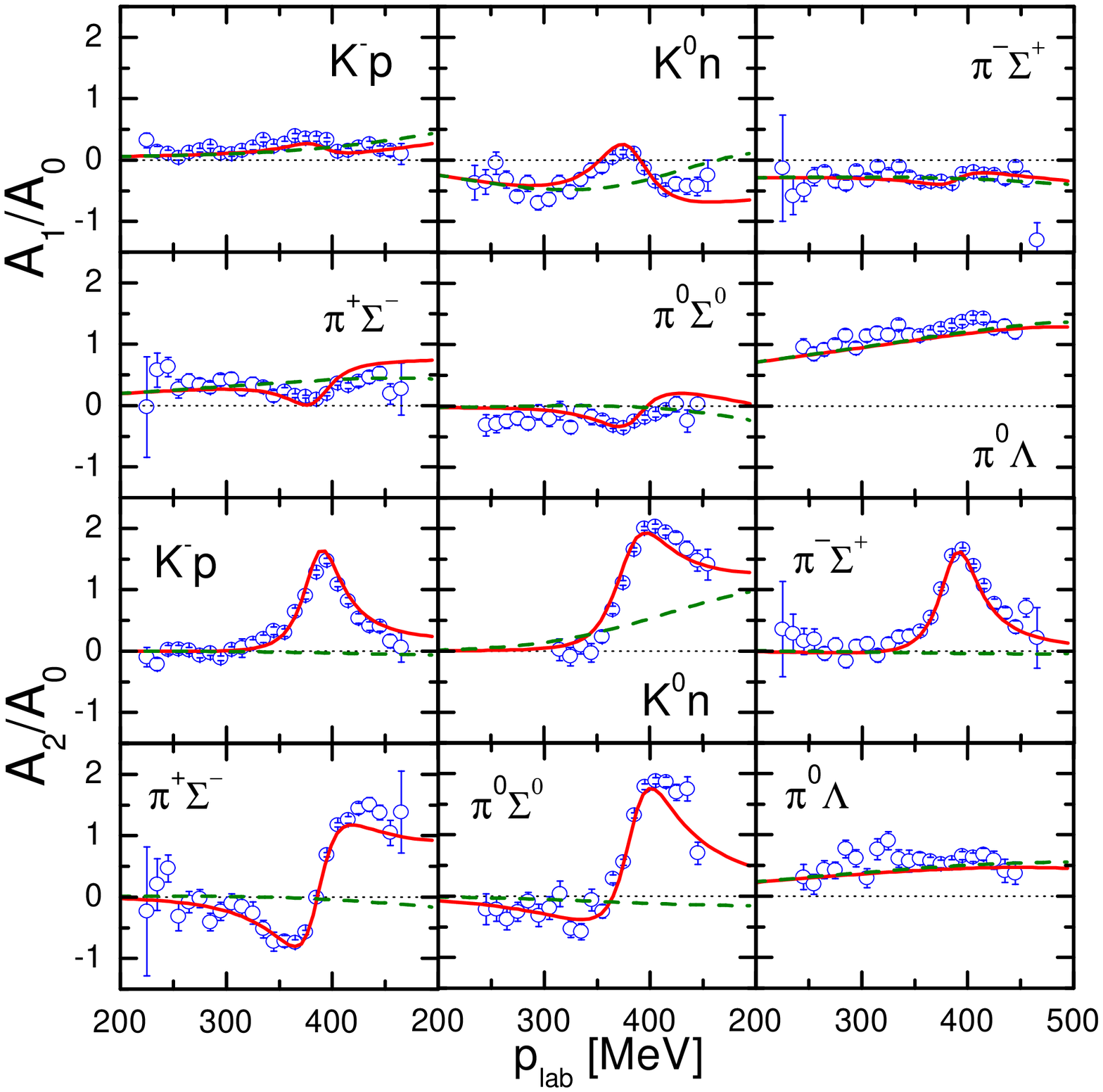} 
\end{center}
\caption{Coefficients $A_1$ and $A_2$ for the $K^-p\to \pi^0 \Lambda$,
$K^-p\to \pi^\mp \Sigma^\pm$  
and $K^-p\to \pi^0 \Sigma$ differential cross sections. The data are
taken from \cite{mast-pio,bangerter-piS}. The solid lines are the result of the $\chi$-BS(3) approach 
with inclusion of the d-wave resonances. The dashed lines show the effect of switching off d-wave contributions.} 
\label{fig:a}
\end{figure}

\begin{figure}[t]
\begin{center}
\includegraphics[width=11.0cm,clip=true]{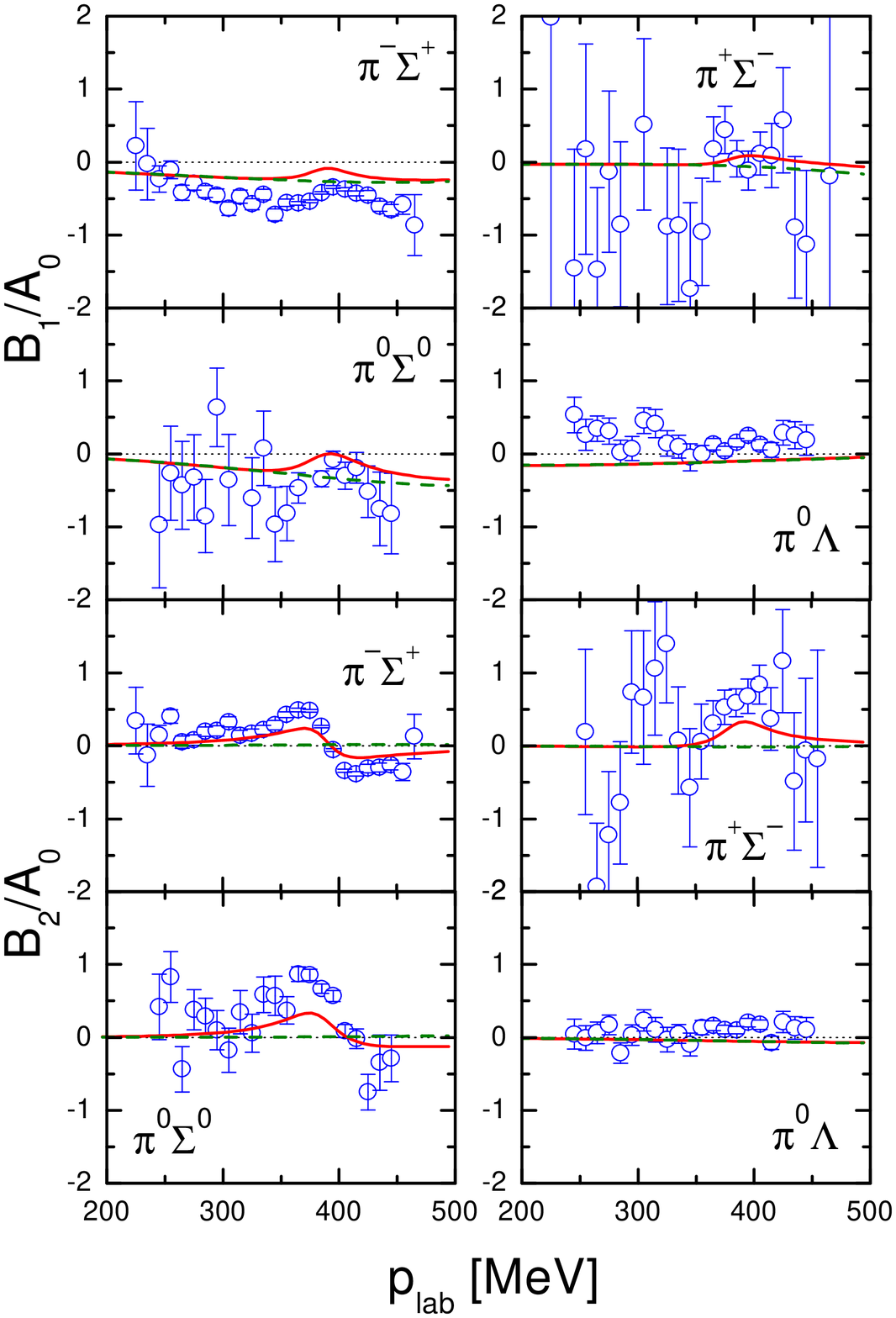}
\end{center}
\caption{Coefficients $B_1$ and $B_2$ for the $K^-p\to \pi^\mp \Sigma^\pm$, 
$K^-p\to \pi^0 \Sigma$ and $K^-p\to \pi^0 \Lambda$ differential cross sections. The data are
taken from \cite{mast-pio,bangerter-piS}. The solid lines are the result of the $\chi$-BS(3) approach 
with inclusion of the d-wave resonances. The dashed lines show the effect of switching off d-wave contributions.} 
\label{fig:b}
\end{figure}

Further important information on the p-wave dynamics is provided by angular distributions for 
the inelastic $K^-p$ reactions. The available data are represented in terms of 
coefficients $A_n$ and $B_n$ characterizing the differential cross section 
$d\sigma(\cos \theta , \sqrt{s}\,) $
and the polarization $P(\cos \theta , \sqrt{s}\,)$ as functions of the center of mass 
scattering angle $\theta $ and the total energy $\sqrt{s}$:
\begin{eqnarray}
\frac{d\sigma (\sqrt{s}, \cos \theta )}{d\cos \theta }  &=&
\sum_{n=0}^\infty A_n(\sqrt{s}\,)\,P_n(\cos \theta ) \,, 
\nonumber\\
\frac{d\sigma (\sqrt{s}, \cos \theta )}{d\cos \theta } \,P (\sqrt{s}, \cos \theta ) 
&=& \sum_{n=1}^\infty B_l(\sqrt{s}\,)\,P^1_n(\cos \theta ) \;.
\label{a-b-def}
\end{eqnarray}
In Fig.~\ref{fig:a} we compare the empirical ratios $A_1/A_0$ and $A_2/A_0$ with the results of 
the $\chi$-BS(3) approach. Note that for $p_{\rm lab} < 300$ MeV the empirical ratios with $n\geq 3$ are 
compatible with zero within their given errors.  A large $A_1/A_0$ ratio 
is found only in the $K^-p\to \pi^0 \Lambda$ channel demonstrating again the importance of 
p-wave effects in the isospin one channel. The dashed lines of Fig.~\ref{fig:a}, which are obtained when  
switching off d-wave contributions, confirm the importance of this resonance for the angular 
distributions in the isospin zero channel. The fact that the $\Lambda(1520)$ resonance appears 
more important in the differential cross sections than in the total cross sections follows simply 
because the tail of the resonance is enhanced if probed via an interference term. In the differential 
cross section the $\Lambda(1520)$ propagator enters linearly whereas the total cross section probes 
the squared propagator only. Note also the sizeable p-wave contributions at somewhat larger momenta  
seen in the charge-exchange reaction of Fig. \ref{fig:a} and also in Fig. \ref{fig:totcross}.

The constraint from the ratios $B_1/A_0$ and $B_2/A_0$, presented in Fig. \ref{fig:b}, is weak due to rather large 
empirical errors. New polarization data, possibly with polarized hydrogen targets, would be highly desirable.

\tabcolsep=0.9mm
\begin{table}[t]\begin{center}
\begin{tabular}{|c|c|c||c|c|c|}
\hline 
& $a_{K^-p }$ [fm] & $a_{K^-n }$ [fm] & $\gamma  $ & $R_c$ & $R_n $   \\
\hline \hline
Exp. & -0.78$\pm$0.18
 & - & 2.36$\pm$ 0.04 & 0.664$\pm$0.011 & 0.189$\pm$0.015 \\
  & +$i$\,(0.49$\pm$0.37)  & & & & \\
\hline
$\chi$-BS(3) & -1.09+$i$ 0.82 & 0.29+$i$ 0.54 & 2.42 & 0.65 & 0.19  \\
\hline
 SU(2) & -0.79+$i$ 0.95 & 0.30+$i$ 0.49 & 4.58 & 0.63 & 0.32 \\
\hline
\hline 
& 
$a^{(\bar K N)}_{P_{01}}$ $[m_\pi^{-3}]$ & $a^{(\bar K N)}_{P_{03}}$ $[m_\pi^{-3}]$ & &
$a^{(\bar K N)}_{P_{21}}$ $[m_\pi^{-3}]$ & $a^{(\bar K N)}_{P_{23}}$ $[m_\pi^{-3}]$  \\
\hline \hline
$\chi$-BS(3)   & 0.025+$i$ 0.001 & 0.002+$i$ 0.001 & & -0.004+$i$ 0.001 & -0.055+$i$ 0.021 \\
\hline
\end{tabular}
\end{center}
\caption{$K^-$-nucleon threshold parameters. The row labelled by $SU(2)$ gives the results in the isospin limit with 
$m_{K^-}=m_{\bar K^0}= 493.7$ MeV and $m_p = m_n = 938.9$ MeV. }
\label{tab-r-km}
\end{table}

We turn to the threshold characteristics of the $K^-p$ reaction which is 
constrained by experimental data for the threshold branching ratios 
$\gamma, R_c$ and $R_n$ where
\begin{eqnarray}
\gamma &=& \frac{\sigma (K^-\,p\rightarrow \pi^+\,\Sigma^-)}
{\sigma (K^-\,p\rightarrow \pi^-\,\Sigma^+)}  \;, \quad 
R_c = \frac{\sigma (K^-\,p\rightarrow \mbox{charged\,particles })}
{\sigma (K^-\,p\rightarrow \mbox{all})} \;,
\nonumber\\
R_n &=& \frac{\sigma (K^-\,p\rightarrow \pi^0\,\Lambda)}
{\sigma (K^-\,p\rightarrow \mbox{all\,neutral \,channels})}  \;.
\label{km:thres}
\end{eqnarray}
A further important piece of information  
is provided by the recent measurement of the $K^-$ hydrogen atom state
which leads  to a value for the $K^-p$ scattering length \cite{Iwasaki}. 
In Tab. \ref{tab-r-km} we confront the empirical numbers with our analysis.  
All threshold parameters are well described within the $\chi$-BS(3) 
approach. We confirm the result of \cite{Kaiser,Ramos} that the branching ratios
are rather sensitive to isospin breaking effects. However, note that 
it is sufficient to include isospin breaking effects only in the $\bar
KN$ channel to good accuracy. The empirical branching ratios are taken from
\cite{branch-rat}.  The real part of our $K^-n$ scattering length with 
$\Re \,a_{K^-n \to K^- n} \simeq 0.29$ fm turns out considerably smaller than the value of 
$ 0.53$ fm found in the recent analysis \cite{Ramos}. 
In Tab. \ref{tab-r-km} we present also our results for the p-wave scattering volumes. 
Here isospin breaking effects are negligible. All scattering volumes but the one in the 
$P_{23}$ channel are found to be small. The not too small and repulsive scattering volume 
$a_{P_{23}} \simeq (-0.16+i\,0.06)$ fm$^3$ reflects the presence of the $\Sigma (1385)$ resonance 
just below the $\bar K N$ threshold.  The precise values of the threshold parameters 
are of crucial importance when describing  $K^-$-atom data which constitute a rather sensitive 
test of the in-medium dynamics of antikaons. In particular one expects a strong sensitivity 
of the level shifts to the s-wave scattering lengths.

\begin{figure}[t]
\begin{center}
\includegraphics[width=12cm,clip=true]{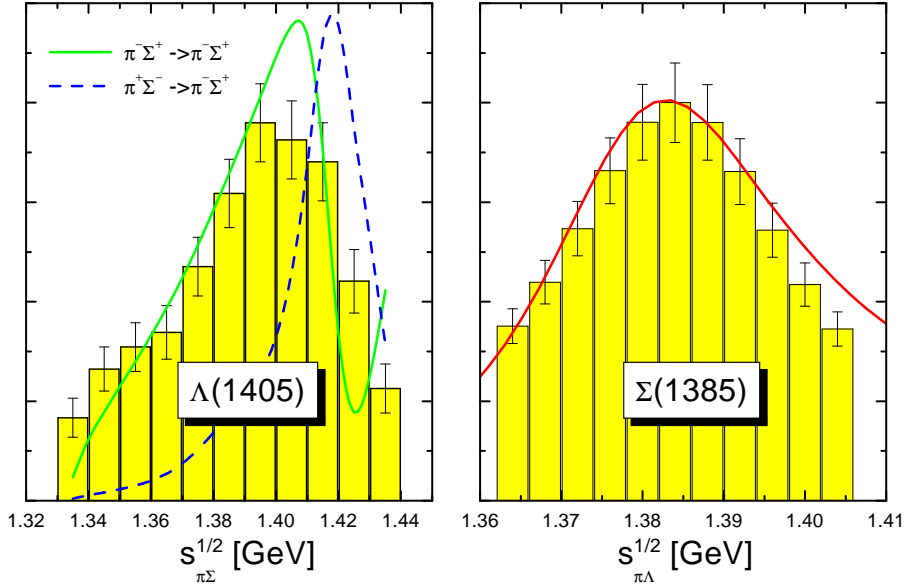}
\end{center}
\caption{$\Lambda(1405)$ and $\Sigma (1385)$ resonance mass
distributions in arbitrary units.}
\label{fig:massspec}
\end{figure}

In Fig.~\ref{fig:massspec} we show the $\Lambda(1405)$ and $\Sigma (1385)$ spectral 
functions measured in the reactions $K^-p\rightarrow \Sigma^+\pi^-\pi^+\pi^-$ \cite{lb-spec}
and $K^-p\rightarrow \Lambda \pi^+\pi^-$ \cite{sig-spec}  respectively. 
We did not include the $\Lambda(1405)$ spectrum of \cite{lb-spec} in our global fit.
Since the $\Lambda(1405)$-spectrum shows a strong energy dependence, incompatible with a 
Breit-Wigner resonance shape, the spectral form depends rather strongly on the initial and
final states through which it is measured. The empirical spectrum of  \cite{lb-spec}
describes the reaction $\Sigma^+(1600) \,\pi^- \to \Lambda(1405) \to \Sigma^+ \,\pi^-$
rather than the reactions $\Sigma^\pm \,\pi^\mp \to \Lambda(1405) \to \Sigma^+ \,\pi^-$
accessible in our present scheme. In Fig. 6 the spectral form of the $\Lambda(1405)$ 
resulting from two different initial states $\Sigma^\pm \,\pi^\mp$ are confronted with the 
empirical spectrum of \cite{lb-spec}. While the spectrum defined with respect to the initial 
state $\Sigma^+ \,\pi^-$ represents the empirical spectrum reasonably well the 
other choice of initial state $\Sigma^+ \,\pi^-$ leads to a significantly altered spectral 
form. We therefore conclude that in a scheme not including the $\Sigma(1600) \pi$ state 
explicitly it is not justified to use the $\Lambda(1405)$ spectrum of \cite{lb-spec} as a 
quantitative constraint for the kaon-nucleon dynamics.

We turn to the mass spectrum of the decuplet $\Sigma (1385)$ state. The spectral 
form, to good accuracy of Breit-Wigner form, is 
reproduced reasonably well. Our result for the ratio of $\Sigma (1385) \to \pi \Lambda $
over $\Sigma (13 85) \to \pi \Sigma $ of about $17\, \% $ compares well with the most recent empirical 
determination. In \cite{Dionisi} that branching ratio was extracted from the 
$K^-p \to \Sigma (1385) \,K\,\bar K$
reaction and found to be $20 \pm 6 \,\%$. Note that we obtained our ratio from the two reaction amplitudes 
$\pi \Lambda \to \pi \Lambda$ and $\pi \Lambda \to \pi \Sigma $ evaluated at the $\Sigma (1385)$ pole. 
The schematic expression (\ref{decuplet-decay}) would give a smaller value of about $ 15 \, \%$. Finally 
we mention that the value for our $\Xi (1530)$ total width of $10.8 $ MeV comfortably meets the 
empirical value of $9.9^{+1.7}_{-1.9} $ MeV given in \cite{fpi:exp}.

\subsection{Analyticity and crossing symmetry}

It is important to investigate to what extent our multi-channel scattering amplitudes 
are consistent with the expectations from analyticity and crossing symmetry. As discussed in 
detail in section 4.3 the crossing symmetry constraints should not be considered in terms of 
partial-wave amplitudes, but rather in terms of the forward scattering amplitudes only. 
We expect crossing symmetry to be particularly important in the strangeness sector, because that sector 
has a large subthreshold region not directly accessible and constrained by data. 
In the following we reconstruct the forward $\bar KN$ and $KN$ scattering amplitudes in terms of their 
imaginary parts by means of dispersion integrals. We then confront the reconstructed scattering amplitudes
with the original ones. It is non-trivial that those amplitudes match even though our loop functions and 
effective interaction kernels are analytic functions. One can not exclude that the coupled channel dynamics
generates unphysical singularities off the real axis which would then spoil the representation of the 
scattering amplitudes in terms of dispersion integrals. Note that within effective field theory unphysical 
singularities are acceptable, however, only far outside the applicability domain of the approach.

Our analysis is analogous to that of Martin \cite{A.D.Martin}. However, 
we consider the dispersion integral as a consistency check of our theory rather than as a 
predictive tool to derive the low-energy kaon-nucleon scattering amplitudes in terms 
of the more accurate scattering data at $E_{\rm lab}>300$ MeV \cite{Queen}. This way we avoid any subtle 
assumptions on the number of required subtractions in the dispersion integral.
Obviously the dispersion integral, if evaluated for small energies, must be dominated by the low-energy total 
cross sections which are not known empirically too well. 
We write a subtracted dispersion integral
\begin{eqnarray}
&& \! \!T^{(0)}_{\bar K N}(s) = \frac{f^2_{KN \Lambda }}
{s-m^2_\Lambda} +  \sum_{k=1}^n\,c_{\bar K N}^{(0,k)}\,(s-s_0)^k
+\!\!\!\int_{(m_\Sigma+m_\pi)^2}^\infty \!\!\! \! \frac{d \,s'}{\pi }\,
\frac{(s-s_0)^n}{(s'-s_0)^n}\,\frac{\Im \,T^{(0)}_{\bar K N} (s')}{s'-s -i\,\epsilon}\;,
\nonumber\\
&& \! \! T^{(1)}_{\bar K N}(s) = \frac{f_{KN \Sigma }^2}
{s-m^2_\Sigma} +  \sum_{k=1}^n\,c_{\bar K N}^{(1,k)}\,(s-s_0)^k
+\!\!\! \int_{(m_\Lambda+m_\pi)^2}^\infty \!\!\! \!\frac{d \,s'}{\pi }\,
\frac{(s-s_0)^n}{(s'-s_0)^n}\,\frac{\Im \,T^{(1)}_{\bar K N} (s')}{s'-s -i\,\epsilon} \;,
\nonumber\\
&&\qquad \quad f_{KN Y } = \sqrt{\frac{m_K^2-(m_Y-m_N)^2}{2\,m_N}}\,
\frac{m_N+m_Y}{2\,f}\,G^{(Y)}_{\bar  K N } \,,
\label{disp-check}
\end{eqnarray}
where we identify $s_0 = \Lambda_{\rm opt.}^2= m_N^2+m_K^2$ with the optimal matching point
of (\ref{opt-match}). We recall that the  kaon-hyperon coupling constants 
$G^{(Y)}_{\bar K N}=A^{(Y)}_{\bar K N}+P^{(Y)}_{\bar K N}$ with $Y=\Lambda, \Sigma $ receive 
contributions from pseudo-vector and pseudo-scalar vertices as specified 
in Tab. \ref{meson-baryon:tab}. The values $f_{K N \Lambda } \simeq -12.8\,m^{1/2}_{\pi^+}$ and 
$f_{K N \Sigma } \simeq 6.1\,m^{1/2}_{\pi^+}$ follow.

The forward scattering amplitude
$T_{\bar KN}^{}(s)$  reconstructed in terms of the partial-wave 
amplitudes $f_{\bar K N, J}^{(L)}(\sqrt{s}\,)$ of (\ref{match}) reads
\begin{eqnarray}
T^{}_{\bar KN}(s) &=&4\,\pi\,\frac{\sqrt{s}}{m_N}\,
\sum_{n=0}^\infty \,\Big(n+1 \Big)\,\Big( f_{\bar K N,n+\half}^{(n)}(\sqrt{s}\,)+
f_{\bar K N,n+\half}^{(n+1)}(\sqrt{s}\,)
\Big) \;.
\label{forward-amplitude}
\end{eqnarray}
The subtraction coefficients $c_{\bar K N}^{(I,k)}$ are 
adjusted to reproduce the scattering amplitudes close to the kaon-nucleon threshold. 
One must perform a sufficient number of subtractions so that 
the dispersion integral in (\ref{disp-check}) is dominated by energies still within the 
applicability range of our theory. With $n=4$ in (\ref{disp-check}) we indeed find that 
we are insensitive to the scattering amplitudes for $\sqrt{s} > 1600$ MeV to good 
accuracy. Similarly we write a subtracted dispersion integral for the 
amplitudes $T^{(I)}_{K N}(s)$ of the strangeness plus one sector
\begin{eqnarray}
 T^{(0)}_{K N}(s) &=& -\frac{1}{2}\,\frac{f^2_{KN \Lambda }}{2\,s_0-s-m^2_\Lambda}
+\frac{3}{2}\,\frac{f^2_{KN \Sigma }}{2\,s_0-s-m^2_\Sigma}
\nonumber\\
&+&\sum_{k=1}^n\,c_{K N}^{(0,k)}\,(s-s_0)^k+\int_{(m_N+m_K)^2}^\infty \frac{d \,s'}{\pi }\,
\frac{(s-s_0)^n}{(s'-s_0)^n}\,\frac{\Im \,T^{(0)}_{K N} (s')}{s'-s -i\,\epsilon}\;,
\nonumber\\
T^{(1)}_{K N}(s) &=& \frac{1}{2}\, \frac{f^2_{KN \Lambda }}{2\,s_0-s-m^2_\Lambda}
+\frac{1}{2}\, \frac{f^2_{KN \Sigma }}{2\,s_0-s-m^2_\Sigma}
\nonumber\\
&+&\sum_{k=1}^n\,c_{K N}^{(1,k)}\,(s-s_0)^k+\int_{(m_N+m_K)^2}^\infty \frac{d \,s'}{\pi }\,
\frac{(s-s_0)^n}{(s'-s_0)^n}\,\frac{\Im \,T^{(1)}_{K N} (s')}{s'-s -i\,\epsilon} \;.
\label{disp-check-2}
\end{eqnarray}
The subtraction coefficients $c_{K N}^{(I,k)}$ are 
adjusted to reproduce the scattering amplitudes close to the kaon-nucleon threshold. The 
choice $n=4$ leads to a sufficient emphasis of the low-energy $KN$-amplitudes.

\begin{figure}[t]
\begin{center}
\includegraphics[width=13cm,clip=true]{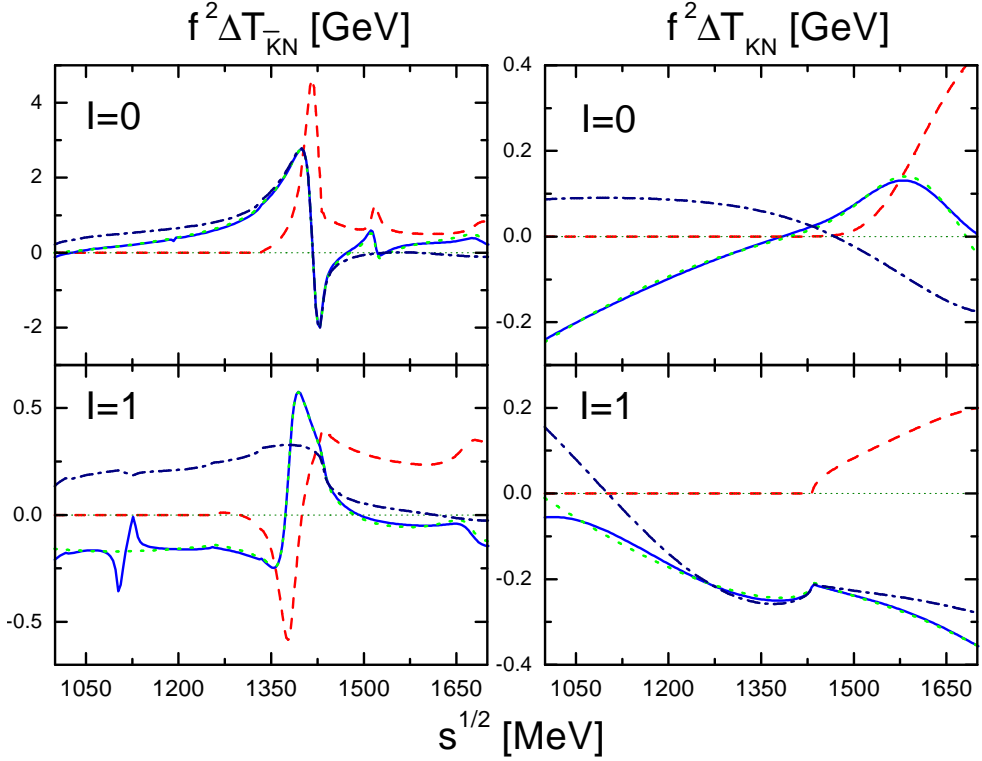}
\end{center}
\caption{Causality check of the pole-subtracted kaon-nucleon scattering amplitudes $f^2\,\Delta T^{(I)}_{K N}$
and $f^2\,\Delta T^{(I)}_{\bar K N}$. The full lines represent the real part of the forward scattering amplitudes. 
The dotted lines give the amplitudes as obtained from their imaginary parts (dashed lines) in terms of the 
dispersion integrals (\ref{disp-check}) and (\ref{disp-check-2}). The dashed-dotted lines give the s-wave 
contribution to the real part of the forward scattering amplitudes only.}
\label{fig:causality}
\end{figure}

In Fig.~\ref{fig:causality} we compare the real part of the pole-subtracted amplitudes 
$\Delta T^{(I)}_{\bar KN}(s)$ and $\Delta T^{(I)}_{KN}(s)$ with the corresponding amplitudes 
reconstructed via the dispersion integral (\ref{disp-check})
\begin{eqnarray}
\Delta T_{\bar KN}^{(I)}(s) &=&  T_{\bar KN}^{(I)}(s)- \frac{f^2_{K N Y}}{s-m_H^2}
-\left( A_{\bar K N}^{(Y)}\right)^2 \frac{2\,m_N^2+m_K^2-s-m_H^2}{8\,f^2\,m_N} 
\nonumber\\
&+&\left( P_{\bar K N}^{(Y)}\right)^2 \frac{(m_N+m_H)^2}{8\,f^2\,m_N}
+ P_{\bar K N}^{(Y)}\,A_{\bar K N}^{(Y)}\frac{m^2_H-m_N^2}{4\,f^2\,m_N}  \,,
\label{pole-subtr}
\end{eqnarray}
with $Y=\Lambda$ for $I=0$ and $Y= \Sigma $ for $I=1$. Whereas it 
is straightforward to subtract the complete hyperon pole contribution in the $\bar K N$ amplitudes 
(see (\ref{pole-subtr})), it is less immediate how to do so for the $KN$ amplitudes. Since we 
do not consider all partial wave contributions in the latter amplitudes the u-channel pole 
must be subtracted in its approximated form as given in (\ref{u-approx-1}). 
The reconstructed amplitudes agree rather well with the original amplitudes. For 
$\sqrt{s} > 1200$ MeV the solid and dotted lines in Fig. \ref{fig:causality} can hardly be discriminated. 
This demonstrates that our amplitudes are causal too good accuracy. Note that the 
discrepancy for $\sqrt{s} < 1200$ MeV is a consequence of the approximate treatment 
of the non-local u-channel exchanges  which violates analyticity at subthreshold energies to some extent 
(see (\ref{prescription})). With Fig.~\ref{fig:causality} we confirm that such effects are well controlled 
for $\sqrt{s} > 1200$ MeV. In any case, close to $\sqrt{s}\simeq 1200$ MeV the complete forward scattering amplitudes 
are largely dominated by the s- and u-channel hyperon pole contributions absent in $\Delta T_{\bar KN}^{(I)}(s)$.

As can be seen from Fig.~\ref{fig:causality} also, we find sizeable p-wave contributions in the pole-subtracted 
amplitudes at subthreshold energies. This follows from comparing the dashed-dotted lines, which give the s-wave 
contributions only, with the solid lines which represent the complete real part of the pole-subtracted forward 
scattering amplitudes. The p-wave contributions are typically much larger below threshold than above threshold. 
The fact that this is not the case in the isospin zero $KN$ amplitude reflects a subtle 
cancellation mechanism of hyperon exchange contributions and quasi-local two-body interaction terms. In the 
$\bar K N$ amplitudes the subthreshold effects of p-waves are most dramatic in the isospin one channel. Here 
the amplitude is dominated by the $\Sigma(1385)$ resonance. Note that p-wave channels contribute with a 
positive imaginary part for energies larger than the kaon-nucleon threshold but with a negative imaginary part 
for subthreshold energies. A negative imaginary part of a subthreshold amplitude is consistent with 
the optical theorem which only relates the imaginary part of the forward scattering 
amplitudes to the total cross section for energies above threshold. Our analysis may shed 
some doubts on the quantitative results of the analysis by Martin, which attempted 
to constrain the forward scattering amplitude via a dispersion analysis \cite{A.D.Martin}. 
An implicit assumption of Martin's analysis 
was, that the contribution to the dispersion integral from the subthreshold region, which is not directly determined 
by the data set, is dominated by s-wave dynamics. 
As was pointed out in \cite{Hirschegg} strong subthreshold p-wave 
contributions should have an important effect for the propagation properties of antikaons in dense 
nuclear matter. 

We turn to the approximate crossing symmetry of our scattering amplitudes. 
Crossing symmetry relates the subthreshold $\bar K N$ and $K N$ scattering amplitudes. 
As a consequence the exact amplitude ${\rm T}^{(0)}_{\bar K N}(s)$ shows unitarity cuts not only for 
$\sqrt{s}>m_\Sigma+m_\pi$ but also for $\sqrt{s}<m_N-m_K$ representing the elastic $K N$ 
scattering process. Consider for example the isospin zero amplitude for which one expects 
the following representation:

\begin{figure}[t]
\begin{center}
\includegraphics[width=12cm,clip=true]{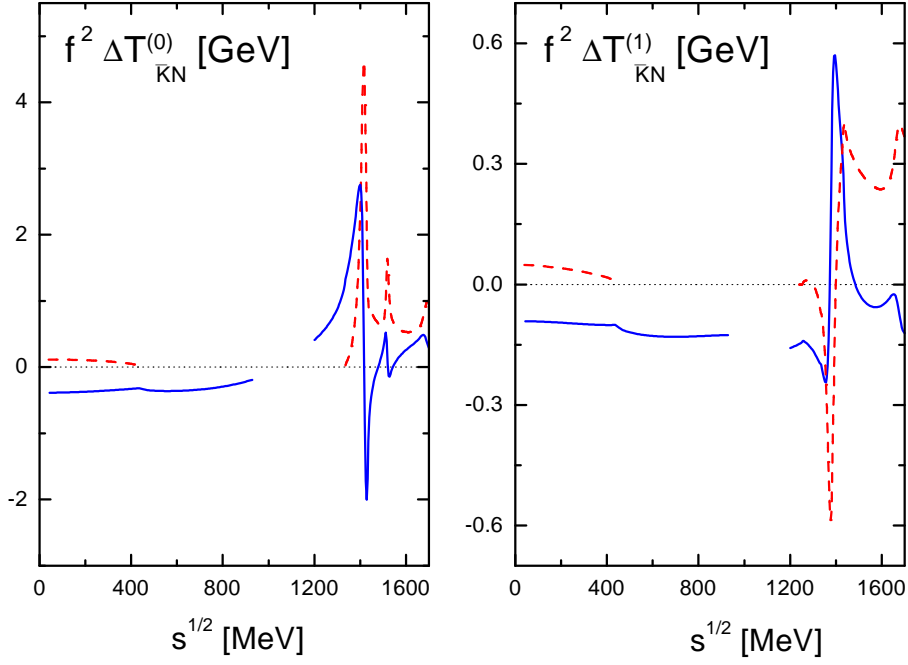}
\end{center}
\caption{Approximate crossing symmetry of the pole subtracted kaon-nucleon forward scattering amplitudes. The lines in 
the left hand parts of the figures result from the $KN$ amplitudes. The lines in the right hand side of 
the figures give the $\bar K N$ amplitudes.}
\label{fig:crossing}
\end{figure}

\begin{eqnarray}
\!\! {\rm T}^{(0)}_{\bar K N}(s)-{\rm T}^{(0)}_{\bar K N}(s_0) &=& 
\frac{f^2_{KN \Lambda }}{s-m^2_\Lambda}-\frac{f^2_{KN \Lambda }}{s_0-m^2_\Lambda} +  
\! \! \!\int_{-\infty}^{(m_N-m_K)^2} \frac{d \,s'}{\pi }\,\frac{s-s_0}{s'-s_0}\,
\frac{\Im \,{\rm T}^{(0)}_{\bar K N} (s')}{s'-s -i\,\epsilon}
\nonumber\\
&+&\int_{(m_\Sigma+m_\pi)^2}^{+\infty} \frac{d \,s'}{\pi }\,\frac{s-s_0}{s'-s_0}\,
\frac{\Im \,{\rm T}^{(0)}_{\bar K N} (s')}{s'-s -i\,\epsilon}
\;,
\label{cross-disp-check}
\end{eqnarray}
where we performed one subtraction to help the convergence of the dispersion integral. 
Comparing the expressions for $T^{(0)}_{\bar K N}(s)$  in (\ref{disp-check}) and 
${\rm  T}^{(0)}_{\bar K N}(s)$ in (\ref{cross-disp-check}) demonstrates that the contribution 
of the unitarity cut at $\sqrt{s}< m_N-m_K$ in (\ref{cross-disp-check}) is effectively 
absorbed in the subtraction coefficients $c_{\bar K N}^{(I,k)}$ of (\ref{disp-check}). 
Similarly the subtraction coefficients $c_{K N}^{(I,k)}$ in (\ref{disp-check-2}) represent the 
contribution to $T^{(I)}_{K N}(s)$ from the inelastic $\bar K$N scattering process. Thus 
both model amplitudes $T^{(I)}_{K N}(s)$ and $T^{(I)}_{\bar K N}(s)$ 
represent the exact amplitude ${\rm T}^{(I)}_{\bar K N}(s)$ within their validity 
domains and therefore we expect approximate crossing symmetry 
\begin{eqnarray}
&& \Delta T^{(0)}_{\bar K N}(s) \simeq   
-\frac{1}{2}\, \Delta T^{(0)}_{K N}(2\,s_0-s)+\frac{3}{2}\, \Delta T^{(1)}_{ K N}(2\,s_0-s) \;,
\nonumber\\
&&  \Delta T^{(1)}_{\bar K N}(s) \simeq  
+ \frac{1}{2}\, \Delta T^{(0)}_{K N}(2\,s_0-s)+\frac{1}{2}\, \Delta T^{(1)}_{ K N}(2\,s_0-s) \;,
\label{exp-cross}
\end{eqnarray}
close to the optimal matching point $s_0=m_N^2+m_K^2$ only.  In Fig.~\ref{fig:crossing} we 
confront the pole-subtracted $ \Delta T^{(I)}_{K N}$ and $  \Delta T^{(I)}_{\bar KN} $ amplitudes with the 
expected approximate crossing identities (\ref{exp-cross}). Since the optimal matching point 
$s_0=m_N^2+m_K^2 \simeq (1068)^2$ MeV$^2$ is slightly below the respective validity range of the 
original amplitudes, we use the reconstructed amplitudes of (\ref{disp-check}) and
(\ref{disp-check-2}) shown in Fig.~\ref{fig:causality}. This is justified, because the 
reconstructed amplitudes are based on the imaginary parts of the amplitude which have support 
within the validity domain of our theory only. Fig.~\ref{fig:crossing} indeed confirms that 
the kaon-nucleon scattering amplitudes are approximatively crossing symmetric. Close to the 
the point $s\simeq m_N^2+m_K^2$ the $K N$ and $\bar K N$ amplitudes match. 
We therefore expect that our subthreshold kaon-nucleon scattering amplitudes are determined rather 
reliably and well suited for an application to the nuclear kaon dynamics.

\begin{figure}[t]
\begin{center}
\includegraphics[width=14cm,clip=true]{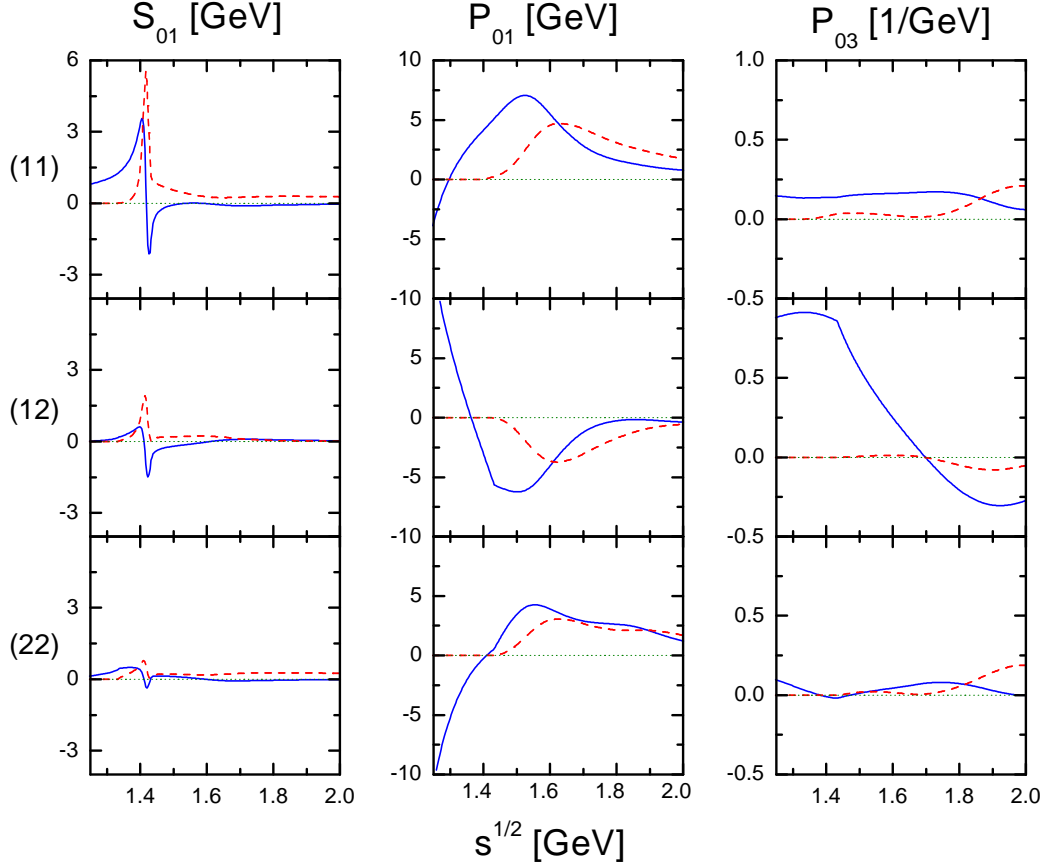}
\end{center}
\caption{Real and imaginary parts of the s- and p-wave partial-wave amplitudes $f^2\,M$ in the 
isospin  zero channel. The labels (ab) refer to our channel convention of (\ref{r-def}) 
with (11): $\bar K N \to \bar K N$, (12): $\bar K N \to \pi \Sigma $ and (22):  $\pi \Sigma \to \pi \Sigma $.}
\label{fig:amplitudes:zero}
\end{figure}

\subsection{Scattering amplitudes}

We discuss now the s- and p-wave partial-wave amplitudes for the various 
$\bar K N, \pi \Sigma $ and $\pi \Lambda$ reactions. Since we are interested in part also in subthreshold 
amplitudes we decided to present the invariant amplitudes $M^{(\pm )}(\sqrt{s},n)$ rather than the more 
common $f^{(l)}_{J}(\sqrt{s}\,)$ amplitudes. Whereas the latter amplitudes lead to convenient  
expressions for the cross sections
\begin{eqnarray}
\sigma_{i\,\to f} (\sqrt{s}\,) &=& 4\,\pi \,\frac{p^{(f)}_{\rm cm}}{p^{(i)}_{\rm cm}}\,
\sum_{l=0}^\infty \,\Bigg(
l\,|f^{(l)}_{J=l-\frac{1}{2}}(\sqrt{s}\,) |^2 
+ (l+1)\,|f^{(l)}_{J=l+\frac{1}{2}}(\sqrt{s}\,)|^2 
\Bigg)\;,  
\label{}
\end{eqnarray}
the former amplitudes $M^{(\pm )}(\sqrt{s},n)$ provide a more detailed picture of the 
higher partial-wave amplitudes, simply because the trivial phase-space factor 
$(p^{(f)}_{\rm cm}\,p^{(i)}_{\rm cm})^{l}$ is taken out

\begin{figure}[t]
\begin{center}
\includegraphics[width=13cm,clip=true]{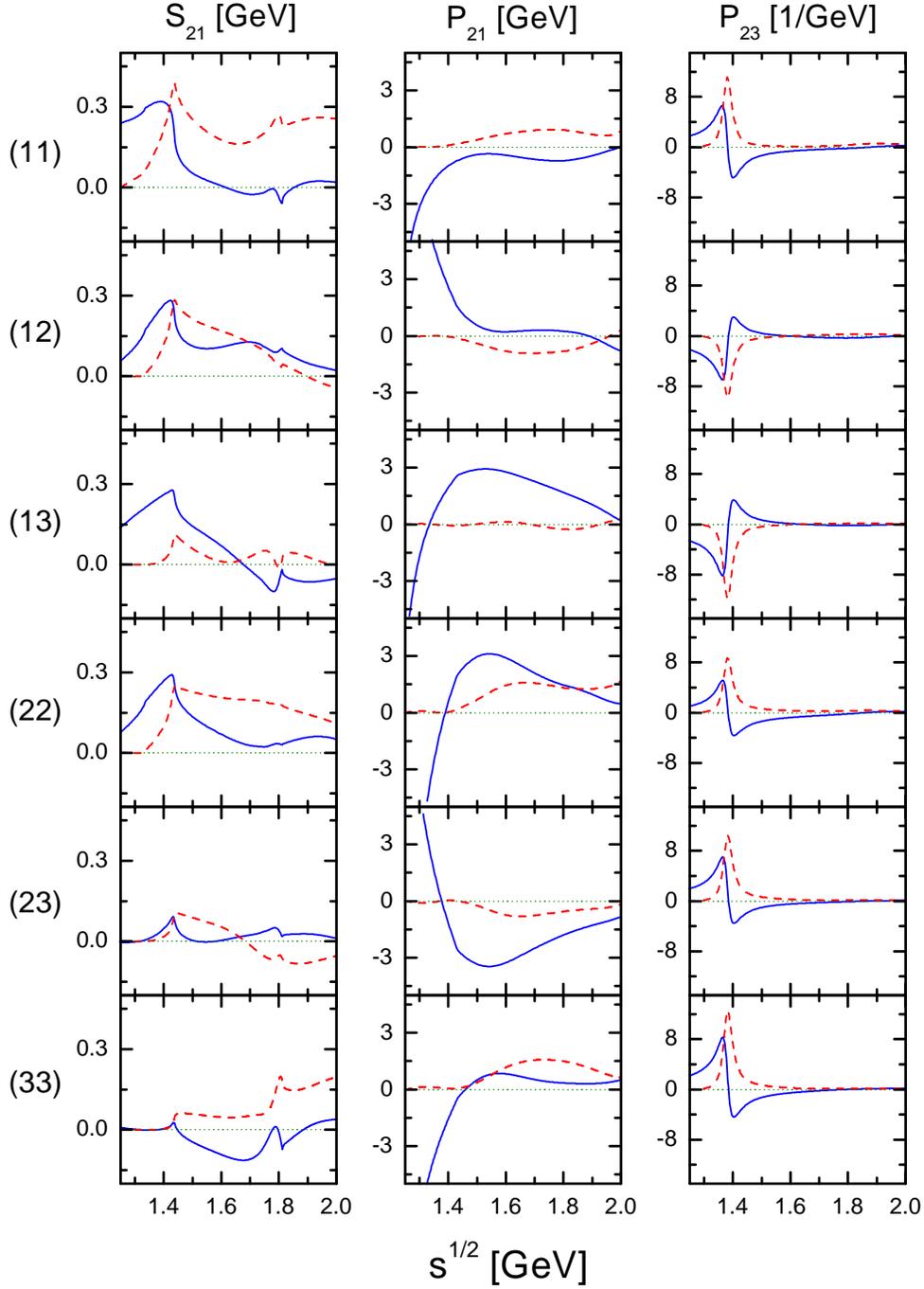}
\end{center}
\caption{Real and imaginary parts of the s- and p-wave partial-wave amplitudes $f^2\,M$ in the 
isospin one channel.
The labels (ab) refer to our channel convention of (\ref{r-def}) with 
(11): $\bar K N \to \bar K N$, (12): $\bar K N \to \pi \Sigma $ and (13):  $\bar K N \to \pi \Lambda $.}
\label{fig:amplitudes:one}
\end{figure}

\begin{eqnarray}
f^{(l)}_{J=l\pm \frac{1}{2}}(\sqrt{s}\,) &=& 
\frac{(p^{(f)}_{\rm cm}\,p^{(i)}_{\rm cm})^{J-\frac{1}{2}}}{8\,\pi\,\sqrt{s}} \;
\sqrt{E_i\pm m_i}\,\sqrt{E_f\pm m_f}\,M^{(\pm )}(\sqrt{s},J-{\textstyle{1\over 2}})\;.
\label{}
\end{eqnarray}
Here $p^{(i,f)}_{\rm cm}$ denote the relative momenta and $E_{i,f}$
the baryon energies  in the center of mass frame. Also $m_{i,f}$ are the baryon masses of the initial and 
final baryons. 

In Fig.~\ref{fig:amplitudes:zero} and in Fig.~~\ref{fig:amplitudes:one} the isospin zero and isospin one amplitudes 
are shown up to rather high energies $\sqrt{s} =2$ GeV. We emphasize that we trust our amplitudes quantitatively 
only up to $\sqrt{s} \simeq 1.6$ GeV. Beyond that energy one should consider our results as  
qualitative only. The amplitudes reflect the presence of the s-wave $\Lambda(1405)$ and p-wave $\Sigma (1385)$ 
resonances. From the relative height of the peak structures in the figures one can read off the branching 
ratios of those resonances. For instance it is clearly seen from Fig.~\ref{fig:amplitudes:zero} that the $\Lambda (1405)$ 
has a rather small coupling to the $\pi \Sigma $ channel. It is gratifying to find precursor effects for the 
p-wave $\Lambda (1600)$  and $\Lambda (1890)$ resonances in that figure also. Note that the estimates for the 
widths of those resonances range up to 200 MeV. Similarly Fig.~\ref{fig:amplitudes:one}
indicates attractive strength in the s- and p- wave channel where one would expect the s-wave $\Sigma (1750)$
and p-wave $\Sigma (1660)$ resonances. 

It is a highly non-trivial but nevertheless expected result, that all resonances but the p-wave baryon decuplet 
and the d-wave baryon nonet resonances were  generated dynamically by the chiral coupled channel dynamics once 
agreement with the low-energy data set with $p_{\rm lab.}< 500$ MeV was achieved. A more accurate description 
of the latter resonances requires the extension of the $\chi$-BS(3) approach including more inelastic channels. 
These finding strongly support the conjecture that all baryon resonances 
but the decuplet ground states are a consequence of coupled-channels dynamics.

\subsection{Predictions  for cross sections}

\begin{figure}[t]
\begin{center}
\includegraphics[width=14cm,clip=true]{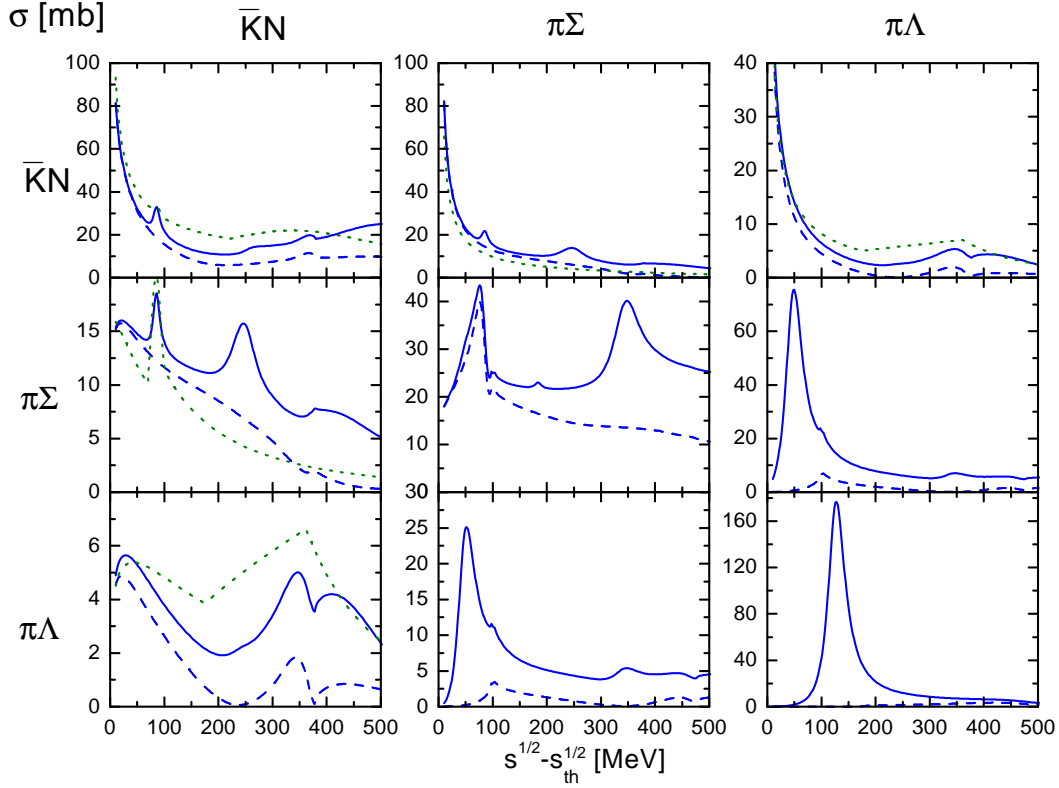}
\end{center}
\caption{Total cross sections
$\bar K N \to \bar K N$, $\bar K N \to \pi \Sigma $, $\bar K N \to \pi \Lambda $ etc relevant for subthreshold 
production of kaons in heavy-ion reactions. The solid and dashed lines give the results of the $\chi$-BS(3) approach
with and without p- and d-wave contributions respectively. The 
dotted lines correspond to the parameterizations given in \cite{Brown-Lee}.}
\label{fig:cross-pred}
\end{figure}

We close the result section by a presentation of total cross sections relevant for transport model 
calculation of heavy-ion reactions. We believe that the $\chi$-BS(3) approach is particularly well suited 
to determine some cross sections not directly accessible in scattering experiments. Typical examples would be 
the $\pi \Sigma \to \pi \Sigma , \pi \Lambda$ reactions. Here the quantitative realization of the chiral 
SU(3) flavor symmetry including its important symmetry breaking effects are an extremely useful constraint 
when deriving cross sections not accessible in the laboratory directly. It is common to consider 
isospin averaged cross sections \cite{Cugnon,Brown-Lee}
\begin{eqnarray}
\bar \sigma_{}(\sqrt{s}\,) = \frac{1}{N}\,\sum_{I} \,(2\,I+1)\,\sigma_{I}(\sqrt{s}\,) \,.
\label{}
\end{eqnarray}
The reaction dependent normalization factor is determined by $ N= \sum \,(2\,I+1)$ where the sum extends over 
isospin channels which contribute in a given reaction. In Fig. \ref{fig:cross-pred} we confront the 
cross sections of the channels $\bar K N, \pi \Sigma $ and $\pi \Lambda$. The results   
in the first row repeat to some extent the presentation of Fig. \ref{fig:totcross} only that here we confront 
the cross sections with typical parameterizations used in transport model calculations. The cross sections 
in the first column are determined by detailed balance from those of the first row. Uncertainties are present 
nevertheless, reflecting the large empirical uncertainties of the antikaon-nucleon cross sections close to 
threshold. The remaining four cross sections in Fig. \ref{fig:cross-pred} are true predictions of the 
$\chi$-BS(3) approach. Again we should emphasize that we trust our results quantitatively only for 
$\sqrt{s}< 1600$ MeV. It is remarkable that nevertheless our cross sections agree with the parameterizations 
in \cite{Brown-Lee} qualitatively up to much higher energies except in the 
$\bar K N  \leftrightarrow \pi \Sigma $ reactions where we overshoot those parameterizations somewhat. 
Besides some significant deviations of our results from \cite{Cugnon,Brown-Lee} at $\sqrt{s}-\sqrt{s_{\rm th}} <$ 200 MeV,
an energy range where we trust our results quantitatively, we find most interesting the sizeable 
cross section of about 30 mb for the $\pi \Sigma \to \pi \Sigma $ reaction. Note that here we include 
the isospin two contribution as part of the isospin averaging. As demonstrated by the dotted line in  
Fig. \ref{fig:cross-pred}, which represent the $\chi$-BS(3) approach with s-wave contributions only, 
the p- and d-wave amplitudes are of considerable importance for the $\pi \Sigma \to \pi \Sigma $ reaction.

\section{Summary and  outlook}

In this work we successfully used the relativistic chiral SU(3) Lagrangian to describe 
meson-baryon scattering. Within our $\chi$-BS(3) approach we established a unified description of pion-nucleon, 
kaon-nucleon and antikaon-nucleon scattering describing a large amount of empirical scattering data including the 
axial vector coupling constants for the baryon octet ground states. 
We derived the Bethe-Salpeter interaction kernel to chiral order $Q^3$ and then computed the scattering 
amplitudes by solving the Bethe-Salpeter equation. This leads to results consistent with covariance and unitarity. 
Moreover we consider the number of colors ($N_c$) in QCD as a large parameter 
performing a systematic $1/N_c$ expansion of the interaction kernel. This establishes a significant 
reduction of the number of parameters. Our analysis 
provides the first reliably estimates of previously poorly known s- and p-wave parameters. It is a highly 
non-trivial and novel result that the strength of all quasi-local 2-body interaction terms are 
consistent with the expectation from the large $N_c$ sum rules of QCD. Further intriguing results 
concern the meson-baryon coupling constants. The chiral $SU(3)$ flavor symmetry is found to be an extremely 
useful and accurate tool. Explicit symmetry breaking effects are quantitatively important but sufficiently 
small to permit an evaluation within chiral perturbation theory. We established two essential ingredients 
in a successful application of the chiral Lagrangian to the meson-baryon dynamics. 
First it is found that the explicit s- and u-channel decuplet contributions are indispensable for 
a good fit. Second, we find that it is crucial to employ the relativistic chiral Lagrangian. 
It gives rise to well defined kinematical structures in the quasi-local 4-point interaction terms which 
leads to a mixing of s-wave and p-wave parameters. Only in the heavy-baryon mass limit, not applied in this 
work, the parameters decouple into the s-wave and p-wave sector. In the course of developing our scheme we 
constructed a projector formalism which decouples in the Bethe-Salpeter equation covariant partial 
wave amplitudes and also suggested a minimal chiral subtraction scheme within dimensional regularization which 
complies  manifestly with the chiral counting rules. An important test of our analysis could be provided by 
new data on kaon-nucleon scattering from the DA$\Phi$NE facility \cite{DAPHNE}. In particular additional  
polarization data, possibly with a polarized hydrogen or deuteron target, would be extremely useful.

We performed a consistency check of our forward scattering amplitudes by confronting 
them with their dispersion-integral representations. Our analysis shows that  
the scattering amplitudes are compatible with their expected analytic structure. 
Moreover we demonstrate that the kaon-nucleon and antikaon-nucleon scattering amplitudes are approximatively 
crossing symmetric in the sense that the $K N$ and $\bar K N$ amplitudes match at subthreshold 
energies. Our results for the $\bar K N$ amplitudes have interesting consequences for kaon 
propagation in dense nuclear matter as probed in heavy ion collisions \cite{Senger}. 
According to the low-density theorem 
\cite{dover,njl-lutz} an attractive in-medium kaon spectral function 
probes the kaon-nucleon scattering amplitudes at subthreshold energies. The required
amplitudes are well established in our work. In particular we find sizeable 
contributions from p-waves not considered systematically so far 
\cite{Kolomeitsev,ml-sp,ramos-sp}.

We expect our scattering amplitudes to lead to an improved description of the spectral 
functions of antikaons in nuclear matter and pave the way for a microscopic 
description of kaonic atom data. The latter are known to be a rather sensitive test of 
the antikaon-nucleon dynamics \cite{Gal}.

\newpage 

\appendix
\noindent
{\bf\large  Appendices}

\section{Isospin in $SU(3)$ }

The $SU(3)$ meson and baryon octet fields $\Phi=\sum \Phi_i\,\lambda_i$ and the 
baryon octet field $B= \sum B_i\,\lambda_i/\sqrt{2}$ with the Gell-Mann matrices
$\lambda_i$ normalized by $\tr \lambda_i \,\lambda_j = 2\,\delta_{ij}$
are decomposed  into their isospin symmetric components
\begin{eqnarray}
\Phi &=& \tau \cdot  \pi
+ \alpha^\dagger \cdot  K +  K^\dagger \cdot \alpha  +
\eta \,\lambda_8 \;,
\nonumber\\
B &=&  {\textstyle{1\over\sqrt{2}}}\,
\left( \tau \cdot \Sigma +
\alpha^\dagger \cdot  N +\Xi^t\,i\,\sigma_2 \cdot \alpha +\lambda_8 \,\Lambda \right) \, ,
\nonumber\\
\alpha^\dagger &=&
 {\textstyle{1\over\sqrt{2}}}\left( \lambda_4+i\,\lambda_5 ,
\lambda_6+i\,\lambda_7 \right)
\;,\;\;\;\tau = (\lambda_1,\lambda_2,\lambda_3)\;,
\label{field-decomp-app}
\end{eqnarray}
with the isospin singlet fields $\eta, \Lambda$, the isospin doublet fields 
$K =(K^+,K^0)^t $, $N=(p,n)^t$, $\Xi = (\Xi^0,\Xi^-)^t$ and the isospin triplet 
fields $\vec \pi = (\pi^{(1)},\pi^{(2)},\pi^{(3)})$, 
$\vec \Sigma =(\Sigma^{(1)},\Sigma^{(2)},\Sigma^{(3)})$.
Similarly we derive the isospin decomposition of  $(\bar \Delta_\mu \cdot \Phi)$,
$(\Phi \cdot \Delta_\mu )$ and $(\bar \Delta_\mu \cdot \Delta_\nu)$ as defined in 
(\ref{dec-prod}). The latter objects are expressed in terms of the isospin singlet field
$\Omega^-$, the isospin doublet fields 
$\Xi_\mu = (\Xi^0_\mu, \Xi^-_\mu)^t$, the isospin triplet field
$\vec \Sigma_\mu =(\Sigma^{(1)}_\mu,\Sigma^{(2)}_\mu,\Sigma^{(3)}_\mu)$ 
and the isospin 3/2 field 
$\underline \Delta_\mu = (\Delta_\mu^{++},\Delta_\mu^+,\Delta_\mu^0,\Delta_\mu^-)^t$. 
We find
\begin{eqnarray}
\Big(\bar \Delta_\mu \cdot \Phi \Big)_{b}^{a}
&=&\Big(\Big(\bar
{\underline  \Delta}_\mu \, S \, K \Big)\,\cdot \tau
-\Big(\bar {\underline  \Delta}_\mu
\, S \, \alpha  \Big)\,\cdot  \pi
-\frac{1}{\sqrt{6}}\,K^\dagger
\,\Big(\bar \Sigma_\mu \cdot \sigma \Big) \,\alpha
\nonumber\\
&+&\frac{1}{\sqrt{6}}\,\alpha^\dagger\, \Big(\bar \Sigma_\mu \cdot
\sigma \Big)\,K
+\frac{1}{\sqrt{2}}\,\Big(\bar \Sigma_\mu \cdot \pi \Big)\,\lambda_8
-\frac{1}{\sqrt{2}}\,\Big(\bar
\Sigma_\mu \cdot \tau \Big)\,\eta
\nonumber\\
&-&\frac{i}{\sqrt{6}}\, \Big( \bar \Sigma_\mu \times \pi \Big)\,\cdot \tau
+\frac{1}{\sqrt{2}}\,\Big( K^\dagger\,i\,\sigma_2\,\bar
\Xi_\mu  ^t\Big) \,\lambda_8
- \frac{1}{\sqrt{6}}\,\Big(K^\dagger \,\sigma\,i\,\sigma_2\,\bar
\Xi^t_\mu\Big)\,\cdot \tau
\nonumber\\
&+&\frac{1}{\sqrt{6}}\,\alpha^\dagger
\,\Big(\sigma\cdot \pi \Big)\,i\,\sigma_2\,\bar \Xi^t_\mu
-\frac{1}{\sqrt{2}}\,\Big(\alpha^\dagger\,i\,\sigma_2\,\bar
\Xi^t_\mu
\Big)\,\eta
\nonumber\\
&-&\bar \Omega^-_\mu\,\Big(\alpha^\dagger \,i\,\sigma_2\,\Big(K^\dagger \Big)^t
\Big)
\Big)_{b}^{a} \;,
\nonumber\\
\Big(\Phi \cdot \Delta_\mu \Big)_{b}^a
&=&\Big(\Big( K^\dagger \, S^\dagger \, {\underline  \Delta}_\mu  \Big)\,\cdot
\tau
-\Big(\alpha^\dagger
\, S^\dagger \,{\underline  \Delta}_\mu   \Big)\, \cdot \pi
-\frac{1}{\sqrt{6}}\,\alpha^\dagger
\,\Big( \Sigma_\mu  \cdot \sigma \Big) \,K
\nonumber\\
&+&\frac{1}{\sqrt{6}}\, K^\dagger\,\Big(\Sigma \cdot \sigma  \Big)
\,\alpha+\frac{1}{\sqrt{2}}\,
\Big( \Sigma_\mu  \cdot \pi \Big)\,\lambda_8
-\frac{1}{\sqrt{2}}\,\Big(\Sigma \cdot \tau \Big)\,\eta
\nonumber\\
&+&\frac{i}{\sqrt{6}}\, \Big( \Sigma_\mu  \times \pi \Big)\,\cdot \tau
-\frac{1}{\sqrt{2}}\Big( \Xi^t_\mu \,i\,\sigma_2\,K
\Big) \, \lambda_8
+\frac{1}{\sqrt{6}}\,\Big(\Xi^t_\mu \,i\,\sigma_2\,\sigma \,K\Big)\,\cdot \tau
\nonumber\\
&-& \frac{1}{\sqrt{6}}\,\Xi^t\,\,i\,\sigma_2
\,\Big(\sigma\cdot \pi \Big)\,\alpha
+\frac{1}{\sqrt{2}}\,\Big(\Xi^t_\mu \,i\,\sigma_2\,\alpha
\Big)\,\eta
\nonumber\\
&+&\Omega^-_\mu \,\Big(K^t\,i\,\sigma_2\, \alpha \Big)
\Big)_{b}^a
\label{dec-ap-1}
\end{eqnarray}
and
\begin{eqnarray}
\Big(\bar \Delta_\mu \cdot \Delta_\nu \Big)_{b}^a
&=&
\Big(\frac{1}{2}\,\Big( \sum_j\,\bar {\underline  \Delta}_\mu \,S_j\,\sigma\, S^\dagger_j\,{\underline  \Delta}_\nu \Big)\,\cdot \tau
+\frac{1}{6}\,\bar {\underline  \Delta}_\mu \cdot {\underline  \Delta}_\nu \,\Big(
2+\sqrt{3}\,\lambda_8
\Big)
\nonumber\\
&+&\frac{1}{3}\,\bar
\Sigma_\mu \cdot \Sigma_\nu
+\frac{i}{3}\, \tau \cdot \Big( \bar \Sigma_\mu \times \Sigma_\nu \Big)
\nonumber\\
&+& \frac{1}{6}\,\bar \Xi_\mu \,\Xi_\nu \,\Big( 2-\sqrt{3}\,\lambda_8\Big)
+\frac{1}{6}\,\Big(\bar \Xi_\mu \,\sigma \, \Xi_\nu \Big)\,\cdot \tau
\nonumber\\
&+&
\frac{1}{3}\,\bar \Omega^-_\mu\,\Omega^-_\nu \,\Big(1-\sqrt{3}\,\lambda_8
\Big)
\nonumber\\
&+&\frac{1}{\sqrt{6}}\,\bar {\underline  \Delta}_\mu \,\Big(S \cdot \Sigma_\nu
\Big)\,\alpha + \frac{1}{\sqrt{6}}\,
\alpha^\dagger \,\Big(\bar \Sigma_\mu \cdot S^\dagger \Big)\,{\underline  \Delta}_\nu
\nonumber\\
&-&\frac{1}{3}\,
\bar \Sigma_\mu  \cdot \, \Big( \Xi^t_\nu \,i\,\sigma_2\,\sigma \,\alpha  \Big)
+\frac{1}{3}\, \Big( \alpha^\dagger \,\sigma \,i\,\sigma_2\,\bar
\Xi^t_\mu\Big) \,\cdot \Sigma_\nu
\nonumber\\
&+& \frac{1}{\sqrt{6}}\,\bar \Omega^-_\mu\,\Big(\alpha^\dagger \,\Xi_\nu
\Big) +\frac{1}{\sqrt{6}}\,
\Big(\bar \Xi_\mu \, \alpha \Big) \, \Omega^-_\nu \Big)_{b}^a \;.
\label{dec-ap-2}
\end{eqnarray}
The isospin transition matrices $S_i$ are normalized by  
$S^\dagger_i\,S_j = \delta_{ij}-\sigma_i\,\sigma_j/3$. Note that the isospin Pauli 
matrices $\sigma_i$ act exclusively in the space of isospin doublet fields 
$K,N,\Xi,\Xi_\mu$ and the matrix valued isospin doublet $\alpha$. 
Expressions analogous to (\ref{dec-ap-1}) 
hold for $(\bar \Delta_\mu \cdot B)$ and $(\bar B \cdot \Delta_\mu )$. 

The isospin reduction of the $SU(3)$ symmetric interaction terms is
most conveniently derived applying the set of identities
\begin{eqnarray}
\tau \cdot \alpha^\dagger &=&   \alpha^\dagger  \cdot \sigma
\;,\;\;\;\;\;\alpha^\dagger \cdot  \tau =0\;,\;\;\;\;\;
\alpha \cdot \tau = \sigma \cdot \alpha
\;,\;\;\;\;\;  \tau \cdot \alpha =0 \;,
\nonumber\\
\tau \, \lambda_8 &=& \lambda_8\, \tau
\;,\;\;\;\;\;
\tr \big(\tau_i\,\tau_j \big)= 2\,\delta_{ij}
\;,\;\;\;\;
\tr \big(\alpha_i\,\alpha^\dagger_j \big)= 2\,\delta_{ij} \,,
\label{isospin-reduction-1}
\end{eqnarray}
where the $SU(2)$ Pauli matrices $ \vec \sigma $ act exclusively on
the isospin doublet fields. For example
$\big(\vec \sigma \cdot \alpha \big)_a=\sum_b\,\vec \sigma_{ab}\,\alpha_b$.
The algebra (\ref{isospin-reduction-1})
is completed with
\begin{eqnarray}
\Big[ \alpha  , \alpha^\dagger \Big]_- &=&-\tau \cdot \sigma
-\sqrt{3}\,\lambda_8
\;,\;\;\;\;\;\alpha \cdot \alpha=0\;,
\nonumber\\
\Big[ \alpha \,,\alpha^\dagger \Big]_+ &=& {\textstyle{4\over 3}}\,1
-{\textstyle{1\over \sqrt{3}}}\, \lambda_8 + \tau \cdot  \sigma
\;,\;\;\;\;\;
\alpha^\dagger \cdot \alpha^\dagger =0 \;,
\nonumber\\
\Big[ \tau_i\,, \tau_j \Big]_- &=&2\,i\,\epsilon_{ijk}\,\tau_k
\;,\;\;\;
\Big[ \tau_i\,, \tau_j \Big]_+ =
\left( {\textstyle{2\over \sqrt{3}}}\,\lambda_8
+{\textstyle{4\over 3}}\,1 \right) \delta_{ij}\;,
\nonumber\\
\Big[ \lambda_8\,, \alpha^\dagger \Big]_- &=&
\textstyle{\sqrt{3}}\,\alpha^\dagger
\;,\;\;\;\;
\Big[ \lambda_8 \, , \alpha^\dagger \Big]_+
=-{\textstyle{1\over \sqrt{3}}}\,\alpha^\dagger
\;,\;\;\;\;
\Big[ \lambda_8\, , \lambda_8 \Big]_+ = {\textstyle{4\over 3}}\,1
-{\textstyle{2\over \sqrt{3}}}\,\lambda_8 \;,
\nonumber\\
\Big[ \lambda_8\,, \alpha \,\Big]_- &=&-\textstyle{\sqrt{3}}\,\alpha
\;,\;\;\;
\Big[ \lambda_8\,, \alpha \,\Big]_+ =-{\textstyle{1\over \sqrt{3}}}\,\alpha
\;,\;\;\;
\Big[\lambda_8 \,,  \tau \Big]_+ ={\textstyle{2\over \sqrt{3}}}\,\tau \;
\label{isospin-reduction-2}
\end{eqnarray}
where $[A,B]_\pm= A\,B\pm B\,A$. By means of 
(\ref{isospin-reduction-1}, \ref{isospin-reduction-2}) it is straightforward to 
derive the isospin structure of the chiral interaction terms. Note a typical intermediate 
result 
\begin{eqnarray}
\Big[\Phi ,B\Big]_+
&=&\sqrt{\frac{2}{3}}\, \Big(\lambda_8 + \frac{2}{\sqrt{3}}\Big)
\,\Big( \pi \cdot  \Sigma \Big)
+\frac{1}{\sqrt{2}}\, \pi \cdot \Big(  \alpha^\dagger \, \sigma  \,N \Big)
+\sqrt{\frac{2}{3}}\, \tau \cdot \Big(  \pi \,\Lambda \Big)
\nonumber\\
&-&\frac{1}{\sqrt{6}}\, \Big(\lambda_8 - \frac{4}{\sqrt{3}}\Big)\,
\Big( K^\dagger \,N \Big) +\frac{1}{\sqrt{2}}\,  \tau\cdot \Big( K^\dagger \,
\sigma \,N \Big)
-\frac{1}{\sqrt{6}}\,\Big( K^\dagger\,\alpha  \Big)\,\Lambda
\nonumber\\
&+&\frac{1}{\sqrt{2}}\,\Big( K^\dagger  \sigma \,\alpha \Big) \cdot \Sigma
-\frac{1}{\sqrt{6}}\,\Big(\alpha^\dagger\,K \Big) \,\Lambda
+\frac{1}{\sqrt{2}}\,\Big( \alpha^\dagger \, \sigma \,K \Big) \cdot \Sigma
\nonumber\\
&-&\sqrt{\frac{2}{3}}\, \Big(\lambda_8 - \frac{2}{\sqrt{3}}\Big)\,
\,\Big( \eta \,\Lambda \Big)
-\frac{1}{\sqrt{6}}\,\eta\,\Big( \alpha^\dagger\,N \Big)
+\sqrt{\frac{2}{3}}\, \tau \cdot  \Big( \eta \,\Sigma \Big)
\nonumber\\
&+&\frac{1}{\sqrt{2}}\, \pi  \cdot \Big( \Xi^t\,i\,\sigma_2\,\sigma 
\,\alpha \Big)
-\frac{1}{\sqrt{6}}\, \eta \,\Big(\Xi^t\,i\,\sigma_2\,\alpha\Big)
\nonumber\\
&+& \frac{1}{\sqrt{2}}\, \tau \cdot \Big(\Xi^t\,i\,\sigma_2\,
\sigma \,K \Big)
-\frac{1}{\sqrt{6}}\, \Big(\lambda_8 - \frac{4}{\sqrt{3}}\Big)\, \Big(\Xi^t\,i\,\sigma_2\,
K \Big) \;.
\label{ex-app}
\end{eqnarray}

\section{Local interaction terms of chiral order $Q^3$}

For heavy-baryon chiral $SU(3)$ perturbation theory 102 terms of chiral order $Q^3$ 
are displayed in \cite{q3-meissner}. We find that only 10 chirally symmetric terms are 
relevant for elastic meson-baryon scattering. Following the constructing rules of 
Krause \cite{Krause} one writes down 19 interaction terms for the relativistic chiral 
Lagrangian where terms which are obviously redundant by means of the equation of motion 
or $SU(3)$ trace identities \cite{Fearing} are 
not displayed. The Bethe-Salpeter interaction kernel (see (\ref{BS-eq})) 
receives the following contributions
\begin{eqnarray}
&&K^{(I)}_{[8][8]}(\bar k, k; w)= 
\frac{1}{8\,f^2}\,\Big( \big( \bar p \cdot \bar q \big)\big( p \cdot q\big)
+\big( \bar p \cdot q \big)\big( p \cdot \bar q\big)\Big)\,C^{(I)}[h^{(1)}]
\\
&&\qquad +\frac{1}{16\,f^2}\,\Big( \qslash\,\big( p-\bar p\big) \cdot \bar q
-\barqslash \,\big(p- \bar p\big) \cdot  q\Big)\,\bar C^{(I)}[h^{(2)}]
\nonumber\\
&&\qquad +\frac{1}{8\,f^2}\,\big( \barqslash +\qslash \big)\,(\bar q \cdot q) \,
\,\bar C^{(I)}[h^{(3)}]
\nonumber\\
&&\qquad +\frac{1}{8\,f^2}\,i\,\gamma_5\,\gamma^\mu \, (p+\bar p)^\nu\,\epsilon_{\mu \nu \alpha \beta}\,
\bar q^\alpha\,q^\beta \,\bar C^{(I)}[h^{(4)}]
\nonumber\\
&&\qquad +\frac{1}{16\,f^2}\,i\,\gamma_5\,\gamma^\mu\,
\Big((\bar p \cdot (\bar q-q) )\, p^\mu +(p \cdot (\bar q-q) )\,\bar p^\mu \Big)\,
\epsilon_{\mu \nu \alpha \beta}\,\bar q^\alpha\,q^\beta \,\bar C^{(I)}[h^{(5)}]
\nonumber\\
&&\qquad +\frac{1}{16\,f^2}\,i\,\gamma_5\,\gamma^\mu\,
\Big((\bar p \cdot (\bar q+q) )\, p^\mu -(p \cdot (\bar q+q) )\,\bar p^\mu \Big)\,
\epsilon_{\mu \nu \alpha \beta}\,\bar q^\alpha\,q^\beta \,\bar C^{(I)}[h^{(6)}]
\nonumber \;,
\label{k-q3}
\end{eqnarray}
where the interaction terms are already presented in momentum space for notational 
convenience. Their SU(3) structure is expressed in terms of the matrices $C_{0,1,D,F}$ and 
$\bar C_{1,D,F}$ introduced in (\ref{def-local-1}). In (\ref{k-q3}) we applied the convenient notation:
\begin{eqnarray}
\bar C^{(I)}[ h^{(i)}]&=&h_{1}^{(i)}\,\bar C_1^{(I)}
+h_{D}^{(i)}\,C_D^{(I)}+h_{F}^{(i)}\,C_F^{(I)} \,,
\nonumber\\
C^{(I)}[h^{(i)}]&=&h_{0}^{(i)}\,C_0^{(I)}+h_{1}^{(i)}\,C_1^{(I)}
+h_{D}^{(i)}\,C_D^{(I)}+h_{F}^{(i)}\,C_F^{(I)} \,.
\label{}
\end{eqnarray}
Explicit evaluation of the terms in (\ref{k-q3}) demonstrates that in fact only 
10 terms contribute to the s and p- wave interaction kernels to chiral order $Q^3$.
Obviously there are no quasi-local counter terms contributing to higher 
partial waves to this order. The terms of chiral order $Q^2$ are
\begin{eqnarray}
V^{(I,+)}_{[8][8],2}(s;0)&=&  \frac{s}{4\,f^2}\,
\Big( \sqrt{s}-M^{(I)}\Big)\,C^{(I)}[h^{(1)}]\,\Big( \sqrt{s}-M^{(I)}\Big)\,,
\nonumber\\
V^{(I,-)}_{[8][8],2}(s;0)&=&  \frac{1}{6\,f^2}\,
M^{(I)}\,\Big[ \,\bar C^{(I)}[h^{(4)}]\, , M^{(I)} \Big]_+\,M^{(I)}\,,
\nonumber\\
&-& \frac{1}{8\,f^2}\,
\Big( \sqrt{s}+M^{(I)}\Big)\,\Big[\, \bar C^{(I)}[h^{(4)}] \,, M^{(I)} \Big]_+ \,
\Big( \sqrt{s}+M^{(I)}\Big)\,,
\nonumber\\
V^{(I,+)}_{[8][8],2}(s;1)&=&  \frac{1}{24\,f^2}\,
\Big[ \,\bar C^{(I)}[h^{(4)}] \,, M^{(I)} \Big]_+ \, .
\label{}
\end{eqnarray}
We observe that the $h^{(1)}_{0,1,D,F}$ parameters can be absorbed into the $g^{(V)}_{0,1,D,F}$ to  
order $Q^2$ by the replacement $ g^{(V)} \to g^{(V)}- M \, h^{(1)}$. Similarly 
the replacement $g^{(T)} \to g^{(T)} + M \,h^{(4)} $ cancels the dependence on 
$h^{(4)}$ at order $Q^2$ (see (\ref{local-v})). This mechanism illustrates the 
necessary regrouping of interaction terms required for the relativistic chiral 
Lagrangian as discussed in \cite{nn-lutz}. We turn to the quasi-local interaction terms of 
chiral order $Q^3$:
\begin{eqnarray}
\nonumber\\
V^{(I,+)}_{[8][8],3}(s;0)&=& \frac{1}{8\,f^2}\,
\,\Bigg(
\Big(\phi^{(I)}+m_{\Phi(I)}^{2}\Big)\, \bar C^{(I)}[h^{(3)}]\,\Big( \sqrt{s}-M^{(I)}\Big)
\nonumber\\
&& \qquad + 
\Big( \sqrt{s}-M^{(I)}\Big)\,\bar C^{(I)}[h^{(3)}]\,
\Big(\phi^{(I)}+m_{\Phi(I)}^{2}\Big)\Bigg) 
\nonumber\\
-\frac{1}{4\,f^2} \!\!\!&&\!\!\!
\Bigg( \frac{\phi^{(I)}}{2 \,\sqrt{s}}\,\Big( C^{(I)}[ g^{(S)}]
+{\textstyle{1\over 2}} \,\sqrt{s}\,\bar C^{(I)}[ h^{(2)}+2\,h^{(4)}] \Big)
\,\Big( \sqrt{s}-M^{(I)}\Big)
\nonumber\\
 + \!\!\!&&\!\!\!\Big( \sqrt{s}-M^{(I)}\Big)\,\Big( C^{(I)}[ g^{(S)}]
+{\textstyle{1\over 2}} \,\sqrt{s}\,\bar C^{(I)}[ h^{(2)}+2\,h^{(4)}] \Big)
\,\frac{\phi^{(I)}}{2 \,\sqrt{s}} \Bigg)
\nonumber\\
-\frac{1}{4\,f^2} \!\!\!&&\!\!\!
\Bigg( \frac{\phi^{(I)}}{2 \,\sqrt{s}}\,\Big( \bar C^{(I)}[ g^{(T)}]
-{\textstyle{1\over 2}} \,\Big[\, \bar C^{(I)}[ h^{(4)}]\,, M^{(I)} \Big]_+ \Big)
\,\Big( \sqrt{s}-M^{(I)}\Big)
\nonumber\\
+ \!\!\!&&\!\!\! \Big( \sqrt{s}-M^{(I)}\Big)\,\Big( \bar C^{(I)}[ g^{(T)}]
-{\textstyle{1\over 2}} \,\Big[\, \bar C^{(I)}[ h^{(4)}]\,, M^{(I)} \Big]_+  \Big)
\,\frac{\phi^{(I)}}{2 \,\sqrt{s}} \Bigg)
\nonumber\\
&-&\frac{1}{8\,f^2}\,
\,\Bigg(
\Big(\phi^{(I)}+m_{\Phi(I)}^{2}\Big)\, C^{(I)}[g^{(V)}\!+\!2\,\sqrt{s}\,h^{(1)}]\,\Big( \sqrt{s}-M^{(I)}\Big)
\nonumber\\
&& \qquad + 
\Big( \sqrt{s}-M^{(I)}\Big)\,C^{(I)}[g^{(V)}\!+\!2\,\sqrt{s}\,h^{(1)}]\,
\Big(\phi^{(I)}+m_{\Phi(I)}^{2}\Big)\Bigg) \,,
\nonumber\\
V^{(I,-)}_{[8][8],3}(s;0)&=&
\frac{1}{8\,f^2} \,\sqrt{s}\,\Bigg(
\Big( \sqrt{s}+M^{(I)}\Big)\,C^{(I)}[g^{(V)}]\,\Big( \sqrt{s}-M^{(I)}\Big)
\nonumber\\
&& \qquad \quad +\Big( \sqrt{s}-M^{(I)}\Big)\,C^{(I)}[g^{(V)}]\,\Big( \sqrt{s}+M^{(I)}\Big) \Bigg)
\nonumber\\
&+&\frac{1}{12\,f^2}\,M^{(I)}\,
\Big[\,C^{(I)}[g^{(V)}+2\,\sqrt{s}\,h^{(1)}]\,,\sqrt{s}-M^{(I)}\Big]_+\,M^{(I)}
\nonumber\\
&+&\frac{1}{12\,f^2}\,M^{(I)}\,
\Big[\,\bar C^{(I)}[h^{(2)}+2\,h^{(4)}-2\,h^{(3)}]\,,\sqrt{s}-M^{(I)}\Big]_+\,M^{(I)}
\,,
\nonumber\\
V^{(I,+)}_{[8][8],3}(s;1)&=& \frac{1}{48\,f^2}\,
\Big[\,C^{(I)}[g^{(V)}+2\,\sqrt{s}\,h^{(1)}]\,,\sqrt{s}-M^{(I)}\Big]_+ 
\nonumber\\
&+&\frac{1}{48\,f^2}\,
\Big[\,\bar C^{(I)}[h^{(2)}+2\,h^{(4)}-2\,h^{(3)}]\,,\sqrt{s}-M^{(I)}\Big]_+\,.
\label{local-v-q3}
\end{eqnarray}
We observe that neither the $h^{(5)}$ nor the $h^{(6)}$ coupling constants enter  
the interaction kernel to chiral order $Q^3$. Furthermore the structure $h^{(4)}$ is 
redundant, because it disappears with the replacements to $g^{(T)} \to g^{(T)} + M \,h^{(4)} $  
and $h^{(2)} \to h^{(2)}  -2\,h^{(4)} $.  Thus at chiral order $Q^3$ we find 
10 relevant chirally symmetric parameters $h^{(1)}$, $h^{(2)}$ and $h^{(3)}$.

\section{Projector algebra}

We establish the loop-orthogonality of the projectors 
${Y}^{(\pm )}_n(\bar q,q;w)$ introduced in (\ref{cov-proj}).
In order to facilitate our derivations we rewrite the projectors in terms 
of the convenient building objects $P_\pm$ and $V_\mu$ as 
\begin{eqnarray}
&&{Y}^{(\pm )}_n(\bar q,q;w)=\pm \,P_\pm\,\bar Y_{n+1}(\bar q,q;w)
\pm 3\,(\bar q \cdot V)\,P_\mp\, (V \cdot q)\,\bar Y_{n}(\bar q,q;w)\;,
\nonumber\\
&&\bar Y_{n}(\bar q,q;w)= \sum_{k=0}^{[(n-1)/2]}\,\frac{(-)^k\,(2\,n-2\,k) !}{2^n\,k !\,(n-k)
!\,(n-2\,k -1) !}\,Y_{\bar q \bar q}^{k}\,Y_{\bar q q}^{n-2\,k-1}\,Y_{q q}^{k}\;,
\nonumber\\
&&Y_{a b}=\frac{(w\cdot a )\,(b\cdot w)}{w^2} -a \cdot b \,, \quad 
\label{cov-proj:a}
\end{eqnarray}
where 
\begin{eqnarray}
&& P_\pm = \frac{1}{2} \left(1\pm \frac{\wslash}{\sqrt{w^2}} \right)\,,\quad 
V_\mu = \frac{1}{\sqrt{3}}\left(\gamma_\mu - \frac{\wslash }{w^2}\,w_\mu \right)\,,
\quad P_\pm \,P_\pm =P_\pm \,,
\nonumber\\
&& P_\pm \,P_\mp =0\,,\quad P_\pm \,\lslash  = \lslash \,P_\mp 
\pm (l \cdot w)/\sqrt{w^2}\,,\quad 
P_\pm \,V_\mu = V_\mu \,P_\mp \,.
\label{cov-proj:b}
\end{eqnarray}
In this appendix we will derive the identities:
\begin{eqnarray}
&&\Im \ll n \,\pm | m \,\pm \gg  (\bar q,q;w) = \delta_{nm}\,Y^{(\pm)}_n(\bar q,q;w)
\,\Im J_{ab}^{(\pm)}(w;n) \;,
\nonumber\\
&&\Im \ll n \,\pm | m \,\mp \gg  (\bar q,q;w) =0 \,,
\label{des-result}
\end{eqnarray}
with the convenient notation 
\begin{eqnarray}
&&\ll n \,\pm | m \,\pm \gg  (\bar q,q;w)= 
\int \frac{d^4l}{(2\pi)^4}\,
Y^{(\pm)}_n(\bar q,l;w)\,G(l;w)\,Y^{(\pm)}_m(l,q;w)\,,
\nonumber\\
&&G(l;w)=\frac{-i}{(w-l)^2-m_a^2+i\,\epsilon}\,\frac{\lslash +m_b}{l^2-m_b^2+i\,\epsilon}\,,
\nonumber\\
&& \Im\,J^{(\pm)}_{ab}(w;n) = \frac{p^{2\,n+1}_{ab}}{8\,\pi\,\sqrt{w^2}}
\left(  \frac{\sqrt{w^2}}{2}+ \frac{m_b^2-m_a^2}{2\,\sqrt{w^2}}\pm m_b \right) \,, 
\nonumber\\
&& \sqrt{w^2}= \sqrt{m_a^2+p^2}+\sqrt{m_b^2+p^2}\,.
\label{im-loop}
\end{eqnarray}
The real parts of the loop functions $J_{ab}^{(\pm )}(w;n)$ are readily reconstructed by 
means of a dispersion integrals in terms of their imaginary parts.
Since the loop functions are highly divergent they require a finite number of subtractions 
which are to be specified 
by the renormalization scheme. These terms are necessarily real and represent typically 
power divergent tadpole terms (see (\ref{tadpole:a})).
According to our renormalization condition (\ref{ren-v}) such terms must be moved into 
the effective interaction kernel.

In order to arrive at the desired result (\ref{des-result}) we introduce further notation
streamlining our derivation. For any covariant function $F(\bar q,l,q;w)$ we write
\begin{eqnarray}
&&\langle F \rangle_{n,m} (\bar q,q;w)=
\int \frac{d^4l}{(2\pi)^4}\,F(\bar q,l,q;w)\,
\bar Y_n(\bar q,l;w)\,\bar G(l;w)\,\bar Y_m(l,q;w)\,,
\nonumber\\
&& \bar G(l;w)= \frac{1}{(w-l)^2-m_a^2+i\,\epsilon}\,
\frac{-i}{l^2-m_b^2+i\,\epsilon}\,.
\label{f-def}
\end{eqnarray}
The result (\ref{des-result}) is now derived in two steps. First the expressions in 
(\ref{des-result}) are simplified by standard Dirac algebra methods. In 
the convenient notation of (\ref{f-def}) we find
\begin{eqnarray}
\ll n \,\pm | m \,\pm \gg  &=& 
\langle m_N \pm (l \cdot \hat w)\rangle_{n+1,m+1}\,P_\pm
\nonumber\\
&+&3\, \langle \big(m_N \pm (l \cdot \hat w)\big) \,
(l \cdot V)\rangle_{n+1,m}\,P_\mp\,(V \cdot q)
\nonumber\\
&+& 3\,(\bar q \cdot V )\,P_\mp \,\langle (V \cdot l)\,\big(m_N \pm (l \cdot \hat w)\big) 
\rangle_{n,m+1}
\nonumber\\
&-& 3\,(\bar q \cdot V )\,P_\mp \,\langle Y_{ll}\,\big(m_N \pm (l \cdot \hat w)\big) 
\rangle_{n,m}\,(V \cdot q)\,,
\nonumber\\
\ll n \,\pm | m \,\mp \gg  &=& -P_\pm
\langle \lslash \rangle_{n+1,m+1}\,P_\mp
+ 3\,(\bar q \cdot V )\,P_\mp \,\langle \lslash\,Y_{ll} \rangle_{n,m}\,P_\pm\,(V \cdot q)
\nonumber\\
&+& P_\pm\,\langle \barqslash \,Y_{ll} \rangle_{n,m+1}\,P_\mp
+P_\pm\, \langle \qslash \, Y_{ll} \rangle_{n+1,m}\,P_\mp \,,
\label{first-step}
\end{eqnarray}
where $\hat w_\mu = w_\mu/\sqrt{w^2}$. 
Next we observe that the terms $2\,(l \cdot w)=l^2-m_b^2-(w-l)^2+m_a^2+m_b^2-m_a^2+w^2$ 
and $Y_{ll}$ in (\ref{first-step}) can be replaced by
\begin{eqnarray}
l \cdot \hat w \to \frac{\sqrt{w^2}}{2}+ \frac{m_b^2-m_a^2}{2\,\sqrt{w^2}} \,, \quad 
Y_{ll} \to p^2 \,,
\label{rep}
\end{eqnarray}
if tadpole contributions are neglected. The first replacement rule in (\ref{rep}) 
generates the typical structure occurring in (\ref{im-loop}). 
The final step consists in evaluating the remaining integrals. 
Consider for example the identities:
\begin{eqnarray}
&&\langle 1 \rangle_{n,m} (\bar q,q;w) = \langle 1 \rangle_{m,n} (q,\bar q;w)=
Y_{\bar q \bar q}^{(n-m)/2}\,\bar Y_{m}(\bar q,q;w)\,J_{n+m-1}(w)\,,
\nonumber\\
&&\langle 1 \rangle_{n+1,m} (\bar q,q;w) = \langle 1 \rangle_{m,n+1} (q,\bar q;w)= 0\,, 
\quad 
\Im \,J_n(w) = \frac{p^{n}}{8\pi \,\sqrt{w^2}}\,,
\label{ex:1}
\end{eqnarray}
which hold modulo some subtraction polynomial for $n\geq m$ and both $n,m$ either even or odd. The 
result (\ref{ex:1}) is readily confirmed in the particular frame where $\vec w =0 $
\begin{equation}
\bar Y_{n}(l,q;w) = \big(|\vec l\, |\,|\vec q\,|\big)^{n-1}\,
P_n'\big(\cos (\vec l\,,\vec q\,)\big)\,,
\label{intp}
\end{equation}
by applying the Cutkosky cutting rule in conjunction with standard properties of the Legendre 
polynomials\footnote{Note the convenient identities: 
$P_{n+1}'(x)=\sum_{l={\rm even}}^n\,(2\,l+1)\,P_l(x)$ for n even and 
$P_{n+1}'(x)=\sum_{l={\rm odd}}^n\,(2\,l+1)\,P_l(x)$ for n odd.}. Here it is crucial 
to observe that the object $\bar Y_{n}(q,l;w)=\bar Y_{n}(l,q;w)$ does not exhibit
any singularity in $l_\mu$. For example a square root term $\sqrt{Y_{\bar q l}}$ in 
$\bar Y(\bar q, l;w)$ would invalidate our derivation. We point out that this observation 
leads to the unique interpretation of $P_n'$ in terms of $Y_{qq}$, $Y_{\bar q \bar q}$ 
and $Y_{\bar q q}$ (see (\ref{intp})) and thereby defines the unambiguous 
form of our projectors (\ref{cov-proj:a}). Similarly one derives the 
identities:
\begin{eqnarray}
&&\langle Y_{\bar q l} \rangle_{n+1,m} (\bar q,q;w) =
\langle Y_{\bar q l } \rangle_{m,n+1} (q, \bar q;w)
\nonumber\\
&& \qquad \qquad \qquad \quad \quad = Y_{\bar q \bar q}\,
Y_{\bar q \bar q}^{(n-m)/2}\,\bar Y_{m}(\bar q,q;w)\,J_{n+m+1}(w)\,,
\nonumber\\
&&\langle Y_{l q} \rangle_{n+1,m} (\bar q,q;w) =
\langle Y_{l q } \rangle_{m,n+1} (q, \bar q;w)
\nonumber\\
&& \qquad \qquad \qquad \quad \quad = Y_{\bar q q}\,
Y_{\bar q \bar q}^{(n-m)/2}\,\bar Y_{m}(\bar q,q;w)\,J_{n+m+1}(w)\,,
\nonumber\\
&&\langle Y_{\bar q l} \rangle_{n,m} (\bar q,q;w) =
\langle Y_{\bar q l } \rangle_{m,n} (q, \bar q;w) =0 \,, 
\nonumber\\
&&\langle Y_{l q} \rangle_{n,m} (\bar q,q;w) =
\langle Y_{l q } \rangle_{m,n} (q, \bar q;w) =0\,,
\label{ex:2}
\end{eqnarray}
which again hold modulo some subtraction polynomial for $n\geq m$ and both $n,m$ either 
even or odd. Our proof of (\ref{des-result}) is completed with the convenient identities:
\begin{eqnarray}
&&\langle l_\mu \rangle_{n+1,m} = 
\hat w_\mu\,\langle Y_{ll}\,(l-\bar q )\cdot \hat w \rangle_{n,m} 
+\bar q_\mu \,\langle Y_{ll} \rangle_{n,m} \quad 
{\rm if\;n\geq m } \;,
\nonumber\\
&&Y_{qq}\,\langle l_\mu \rangle_{n+1,m} = 
\hat w_\mu\,\langle (l-q )\cdot \hat w \rangle_{n+1,m+1} 
+q_\mu \,\langle 1 \rangle_{n+1,m+1} \quad 
{\rm if\;n < m } \;,
\nonumber\\
&&\langle l_\mu \rangle_{n,m+1} = 
\hat w_\mu\,\langle Y_{ll}\,(l-q )\cdot \hat w \rangle_{n,m} 
+q_\mu \,\langle Y_{ll} \rangle_{n,m} \quad 
{\rm if\;n\leq m } \;,
\nonumber\\
&&Y_{\bar q \bar q}\,\langle l_\mu \rangle_{n,m+1} = 
\hat w_\mu\,\langle (l-\bar q )\cdot \hat w \rangle_{n+1,m+1} 
+\bar q_\mu \,\langle 1 \rangle_{n+1,m+1} \quad 
{\rm if\;n > m } \;,
\label{help:a}
\end{eqnarray}
which follow from (\ref{ex:2}) and covariance which implies the replacement rule
\begin{eqnarray}
&&l_\mu \to  \frac{Y_{q \bar q}\,Y_{\bar q l}-Y_{\bar q \bar q}\,Y_{q l}}
{Y_{q \bar q}^2-Y_{q q}\,Y_{\bar q \bar q}}\,\left( q_\mu-\frac{w \cdot q}{w^2}\,w_\mu \right) 
\nonumber\\
&&\quad  +\frac{Y_{q \bar q}\,Y_{q l}-Y_{q q}\,Y_{\bar q l}}
{Y_{q \bar q}^2-Y_{q q}\,Y_{\bar q \bar q}}\,
\left( \bar q_\mu-\frac{w \cdot \bar q}{w^2}\,w_\mu \right)+ \frac{l \cdot w}{w^2}\,w_\mu \,.
\label{}
\end{eqnarray}
The identities (\ref{des-result}) follow now from (\ref{first-step}) and (\ref{help:a}). 
For example, the first identity follows for $n=m$, because the first two terms in 
(\ref{first-step}) lead to the loop function $J_{ab}^{(\pm)}(w;n)$ and the last two terms
cancel. Similarly for $n> m$ the first and second terms are canceled by 
the third and fourth terms respectively whereas for $n< m$ both, the first two and last 
two terms in (\ref{first-step}) cancel separately.

\section{Isospin breaking effects}

Isospin breaking effects are easily incorporated by 
constructing super matrices $V^{(I'I)}, J^{(I'I)}$ and $T^{(I'I)}$ which couple different 
isospin states. Here we only consider isospin breaking effects induced by the loop 
functions, i.e. the interaction kernel $V^{(I'I)} = \delta_{I'I}\,V^{(I)}$ is assumed 
isospin diagonal. 

Furthermore we neglect isospin breaking effects in all but the 
s-wave $\bar KN$-channels. This leads to
\begin{eqnarray}
J_{\bar K N}^{(00)} =J_{\bar K N}^{(11)} = \frac{1}{2}\Big(J_{K^- p }+ J_{\bar{K^0} n }
\Big)
\,,\,\,
J_{\bar K N}^{(01)} =J_{\bar K N}^{(10)} = \frac{1}{2}\Big(J_{K^- p }-J_{\bar{K^0} n }
\Big)
\,.
\label{}
\end{eqnarray}
The remaining channels $X$ are defined via $J^{(I'I)}_X=\delta_{I'I}\,J^{(I)}_X$
with isospin averaged masses in the loop functions. We use also isospin averaged 
meson and baryon masses in $V^{(I)}$. Note that there is an ambiguity in the 
subtraction point of the isospin transition loop function $J_{\bar KN}^{(01)} $. We
checked that taking $m_\Lambda $ as used for $J_{\bar KN}^{(00)} $
or $m_\Sigma $ as used for $J_{\bar K N}^{(11)} $ makes little difference. We use the 
average hyperon mass $(m_\Lambda+m_\Sigma)/2$. 
The $K^-p$-reaction matrices can now be linearly combined in terms of appropriate matrix
elements of $M^{(I'I)}$
\begin{eqnarray}
M_{K^-\,p \,\rightarrow K^-\, p\;} &=& \frac{1}{2} \,M^{I=(0 ,0)}_{\bar K N
\rightarrow \bar K N} +\frac{1}{2}\,M^{I=(1,1)}_{\bar K N \rightarrow \bar K N}
+\frac{1}{2}\,M^{I=(0,1)}_{\bar K N \rightarrow \bar K N}
+\frac{1}{2}\,M^{I=(1,0)}_{\bar K N \rightarrow \bar K N} \;,
\nonumber\\
M_{K^-\,p \,\rightarrow \bar K^0\; n\;} &=& \frac{1}{2} \,M^{I=(0,0)}_{\bar K N
\rightarrow \bar K N} -\frac{1}{2}\,M^{I=(1,1)}_{\bar K N \rightarrow \bar K N}
-\frac{1}{2}\,M^{I=(0,1)}_{\bar K N \rightarrow \bar K N}
+\frac{1}{2}\,M^{I=(1,0)}_{\bar K N \rightarrow \bar K N}\;,
\nonumber\\
M_{K^-\,p \,\rightarrow \pi^- \Sigma^+\!} &=& \frac{1}{\sqrt{6}}
\,M^{I=(0,0)}_{\bar K N
\rightarrow \pi \Sigma } +\frac{1}{2}\,M^{I=(1,1)}_{\bar K N \rightarrow \pi \Sigma}
+\frac{1}{\sqrt{6}} \,M^{I=(1,0)}_{\bar K N
\rightarrow \pi \Sigma } +\frac{1}{2}\,M^{I=(0,1)}_{\bar K N \rightarrow \pi \Sigma} \;,
\nonumber\\
M_{K^-\,p \,\rightarrow \pi^+ \Sigma^-\!} &=& \frac{1}{\sqrt{6}}
\,M^{I=(0,0)}_{\bar K N \rightarrow \pi \Sigma }
-\frac{1}{2}\,M^{I=(1,1)}_{\bar K N \rightarrow \pi \Sigma}
+\frac{1}{\sqrt{6}} \,M^{I=(1,0)}_{\bar K N \rightarrow \pi \Sigma }
-\frac{1}{2}\,M^{I=(0,1)}_{\bar K N \rightarrow \pi \Sigma} \;,
\nonumber\\
M_{K^-\,p \,\rightarrow \pi^0\, \Sigma\;} &=&
\frac{1}{\sqrt{6}} \,M^{I=(0,0)}_{\bar K N \rightarrow \pi \Sigma }
+\frac{1}{\sqrt{6}} \,M^{I=(1,0)}_{\bar K N \rightarrow \pi \Sigma } \;,
\nonumber\\
M_{K^-\,p \,\rightarrow \,\pi^0\, \Lambda\;} &=& \frac{1}{\sqrt{2}}
\,M^{I=(0,1)}_{\bar K N \rightarrow \pi \Lambda }
+\frac{1}{\sqrt{2}}\,M^{I=(1,1)}_{\bar K N \rightarrow \pi \Lambda } \;.
\label{}
\end{eqnarray}

The $K^+p \to K^+p$ reaction of the strangeness  +1 channel remains a single channel 
problem due to charge conservation. One finds
\begin{eqnarray}
M_{K^+\,p \,\rightarrow K^+\, p\;} &=& 
M^{I=(1,1)}_{K N \rightarrow K N} \;, \quad 
J_{K N}^{(11)} =J_{K^+ p}  \;.
\label{}
\end{eqnarray}
The charge exchange reaction $K^+n \to K^0 p$, on the other hand,
turns into a 2 channel problem with ($K^+n$, $K^0 p$). The proper matrix 
structure in the isospin basis leads to
\begin{eqnarray}
&&M_{K^+\,n \,\rightarrow K^0\, p\;} = \frac{1}{2} \,M^{I=(1 ,1)}_{K N
\rightarrow K N} -\frac{1}{2}\,M^{I=(0,0)}_{K N \rightarrow K N}
+\frac{1}{2}\,M^{I=(0,1)}_{K N \rightarrow K N}
+\frac{1}{2}\,M^{I=(1,0)}_{K N \rightarrow K N} \;,
\nonumber\\
&&J_{K N}^{(00)} =J_{K N}^{(11)}=
 \frac{1}{2}\,\Big( J_{K^+ n}+  J_{K^0 p} \Big) \;,\;\; 
J_{K N}^{(01)} =J_{K N}^{(10)} =\frac{1}{2}\,\Big( J_{K^+ n}-  J_{K^0 p} \Big)\;.
\label{}
\end{eqnarray}
The subtraction point $\mu^{(I)}$ in the loop functions $J_{K^+ p}, J_{K^+ n}$ and
$J_{K^0 p}$ is identified with the average hyperon mass 
$\mu^{(0)}=\mu^{(1)}=(m_\Lambda+m_\Sigma)/2$.

\section{Strangeness minus one channel}

We collect the coefficients $C^{(I)}_{...}$ defined 
in (\ref{k-nonlocal},\ref{G-explicit},\ref{def-local-1}) in Tab. \ref{tabkm-1} and 
Tab. \ref{tabkm-2}. For completeness we also include the $I=2$ channel. All coefficients are defined 
within the natural extension of our notation in (\ref{r-def}). The appropriate isospin states $R^{(2)}$ are
introduced as 
\begin{eqnarray}
R^{(2)}_{[n]} = \pi_c  \cdot S_{[n]}\cdot \Sigma  \,, \qquad 
\label{I=2:def}
\end{eqnarray}
where the matrix valued vector $S_{[n]}$ satisfies 
\begin{eqnarray}
\sum_{n=1}^5\,S^\dagger_{[n],\,ac}\,S^{\,}_{[n],\,bd} ={\textstyle{1\over 2}}\, \delta_{ab}\,\delta_{cd}+
{\textstyle{1\over 2}}\,\delta_{ad}\,\delta_{cb} -{\textstyle{1\over 3}}\,\delta_{ac}\,\delta_{bd} \,.
\label{}
\end{eqnarray}

\newpage

\tabcolsep=0.7mm
\renewcommand{\arraystretch}{1.05}
\begin{table}[t]
\begin{tabular}{|r||c||c|c|c||c|c||c|c|c|c||c|c|c|c|}
 \hline   
& $ C_{WT}^{(0)} $  &  $ C_{N_{[8]}}^{(0)}$  & $C_{\Lambda_{[8]}}^{(0)}$  &  
$C_{\Sigma_{[8]}}^{(0)}$ 
& $C_{\Delta_{[10]}}^{(0)}$ & $C_{\Sigma_{[10]} }^{(0)}$ 
    &$\widetilde{C}_{N_{[8]}}^{(0)}$ & $\widetilde{C}_{\Lambda_{[8]} }^{(0)}$ &
    $\widetilde{C}_{\Sigma_{[8]}}^{(0)}$ & $\widetilde {C}_{\Xi_{[8]} }^{(0)}$   &
$\widetilde {C}_{\Delta_{[10]} }^{(0)}$ &
    $\widetilde {C}_{\Sigma_{[10]}}^{(0)}$& $\widetilde C_{\Xi_{[10]}}^{(0)}$
\\ \hline \hline 
$11$& $3$ & 0& $1$  & $0$ & 0 & 0  &0  &0  & 0 &0 &0 &0 & 0\\ \hline 
$12$& $\sqrt{\frac{3}{2}}$ & 0& $1$  & $0$ & 0 & 0  &$\sqrt{\frac{2}{3}}$  &0  & 0 &0 &$2\sqrt{\frac{2}{3}}$ &0 & 0\\ \hline 
$13$& $\frac{3}{\sqrt{2}}$ & 0& $1$  & $0$  & 0& 0  & $\sqrt{2}$  &0  & 0 &0 &0 &0 & 0\\ \hline 
$14$& $0$ & 0& $1$  & $0$  & 0 & 0  &0  & $\frac{1}{2}$  & -$\frac{3}{2}$ &0 &0 &-$\frac{3}{2}$ & 0\\ \hline \hline

$22$& $4$ & 0& $1$  & $0$  & 0 & 0  &0  &$\frac{1}{3}$  & -1 &0 &0 &-1 & 0\\ \hline 
$23$& $0$& 0 & $1$  & $0$  & 0& 0  &0  &0  & $\sqrt{3}$ &0 &0 &$\sqrt{3}$ & 0\\ \hline 
$24$& -$\sqrt{\frac{3}{2}}$ & 0 & $1$  & $0$ & 0 & 0  &0  &0  & 0 &-$\sqrt{\frac{2}{3}}$ &0 &0 & -$\sqrt{\frac{2}{3}}$\\ \hline  \hline

$33$& $0$ & 0 & $1$  & $0$  & 0& 0  &0  &1  & 0 &0 &0 &0 & 0\\ \hline 
$34$& -$\frac{3}{\sqrt{2}}$ & 0 & $1$  & $0$ & 0 & 0  &0  &0  & 0 &-$\sqrt{2}$ &0 &0 & -$\sqrt{2}$\\ \hline \hline

$44$& $3$& 0  & $1$  & $0$  & 0 & 0  &0  &0  & 0 &0 &0 &0 & 0\\ \hline \hline

\hline   
& $ C_{WT}^{(1)} $  &  $ C_{N_{[8]}}^{(1)}$  & $C_{\Lambda_{[8]}}^{(1)}$  &  
$C_{\Sigma_{[8]}}^{(1)}$ 
& $C_{\Delta_{[10]}}^{(1)}$ & $C_{\Sigma_{[10]} }^{(1)}$ 
    &$\widetilde{C}_{N_{[8]}}^{(1)}$ & $\widetilde{C}_{\Lambda_{[8]} }^{(1)}$ &
    $\widetilde{C}_{\Sigma_{[8]}}^{(1)}$ & $\widetilde {C}_{\Xi_{[8]} }^{(1)}$   &
$\widetilde {C}_{\Delta_{[10]} }^{(1)}$ &
    $\widetilde {C}_{\Sigma_{[10]}}^{(1)}$& $\widetilde C_{\Xi_{[10]}}^{(1)}$
\\ \hline \hline 
$11$& 1 & 0& 0 & 1 & 0 & 1  &0  &0  & 0 &0 &0 &0 & 0\\ \hline 
$12$& 1 & 0& 0 & 1 & 0 & 1  &-$\frac{2}{3}$  &0  & 0 &0 &$\frac{2}{3}$ &0 & 0\\ \hline 
$13$& $\sqrt{\frac{3}{2}}$ & 0 & $0$  & $1$ & 0 & 1  &$\sqrt{\frac{2}{3}}$  &0  & 0 &0 &0 &0 & 0\\ \hline 
$14$& $\sqrt{\frac{3}{2}}$ & 0 & $0$  & $1$ & 0 & 1  &$\sqrt{\frac{2}{3}}$  &0  & 0 &0 &0 &0 & 0\\ \hline 
$15$& $0$ & 0 & $0$  & $1$  & 0 & 1  &0  &-$\frac{1}{2}$  & -$\frac{1}{2}$ &0 &0 &-$\frac{1}{2}$ & 0\\ \hline \hline

$22$& $2$ & 0 & $0$  & $1$ & 0 & 1  &0  &-$\frac{1}{3}$  & $\frac{1}{2}$ &0 &0 &$\frac{1}{2}$ & 0\\ \hline 
$23$& $0$ & 0 & $0$  & $1$ & 0 & 1  &0  &0  & -1 &0 &0 &-1 & 0\\ \hline 
$24$& $0$ & 0 & $0$  & $1$ & 0 & 1  &0  &0  & 1 &0 &0 &1 & 0\\ \hline 
$25$& $-1$ & 0 & $0$  & $1$ & 0 & 1  &0  &0  & 0 &$\frac{2}{3}$ &0 &0 & $\frac{2}{3}$\\ \hline  \hline

$33$& $0$ & 0& $0$  & $1$ & 0 & 1  &0  &0  & 1 &0 &0 &1 & 0\\ \hline 
$34$& $0$ & 0& $0$  & $1$  & 0& 1  &0  &$\frac{1}{\sqrt{3}}$ & 0 &0 &0 &0 & 0\\ \hline 
$35$& $\sqrt{\frac{3}{2}}$ & 0 & $0$  & $1$& 0  & 1  &0  &0  & 0 &-$\sqrt{\frac{2}{3}}$ &0 &0 & -$\sqrt{\frac{2}{3}}$\\ \hline \hline

$44$& $0$ & 0 & $0$  & $1$ & 0 & 1  &0  &0  & 1 &0 &0 &1 & 0\\ \hline 
$45$& $\sqrt{\frac{3}{2}}$ & 0 & $0$  & $1$ & 0 & 1  &0  &0  & 0 &-$\sqrt{\frac{2}{3}}$ &0 &0 & -$\sqrt{\frac{2}{3}}$\\ \hline  \hline

$55$& $1$ & 0 & $0$  & $1$ & 0 & 1  &0  &0  & 0 &0 &0 &0 & 0\\ \hline \hline   

& $ C_{WT}^{(2)} $  &  $ C_{N_{[8]}}^{(2)}$  & $C_{\Lambda_{[8]}}^{(2)}$  &  
$C_{\Sigma_{[8]}}^{(2)}$ 
& $C_{\Delta_{[10]}}^{(2)}$ & $C_{\Sigma_{[10]} }^{(2)}$ 
    &$\widetilde{C}_{N_{[8]}}^{(2)}$ & $\widetilde{C}_{\Lambda_{[8]} }^{(2)}$ &
    $\widetilde{C}_{\Sigma_{[8]}}^{(2)}$ & $\widetilde {C}_{\Xi_{[8]} }^{(2)}$   &
$\widetilde {C}_{\Delta_{[10]} }^{(2)}$ &
    $\widetilde {C}_{\Sigma_{[10]}}^{(2)}$& $\widetilde C_{\Xi_{[10]}}^{(2)}$
\\ \hline \hline 
$11$& -2 & 0& 0 & 0 & 0 & 0  &0  &$\frac{1}{3}$ & $\frac{1}{2}$ &0 &0 &$\frac{1}{2}$ & 0\\ \hline

\end{tabular}
\vspace*{2mm} \caption{Weinberg-Tomozawa interaction strengths and baryon exchange coefficients in the 
strangeness minus one channels as defined in (\ref{k-nonlocal}) and (\ref{I=2:def}).} 
\label{tabkm-1}
\end{table}

\tabcolsep=1.05mm
\renewcommand{\arraystretch}{1.05}
\begin{table}[t]
\begin{tabular}{|r||c||c|c||c|c|c||c|c|c|c||c|c|c|}
\multicolumn{14}{c}{}\\ \multicolumn{14}{c}{}\\
 \hline
    & $C_{\pi,0}^{(0)}$ &  $C_{\pi,D}^{(0)}$ &  $C_{\pi,F}^{(0)}$ 
    & $C_{K,0}^{(0)}$   & ${C}_{K,D}^{(0)}$  & ${C}_{K,F}^{(0)}$ 
    & ${C}_{0}^{(0)}$   &  ${C}_{1}^{(0)}$  & 
    ${C}_{D}^{(0)}$ & ${C}_{F}^{(0)}$   
    & $\bar C_{1}^{(0)}$ & $\bar C_{D}^{(0)}$ & $\bar C_{F}^{(0)}$\\
\hline \hline 
$11$& 0 & 0 & 0
    & -4 &-6 & -2
    & 2 & 2 & 3& 1  
    & 2 &1& 3   \\ \hline 
$12$& 0& -$\sqrt{\frac32}$ &  $\sqrt{\frac32}$ 
    & 0& -$\sqrt{\frac32}$ & $\sqrt{\frac32}$  
    & 0 & $\sqrt 6$ & $\sqrt{\frac32}$ &-$\sqrt{\frac32}$   
    & $\sqrt 6$ & -$\sqrt{\frac32}$ &$\sqrt{\frac32}$      \\ \hline 
$13$& 0& $\frac{1}{\sqrt 2}$ & $\frac{3}{\sqrt 2}$ 
    & 0& -$\frac{5}{3\,\sqrt 2}$  & -$\frac{5}{\sqrt 2}$ 
    & 0  & $\sqrt 2$ & $\frac{1}{3\sqrt 2}$ & $\frac{1}{\sqrt 2}$ 
    & $\sqrt 2$ &$\frac{1}{\sqrt 2}$&$\frac{3}{\sqrt 2}$   \\  \hline
$14$& 0 & 0  & 0
    & 0 & 0  & 0
    & 0  & -3& 0 & 0  
    & -1& 0 & 0   \\ \hline\hline
$22$& -4 & -4 & 0
    & 0  & 0 & 0
    & 2  & 4& 2 & 0  
    & 2 & 0& 4    \\ \hline
$23$& 0 &  -$\frac{4}{\sqrt 3}$ & 0  
    &  0  & 0  & 0
    & 0  & $\sqrt 3$& $\frac{2}{\sqrt 3}$ &  0 
    & $\sqrt 3$ &  0 &  0  \\ \hline
$24$& 0&  $\sqrt{\frac32}$ & $\sqrt{\frac32}$ 
    & 0&  $\sqrt{\frac32}$ & $\sqrt{\frac32}$ 
    & 0 & -$\sqrt 6$& -$\sqrt{\frac32}$ &  -$\sqrt{\frac32}$ 
    & -$\sqrt 6$ &-$\sqrt{\frac32}$&  -$\sqrt{\frac32}$   \\ \hline\hline
$33$& $\frac43$ & $\frac{28}{9}$ & 0 
    & -$\frac{16}{3}$ & -$\frac{64}{9}$& 0 
    & 2 & 2 & 2 & 0  
    & 0 & 0  & 0 \\ \hline
$34$& 0&  -$\frac{1}{\sqrt 2}$ & $\frac{3}{\sqrt{2}}$ 
    & 0& $\frac{5}{3\sqrt 2}$  & -$\frac{5}{\sqrt 2}$ 
    & 0 & -$\sqrt 2$ & -$\frac{1}{3\sqrt 2}$ & $\frac{1}{\sqrt 2}$  
    & -$\sqrt 2$ &$\frac{1}{\sqrt 2}$ & -$\frac{3}{\sqrt 2}$    \\ \hline\hline
$44$& 0 & 0 & 0
    & -4 &-6 & 2 
    & 2 & 2 & 3&-1  
    & 2 & -1 &3  \\ \hline \hline

  & $C_{\pi,0}^{(1)}$    &  $C_{\pi,D}^{(1)}$      &  $C_{\pi,F}^{(1)}$ 
  & $C_{K,0}^{(1)}$  &${C}_{K,D}^{(1)}$ & ${C}_{K,F}^{(1)}$ 
  & ${C}_{0}^{(1)}$ & ${C}_{1}^{(1)}$& ${C}_{D}^{(1)}$ & ${C}_{F}^{(1)}$   
  & $\bar C_{1}^{(1)}$ & $\bar C_{D}^{(1)}$ & $\bar C_{F}^{(1)}$\\
\hline \hline 
$11$&0 & 0 & 0  
    & -4 & -2& 2   
    & 2  & 0  & 1& -1 
    & 0& -1 &  1   \\ \hline 
$12$& 0& -1 & 1  
    & 0& -1 & 1  
    & 0 & 0 & 1& -1  
    & 0 &-1& 1    \\ \hline 
$13$& 0 &  $\frac{1}{\sqrt 6}$ & $\sqrt{\frac32}$ 
    & 0 & $\frac{1}{\sqrt 6}$& $\sqrt{\frac32}$     
    & 0 & 0 & -$\frac{1}{\sqrt 6}$  
    &-$\sqrt{\frac32}$  & 0 &$\frac{1}{\sqrt 6}$& $\sqrt{\frac32}$   \\  \hline
$14$& 0 & -$\sqrt{\frac32}$&  $\sqrt{\frac32}$ 
    & 0 & $\frac{5}{\sqrt 6}$ &-$\frac{5}{\sqrt 6}$   
    & 0 & 0 & -$\frac{1}{\sqrt 6}$   & $\frac{1}{\sqrt 6}$   
    &0 & -$\sqrt{\frac32}$& $\sqrt{\frac32}$    \\  \hline
$15$& 0  &  0 & 0 
    & 0  &  0 & 0  
    & 0 & 1 & 0  &  0  
    & -1 & 0  &  0  \\ \hline\hline
$22$& -4&-4 & 0  
    & 0& 0  & 0  
    & 2  & -1 & 2 & 0    
    & 1 & 0  & 2  \\ \hline
$23$& 0  & 0 & 0  
    & 0  & 0 & 0  
    & 0  & 0 & 0  & 0 
    & 0  & -$2\sqrt{\frac23}$& 0   \\ \hline
$24$& 0& 0  &  $4\sqrt{\frac23}$
    & 0  & 0  & 0  
    & 0 & 0 & 0   & -$2\sqrt{\frac23}$   
    & 0  & 0 &0 \\ \hline
$25$& 0 & 1& 1   
    & 0 & 1 & 1    
    &0  & 0 & -1  & -1   
    &0 & -1  & -1  \\ \hline\hline
$33$& -4 & -$\frac43$ & 0    
    & 0    & 0   & 0  
    & 2 & 0 & $\frac23$& 0   
    &0 & 0  & 0  \\ \hline
$34$& 0 & -$\frac43$  & 0  
    & 0   &0  & 0 
    & 0 & 1 & $\frac23$ & 0   
    &-1 & 0   & 0  \\ \hline
$35$& 0 & $\frac{1}{\sqrt 6}$  &  -$\sqrt{\frac32}$    
    & 0 & $\frac{1}{\sqrt 6}$ &-$\sqrt{\frac32}$   
    & 0 & 0 &-$\frac{1}{\sqrt 6}$& $\sqrt{\frac32}$  
    & 0 & -$\frac{1}{\sqrt 6}$ & $\sqrt{\frac32}$    \\ \hline\hline
$44$& $\frac43$   & -$\frac43$  &  0 
    & -$\frac{16}{3}$ & 0  & 0   
    & 2 & 0 & $\frac23$ & 0  
    & 0 & 0 & 0  \\ \hline
$45$& 0  & -$\sqrt{\frac32}$  &  -$\sqrt{\frac32}$  
    & 0  & $\frac{5}{\sqrt 6}$  &  $\frac{5}{\sqrt 6}$  
    & 0 & 0 & -$\frac{1}{\sqrt 6}$  & -$\frac{1}{\sqrt 6}$  
    & 0 & $\sqrt{\frac32}$ & $\sqrt{\frac32}$   \\ \hline\hline
$55$& 0 & 0  & 0  
    & -4 & -2  & -2   
    & 2 & 0 & 1   & 1 
    & 0 & 1   & 1  \\ \hline \hline

  & $C_{\pi,0}^{(2)}$    &  $C_{\pi,D}^{(2)}$      &  $C_{\pi,F}^{(2)}$ 
  & $C_{K,0}^{(2)}$  &${C}_{K,D}^{(2)}$ & ${C}_{K,F}^{(2)}$ 
  & ${C}_{0}^{(2)}$ & ${C}_{1}^{(2)}$& ${C}_{D}^{(2)}$ & ${C}_{F}^{(2)}$   
  & $\bar C_{1}^{(2)}$ & $\bar C_{D}^{(2)}$ & $\bar C_{F}^{(2)}$\\
\hline \hline 
$11$ &-4 & -4 & 0  & 0 & 0& 0 & 2  & 1  & 2 & 0 
    & -1& 0 &  -2   \\ \hline 
\end{tabular}
\vspace*{2mm} \caption{Coefficients of quasi-local interaction terms in the strangeness minus one
channels as defined in (\ref{def-local-1}) and (\ref{I=2:def}).} 
\label{tabkm-2}
\end{table}

\newpage

\section{Differential cross sections}

In this appendix we provide the expressions needed for the evaluation of 
cross sections as required for the comparison with available empirical data. 
Low-energy data are available for the reactions 
$K^-p \to K^-p, \bar K^0 n, \pi^\mp\Sigma^\pm, \pi^0 \Sigma^0$, $\pi^0 \Lambda$
and $K^-p \to \pi^0 \pi^- p, \pi^+ \pi^- n $ in the strangeness minus one channel and for the 
reactions $K^+p \to K^+p $ and $K^+ n \to K^0p $ in the strangeness +1 channel. The 
differential cross sections with a two-body final state can be written in the generic form
\begin{eqnarray}
\frac{d\,\sigma (\sqrt{s}, \cos \theta )}{d\,\cos \theta } &=&
\frac{1}{32\,\pi\,s}\,\frac{p^{(f)}_{\rm cm}}{p^{(i)}_{\rm cm}}\,\Bigg(
| F_+ (\sqrt{s},\theta )|^2\,\Big(m_i+E_i\Big) \,\Big(m_f+E_f\Big)
\nonumber\\
&+&| F_-(\sqrt{s},\theta )|^2\Big(m_i-E_i\Big)
\,\Big(m_f-E_f\Big)
\nonumber\\
&+&2\,\Re  \,\Big( F_+(\sqrt{s},\theta )\,F_-^\dagger(\sqrt{s},\theta )
\Big)\,p_{\rm cm}^{(i)}\,p_{\rm cm}^{(f)}\,\cos
\theta \Bigg)
\label{cross-section}
\end{eqnarray}
where $p_{\rm cm}^{(i)}$ and $p_{\rm cm}^{(f)}$ is the relative momentum in 
the center of mass system of the initial and final channel
respectively. The angle $ \theta $ denotes the scattering angle in the center 
of mass system. Also, $m_{i}$ and $m_{f}$ are the masses of incoming and
outgoing fermions, respectively, and
\begin{eqnarray}
E_i=\sqrt{m^2_i+\Big(p_{\rm cm}^{(i)}\Big)^2}\,, \qquad
E_f=\sqrt{m^2_f+\Big(p_{\rm cm}^{(f)}\Big)^2}
\,.
\label{}
\end{eqnarray}
The amplitudes $F_\pm(\sqrt{s},\theta )$ receive contributions from the 
amplitudes $M^{(\pm)}(\sqrt{s},0)$ and $M^{(\pm)}(\sqrt{s},1)$ as introduced
in (\ref{t-sum}). We derive
\begin{eqnarray}
F_+(\sqrt{s},\theta ) &=& M^{(+)}(\sqrt{s}; 0)
+3 \, p_{\rm cm}^{(f)}\,p_{\rm cm}^{(i)}\,M^{(+)}(\sqrt{s};1)\,\cos \theta
\nonumber\\
&-& \Big(E_f-m_f\Big)\,\Big(E_i-m_i\Big)\,M^{(-)}(\sqrt{s};1) \,,
\nonumber\\
F_- (\sqrt{s},\theta )&=&M^{(-)}(\sqrt{s}; 0)
+3 \, p_{\rm cm}^{(f)}\,p_{\rm cm}^{(i)}\,M^{(-)}(\sqrt{s};1)\,\cos \theta
\nonumber\\
&-&\Big(E_f+m_f\Big)\,\Big(E_i+m_i\Big)\,M^{(+)}(\sqrt{s};1) \;,
\label{mplus-minus}
\end{eqnarray}
where the appropriate element of the coupled channel matrix $M^{(I,\pm)}_{ab}$ is assumed. 
For further considerations it is convenient to introduce reduced partial-wave 
amplitudes $f^{(l)}_{J=l\pm \frac{1}{2}}(s)$ with
\begin{eqnarray}
f^{(l)}_{J=l\pm \frac{1}{2}}(\sqrt{s}\,) &=& 
\frac{(p^{(f)}_{\rm cm}\,p^{(i)}_{\rm cm})^{J-\frac{1}{2}}}{8\,\pi\,\sqrt{s}} \;
\sqrt{E_i\pm m_i}\,\sqrt{E_f\pm m_f}\,M^{(\pm )}(\sqrt{s},J-{\textstyle{1\over 2}})\;,
\label{}
\end{eqnarray}
in terms of which the total cross section reads
\begin{eqnarray}
\sigma_{i\,\to f} (\sqrt{s}\,) &=& 4\,\pi \,\frac{p^{(f)}_{\rm cm}}{p^{(i)}_{\rm cm}}\,
\sum_{l=0}^\infty \,\Bigg(
l\,|f^{(l)}_{J=l-\frac{1}{2}}(\sqrt{s}\,) |^2 
+ (l+1)\,|f^{(l)}_{J=l+\frac{1}{2}}(\sqrt{s}\,)|^2 
\Bigg)\;.  
\label{}
\end{eqnarray}

The measurement of the three-body final states $\pi^0 \pi^- p$ and $\pi^+\pi^- n$
provides more detailed constraints on the kaon induced $\Lambda$ and
$\Sigma$ production matrix elements due to the self-polarizing property
of the hyperons. Particularly convenient is the triple differential cross section
\begin{eqnarray}
\frac{k_{\pi^- n}\,d\sigma_{K^-p\,\to \pi^+ \pi^-n}}{dm_{\Sigma^- }^2\,
d\! \cos \theta \,d k_\perp}
&=&\frac{\Gamma_{\Sigma^{-}\to \pi^- n}}{2\pi\,\Gamma_{\Sigma^-}^{(tot.)}}\,
\frac{d\sigma_{K^-p\,\to \pi^+ \Sigma^- } }{d\!\cos \theta }\,\Big( 1
+\alpha_{\Sigma^-}\,\frac{k_\perp}{k_{\pi- n}} \,P_{K^-p\,\to \pi^+ \Sigma^- }\Big)\;,
\nonumber\\
\frac{k_{\pi^0p}\,d\sigma_{K^-p\,\to \pi^0 \pi^-p}}{dm_{\Sigma^+}^2\,d\! \cos \theta \,
d k_\perp}
&=&\frac{\Gamma_{\Sigma^+\to \pi^0 p}}{2\pi\,\Gamma_{\Sigma^+}^{(tot.)}}\,
\frac{d\sigma_{K^-p\,\to \pi^-\Sigma^+ } }{d\!\cos \theta }\,\Big( 1
+\alpha_{\Sigma^+}\,\frac{k_\perp}{k_{\pi^0 p}}\,P_{K^-p\,\to \pi^-\Sigma^+ }\Big)
\nonumber\\
\frac{k_{\pi^-p}\,d\sigma_{K^-p\,\to \pi^0 \pi^-p}}{dm_\Lambda^2\,d\! \cos \theta \,d k_\perp}
&=&\frac{\Gamma_{\Lambda\to \pi^- p}}{2\pi\,\Gamma_\Lambda^{(tot.)}}\,
\frac{d\sigma_{K^-p\,\to \pi^0\Lambda } }{d\!\cos \theta }\,\Big( 1
+\alpha_\Lambda\,\frac{k_\perp}{k_{\pi^- p}}\,P_{K^-p\,\to \pi^0\Lambda }\Big)\;,
\label{}
\end{eqnarray}
which determines the polarizations $P_{K^-p\to \pi^0\Lambda }$ and
$P_{K^-p\to \pi\mp \Sigma^\pm }$
\begin{eqnarray}
\frac{d\sigma (\sqrt{s},\cos \theta )}{d\!\cos \theta }\,P(\sqrt{s},\theta )
&=& \frac{\Big(p^{(f)}_{\rm cm}\Big)^2\,\sin \theta }{32\,\pi\,s}
\,\Im \,\Big( F_+(\sqrt{s},\theta )\,F_-^\dagger(\sqrt{s},\theta ) \Big)\;.
\label{}
\end{eqnarray}
Here $\alpha_{\Sigma^-} =-0.068\pm$0.013, $\alpha_{\Sigma^+} =-0.980\pm$0.017
and $\alpha_\Lambda =0.642\pm$0.013 are the hyperon polarizeabilties. The 
variable $k_\perp$ is defined in the $K^-p$ center of mass frame via 
$k_\perp\,p_{\rm cm }^{(i)}\,p_{\rm cm }^{(f)}\sin \theta  = 
{\vec k} \cdot ({\vec p}^{\,(i)}_{\rm cm}\! \times {\vec p}^{\,(f)}_{\rm cm})$. In the 
hyperon rest frame it represents the decay angle 
$\beta $ of the $\Sigma $ and $\Lambda $ with $k_\perp =k_{\pi^\pm n}\,\cos \beta$ and 
$k_\perp =k_{\pi^- p}\,\cos \beta$  relative to the final neutron and proton 
three momentum ${\vec k}$ respectively. Furthermore 
$m_{\Sigma^\pm}^2= (m_{\pi^\pm}^2+k_{\pi^\pm n}^2)^{1/2}+(m_n^2+k_{\pi^\pm n}^2)^{1/2}$ 
and $m_\Lambda^2= (m_{\pi^-}^2+k_{\pi^- p}^2)^{1/2}+(m_p^2+k_{\pi^- p}^2)^{1/2}$.

Differential cross sections and polarizations are conveniently parameterized in 
terms of moments $A_n(\sqrt{s}\,)$ and 
$B_n(\sqrt{s}\,)$ \cite{mast-pio,mast-ko,bangerter-piS}
defined as the nth order Legendre weights
\begin{eqnarray}
\sum_{n=0}^\infty A_n(\sqrt{s}\,)\,P_n(\cos \theta )&=&
\frac{d\sigma (\sqrt{s}, \cos \theta )}{d\cos \theta } \,, 
\nonumber\\
-\sum_{l=1}^\infty B_l(\sqrt{s}\,)\,P^1_l(\cos \theta )&=&
\frac{\Big(p^{(f)}_{\rm cm}\Big)^2\,\sin \theta }{32\,\pi\,s}\,
\Im \,\Big( F_+(\sqrt{s},\theta )\,F_-^\dagger(\sqrt{s},\theta ) \Big)\;,
\label{a-distributions}
\end{eqnarray}
where $P_l^1(\cos \theta)=-\sin \theta \,P_l'(\cos \theta)$.
One derives
\begin{eqnarray}
A_1(\sqrt{s}\,)&=& 4\,\pi\,\frac{p^{(f)}_{\rm cm}}{p^{(i)}_{\rm cm}}\,
\Re \Bigg( f^{(S)}_{J=\frac{1}{2}}(\sqrt{s}\,)\,\Big( 
f^{(P,*)}_{J=\frac{1}{2}}(\sqrt{s}\,)+2\,f^{(P,*)}_{J=\frac{3}{2}}(\sqrt{s}\,)
\Big)\,
\nonumber\\
&& \qquad \quad +
\Big(2\,f^{(P)}_{J=\frac{1}{2}}(\sqrt{s}\,)+\frac{2}{5}\, f^{(P)}_{J=\frac{3}{2}}(\sqrt{s}\,)
\Big)\,f^{(D,*)}_{J=\frac{3}{2}}(\sqrt{s}\,) \Bigg)\;,
\nonumber\\
A_2(\sqrt{s}\,)&=&4\,\pi\, \frac{p^{(f)}_{\rm cm}}{p^{(i)}_{\rm cm}}\,
\Re \Bigg(| f^{(P)}_{J=\frac{3}{2}}(\sqrt{s}\,)|^2
+ |f^{(D)}_{J=\frac{3}{2}}(\sqrt{s}\,)|^2
\nonumber\\
&& \qquad \quad +2\,
f^{(S)}_{J=\frac{1}{2}}(\sqrt{s}\,)\,f^{(D,*)}_{J=\frac{3}{2}}(\sqrt{s}\,)
+2\,f^{(P)}_{J=\frac{1}{2}}(\sqrt{s}\,)\,f^{(P,*)}_{J=\frac{3}{2}}(\sqrt{s}\,)
\Bigg) \;,
\nonumber\\
B_1(\sqrt{s}\,)&=& 2\,\pi\,\frac{p^{(f)}_{\rm cm}}{p^{(i)}_{\rm cm}}\,
\Im \Bigg(
f^{(S)}_{J=\frac{1}{2}}(\sqrt{s}\,)\,
\Big(f^{(P,*)}_{J=\frac{1}{2}}(\sqrt{s}\,)- f^{(P,*)}_{J=\frac{3}{2}}(\sqrt{s}\,)
\Big)
\nonumber\\
&& \qquad \quad 
- f^{(D)}_{J=\frac{3}{2}}(\sqrt{s}\,)
\,\Big(f^{(P,*)}_{J=\frac{1}{2}}(\sqrt{s}\,)- f^{(P,*)}_{J=\frac{3}{2}}(\sqrt{s}\,)
\Big) \Bigg)\;,
\nonumber\\
B_2(\sqrt{s}\,)&=& 2\,\pi\,\frac{p^{(f)}_{\rm cm}}{p^{(i)}_{\rm cm}}\,
\Im \Bigg(
f^{(S)}_{J=\frac{1}{2}}(\sqrt{s}\,)\,f^{(D,*)}_{J=\frac{3}{2}}(\sqrt{s}\,)
- f^{(P)}_{J=\frac{1}{2}}(\sqrt{s}\,)\,f^{(P,*)}_{J=\frac{3}{2}}(\sqrt{s}\,)  \Bigg) \;.
\label{}
\end{eqnarray}

We use the empirical mass values for the kinematical factors in (\ref{cross-section}) and 
include isospin breaking effects in the matrix elements $M_\pm $ as described in the 
Appendix D. Coulomb effects which are particularly important in the $K^+ p \to K^+ p$ 
reaction are also considered. They can be generated by the formal replacement 
rule \cite{Coulomb}
\begin{eqnarray}
&&F_{+}(\sqrt{s},\theta) \to F_{+}(\sqrt{s},\theta )
-\frac{\alpha }{v}\,\frac{4 \pi \,\sqrt{s}}{E+m}\,\frac{
\exp \left( -i\,\frac{\alpha}{v}\,\ln \left(\sin^2 (\theta/2) \right) 
\right)}{p^{}_{\rm cm}\,\sin^2 (\theta/2)}\,,
\nonumber\\
&& \quad \quad \quad  \quad  \quad 
v= \frac{p^{}_{\rm cm}\,\sqrt{s}}{E\,(\sqrt{s}-E)} \;,
\quad  \quad  \quad 
\alpha = \frac{e^2\,Z^{}_1\,Z^{}_2}{4\,\pi} \;
\label{Coulomb}
\end{eqnarray}
for the amplitude $F_{+}(\sqrt{s},\theta )$ in 
(\ref{cross-section}). Here $E_i=E_f=E$, $m_f=m_i=m $, $e^2/(4\pi) \simeq 1/137$ 
and $e\,Z_{1,2}^{}$ are 
the charges of the  particles.

\section{Chiral expansion of baryon exchange}

In this appendix we provide the leading and subleading terms of the baryon exchange 
contributions not displayed in the main text. We begin with the s-channel exchange 
contributions for which only the baryon decuplet states induce terms of chiral order 
$Q^3$:
\begin{eqnarray}
V^{(I,+)}_{s-[10],3}(\sqrt{s};0)
&=& \sum_{c=1}^2\,\frac{(2-Z_{[10]})\,\sqrt{s}+3\,m^{(c)}_{[10]}}{3\,(m^{(c)}_{[10]})^2}\,
\Bigg(
\frac{\phi^{(I)}(s)}{2 \,\sqrt{s}}\,\frac{C^{(I,c)}_{[10]}}{4\,f^2 }\,
\Big( \sqrt{s}-M^{(I)}\Big)
\nonumber\\
&& \qquad \quad + 
\Big( \sqrt{s}-M^{(I)}\Big)\,\frac{C^{(I,c)}_{[10]}}{4\,f^2 }\,
\frac{\phi^{(I)}(s)}{2 \,\sqrt{s}} \Bigg)\,,
\nonumber\\
V^{(I,-)}_{s-[10],3}(\sqrt{s};0)&=&\sum_{c=1}^2\, 
\frac{Z_{[10]}\,\sqrt{s}}{3\,(m^{(c)}_{[10]})^2}\,\Bigg(
\Big( \sqrt{s}-M^{(I)}\Big)\,\frac{C^{(I,c)}_{[10]}}{4\,f^2 }\,
\Big( \sqrt{s}+M^{(I)}\Big)
\nonumber\\
&& \qquad \quad +\Big( \sqrt{s}+M^{(I)}\Big)\,\frac{C^{(I,c)}_{[10]}}{4\,f^2 }\,
\Big( \sqrt{s}-M^{(I)}\Big) \Bigg) \,,
\nonumber\\
V^{(I,+)}_{s-[10],3}(\sqrt{s};1)&=&
V^{(I,\pm)}_{s-[8],3}(\sqrt{s};0)=V^{(I,+)}_{s-[8],3}(\sqrt{s};1)=0 \,,
\label{v-result-1:q3}
\end{eqnarray}
where the index '3' in (\ref{v-result-1:q3}) indicates that only the terms of order $Q^3$ 
are shown. Next we give the $Q^3$-terms of the u-channel baryon exchanges characterized
by the subleading moment of the functions $h_{n,\pm}^{(I)}(\sqrt{s},m)$
\begin{eqnarray}
\Big[V^{(I,\pm )}_{u-[8]}(\sqrt{s};n)\Big]_{ab} &=& \sum_{c=1}^4\,
\frac{1}{4\,f^2 }\,\Big[\widetilde C^{(I,c)}_{[8]}\Big]_{ab}\,
\Big[{ h}^{(I)}_{n \pm }(\sqrt{s},m^{(c)}_{[8]})\Big]_{ab} \;.
\label{u-result-8:appendix}
\end{eqnarray}
We derive
\begin{eqnarray}
&&\Big[{h}_{0+}^{(I)} (\sqrt{s},m)\Big]_{ab,3}=
\frac{m+M_{ab}^{(L)}}{\mu_{+,ab}}\,\Bigg(
\frac{2}{3}\,\frac{\phi_a\,\phi_b}{s\,\mu_{-,ab}}
+\frac43\,\frac{\phi_a\,
\Big(\omx{a}+\omx{b}\Big)\,\phi_b }
{\sqrt{s}\,\mu_{+,ab}\,\big(\mu_{-,ab}\big)^2}
\Bigg)\,
\frac{m+M_{ab}^{(R)}}{\mu_{+,ab}}\,,
\nonumber\\
&&\Big[{h}_{0-}^{(I)} (\sqrt{s},m)\Big]_{ab,3}=\sqrt{s}-m
-\tilde R^{(I)}_{L,ab}-\tilde R^{(I)}_{R,ab}
-\frac{m+M_{ab}^{(L)}}{\mu_{+,ab}}\,
\Bigg( \frac23\,\frac{\phi_a\,M_b+M_a\,\phi_b }
{\sqrt s \,\mu_{-,ab}}
\nonumber\\
&& \qquad \quad
+\frac{16}{3}\,\frac{\phi_a\,M_a\,M_b\,\omx{b}}
{\sqrt{s}\,\mu_{+,ab}\,(\mu_{-,ab})^2}
+\frac{16}{3}\,\frac{\omx{a}\,M_a\,M_b\,\phi_b}
{\sqrt{s}\,\mu_{+,ab}\,(\mu_{-,ab})^2} -\frac{8}{3}\,\frac{\phi_a\,\sqrt{s}\,\phi_b}{\mu_{-,ab}^3\,\mu_{+,ab}}
\nonumber\\
&& \qquad \quad
-2\,\frac{\phi_a\,\omx{b}+\omx{a}\,\phi_b}{(\mu_{-,ab})^2}
+\frac{32}{5}\,\frac{\phi_a\,M_a\,M_b\,\phi_b\,}
{(\mu_{+,ab})^2\,(\mu_{-,ab})^3}
\Bigg)\,\frac{m+M_{ab}^{(R)}}{\mu_{+,ab}} \,,
\nonumber\\
&&\Big[{h}_{1+}^{(I)} (\sqrt{s},m)\Big]_{ab,3}=
-\frac{m+M_{ab}^{(L)}}{\mu_{+,ab}}\,
\Bigg( \frac{8}{5}\,\frac{\phi_a\,\phi_b\,}
{(\mu_{+,ab})^2\,(\mu_{-,ab})^3} \left(1-\frac{\mu_{+,ab}}{6\,\sqrt{s}} \right)
\nonumber\\
&& \qquad \quad
+\frac{4}{3}\,\frac{\phi_a\,\omx{b}}
{\sqrt{s}\,\mu_{+,ab}\,(\mu_{-,ab})^2}
+\frac{4}{3}\,\frac{\omx{a}\,\phi_b}
{\sqrt{s}\,\mu_{+,ab}\,(\mu_{-,ab})^2}
\Bigg)\,\frac{m+M_{ab}^{(R)}}{\mu_{+,ab}}
\nonumber\\
&&\Big[{h}_{1-}^{(I)} (\sqrt{s},m)\Big]_{ab,3}=
\frac23\,\frac{(m+M_{ab}^{(L)})\,(m+M_{ab}^{(R)}) }{\mu_{-,ab}\,
\big(\mu_{+,ab}\big)^2}
\label{u-approx-1:q3}\;,
\end{eqnarray}
where we introduced the short hand notations 
$M_a=m_{B(I,a)}$ and $ m_a=m_{\Phi(I,a)}$. Also,  
$\omx{a}= \sqrt{s}-M_a$ and $\phi_a= \omx{a}^2-m_a^2$. We recall here 
$M_{ab}^{(L)}=m_{B(I,a)}+\tilde R^{(I,c)}_{L,ab}$ and 
$M_{ab}^{(R)}=m_{B(I,b)}+\tilde R^{(I,c)}_{R,ab}$ with $\tilde R $ 
specified in (\ref{P-result}).

We turn to the decuplet functions ${p}^{(I)}_{n \pm}(\sqrt{s},m)$
\begin{eqnarray}
\Big[V^{(I,\pm )}_{u-[10]}(\sqrt{s};n)\Big]_{ab} &=& \sum_{c=1}^2\,
\frac{1}{4\,f^2 }\,\Big[\widetilde C^{(I,c)}_{[10]}\Big]_{ab}\,
\Big[{ p}^{(I)}_{n \pm }(\sqrt{s},m^{(c)}_{[10]})\Big]_{ab} \;,
\label{u-result-8:appendix}
\end{eqnarray}
for which we provide the leading moments
\begin{eqnarray}
&&\left[p_{0+}^{(I)}(s,m)\right]_{ab}=
\frac{\omx{a}\, \omx{b}}{\mu_{-,ab}}\,
\left(1-\frac{s}{m^2}+\frac{\sqrt s}{m^2}\, \Big(\omx{a}+\omx{b}\Big)\right)
+ \frac{s}{3\,m^2}\, \frac{\omx{a}\, \omx{b}}{\mu_{+,ab}}
\nonumber\\
&&\qquad+ \frac{(M_a+m)\,(M_b+m)}{3\,\mu_{+,ab}}\,
\Bigg(
1+\frac{\omx{a}\, \phi_b+\omx{b}\,
\phi_a}{\sqrt s\, \mu_{+,ab}\,\mu_{-,ab}}
+ \frac{\phi_a\,(4\,\sqrt{s}-\mu_{+,ab})\,\phi_b}
{3\,\sqrt{s}\,\mu_{+,ab}^2\,\mu_{-,ab}^2} \Bigg)
\nonumber\\
&&\qquad -\frac13\, (\sqrt s +m)
+\left(\frac{M_a+m}{3\,m\, \mu_{+,ab}}-\frac{1}{3\,m}\right)
\left(\sqrt s\, \omx{a}-\frac12\,
\phi_a-\omx{a}\,\omx{b}  -m_a^2\right)
\nonumber\\
&&\qquad +\left(\frac{M_b+m}{3\,m\, \mu_{+,ab}}-\frac{1}{3\,m}\right)
\left(\sqrt s\, \omx{b}-\frac12\,\phi_b-\omx{a}\,\omx{b}- m_b^2\right)
-\frac{2}{3}\,\frac{\phi_a\,\phi_b}{\mu_+ \,\mu_{-,ab}^2}
\nonumber\\
&&\qquad 
+ \sqrt{s}\,\frac{ \omx{a} \, m_b^2 +\omx{b} \, m_a^2}{m^2\, \mu_{-,ab}}
-\frac{\omx{a}\,\omx{b}}{6\,m^2} \,\Big(
M_a+M_b-\sqrt{s}-2\,m \Big)\,Z_{[10]}^2
\nonumber\\
&&\qquad +\frac{\omx{a}\,\omx{b}}{3\,m^2} \, \Big(2\,\sqrt{s}-m \Big)\,Z_{[10]}
+{\mathcal O}\left(Q^3 \right)
-\frac{ \omx{a}\,\phi_b+\phi_a\,\omx{b}}{2\,\sqrt{s}\,\mu_{-,ab}} \left( 1-\frac{s}{m^2}\right)
\nonumber\\
&& \qquad 
+\frac{\phi_a\,\phi_b}{4\,s\mu_{-,ab}}\,\Bigg( 1-\frac{7\,s}{3\,m^2} \Bigg)
+\frac{4}{3}\,\frac{ \omx{a}\,\phi_a\,\phi_b\, \omx{b}}{\mu_{+,ab}^2\,\mu_{-,ab}^3}
-\omx{a}\,\omx{b}\,\frac{(\omx{a}+\omx{b})^2}{m^2\, \mu_{-,ab}}
\nonumber\\
&&\qquad
-\frac{2}{3}\,\frac{\phi_a\,\phi_b}{ \mu^2_{-,ab}}\,\Bigg( 
\frac{\sqrt{s}\,(\omx{a}+\omx{b})}{m^2\,\,\mu_{+,ab}} 
+\frac{2\,s}{m^2}\,\frac{\omx{a}\,\omx{b}}{ \mu^2_+\,\mu_{-,ab}}
\Bigg)
-\frac{\sqrt{s}}{m^2}\,
\frac{\omx{a}^2\,\phi_b\,\omx{b} +\omx{a}\, \phi_a\,\omx{b}^2}{ \mu_{+,ab}\,\mu_{-,ab}^2}
\nonumber\\
&&\qquad
+\frac{1}{6}\left(\frac{\phi_a\,\omx{b}+\omx{a}\,\phi_b}{\sqrt{s}\,m\,\mu_{+,ab}} 
+\frac{\phi_a\,\big(4\,\sqrt{s}-\mu_{+,ab}\big)\phi_b}{3\,\sqrt{s}\,m\,\mu_{+,ab}^2\,\mu_{-,ab}}  \right) 
\Big(M_a+M_b+2\,m \Big)
\nonumber\\
&&\qquad
+\frac{2}{3}\,\frac{\phi_a\,\phi_b}{\sqrt{s}\,\mu_{-,ab}^2}\,\frac{\chi_a+\chi_b}{\mu_{+,ab}}
+\frac{\chi_a\,\chi_b}{\mu_{+,ab}}\left(\frac{\chi_a\,\phi_b+\chi_b\,\phi_a}{\sqrt{s}\,\mu_{-,ab}^2}
-\frac{\sqrt{s}}{m}\,\frac{\chi_a+\chi_b}{3\,m} \right)
\nonumber\\
&&\qquad+
\,\frac{\omx{a}\,(M_a+m )+\omx{b}\,(M_b+m )}{3\,m\,\mu^2_{+,ab}}\, \Bigg(
\frac{\omx{a}\, \phi_b+\omx{b}\, \phi_a}{\mu_{-,ab}}
+\frac{\phi_a\,\big(4\,\sqrt{s}-\mu_{+,ab}\big)\,\phi_b}{3\,\mu_{+,ab}\,\mu^2_{-,ab}}
\Bigg)
\nonumber\\
&&\qquad
-\frac{\sqrt{s}}{m}\,\frac{\chi_a\,m_b^2+\chi_b\,m_a^2}{3\,m\,\mu_{+,ab}}
-\frac{\sqrt{s}}{m}\,\frac{\chi_a\,\phi_b+\chi_b\,\phi_a}{6\,m\,\mu_{+,ab}}
+\frac{\phi_a\,\phi_b}{\mu_{-,ab}} \,\frac{2\,\sqrt{s}+\mu_{+,ab}}{12\,s\,\mu_{+,ab}}
\nonumber\\
&&\qquad-\frac{2}{9}
\frac{\phi_a\,\phi_b}{\mu_{-,ab}} \, 
\frac{(M_a+m)\,(M_b+m)}{s\,\mu^2_{+,ab}}\Bigg(
1+ \frac{2\,\sqrt{s}}{\mu_{+,ab}}\frac{\chi_a+\chi_b}{\mu_{-,ab}} \Bigg)
\nonumber\\
&&\qquad 
+\frac{1}{3\,\sqrt{s}\, m}\, \Big(\omx{a}\,\phi_b+\phi_a \,\omx{b}\Big)\,
\Bigg(Z_{[10]}-1-Z_{[10]}^2\,\left(1+\frac{\sqrt{s}}{2\,m} \right) \Bigg)
\nonumber\\
&&\qquad 
-\frac{Z_{[10]}}{3\,m^2} \,\Bigg(
 \Big({\textstyle{3\over 2}}\,\phi_a+2\,m_a^2 \Big) \,\omx{b}
+\omx{a}\,\Big({\textstyle{3\over 2}}\,\phi_b+2\,m_b^2 \Big)
\Bigg)+{\mathcal O}\left(Q^4 \right) \,,
\end{eqnarray}
and
\begin{eqnarray}
&&\Big[{ p}_{0-}^{(I)} (\sqrt{s},m)\Big]_{ab} =
\frac{M_a\,M_b}{3} \,
\Bigg( 4\,\frac{M_a+M_b+2\,m}{3\,m\,\mu_{+,ab}}
-\frac{4}{\mu_{-,ab}} -\frac{8}{3\,m} \Bigg)
\nonumber\\
&&\qquad 
- \Big(M_a+m \Big)\,
\Bigg(\frac{1}{3\,\mu_{+,ab}} +\frac{2\,\sqrt s}{3\,\mu_{-,ab}\,\mu_{+,ab}}
-\frac{8\,M_a\,M_b} {9\,\mu_{-,ab}\,\mu^2_{+,ab}}
\Bigg)\,\Big( M_b+m\Big)
\nonumber\\
&&\qquad -\frac{\omx{a}}{3\,\mu_{-,ab}}
\Bigg(\frac{2\,s\,(M_a+m)}{m\,\mu_{+,ab}}
+4\,\frac{\sqrt s \,M_a\,M_b}{m^2}
-\frac{8}{3}\,\frac{\sqrt s \,(M_a+m)}{m\,\mu_{+,ab}}
\,\frac{M_a\,M_b}{\mu_{+,ab}}
\Bigg)
\nonumber\\
&&\qquad -\frac{\omx{b}}{3\,\mu_{-,ab}}
\Bigg(\frac{2\,s\,(M_b+m)}{m\,\mu_{+,ab}}
+4\,\frac{\sqrt s \,M_a\,M_b}{m^2}
-\frac{8}{3}\,\frac{\sqrt s \,(M_b+m)}
{m\,\mu_{+,ab}}
\,\frac{M_a\,M_b}{\mu_{+,ab}}
\Bigg)
\nonumber\\
&&\qquad -8\,Z_{[10]}\,\frac{M_a\,M_b}{9\,m} \,\Bigg(
Z_{[10]}-1-3\,Z_{[10]} \,\frac{M_a+M_b}{16\,m} \Bigg)
+\frac{Z_{[10]}^2\,\sqrt s }{6\,m^2}\,
\left(s-\frac{5}{3}\,M_a\,M_b\right)
\nonumber\\
&& \qquad 
+\frac{Z_{[10]}\,(Z_{[10]}-1)}{3\,m}\,\Big( \sqrt s +M_a\Big)\,
\Big( \sqrt s +M_b\Big)
+{\mathcal O}\left( Q\right)
\nonumber\\
&& \qquad
+\Bigg( \frac{2}{9}\,\frac{(M_a+m)\,(M_b+m)}{\mu_{+,ab}^2}-\frac{1}{3}
\Bigg)\,\frac{\phi_a\,M_b+M_a\,\phi_b}{\sqrt{s}\,\mu_{-,ab}}
\nonumber\\
&&\qquad
-\frac{2}{3}\,\frac{(M_a+m)\,(M_b+m)}{\mu_{-,ab}\,\mu_{+,ab}}\,
\left(
\frac{\omx{a}\, \phi_b+\omx{b}\, \phi_a}{ \mu_{+,ab}\,\mu_{-,ab}}
+ \frac43\, \frac{\phi_a\,\sqrt{s}\,\phi_b}{\mu_{+,ab}^2\,\mu_{-,ab}^2} \right)
\nonumber\\
&&\qquad + \frac{M_a\, M_b}{\mu_{+,ab}} \,\Bigg( 
\frac83\,\frac{\omx{a}\,\omx{b}}{\mu_{-,ab}^2} \left(1-\frac{s}{m^2} \right)
-\frac{16\,\phi_a\,\phi_b}{5\,\mu_{+,ab}\,\mu_{-,ab}^3}
-\frac{8}{3}\,\frac{\omx{a}\,\phi_b+\phi_a\,\omx{b}}{\sqrt{s}\,\mu^2_{-,ab}}
\Bigg)
\nonumber\\
&& \qquad +\left(\frac{2}{3}\,\sqrt{s}-\frac{8}{9}\,\frac{M_a\,M_b}{\mu_{+,ab}}\right)
\frac{M_a+m}{\mu_{+,ab}}\,\frac{\frac12\,
\phi_a+\omx{a}\,\omx{b}+m_{a}^2}{m\,\mu_{-,ab}}
\nonumber\\
&& \qquad +\left(\frac{2}{3}\,\sqrt{s}-\frac{8}{9}\,\frac{M_a\,M_b}{\mu_{+,ab}}\right)
\frac{M_b+m}{\mu_{+,ab}}\,\frac{\frac12\,
\phi_b+\omx{a}\,\omx{b}+m_{b}^2}{m\,\mu_{-,ab}}
\nonumber\\
&& \qquad 
-\left(\frac{2}{3}\,\sqrt{s}-\frac{8}{9}\,\frac{M_a\,M_b}{\mu_{+,ab}}\right)
\frac{s}{m^2}\,\frac{\omx{a}\,\omx{b}}{\mu_{+,ab}\,\mu_{-,ab}}
+\frac{4}{9}\,\frac{M_a\,\sqrt{s}\,M_b}{m^2 }\,\frac{\omx{a}+\omx{b}}{\mu_{+,ab}}
\nonumber\\
&& \qquad 
+\frac{m+M_a}{\mu_{+,ab}}\,\Bigg( 
\frac{16}{9}\,\frac{M_a\,M_b}{\sqrt{s}\,\mu_{+,ab}}\,\frac{\chi_a\,\phi_b+\chi_b\,\phi_a}{\mu_{-,ab}^2}
+\frac{32}{15}\,\frac{M_a\,M_b}{\mu_{+,ab}^2}\,\frac{\phi_a\,\phi_b}{\mu_{-,ab}^3}
\Bigg)\,\frac{m+M_b}{\mu_{+,ab}}  
\nonumber\\
&& \qquad 
-\frac{1}{3}\,\frac{\sqrt{s}}{m}\left( \frac{m+M_a}{\mu_{+,ab}}\,\chi_a
+\frac{m+M_b}{\mu_{+,ab}}\,\chi_b \right)
+\frac{4}{3}\,\frac{M_a\,\sqrt{s}\,M_b}{m^2\,\mu_{+,ab}}\,\frac{\chi_a\,m_b^2+\chi_b\,m_a^2}{\mu_{-,ab}^2}
\nonumber\\
&& \qquad 
+\left( \frac{8}{3}\,\frac{\chi_a\,\chi_b}{\mu_{-,ab}}\,
+\frac{4}{3}\,\frac{m_a^2+m_b^2}{\mu_{-,ab}}+\frac{2}{3}\,\frac{\phi_a+\phi_b}{\mu_{-,ab}}
+\frac{4\,\sqrt{s}}{\mu_{+,ab}}\,\frac{\chi_a^2\,\chi_b+\chi_b^2\,\chi_a}{\mu_{-,ab}^2}\
\right) \frac{M_a\,M_b}{m^2}
\nonumber\\
&& \qquad 
-\frac{1}{3}\,\Big( \sqrt{s}-m \Big) +\frac{\sqrt{s}}{3\,m}\,
\Bigg(1-\frac{\sqrt{s}}{m}\,Z_{[10]}^2+\frac{4}{3}\,\frac{M_a\,M_b}{\sqrt{s}\,m}\,Z_{[10]} \Bigg)
\,\Big( \omx{a}+\omx{b}\Big)
\nonumber\\
&& \qquad 
+\frac{2\,\sqrt{s}}{3\,m^2} \,\Big( s-M_a\,M_b \Big)\,Z_{[10]}
+{\mathcal O}\left( Q^2\right)\;,
\label{}
\end{eqnarray}
and
\begin{eqnarray}
&&\Big[{ p}_{1+}^{(I)} (\sqrt{s},m)\Big]_{ab} =
-\frac{1}{3\,\mu_{-,ab}}
+ \frac29\,
\frac{(m+M_a)\,(m+M_b)}
{\mu_{+,ab}^2\,\mu_{-,ab}}
+\frac19\, \frac{M_a+M_b}{m\,\mu_{+,ab}}
\nonumber\\
&& \qquad 
-\frac{\sqrt s }{3\,m}\,\frac{\omx{a}}
{\mu_{-,ab}}\,
\Bigg(\frac{1}{m}-\frac23\,\frac{m+M_a}
{\mu^2_{+,ab}}
\Bigg)
-\frac{\sqrt s }{3\,m}\,\frac{\omx{b}}
{\mu_{-,ab}}\,
\Bigg(\frac{1}{m}-\frac{2}{3}\,\frac{m+M_b}
{\mu_{+,ab}^2} \Bigg)
\nonumber\\
&&\qquad 
+\frac{2}{9\,\mu_{+,ab}}-\frac{2}{9\,m}
-\frac{Z_{[10]}^2}{9\,m^2}\,\Big(\sqrt s +2\,m \Big)
+\frac{2\,Z_{[10]}}{9\,m}+{\mathcal O}\left( Q\right)
\nonumber\\
&& \qquad 
+\frac{2}{3}\,\frac{{\textstyle{3\over 2}}\,\phi_a+2\,m_a^2 }{m^2\,\mu_{-,ab}^2}\,
\frac{\sqrt{s}\,\omx{b}}{\mu_{+,ab}} 
+\frac{2}{3}\,\frac{\sqrt{s}\,\omx{a}}{\mu_{+,ab}}\,
\frac{{\textstyle{3\over 2}}\,\phi_b+2\,m_b^2 }{m^2\,\mu_{-,ab}^2}
-\frac{4}{5}\,\frac{\phi_a\,\phi_b}{\mu_{+,ab}^2\,\mu_{-,ab}^3}
\nonumber\\
&& \qquad 
+\frac{2}{3}\, \frac{\omx{a}\,\omx{b}}{\mu_{+,ab}\,\mu_{-,ab}^2}\left( 1-\frac{s}{m^2}\right)
+\frac{2}{9}\,\frac{\omx{a}\,s\,\omx{b}}{m^2\,\mu^2_{+,ab}\,\mu_{-,ab}}
+\frac{\sqrt{s}\,(\omx{a}+\omx{b})}{9\,m^2\,\mu_{+,ab}} 
\nonumber\\
&& \qquad -\frac{2\,(M_a+m)}{9\,m}\,
\frac{\frac12\,
\phi_a+\omx{a}\,\omx{b}+m_a^2}{\mu_{-,ab}\,\mu_{+,ab}^2}
-\frac{2\,(M_b+m)}{9\,m}\,
\frac{\frac12\,
\phi_b+\omx{a}\,\omx{b}+m_b^2}{\mu_{-,ab}\,\mu_{+,ab}^2}
\nonumber\\
&& \qquad 
-\frac{2}{3}\,
\frac{\omx{a}\,\phi_b+\phi_a\,\omx{b}}{\sqrt{s}\,\mu_+\,\mu^2_-}
+\frac{\phi_a\,\phi_b}{\mu_{-,ab}^3}\,
\frac{(m+M_a)\,(m+M_b)}{\mu_{+,ab}^4}\,\Bigg( \frac{8}{15}-
\frac{4}{45}\,\frac{\mu_{+,ab}}{\sqrt{s}}\Bigg)
\nonumber\\
&& \qquad 
+\frac{2}{3}\,\frac{\chi_a\,\chi_b}{m^2\,\mu_{-,ab}}
+\frac{m_a^2+m_b^2}{3\,m^2\,\mu_{-,ab}}
+\frac{1}{6}\,\frac{\phi_a+\phi_b}{m^2\,\mu_{-,ab}}
\nonumber\\
&& \qquad 
+\frac{4}{9}\,\frac{(m+M_a)\,(m+M_b)}{\sqrt{s}\,\mu_{-,ab}^2\,\mu_{+,ab}^3}\,
\Big(\chi_a\,\phi_b+\chi_b\,\phi_a \Big)
 + \frac{\omx{a}+\omx{b}}{9\,m^2} \,Z_{[10]}+{\mathcal O}\left( Q^2\right)\;,
\label{u-approx-2}
\end{eqnarray}
and 
\begin{eqnarray}
&&\Big[{ p}_{1-}^{(I)} (\sqrt{s},m)\Big]_{ab} =
-\frac{16}{15}\, \frac{M_a\,M_b}{\big(\mu_{-,ab}\big)^2\,\mu_{+,ab}}
+\frac{32}{45}\, \frac{(m+M_a)\,M_a\,M_b\,(m+M_b)}
{\big(\mu_{-,ab}\big)^2\,\big(\mu_{+,ab}\big)^3}
\nonumber\\
&&\qquad -\frac{4}{9}\, \frac{(m+M_a)\,\sqrt{s}\,(m+M_b)}
{\big(\mu_{-,ab}\big)^2\,\big(\mu_{+,ab}\big)^2}
+{\mathcal O}\left( Q^{-1}\right) 
\nonumber\\
&&\qquad +\frac{1}{3}\,\frac{\mu_{+,ab}+2\,\sqrt{s}}{\mu_{+,ab}\,\mu_{-,ab}}
-\frac{8}{15}\,\frac{M_a\,M_b}{m^2\,\mu_{-,ab}}
-\frac{2}{9}\,\frac{(m+M_a)\,(m+M_b)}{\mu_{-,ab}\,\mu_{+,ab}^2}
\nonumber\\
&& \qquad -\frac{16}{15}\,\frac{M_a\,\sqrt{s}\,M_b}{m^2}\,
\frac{\omx{a}+\omx{b}}{\mu_{+,ab}\,\mu_{-,ab}^2}
+\frac{16}{45}\,\frac{M_a\,M_b}{\mu_{-,ab}}\,\frac{M_a+M_b+2\,m}{m\,\mu_{+,ab}^2}
\nonumber\\
&& \qquad -\frac{4}{9}\,\frac{\omx{a}\,s\,(m+M_a)}{m\,\mu_{-,ab}^2\,\mu_{+,ab}^2}
-\frac{4}{9}\,\frac{(m+ M_b)\,s\,\omx{b}}{m\,\mu_{-,ab}^2\,\mu_{+,ab}^2}
-\frac{2}{9}\,\frac{\sqrt{s}}{m } \,\frac{2\,m +M_a+M_b}{\mu_{+,ab}\,\mu_{-,ab}}
\nonumber\\
&& \qquad +\frac{32}{45}\,\frac{M_a\,\sqrt{s}\,M_b}{m\,\mu_{+,ab}^3\,\mu_{-,ab}^2}
\Big( (m +M_a)\,\omx{a}+(m+ M_b)\,\omx{b}\Big)
+{\mathcal O}\left( Q^{0}\right) \;. 
\label{u-approx-2:b}
\end{eqnarray}
This appendix ends with the d-wave resonance functions $q_{n\pm}^{(I)}(\sqrt{s},m)$
\begin{eqnarray}
\Big[V^{(I,\pm )}_{u-[9]}(\sqrt{s};n)\Big]_{ab} &=& \sum_{c=1}^4\,
\frac{1}{4\,f^2 }\,\Big[\widetilde C^{(I,c)}_{[9]}\Big]_{ab}\,
\Big[{ q}^{(I)}_{n \pm }(\sqrt{s},m^{(c)}_{[9]})\Big]_{ab} \;.
\label{u-result-8:appendix}
\end{eqnarray}
One may or may not apply the questionable formal rule $\sqrt{s}-m_{[9]}\sim Q$. The explicit expressions below, 
which rely on $\sqrt{s}-m_{[9]}\sim Q$ and $\mu_- \sim Q$, demonstrate that our total result to order $Q^3$ are 
basically independent on this assumption. Modified results appropriate for an expansion with 
$\sqrt{s}-m_{[9]}\sim Q^0$ and $\mu_- \sim Q^0$ follow upon dropping some terms proportional to $(1/\mu_-)^n$.
We derive
\begin{eqnarray}
&& [q_{0+}^{(I)}(\sqrt{s},m)]_{ab}=
-\frac{ \omx{a}\,\omx{b}}{3\,m^2} \, \Big(2\,\sqrt{s}+m \Big)\,Z_{[9]}
\nonumber\\
&&\qquad
 +\frac{\omx{a}\,\omx{b}}{6\,m^2} \,\Big(
M_a+M_b-\sqrt{s}+2\,m \Big)\,Z_{[9]}^2
+{\mathcal O}\left(Q^3 \right)
\nonumber\\
&& \qquad 
-\frac{\omx{a}\,\omx{b}}{\mu_{+,ab}}\, \left(1-\frac{s}{m^2}\right)
-\frac{\sqrt{s}\,\omx{a}}{m^2}\,
\frac{{\textstyle{3 \over 2}}\,\phi_b+2\,m_b^2}{\mu_{+,ab}}
-\frac{{\textstyle{3 \over 2}}\,\phi_a+2\,m_a^2 }{\mu_{+,ab}}\,
\frac{\sqrt{s}\,\omx{b}}{m^2}
\nonumber\\
&&\qquad
+\frac{\omx{a}\,\phi_b+\phi_a\,\omx{b}}{6\,\sqrt{s}\,\mu_{+,ab}}
+\frac{2}{9}\,\frac{\phi_a\,\phi_b}{\mu^2_+\,\mu_{-,ab}}
\left(1+2\,\frac{\mu_{+,ab}}{m} \right)
-\frac{1}{9}\,\frac{\phi_a\,\phi_b}{m^2\,\mu_{-,ab}}
\nonumber\\
&&\qquad
-\frac{1}{3\,m^2\mu_{-,ab}}\,\left(\omx{a}\,\mu_{-,ab}+{\textstyle{1 \over 2}}\,\phi_a \right)
\left(\omx{b}\,\mu_{-,ab}+{\textstyle{1 \over 2}}\,\phi_b \right)
\nonumber\\
&&\qquad
+\frac{1}{3\,m^2} \,\left(
2\,m_a^2\,\omx{b}+2\,\omx{a}\,m_b^2 
+\Big(\omx{a}\,\phi_b+\phi_a\,\omx{b}\Big)\left(\frac{3}{2}+
\frac{m}{\sqrt{s}}  \right)\right)Z_{[9]}
\nonumber\\
&&\qquad
+\frac{1}{6\,m^2} \,
\Big(\omx{a}\,\phi_b+\phi_a\,\omx{b}\Big)\,\left(1 -2\,\frac{m}{\sqrt{s}} \right)
Z_{[9]}^2
+{\mathcal O}\left(Q^4 \right)\;,
\end{eqnarray}
and
\begin{eqnarray}
&& [q_{0-}^{(I)}(\sqrt{s},m)]_{ab}= \frac{1}{3}\,(\sqrt{s}+m)
+\frac{4\,M_a\,M_b}{9\,\mu_{+,ab}}
\nonumber\\
&& \qquad
+\frac{Z_{[9]}^2}{6\,m^2}\,\Big( 
M_a\,\Big({\textstyle{5\over3}}\,\sqrt{s}-{\textstyle{16\over3}}\,m-M_a-M_b\Big)\,M_b 
-s\,\sqrt{s} \Big)
\nonumber\\
&& \qquad  +\frac{Z_{[9]}\,(Z_{[9]}-1)}{3\,m} \, 
(\sqrt{s}+M_a) \,(\sqrt{s}+M_b)+\frac{8\,Z_{[9]}}{9\,m} \,M_a\,M_b
+{\mathcal O}\left(Q \right)
\nonumber\\
&& \qquad 
+\frac{2\,\omx{a}\,\sqrt{s}\,\omx{b}}{\mu_{+,ab}\,\mu_{-,ab}} \left(1-\frac{s}{m^2}\right)
+\frac{1}{3}\,\left(1+2\,\frac{\sqrt{s}}{\mu_{+,ab}}\right)\,\mu_{-,ab}
+\sqrt{s}\,\frac{\omx{a}+\omx{b}}{3\,m}
\nonumber\\
&& \qquad
+4\,M_a\,M_b\,\frac{ \omx{a}+\omx{b}}{m\,\mu_{+,ab}}\left(
\frac{\sqrt{s}}{3\,m}-\frac{1}{9}\,\frac{m-\sqrt{s}}{m\,\mu_{-,ab}}\,\Big(2\,m-\mu_{+,ab} \Big) \right)
\nonumber\\
&& \qquad 
-\frac{4\,M_a\,\big(2\,m-\mu_{+,ab}\big)\,M_b}{9\,m^2\,\mu_{+,ab}\,\mu_{-,ab}}
\Bigg(2\,\omx{a}\,\omx{b}
+m_a^2+m_b^2+{\textstyle{1\over 2}} \,(\phi_a+\phi_b)\Bigg)
\nonumber\\
&& \qquad
+\frac{\omx{a}+\omx{b}}{3\,m^2} 
\left(s\,Z_{[9]}-\frac{4}{3}\,M_a\,M_b \right)\,Z_{[9]}
-\frac{2\,\sqrt{s}}{3\,m^2} \,\Big( s-M_a\,M_b\Big)\,Z_{[9]}
+{\mathcal O}\left(Q^2 \right) \;,
\nonumber\\
\end{eqnarray}
and
\begin{eqnarray}
&&[q_{1+}^{(I)}(\sqrt{s},m)]_{ab}
=\frac{1}{9\,\mu_{+,ab}} +\frac{Z_{[9]}^2}{9\,m^2} \,\Big( \sqrt{s} -2\,m\Big)
+\frac{2\,Z_{[9]}}{9\,m} 
+{\mathcal O}\left( Q\right)
\nonumber\\
&& \qquad-\frac{2\,m-\mu_{+,ab}}{9\,m^2\,\mu_{+,ab}\,\mu_{-,ab}}
\Bigg( 2\,\omx{a}\,\omx{b}
+m_a^2+m_b^2+{\textstyle{1\over 2}} \,(\phi_a+\phi_b)\Bigg)
-\frac{\sqrt{s}}{m}\,\frac{\chi_a+\chi_b}{9\,m\,\mu_{-,ab}}
\nonumber\\
&& \qquad
+\frac{ \omx{a}+\omx{b}}{m\,\mu_{+,ab}}\left(
\frac{\sqrt{s}}{3\,m}-\frac{2}{9}\,\frac{m-\sqrt{s}}{\mu_{-,ab}} \right)
-\frac{\omx{a}+\omx{b}}{9\,m^2} \,Z_{[9]}
+{\mathcal O}\left( Q^2\right)\;,
\end{eqnarray}
and
\begin{eqnarray}
&&[q_{1-}^{(I)}(\sqrt{s},m)]_{ab}=0+{\mathcal O}\left( Q^{-1}\right)
-\frac{2\,\sqrt{s}}{3\,\mu_{+,ab}\,\mu_{-,ab}}
+\frac{4}{15}\,\frac{4\,M_a\,M_b}{\mu_{-,ab}\,\mu_{+,ab}^2}
\nonumber\\
&& \qquad 
+\frac{8}{45}\,\frac{4\,M_a\,M_b}{\mu_{-,ab}\,\mu_{+,ab}\,m}
-\frac{2}{45}\,\frac{4\,M_a\,M_b}{\mu_{-,ab}\,m^2}
-\frac{32}{45}\,\frac{M_a\,M_b}{\mu_{-,ab}\,\mu_{+,ab}^2}
+{\mathcal O}\left( Q^{0}\right)\,.
\end{eqnarray}

\section{Perturbative threshold analysis}

Close to threshold the scattering phase
shifts $\delta^{(L)}_{J=L\pm \frac{1}{2}} (s)$ are characterized by the
scattering length or scattering volumes $a_{[L_{2I,2J}]}$
\begin{eqnarray}
\Re \,f^{(L)}_{I,J=L\pm \frac{1}{2}} (s) &=&
q^{2\,L}\,\Big(
a_{[L_{2I,2J}]}+b_{[L_{2I,2J}]}\,q^2+{\mathcal Q}\left( q^{4}\right)
\Big) \;,
\nonumber\\
q^{2\,L+1}\,\cot \delta^{(L)}_{I,J=L\pm \frac{1}{2}} (s) &=&
\frac{1}{a_{[L_{2I,2J}]}}+\frac{1}{2}\,r_{[L_{2I,2J}]}\, q^2
+{\mathcal Q}\left( q^{4}\right) \,,
\label{}
\end{eqnarray}
where we include the isospin in the notation for completeness. For s-wave 
channels one finds $b=-r\,a^2/2$. The threshold parameters, $a^{(I,\pm)}_{\pi N,n}$,
can equivalently be extracted from the threshold values of the reduced amplitudes
\begin{eqnarray}
4\,\pi\,\left(1+\frac{m_\pi}{m_N}\right) a^{(I,+)}_{\pi N,n} &=& 
M_{\pi N}^{(I,+ )}(m_N+m_\pi;n) \;,
\nonumber\\
4\,\pi\,\left(1+\frac{m_\pi}{m_N}\right) a^{(I,-)}_{\pi N,n} &=& \frac{1}{4\,m_N^2}\,
M_{\pi N}^{(I,-)}(m_N+m_\pi;n) \;.
\label{}
\end{eqnarray}
We identify the s and p-wave threshold parameters 
$a^{(I,+)}_{\pi N,0}=a^{(\pi N)}_{[S_{2I,1}]}$, 
$a^{(I,-)}_{\pi N,0}=a^{(\pi N)}_{[P_{2I,1}]}$ and 
$a^{(I,+)}_{\pi N,1}=a^{(\pi N)}_{[P_{2I,3}]}$ for which all terms up 
to chiral order $Q^2$ are collected. The s-wave pion-nucleon scattering lengths are
\begin{eqnarray}
&&4\,\pi \left( 1+\frac{m_\pi}{m_N}\right)
a^{(\pi N)}_{[S_-]}=\frac{m_\pi}{2\,f^2}+{\mathcal O}\left(Q^3 \right)\; ,
\nonumber\\
&&4\,\pi \left( 1+\frac{m_\pi}{m_N}\right)
a^{(\pi N)}_{[S_+]}=-\frac{2}{9}\,\frac{C_{[10]}^2}{f^2}\,
\frac{m_\pi^2}{m_{\Delta}}\,
\,\big(2-Z_{[10]} \big) \left(1+Z_{[10]}+\big(2-Z_{[10]}\big)\,\frac{m_N}{2\,m_\Delta
}\right)
\nonumber\\
&&\quad \quad \quad
+\Big(2\,g^{(S)}_0+g^{(S)}_D+g^{(S)}_F \Big)\, \frac{m_\pi^2}{4 f^2}
+m_N\,\Big(2\,g^{(V)}_0+g^{(V)}_D+g^{(V)}_F \Big)\,\frac{m_\pi^2 }{4 f^2}
\nonumber\\
&&\quad \quad \quad
- \frac{g_A^2\,m_\pi^2}{4\,f^2\,m_N}
-2\,\Big(2\,b_0+b_D+b_F\Big)\frac{m_\pi^2}{f^2}
+{\mathcal O}\left(Q^3 \right) \,,
\label{piN-s}
\end{eqnarray}
where $3\,a_{[S_+]}=a_{[S_{11}]}+2\,a_{[S_{31}]}$ and $3\,a_{[S_-]}=a_{[S_{11}]}-a_{[S_{31}]}$ 
and $F_{[8]}+D_{[8]}=g_A$. For the s-wave range parameter we find
\begin{eqnarray}
&&4\,\pi \left( 1+\frac{m_\pi}{m_N}\right)
b^{(\pi N)}_{[S_-]}=\frac{1}{4\,f^2\,m_\pi}
-\frac{2\,g_A^2+1}{4\,f^2\,m_N}
\nonumber\\
&&\quad \quad \quad
+\frac{C_{[10]}^2}{18\,f^2}\,\frac{m_\pi}{m_N\,(\mu_\Delta+m_\pi)}
+{\mathcal O}\left(Q \right)\; ,
\nonumber\\
&&4\,\pi \left( 1+\frac{m_\pi}{m_N}\right)
b^{(\pi N)}_{[S_+]}= \frac{g_A^2}{4\,f^2\,m_N}
+\frac{C_{[10]}^2}{9\,f^2}\,\frac{m_\pi}{m_N\,(\mu_\Delta+m_\pi)}
\nonumber\\
&&\quad \quad \quad
+\frac{1}{4 f^2}\,\Big( 2g^{(S)}_0+g^{(S)}_D +g^{(S)}_F \Big)
+\frac{m_N}{4 f^2}\,\Big(2\,g^{(V)}_0+g^{(V)}_D+g^{(V)}_F \Big)
\nonumber\\
&&\quad \quad \quad
-\frac{2}{9}\,\frac{C_{[10]}^2}{f^2}\,
\frac{1}{m_{\Delta}}\,
\,\big(2-Z_{[10]} \big) \left(1+Z_{[10]}+\big(2-Z_{[10]}\big)\,\frac{m_N}{2\,m_\Delta
}\right)+{\mathcal O}\left(Q \right)
\label{piN-s-range}
\end{eqnarray}

The p-wave scattering volumes are
\begin{eqnarray}
&&4\,\pi \left( 1+\frac{m_\pi}{m_N}\right)
a^{(\pi N)}_{[P_{11}]}= -\frac{2\,g_A^2}{3\,m_\pi\,f^2}
+\frac{8\,C_{[10]}^2}{27\,f^2}\,\frac{1}{\mu_\Delta+m_\pi}
+\frac{3-4\,g_A^2}{6\,f^2\,m_N}
\nonumber\\
&&\quad \quad \quad+\frac{16}{27}\,\frac{C_{[10]}^2}{f^2}\,
\frac{m_\pi}{m_\Delta\,(\mu_\Delta+m_\pi)}
-\frac{C_{[10]}^2}{f^2}\,\frac{2\,Z_{[10]}}{27\,m_{\Delta}}\,
\left(1-Z_{[10]}-Z_{[10]}\,\frac{m_N}{2\,m_\Delta}\right)
\nonumber\\
&&\quad \quad \quad
-\frac{1}{12 f^2}\,\Big( 2\,g_0^{(S)}+g_D^{(S)}+g_F^{(S)} \Big)
+\frac{1}{3 f^2}\,\Big( g_D^{(T)}+g_F^{(T)} \Big)
+{\mathcal O}\left(Q \right)
\nonumber\\
&&4\,\pi \left( 1+\frac{m_\pi}{m_N}\right)
a^{(\pi N)}_{[P_{31}]}= -\frac{g_A^2}{6\,m_\pi\,f^2}
+\frac{2\,C_{[10]}^2}{27\,f^2}\,\frac{1}{\mu_\Delta+m_\pi}
-\frac{2\,g_A^2+3}{12\,f^2\,m_N}
\nonumber\\
&&\quad \quad \quad+\frac{4}{27}\,\frac{C_{[10]}^2}{f^2}\,
\frac{m_\pi} {m_\Delta\,(\mu_\Delta+m_\pi)}
+\frac{C_{[10]}^2}{f^2}\,\frac{4\,Z_{[10]}}{27\,m_{\Delta}}\,
\left(1-Z_{[10]}-Z_{[10]}\,\frac{m_N}{2\,m_\Delta}\right)
\nonumber\\
&&\quad \quad \quad
-\frac{1}{12 f^2}\,\Big( 2\,g_0^{(S)}+g_D^{(S)}+g_F^{(S)} \Big)
-\frac{1}{6 f^2}\,\Big( g_D^{(T)}+g_F^{(T)} \Big)
+{\mathcal O}\left(Q \right) \,,
\label{piN-pi1}
\end{eqnarray}
and
\begin{eqnarray}
&&4\,\pi \left( 1+\frac{m_\pi}{m_N}\right)
a^{(\pi N)}_{[P_{13}]}= -\frac{g_A^2}{6\,m_\pi\,f^2}
+\frac{2\,C_{[10]}^2}{27\,f^2}\,\frac{1}{\mu_\Delta+m_\pi}
-\frac{g_A^2}{6\,f^2\,m_N}
\nonumber\\
&&\quad \quad \quad+\frac{4}{27}\,\frac{C_{[10]}^2}{f^2}\,
\frac{m_\pi}{m_\Delta\,(\mu_\Delta+m_\pi)}
+\frac{C_{[10]}^2}{f^2}\,\frac{4\,Z_{[10]}}{27\,m_{\Delta}}\,
\left(1-Z_{[10]}-Z_{[10]}\,\frac{m_N}{2\,m_\Delta}\right)
\nonumber\\
&&\quad \quad \quad
-\frac{1}{12 f^2}\,\Big( 2\,g_0^{(S)}+g_D^{(S)}+g_F^{(S)} \Big)
-\frac{1}{6 f^2}\,\Big( g_D^{(T)}+g_F^{(T)} \Big)
+{\mathcal O}\left(Q \right)
\nonumber\\
&&4\,\pi \left( 1+\frac{m_\pi}{m_N}\right)
a^{(\pi N)}_{[P_{33}]}= \frac{g_A^2}{3\,m_\pi\,f^2}
+\frac{C_{[10]}^2}{54\,f^2}\,\left( \frac{1}{\mu_\Delta+m_\pi}
+\frac{9}{\mu_\Delta-m_\pi}\right)
\nonumber\\
&&\quad \quad \quad+\frac{1}{27}\,\frac{C_{[10]}^2}{f^2}\,
\frac{m_\pi}{m_\Delta\,(\mu_\Delta+m_\pi)}
+\frac{C_{[10]}^2}{f^2}\,\frac{Z_{[10]}}{27\,m_{\Delta}}\,
\left(1-Z_{[10]}-Z_{[10]}\,\frac{m_N}{2\,m_\Delta}\right)
\nonumber\\
&&\quad \quad \quad
+\frac{g_A^2}{3\,f^2\,m_N}
-\frac{1}{12 f^2}\,\Big( 2\,g_0^{(S)}+g_D^{(S)}+g_F^{(S)} \Big)
+\frac{1}{12 f^2}\,\Big( g_D^{(T)}+g_F^{(T)} \Big)
\nonumber\\
&&\quad \quad \quad 
+{\mathcal O}\left(Q \right) \,,
\label{piN-pi3}
\end{eqnarray}
where $\mu_\Delta =m_\Delta-m_N\sim Q$.
Note the Z-dependence in (\ref{piN-s},\ref{piN-pi1},\ref{piN-pi3}) can be
completely absorbed into the quasi-local 4-point coupling strength. Expressing the threshold
parameters (\ref{piN-s},\ref{piN-pi1},\ref{piN-pi3}) in terms of renormalized coupling
constants $\tilde g$ as 
\begin{eqnarray}
2\,g^{(V)}_0+g^{(V)}_D+g^{(V)}_F&=&
2\,\widetilde{g}^{(V)}_0+\widetilde{g}^{(V)}_D+
\widetilde{g}^{(V)}_F
\nonumber
\\ &+& \frac{8}{9}\,
\frac{2-Z_{[10]}}{m_{\Delta}\,m_N}\,
 \left(1+\frac{m_N}{m_\Delta} +Z_{[10]} \left( 1-\frac{m_N}{2\,m_\Delta}\right) \right) C_{[10]}^2\;,
\nonumber\\
2\,g^{(S)}_0+g^{(S)}_D+g^{(S)}_F&=&
2\,\widetilde{g}^{(S)}_0+\widetilde{g}^{(S)}_D
\nonumber
\\ &+&
\widetilde{g}^{(S)}_F+\frac89\,\frac{Z_{[10]}}{m_{\Delta}}\,
\left(1-Z_{[10]}\left(1+\frac{m_N}{2\,m_\Delta}\right) \right)C_{[10]}^2 \;,
\nonumber\\
g^{(T)}_D+g^{(T)}_F&=&\widetilde{g}^{(T)}_D+
\widetilde{g}^{(T)}_F + \frac49\,\frac{Z_{[10]}}{m_{\Delta}}\,
\left(1-Z_{[10]}\left(1+\frac{m_N}{2\,m_\Delta}\right)\right) C_{[10]}^2\;,
\label{Z-absorb}
\end{eqnarray}
leads to results which do not depend on $Z_{[10]}$ explicitly. Similarly one can absorb the 
decuplet pole terms $1/(\mu_\Delta \pm m_\pi)$ by expanding in the ratio 
$m_\pi/ \mu_\Delta $.

We turn to the strangeness channels. It only makes sense to provide the 
p-wave scattering volumes of the strangeness plus channel, because all 
other threshold parameters are non-perturbative.  
The leading orders expressions are
\begin{eqnarray}
&&4\,\pi\, \left( 1+\frac{m_K}{m_N}\right)\, a^{(KN)}_{[P_{01}]}
=\frac{1}{36\,f^2}\,\left(
\frac{(3\,F_{[8]}+D_{[8]})^2}{\mu^{(\Lambda )}_{[8]}+m_K}
-9\,\frac{(D_{[8]}-F_{[8]})^2}{\mu_{[8]}^{(\Sigma )}+m_K}\right)\,\Big(1+\frac{m_K}{m_N}\Big)
\nonumber\\
&&\qquad\quad+\frac{C_{[10]}^2}{9\,f^2\,m^{(\Sigma )}_{[10]}}
\left( \frac{m_{[10]}^{(\Sigma )}+2\,m_K}{\mu^{(\Sigma )}_{[10]}+m_K}\,
-\frac{Z_{[10]}}{4}\left(1-Z_{[10]}-Z_{[10]}\,\frac{m_N+m_K}{2\,m^{(\Sigma )}_{[10]}}\right) \right)
\nonumber \\
&&\qquad\quad -\frac{1}{12 f^2}\,\Big(2\,g_0^{(S)}-g_1^{(S)}-2\,g_F^{(S)}\Big)
+\frac{1}{6 f^2}\,\left(1+\frac32\,\frac{m_K}{m_N}\right)\,
\Big(g_1^{(T)}+2\,g_D^{(T)}\Big)
\nonumber\\
&&\qquad\quad+ {\mathcal O}\left(Q \right)\,,
\nonumber\\
&&4\,\pi\, \left( 1+\frac{m_K}{m_N}\right)\, a^{(KN)}_{[P_{21}]}
=
-\frac{1}{36\,f^2}\,\left(
\frac{(3\,F_{[8]}+D_{[8]})^2}{\mu_{[8]}^{(\Lambda )}+m_K}
+3\,\frac{(D_{[8]}-F_{[8]})^2}{\mu_{[8]}^{(\Sigma )}+m_K}\right)\,\Big(1+\frac{m_K}{m_N}\Big)
\nonumber\\
&&\qquad\quad+\frac{C_{[10]}^2}{27\,f^2\,m^{(\Sigma )}_{[10]}}
\left( \frac{m^{(\Sigma )}_{[10]}+2\,m_K}{\mu_{[10]}^{(\Sigma )}+m_K}\,
-\frac{Z_{[10]}}{4}\left(1-Z_{[10]}-Z_{[10]}\,\frac{m_N+m_K}{2\,m^{(\Sigma )}_{[10]}}\right) \right)
\nonumber \\
&&\qquad\quad -\frac{1}{12 f^2}\,\Big(2\,g_0^{(S)}+g_1^{(S)}+2\,g_D^{(S)}\Big)
-\frac{1}{6 f^2}\,\left(1+\frac32\,\frac{m_K}{m_N}\right)\,
\Big(g_1^{(T)}+2\,g_F^{(T)}\Big)
\nonumber\\
&&\qquad\quad-\frac{1}{2\,m_N\,f^2}+ {\mathcal O}\left(Q \right) \,,
\label{kn-p-wave-1}
\end{eqnarray}
and
\begin{eqnarray}
&&4\,\pi\, \left( 1+\frac{m_K}{m_N}\right)\, a^{(KN)}_{[P_{03}]}
=\frac{1}{18\,f^2}\,\left(9\,\frac{(D_{[8]}-F_{[8]})^2}{\mu_{[8]}^{(\Sigma )}+m_K}
-\frac{(3\,F_{[8]}+D_{[8]})^2}{\mu_{[8]}^{(\Lambda )}+m_K}
\right)\,\Big(1+\frac{m_K}{m_N}\Big)
\nonumber\\
&&\qquad\quad+\frac{C_{[10]}^2}{36\,f^2\,m^{(\Sigma )}_{[10]}}\left( 
\frac{m^{(\Sigma )}_{[10]}+2\,m_K}{\mu_{[10]}^{(\Sigma )}+m_K}
+2\,Z_{[10]}\left(1-Z_{[10]}\left( 1+ \frac{m_N+m_K}{2\,m^{(\Sigma )}_{[10]}}\right) \right)
\right)
\nonumber \\
&&\qquad\quad -\frac{1}{12 f^2}\,\Big(2\,g_0^{(S)}-g_1^{(S)}-2\,g_F^{(S)}\Big)
-\frac{1}{12 f^2}\,\Big(g_1^{(T)}+2\,g_D^{(T)}\Big)
+{\mathcal O}\left(Q \right)\,,
\nonumber\\
&&4\,\pi\, \left( 1+\frac{m_K}{m_N}\right)\, a^{(KN)}_{[P_{23}]}
=\frac{1}{18\,f^2}\,\left(
\frac{(3\,F_{[8]}+D_{[8]})^2}{\mu_{[8]}^{(\Lambda )}+m_K}
+3\,\frac{(D_{[8]}-F_{[8]})^2}{\mu_{[8]}^{(\Sigma )}+m_K}\right)\,\Big(1+\frac{m_K}{m_N}\Big)
\nonumber\\
&&\qquad\quad+\frac{C_{[10]}^2}{108\,f^2\,m^{(\Sigma )}_{[10]}}\left( 
\frac{m^{(\Sigma )}_{[10]}+2\,m_K}{\mu_{[10]}^{(\Sigma )}+m_K}\,
+2\,Z_{[10]} \left(1-Z_{[10]}\left( 1+\frac{m_N+m_K}{2\,m^{(\Sigma )}_{[10]}}\right) \right) \right)
\nonumber \\
&&\qquad\quad -\frac{1}{12 f^2}\,\Big(2\,g_0^{(S)}+g_1^{(S)}+2\,g_D^{(S)}\Big)
+\frac{1}{12 f^2}\,
\Big(g_1^{(T)}+2\,g_F^{(T)}\Big)
+{\mathcal O}\left(Q \right)\,,
\label{kn-p-wave-2}
\end{eqnarray}
where $\mu_{[8,10]}^{(Y)}=m_{[8,10]}^{(Y)}-m_N$ with $H=\Lambda ,\, \Sigma $. Note that we included 
in (\ref{kn-p-wave-1},\ref{kn-p-wave-2}) large kinematical correction terms $\sim m_K$ 
of formal order $Q$ induced by the covariant chiral counting assignment scheme at 
leading order.

Note that the p-wave scattering volumes of the kaon-nucleon sector probe four independent 
combinations of background terms as compared to the p-wave scattering volumes  of the 
pion-nucleon sector which probe only two combinations. It is possible to form a particular 
combination which does not depend on the hyperon u-channel exchange dynamics
\begin{eqnarray}
&&4\,\pi\, \left( 1+\frac{m_K}{m_N}\right)\, \left(
2\,a^{(KN)}_{[P_{01}]} +a^{(KN)}_{[P_{03}]}-6\,a^{(KN)}_{[P_{21}]}-3\,a^{(KN)}_{[P_{23}]}\right)
\nonumber\\
&&\quad \quad =2\,\Big(2\,g_0^{(S)}-g_1^{(S)}-2\,g_F^{(S)}\Big)
+4\,\Big(g_1^{(T)}+2\,g_D^{(T)}\Big)
+{\mathcal O}\left(Q \right) \;.
\label{}
\end{eqnarray}

\end{document}